\documentclass{elsearticle}
\usepackage[utf8]{inputenc}
\usepackage{amsmath, mathtools, amssymb, amsthm}
\usepackage[margin=0.6in]{geometry}
\usepackage{soul}
\usepackage{bm}
\usepackage{subfigure}
\usepackage{stmaryrd}
\usepackage{caption}
\usepackage{empheq}
\usepackage{enumitem}
\usepackage{float}
\usepackage{appendix}

\newtheorem{theorem}{Theorem}[section]
\newtheorem{corollary}{Corollary}[theorem]

\newtheorem{definition}{Definition}[section]

\renewcommand*{\appendixname}{}

\makeatletter
\renewcommand*\env@matrix[1][\arraystretch]{%
  \edef\arraystretch{#1}%
  \hskip -\arraycolsep
  \let\@ifnextchar\new@ifnextchar
  \array{*\c@MaxMatrixCols c}}
\makeatother

\journal{}

\begin{document}

\begin{frontmatter}

\title{Entropy-Stable Schemes in the Low-Mach-Number Regime:
Flux-Preconditioning, Entropy Breakdowns, and Entropy Transfers}

\author{Ayoub Gouasmi}
\author{Scott M. Murman}
\address{NASA Ames Research Center, NASA Advanced Supercomputing Division, Moffett field, CA, USA}
\author{Karthik Duraisamy}
\address{University of Michigan, Department of Aerospace Engineering, Ann Arbor, MI, USA}


\begin{abstract}
Entropy-Stable (ES) schemes, specifically those built from [Tadmor \textit{Math. Comput.} 49 (1987) 91], have been gaining interest over the past decade, especially in the context of under-resolved simulations of compressible turbulent flows using high-order methods. These schemes are attractive because they can provide stability in a global and nonlinear sense (consistency with thermodynamics). However, fully realizing the potential of ES schemes requires a better grasp of their local behavior. Entropy-stability itself does not imply good local behavior [Gouasmi \textit{et al.} \textit{J. Sci. Comp.} 78 (2019) 971, Gouasmi \textit{et al.} \textit{Comput. Methd. Appl. M.} 363 (2020) 112912]. In this spirit, we studied ES schemes in problems where \textit{global stability is not the core issue}. In the present work, we consider the accuracy degradation issues typically encountered by upwind-type schemes in the low-Mach-number regime [Turkel \textit{Annu. Rev. Fluid Mech.} 31 (1999) 285] and their treatment using \textit{Flux-Preconditioning} [Turkel \textit{J. Comput. Phys.} 72 (1987) 277, Miczek \textit{et al.} \textit{A \& A} 576 (2015) A50]. ES schemes suffer from the same issues and Flux-Preconditioning can improve their behavior without interfering with entropy-stability. This is first demonstrated analytically: using similarity and congruence transforms we were able to establish conditions for a preconditioned flux to be ES, and introduce the ES variants of the Miczek's and Turkel's preconditioned fluxes. This is then demonstrated numerically through first-order simulations of two simple test problems representative of the incompressible and acoustic limits, the Gresho Vortex and a right-moving acoustic wave. The results are overall consistent with previous studies. For instance, we observe that Turkel's preconditioner improves accuracy in the incompressible limit with the downside of overly damping acoustic waves [Bruel \textit{et al.} \textit{J. Comput. Phys.} 378 (2019) 723]. For Miczek's matrix however, we came across unexpected spurious transients in both problems (a small left-moving acoustic wave in the latter), motivating further analysis. We revisited the pressure fluctuation argument of [Guillard \& Viozat \textit{Comput. Fluids} 28 (1999) 63] in terms of entropy, showing how the standard ES dissipation operator [Ismail \& Roe \textit{J. Comput. Phys.} 228 (2009) 5410] can introduce inconsistent discrete entropy fluctuations in space. These analytical results are achieved by introducing mode-by-mode decompositions of the dissipation operator, similar to [Roe \& Pike \textit{Computing Methods in Applied Science and Engineering} (1984) 499, Tadmor \textit{Acta Numer.} 12 (2003) 482], leading to what we call discrete Entropy Production Breakdowns (EPBs) in space. These EPBs outline the contributions of convective and acoustic modes to the discrete entropy production, both locally and globally. Ultimately, these EPBs enable us to single-out in the ES Miczek flux a skew-symmetric matrix component which we believe causes \textit{entropy transfers between acoustic waves}. Removing this contribution eliminates the spurious transients without interfering with entropy-stability. This conjecture is explored numerically and analytically. 
\end{abstract}

\end{frontmatter}


\section{Introduction}
    Entropy-Stable (ES) schemes have been gaining interest over the past decade, especially in the context of under-resolved simulations of compressible turbulent flows using high-order methods \cite{HO, ES_Diosady, ES_Pazner, ES_Fernandez}. ES schemes are attractive because they can provide stability in both an integral and a nonlinear sense. These schemes are based on the mathematical structure that some systems of Partial Differential Equations (PDEs), such as the compressible Euler equations, possess. These systems, which we write:
\begin{equation}\label{eq:PDE_base}
\frac{\partial \mathbf{u}}{\partial t}  + \sum_{j = 1}^{d} \frac{\partial \mathbf{f}_j}{\partial x_j}  = 0,
\end{equation}
where $\mathbf{u}$ and $(\mathbf{f}_j)_{1 \leq d \leq 3}$ are the state and flux vectors, respectively, and $d$ denotes the number of spatial dimensions, are known to admit a convex extension \cite{ES_Friedrichs, ES_Harten} in the sense that they imply an additional conservation equation for a convex scalar function $U$ (commonly referred to as a \textit{mathematical entropy}):
\begin{equation}\label{eq:PDE_Entropy}
    \frac{\partial U}{\partial t} + \sum_{j = 1}^{d} \frac{\partial F_j}{\partial x_j} = 0.
\end{equation}
The pair $(U, \ F_j) = (U(\mathbf{u}), F_j(\mathbf{u})) \in \mathbb{R} \times \mathbb{R}^{d}$ must satisfy compatibility relations for the baseline system  (\ref{eq:PDE_base}) to imply equation (\ref{eq:PDE_Entropy}). Mathematical entropies have proven to be a key tool in the analysis and discretization of systems of conservation laws \cite{ES_Tadmor_2003}. The existence of a mathematical entropy implies that the PDE system can be symmetrized, which implies its hyperbolicity, and some local existence results. Most notably, a regularization argument by Lax leads to the requirement that admissible weak solutions must satisfy, in the sense of distributions, the \textit{entropy inequality}:
\begin{equation}\label{eq:entropy_ineq}
    \frac{\partial U}{\partial t} + \sum_{j = 1}^{d} \frac{\partial F_j}{\partial x_j} < 0.
\end{equation}
Integrating this inequality over a spatial domain $\Omega$ of trace $\Gamma$, one gets:
\begin{equation}\label{eq:entropy_ineq_integral}
    \frac{d}{dt} \bigg( \int_{\Omega} U dV \bigg)  \ + \ \oint_{\Gamma} F_n dS \ \leq \ 0,
\end{equation}
where $F_n$ denotes the entropy flux projected onto the normal to the local surface element $dS$ of $\Gamma$. If the boundary term is positive or zero (periodic flow), inequality (\ref{eq:entropy_ineq}) becomes a nonlinear integral bound on the solution. This entropy stability property has been actively sought in the development of robust high-order schemes. According to the CFD 2030 Vision Study \cite{HO}: "\textit{Longer term, high-risk research should focus on the development of truly enabling technologies such as monotone or entropy stable schemes}". Efforts are being spent on the development of code infrastructures that can make the most out of ES schemes: the \textit{eddy} solver \cite{eddy0, eddy1, ES_Diosady} under development at NASA Ames Research Center for the simulation of turbulent separated flows and more recently for multi-physics applications is one of several examples. \\
    \indent A variety of ES scheme formulations can be found in the literature. Perhaps the most well-known ones are first-order finite-volume Godunov-type schemes \cite{ES_Harten}, which consist in using solutions of the Riemann problem to compute flux contributions at interfaces. The Godunov scheme, which uses the exact solution to the Riemann problem, is ES as long as the exact Riemann solution exists and satisfies (\ref{eq:entropy_ineq}). Harten \textit{et al.} \cite{ES_Harten} showed that Godunov-type schemes using approximate Riemann solutions (the HLL scheme for instance) are ES if the approximate Riemann solution satisfies (\ref{eq:entropy_ineq}). The Lax-Friedrichs scheme is another well-known ES scheme (an elegant algebraic proof was given by Lax \cite{ES_Lax}). These first-order ES schemes can be used as building blocks for high-order fully-discrete ES schemes by leveraging the convexity of $U$ through convex combinations (see Gottlieb \textit{et al.} \cite{Gottlieb} and Guermond \textit{et al.} for \cite{Guermond_0, Guermond_1}). Some finite-element high-order ES discretizations stem from the seminal work of Hughes \textit{et al.} \cite{ES_Hughes} who showed that a continuous-Galerkin discretization of systems such as (\ref{eq:PDE_base}) of arbitrary order can be made ES \textit{a priori} if the local polynomial representation of the finite-element solution is assigned to the so-called entropy variables instead of the conserved variables. Barth extended these ideas to the discontinuous-Galerkin method \cite{ES_Barth}. \\
    \indent In a seminal paper \cite{ES_Tadmor_1987}, Tadmor introduced finite-volume/finite-difference schemes that achieve entropy-stability in a way that sets them apart from all of the aforementioned ES schemes. Rather than seeking to meet the entropy inequality directly (\ref{eq:entropy_ineq}), Tadmor first developed Entropy Conservative (EC) schemes, namely schemes that are consistent with the conservation equation (\ref{eq:PDE_Entropy}). These require EC numerical flux functions, characterized by a scalar entropy conservation jump condition. Entropy-Stability is then achieved by adding an appropriate dissipation term to the EC flux (the resulting flux is termed ES). From there, a number a developments followed, mostly focused on high-order discretizations \cite{ES_Fisher, ES_Fried, ES_Fjordholm, ES_Pazner, ES_Fernandez}. The present work is exclusively concerned with ES formulations building from Tadmor's ground ideas. \\ 
This focus follows from the authors' stance \cite{Gouasmi_Thesis} that ES schemes could be developed into a solid numerical \textit{foundation} in the simulation of compressible turbulent flows. 
 A key undertaking to fully realize this vision is to \textit{better grasp the local behavior of ES schemes}. In this endeavor, the authors have been studying ES schemes in problems where \textit{global stability is not the core issue} \cite{Gouasmi_0, Gouasmi_2}. A rationale around this stance and approach is given in Gouasmi's PhD thesis \cite{Gouasmi_Thesis}. \\ 
    \indent Numerical schemes designed for compressible flows are known to perform poorly in the low-Mach-number regime \cite{Volpe, Turkel0, Merkle, Turkel2005}, more specifically in the incompressible limit, despite the fact that the incompressible Euler equations are a particular occurrence of the more general compressible Euler equations \cite{Schochet, Majda}.  As there are many flow configurations of engineering interest that exhibit both compressible and incompressible flow phenomena (transonic flow, subsonic combustion, nozzle flows and shock-induced shear instabilities among others), significant research effort has been dedicated to adapting compressible flow codes to handle incompressible flows better. Steady state calculations, which are typically carried out by evolving the unsteady system until a stationary solution is found, require a number of iterations which dramatically increases as the Mach number decreases. \textit{Preconditioning} methods \cite{Turkel0, Turkel1, Turkel1994, Turkel2005, Guillard1, Guillard2, Weiss, Lee, Merkle} have been developed to address this stiffness issue. The idea is to modify the temporal scales of the unsteady system that is iterated by pre-multiplying the time derivative by a well-chosen preconditioning matrix $P$, in order to accelerate convergence. In addition to stiffness, the accuracy of the solution is also known to degrade. The root cause of this issue lies in the artificial viscosity introduced by upwind fluxes. Turkel \cite{Turkel0, Turkel1, Turkel1994, Turkel2005} and Guillard \& Viozat \cite{Guillard1} showed in different ways that the dissipation term $|A|[\mathbf{u}] = R|\Lambda| R^{-1}[\mathbf{u}]$ of upwind fluxes contains terms which prevent the discrete equations solving the compressible system to converge to a set of discrete equations solving the incompressible system. The accuracy degradation problems can be alleviated with \textit{Flux-Preconditioning}, which consists in modifying the upwind dissipation using matrix operations ($P^{-1}|PA| [\mathbf{u}]$ - this operation arises naturally when introducing upwinding in the preconditioned system). For unsteady flows, similar stiffness (stringent CFL condition warrants implicit temporal schemes, whose solution using Newton/GMRES approaches requires efficient preconditioning) and accuracy (excessive damping of vortical structures \cite{Miczek_T, Miczek, Barsukow, Thornber1, Thornber2} due to upwinding) issues arise and can be dealt with similarly. \\ 
     \indent The present work \cite{Gouasmi_USNCCM} considers the accuracy degradation problem in the context of ES schemes. Our first goal was to establish whether Flux-Preconditioning and Entropy-Stability are compatible. We considered two preconditioning matrices: one of the earliest ones by Turkel \cite{Turkel0, Turkel1, Turkel1994} and a more recent one by Miczek \cite{Miczek, Miczek_T}. ES schemes are subject to the same accuracy issues as standard Roe-type schemes because the dissipation terms they use also involves some form of upwinding ($R |\Lambda| R^T [\mathbf{v}]$). We posed the compatibility problem as a linear algebra problem involving $P$. Using similarity and congruence transforms, we established a sufficient condition for compatibility using Barth's eigenscaling theorem. This condition is met by Turkel's preconditioner but does not allow a statement to be made regarding Miczek's preconditioner, whose compatibility is eventually proved using different arguments. From there, we compared four different numerical fluxes in space with Backward Euler in time on two simple periodic flow problems: the Gresho Vortex (incompressible limit) in two dimensions and a right-moving acoustic wave in one dimension (acoustic limit). They use the same EC flux \cite{ES_Chandra}, but differ in their dissipation components. The first flux uses the standard ES dissipation operator \cite{Roe} (ES Roe), the second and third are its preconditioned variants (ES Turkel and ES Miczek), the last one does not use any dissipation operator (EC flux alone). All four flux choices result in fully-discrete ES schemes following \cite{ES_Tadmor_2003}. The numerical results on the two test problems reflected known trends for the most part. The ES Roe flux overly dissipates incompressible vortical structures and behaves consistently in the acoustic limit. The ES Turkel flux has consistent low-Mach behavior in the incompressible limit, but at the expense of overdamping acoustic waves. The ES Miczek flux appears to handle both limits correctly, but upon closer examination, we observed a small spurious transient in both flow configurations. In the sound wave problem, this transient manifests as a small left-going acoustic wave propagating at the same speed as the correct one. To the authors' knowledge, this anomaly has not been reported in the past. The EC flux configuration produced the best results in both limits, hinting that the most simple and effective fix may be to discard the dissipation component in low-Mach-number regimes. The assessment of these low-Mach strategies in a high-order ES setting, on more compelling problems and with a focus on both accuracy and stiffness challenges will be carried out in a follow-up paper. \\ 
    \indent Our desire to better understand discrete local behavior drove us to continue this first-order exploration and further investigate these anomalies, reminiscent of those encountered in previous work \cite{Gouasmi_0, Gouasmi_2}.  In the same spirit as in \cite{Gouasmi_0, Gouasmi_1, Gouasmi_2}, we began our final stretch by looking for a way to explain the accuracy degradation issues in terms of entropy (since it is what ES schemes have a discrete handle on). We chose to revisit Guillard and Viozat's \cite{Guillard1} pressure fluctuation argument in entropy terms. If the incompressible limit can be characterized by a constant density and pressure fluctuations scaling as the square of a reference Mach number $M_r$, then we can argue that fluctuations in entropy $\rho s = \rho (\ln p - \gamma \ln \rho)$ should be of the same order. Likewise, $\mathcal{O}(M_r)$ fluctuations in density and pressure in the acoustic limit amount to $\mathcal{O}(M_r)$ fluctuations in entropy in the acoustic limit \cite{Guillard2, Bruel}. As no accuracy degradation is observed with an EC flux in space, we posit that the discrete entropy production in space, which we denote $\mathcal{E}$ and can express analytically, is responsible for the anomalies. This intuition is confirmed by a dimensional analysis of $\mathcal{E}$ assuming different scalings for the discrete variations in pressure and density. For instance, we find that for the ES Roe flux, $\mathcal{E} = \mathcal{O}(M_r)$ in both limits (inconsistent in the incompressible limit) and that for the ES Turkel flux, $\mathcal{E} = \mathcal{O}(M_r^2)$ in the incompressible limit (consistent behavior), but $\mathcal{E} = \mathcal{O}(1)$ in the acoustic limit (one order of magnitude too high, hence the excess damping). \\ 
    \indent While developing the expression of $\mathcal{E}$ for the ES Roe flux ($\mathcal{E} = [\mathbf{v}]^{T} R |\Lambda| R^{T}[\mathbf{v}]$), we noted that by using a Roe-Pike \cite{Pike}  representation of the dissipation operator, $\mathcal{E}$ can be broken down into distinct positive contributions from each eigenvector of $R$. Without much surprise, we found that the acoustic ($\lambda = u \pm a$) contributions are behind the scaling discrepancies. A similar breakdown was easily achieved for the ES Turkel flux thanks to Barth's eigenscaling theorem \cite{ES_Barth}. For the ES Miczek flux, further manipulations were needed, mainly because of its lack of symmetry, but we were eventually able to dig out a skew-symmetric matrix component that is at the root of the numerical anomalies observed (discarding it removes the anomalies, amplifying it amplifies them). Using these discrete Entropy Production Breakdowns (EPBs), we argue that this skew-symmetric component causes \textit{discrete entropy transfers between acoustic waves}. These findings shed new lights on the local behavior of ES \textit{and} EC schemes. \\
\indent The present work is organized as follows: Section 2 introduces the compressible Euler equations, its underlying entropy structure, and its two low-Mach-number limits following \cite{Guillard1, Guillard2}. Section 3 recaps the root of the accuracy degradation problems and introduces the flux-preconditioning technique. In section 4, we begin analyzing ES schemes in the low-Mach context. We seek to establish whether the flux-preconditioning approach, taking the preconditioner of Miczek \textit{et. al} \cite{Miczek_T, Miczek} and Turkel's \cite{Turkel1994, Guillard1}, is compatible with entropy-stability. Numerical experiments are carried out in section 5 and further analyzed in section 6, where the ideas of Guillard \& Viozat \cite{Guillard1} are used to revisit the accuracy problems from the angle of entropy production. This is where EPBs are introduced and used to develop our discrete entropy transfer argument regarding the anomalies observed with the ES Miczek flux. Section 7 further discusses these developments. 


\section{The Compressible Euler Equations}
\label{LMES:sec:lowMach}
\hspace*{0.1cm} The 3D compressible Euler equations are given by:
\begin{align}\label{eq:Euler}
    \frac{\partial }{\partial t} & \ (\rho) \ + \ \nabla \cdot (\rho \bm{u}) \ = \ 0, \nonumber \\
    \frac{\partial}{\partial t} & \ (\rho \bm{u}) \ + \ \nabla \cdot (\rho \bm{u} \otimes \bm{u} \ + \ p) \ = \ 0, \\
    \frac{\partial}{\partial t} & \ (\rho e^t) \ + \ \nabla \cdot (\bm{u} (\rho e^t \ + \ p)) \ = \ 0. \nonumber
\end{align}
$\rho$ is the density, $\bm{u} := [u,\ v, \ w] \in \mathbb{R}^{3}$ is the velocity vector, $e^{t} := e + k$ is the total energy ($e$ is the internal energy, $k := \frac{1}{2}|\bm{u}|^2$ is the kinetic energy) and $p$ is the pressure. We assume a calorically perfect gas with equation of state $p := (\gamma - 1) \rho e$, where $\gamma = 1.4$ is the adiabatic index. This system can be cast in the form (\ref{eq:PDE_base}) with:
\begin{gather*}
    \mathbf{u} := \begin{bmatrix} \rho & \rho u & \rho v& \rho w & \rho e^t \end{bmatrix}, \ \mathbf{f}_1 := \begin{bmatrix} \rho u & \rho u^2 + p & \rho u v& \rho u w & u \big(\rho e^t + p \big) \end{bmatrix}, \\ 
    \mathbf{f}_2 :=  \begin{bmatrix} \rho v & \rho u v  & \rho v^2 + p & \rho v w & v \big(\rho e^t + p \big) \end{bmatrix}, \ \mathbf{f}_3 :=  \begin{bmatrix} \rho w & \rho u w & \rho v w& \rho w^2 + p & w \big(\rho e^t + p \big) \end{bmatrix}.
\end{gather*}
In quasi-linear form, the system writes:
\begin{equation}\label{eq:Euler_quaslinear}
    \frac{\partial \mathbf{u}}{\partial t} \ + \ \sum_{j = 1}^{3} A_j \frac{\partial \mathbf{u}}{\partial x_j} = 0, \ A_j := \frac{\partial \mathbf{f}_i}{\partial \mathbf{u}}.
\end{equation}
$A_j$ is the flux Jacobian in the direction $x_j$. Let $\mathbf{n}$ be a normal vector of components $(n_1, n_2, n_3)$. Throughout this work, we will denote $A$ the flux jacobian projected along $\mathbf{n}$:
\begin{equation*}
    A \ := \ \sum_{j=1}^{3} n_j A_j \ = \ 
   \begin{bmatrix} 
    0 & n_1 & n_2 & n_3 & 0 \\
    n_1 (\gamma - 1) k - u u_n & (2-\gamma) n_1 u + u_n & (1-\gamma) n_1 v + u n_2 & (1 - \gamma) n_1 w + u n_3 & n_1 (\gamma - 1) \\
    n_2 (\gamma - 1) k - v u_n & (1-\gamma)n_2 u + v n_1 & (2-\gamma) n_2 v + u_n & (1-\gamma) n_2 w + v n_3 & n_2 (\gamma - 1) \\
    n_3 (\gamma - 1) k - w u_n & (1-\gamma)n_3 u + w n_1 & (1-\gamma) n_3 v + w n_2 & (2-\gamma) n_3 w + u_n & n_2 (\gamma - 1) \\
    u_n (-h^t + (\gamma - 1)k) & n_1 h^t - (\gamma - 1) u u_n & n_2 h^t - (\gamma - 1) v u_n & n_3 h^t - (\gamma - 1) w u_n & u_n \gamma
   \end{bmatrix}.
\end{equation*}
By virtue of the hyperbolicity of the compressible Euler equations, $A$ is diagonalizable. We have $A = R \Lambda R^{-1}$ with:
\begin{gather*}
    \Lambda = diag \big( \begin{bmatrix} u_n & u_n & u_n & u_n + a & u_n - a \end{bmatrix}\big), \ u_n := \sum_{i=1}^{3} n_i u_i, \ a := \sqrt{\gamma p / \rho}, \\
    R = \begin{bmatrix}
    n_1                                   & n_2                                    & n_3                                  & 1                  & 1 \\
    n_1 u                                & n_2 u + n_3 a                   & n_3 u - n_2 a                   & u_n + n_1 a & u_n - n_1 a \\
    n_1 v - n_3 a                    & n_2 v                                 & n_3 v + n_1 a                 & u_n + n_2 a & u_n - n_2 a \\
    n_1 w + n_2 a                  & n_2 w - n_1 a                    & n_3 w                              & u_n + n_3 a & u_n - n_3 a \\
    n_1 k  - a (n_3 v - n_2 w) & n_2 k  + a(n_3 u - n_1 w) & n_3 k - a (n_2 u - n_1 v) & h^t + u_n a & h^t - u_n a
    \end{bmatrix},  
\end{gather*} 
where $h^{t} := h + k$ is the total enthalpy ($h := e + p/\rho$ is the enthalpy).
\subsection{Entropy Structure}
The compressible Euler system can be rewritten in terms of the total derivatives of density, velocity and internal energy:
\begin{equation}\label{eq:Euler_total}
    \frac{D\rho}{Dt} \ = \ - \rho \nabla \cdot \bm{u}, \ \frac{D \bm{u}}{Dt} = - \frac{1}{\rho} \nabla p, \ \frac{D e}{Dt} = \ \frac{p}{\rho}\nabla \cdot \bm{u}, 
\end{equation}
with the total derivative operator defined as:
\begin{equation*}
\frac{D}{Dt}  \ := \ \frac{\partial }{\partial t} \ + \ \bm{u} \cdot \nabla .
\end{equation*}
The specific entropy $s = \ln p - \gamma \ln \rho = \ln e - (\gamma - 1) \ln \rho + \ln (\gamma - 1)$ satisfies the differential Gibbs relation:
\begin{equation}\label{eq:Gibbs}
   ds \ = \ \frac{1}{e} \bigg( de - \frac{p}{\rho^2} d\rho\bigg)
\end{equation}
Combining equations (\ref{eq:Euler_total}) and (\ref{eq:Gibbs}) leads to a transport equation for $s$:
\begin{equation}\label{eq:transport_s}
    \frac{\partial s}{\partial t} + \bm{u} \cdot \nabla s = 0,
\end{equation}
which combined with conservation of mass leads to the conservation of entropy:
\begin{equation}\label{eq:Euler_entropy}
    \frac{\partial \rho s}{\partial t} + \nabla \cdot \big( \rho \bm{u} s \big)  = 0.
\end{equation}
It can be shown that $\rho s$ is a concave function of $\mathbf{u}$, hence $U = - \rho s / (\gamma - 1)$ is a mathematical entropy with fluxes $(F_1, F_2, F_3) = - \rho s (u, v, w)$. The entropy variables are defined by:
\begin{equation}\label{eq:entropy_var}
    \mathbf{v} \ := \ \bigg(\frac{\partial U}{\partial \mathbf{u}}\bigg)^T,
\end{equation}
and with the present choice of $U$, it is given by:
\begin{equation*}
    \mathbf{v} \ = \ \begin{bmatrix} \frac{\gamma - s}{\gamma - 1} - \frac{\rho k}{p} & \frac{\rho}{p} \bm{u}^T & -\frac{\rho}{p}\end{bmatrix}^T. 
\end{equation*}
An important result that is central in the construction of an ES scheme and will help us later in section 4 is stated below:
\begin{theorem}[Mock \cite{ES_Mock}]
    The existence of a mathematical entropy $U$ for the system of conservation laws (\ref{eq:PDE_base}) implies that the change of variables $\mathbf{u} \xrightarrow{} \mathbf{v}$ (defined by eq. (\ref{eq:entropy_var})) symmetrizes the system. The matrix $H$ defined by:
\begin{equation}\label{eq:H_def}
    H  \ := \ \frac{\partial \mathbf{u}}{\partial \mathbf{v}},
\end{equation}
is symmetric positive definite and $(A_j H)_{1 \leq j \leq d}$ is symmetric. 
\end{theorem}
For the compressible Euler system and $U = -\rho s / (\gamma - 1)$, the matrix $H$ is given by:
\begin{equation*}
    H = \begin{bmatrix}
    \rho & \rho u & \rho v & \rho w & \rho e^t  \\
      &  \rho u^2 + p & \rho u v & \rho u w & \big(\rho e^t + p \big) u  \\
     &  &  \rho v^2 + p & \rho v w & \big(\rho e^t + p \big) v \\
     &  &  & \rho w^2 + p & \big(\rho e^t + p \big) w \\
    sym & &  &  & \rho (e^t)^2 + p \big( \frac{p}{(\gamma-1)\rho} + (u^2 + v^2 + w^2)\big)
    \end{bmatrix}.
\end{equation*}
\indent $U = -\rho s / (\gamma - 1)$ is not the only mathematical entropy for the compressible Euler system. Harten \cite{ES_Harten} introduced a family of mathematical entropies $U = - \rho h(s)$ with $h' > 0, h^{'} - \gamma h^{''} > 0$  and similar characterizations have been introduced for more complex versions of this system (general equations of state \cite{ES_SuperHarten}, multicomponent \cite{Gouasmi_3}). Throughout this manuscript, we work with the opposite of the thermodynamic entropy as it is the only non-trivial member of Harten's family which, for the more general compressible Navier-Stokes equations, both symmetrizes the system and leads to a stability result \cite{ES_Hughes}. \\ 
\indent The $1/(\gamma - 1)$ factor in the choice of $U$ is such that the entropy flux potentials $\mathcal{F}_j$ and $\mathcal{F}$ defined by:
\begin{equation}\label{eq:potential_def}
    \mathcal{F}_j \ := \ \mathbf{v} \cdot \mathbf{f}_j - F_j, \ \mathcal{F} := \sum_{j=1}^{3} n_j \mathcal{F}_j,
\end{equation}
simplify to $\mathcal{F}_j = \rho u_j$ and $\mathcal{F} = \rho u_n$. These quantities are involved in the construction of EC fluxes.
\subsection{Non-dimensionalization and Low-Mach-Number Limits}  
\indent Here we first introduce the incompressible and acoustic limits of the compressible Euler system following Guillard \& Viozat \cite{Guillard1} and Guillard \& Nkonga \cite{Guillard2}.  We then define the scaled system (together with its entropy) that we will work with in throughout our study of ES schemes in the low-Mach regime. \\ 
\indent Let $\rho_{r}, p_{r}$ and $u_{r}$ be reference values for density, pressure and velocity magnitude, respectively, and let us define a reference speed of sound $a_{r} := \sqrt{p_{r}/\rho_{r}}$. Introduce the non-dimensional variables and operators:
\begin{equation*}
    \tilde{\rho} \ := \ \frac{\rho}{\rho_{r}}, \ \tilde{\bm{u}} \ := \ \frac{\bm{u}}{u_{r}}, \ \tilde{p} \ := \ \frac{p}{p_r}, \ \tilde{e} \ := \ \frac{e}{a_r^2}, \ \tilde{k} \ := \ \frac{k}{\frac{1}{2}u_r^2}, \ \tilde{t} \ := \ \frac{t}{t_r}, \ \tilde{\nabla} \ := \ l_r \nabla,
\end{equation*}
with $l_{r}$ and $t_{r}$ are the reference length scale and time scales, respectively. The reference Mach number is defined as:
\begin{equation*}
    M_r \ := \ \frac{u_r}{a_r}.
\end{equation*} 
The vast majority of the derivations made from here involve the non-dimensional flow variables. For simplicity, we therefore drop the tilde notation. Unless otherwise stated, the flow variables ($\rho, \bm{u}, p, ...$) are dimensionless. \\ \\
\indent \textbf{Incompressible limit.} Setting the reference time scale as $t_r \ := \ u_r/l_r$, the scaled system writes:
\begin{align}\label{LMES:EulerM1}
    \frac{\partial }{\partial t} & \ \big(\rho \big) \ + \ \nabla \cdot \big( \rho \bm{u} \big) \ = \ 0, \nonumber \\
    \frac{\partial}{\partial t} & \ \big( \rho \bm{u} \big) \ + \ \nabla \cdot \big( \rho \bm{u} \otimes \bm{u} \big) \ + \ \frac{1}{M_r^2} \nabla p \ = \ 0, \\
    \frac{\partial}{\partial t} & \ \big( \rho (e \ + \ M_r^2 k)\big) \ + \ \nabla \cdot \big( \bm{u} (\rho (e + M_r^2 k) \ + \ p) \big) \ = \ 0. \nonumber 
\end{align}
The scaled equation of state writes $p := (\gamma - 1) \rho e$. The second step is to consider asymptotic expansions (Klein \cite{Klein}) of the flow variables in powers of the reference Mach number:
\begin{align}
    p \ =& \ p_0 \ + \ M_r p_1 \ + \ M_r^2 p_2 \ + \ \mathcal{O}(M_r^3), \label{LMES:powerM_p} \\
    \bm{u} \ =& \ \bm{u}_0 \ + \ M_r \bm{u}_1 \ + \ M_r^2 \bm{u}_2 \ + \ \mathcal{O}(M_r^3), \label{LMES:powerM_u} \\
    \rho \ =& \ \rho_0 \ + \ M_r \rho_1 \ + \ M_r^2 \rho_2 \ + \ \mathcal{O}(M_r^3). \label{LMES:powerM_r}
\end{align}
Injecting these expansions into (\ref{LMES:EulerM1}) and collecting terms of same order, one gets:
\begin{enumerate}
    \item Order $1/M_r^2$:
    \begin{equation}\label{LMES:p0}
        \nabla p_0 \ = \ 0.
    \end{equation}
    \item Order $1/M_r$:
    \begin{equation}\label{LMES:p1}
        \nabla p_1 \ = \ 0.
    \end{equation}
    \item Order 1:
    \begin{align}
    \frac{\partial }{\partial t} & \ (\rho_0) \ + \ \nabla \cdot (\rho_0 \mathbf{\bm{u}}_0) \ = \ 0, \label{LMES:EulerM_1_1}\\
    \frac{\partial}{\partial  t} & \ (\rho_0 \mathbf{\bm{u}}_0) \ + \ \nabla \cdot (\rho_0 \mathbf{\bm{u}}_0 \otimes \mathbf{\bm{u}}_0) \ +  \ \nabla p_2 \ = \ 0, \label{LMES:EulerM_1_2}\\
    \frac{\partial}{\partial t} & \ p_0 \ + \ \bm{u}_0 \cdot \nabla p_0 \ + \ \rho_0 a_0^2 \nabla \cdot \bm{u}_0 \ = \ 0. \label{LMES:EulerM_1_3}
\end{align}
\end{enumerate}
Equations (\ref{LMES:p0}) and (\ref{LMES:p1}) imply that pressure variations in space scale as $M_r^2$ at least: $p(x,t) = p_0 + M_r p_1(t) + M_r^2 p_2(x,t) = P_0(t) + M_r^2 p_2(x,t)$. If $P_0$ is constant then equation (\ref{LMES:EulerM_1_3}) implies the divergence constraint $\nabla \cdot \mathbf{\bm{u}}_0 = 0$. Injecting it into equation (\ref{LMES:EulerM_1_1}) implies that the material derivative of density is zero. Assuming that all particle paths come from regions of same density $\rho_0$, we get that density is constant everywhere and equations (\ref{LMES:EulerM_1_1}), (\ref{LMES:EulerM_1_2}) and (\ref{LMES:EulerM_1_3}) finally reduce to the incompressible system:
\begin{gather}\label{LMES:EulerM_r}
    \rho_0 \ = \ cte, \\
    \rho_0 \bigg(\frac{\partial}{\partial t}  ( \mathbf{\bm{u}}_0) \ + \ \nabla \cdot ( \mathbf{\bm{u}}_0 \otimes \mathbf{\bm{u}}_0) \bigg) \ + \ \nabla p_2 \ = \ 0, \\
    \nabla \cdot \mathbf{\bm{u}}_0 \ = \ 0. 
\end{gather}
The divergence constraint also implies that the kinetic energy is conserved. \\ \\
\indent \textbf{Acoustic limit.} If the time scale is defined in terms of the reference speed of sound $a_{r}$, that is $t_{r} = l_{r}/a_{r}$, then instead of (\ref{LMES:EulerM1}), we have:
\begin{align}\label{LMES:EulerM2}
     \frac{1}{M_r} \frac{\partial }{\partial t} & \ \big(\rho \big) \ + \ \nabla \cdot \big( \rho \bm{u} \big) \ = \ 0, \nonumber \\
     \frac{1}{M_r}\frac{\partial}{\partial t} & \ \big( \rho \bm{u} \big) \ + \ \nabla \cdot \big( \rho \bm{u} \otimes \bm{u} \big) \ + \ \frac{1}{M_r^2} \nabla p \ = \ 0, \\
     \frac{1}{M_r}\frac{\partial}{\partial t} & \ \big( \rho (e \ + \ M_r^2 k)\big) \ + \ \nabla \cdot \big( \bm{u} (\rho (e + M_r^2 k) \ + \ p) \big) \ = \ 0. \end{align}
Introducing the expansions (\ref{LMES:powerM_p}) - (\ref{LMES:powerM_r}) into (\ref{LMES:EulerM2}) and collecting terms of the same order, one gets:
\begin{enumerate}
    \item Order $1/M_r^2$:
    \begin{equation}\label{LMES:p0_2}
        \nabla p_0 \ = \ 0.
    \end{equation}
    \item Order $1/M_r$:
    \begin{align}
    \frac{\partial }{\partial t} & \ (\rho_0) \ = \ 0, \label{LMES:EulerM_2_1}\\
    \frac{\partial}{\partial t} & \ (\rho_0 \mathbf{\bm{u}}_0) \ + \ \nabla p_1 \ = \ 0, \label{LMES:EulerM_2_2}\\
    \frac{\partial}{\partial t} & \ p_0 \ = \ 0. \label{LMES:EulerM_2_3}
    \end{align}
\end{enumerate}
Equations (\ref{LMES:p0_2}) and (\ref{LMES:EulerM_2_3}) imply that the pressure variations in space scale as $M_r$, that is one order of magnitude bigger than those in the incompressible limit. With further manipulations (see Guillard \& Nkonga \cite{Guillard2} for more details), it can be shown that the first order pressure $p_1$ satisfies the wave equation with propagation speed $a_0$ following:
\begin{equation*}
    \frac{\partial^2}{\partial t^2} p_1 \ - \ a_0^2 \nabla \cdot \big( \nabla p_1 \big) \ = \ 0.
\end{equation*}
\indent \textbf{Non-Dimensional Entropy Structure}. The above analysis shows two distinct scaled versions of the compressible Euler system, namely (\ref{LMES:EulerM1}) and (\ref{LMES:EulerM2}). For the purpose of our work, it is enough to work with the first system because:
\begin{enumerate}
\item As we will see later, the accuracy issues stem from some algebraic implications of upwinding ($|A|$ vs $A$). The flux Jacobian matrices of each scaled system differ by a constant factor $M_r > 0$ only ($|M_r A| = M_r |A|$).
\item It is easy to show that system (\ref{LMES:EulerM1}) implies the exact same entropy conservation as the original system (\ref{eq:Euler}) (with $s$ being the same function of the non-dimensional density and pressure), and that system (\ref{LMES:EulerM2}) implies 
\begin{equation}\label{LMES:entropyM2}
    \frac{1}{M_r}\frac{\partial (\rho s)}{\partial t} + \nabla \cdot (\rho \bm{u} s) = 0.
\end{equation}
The corresponding definitions of the entropy variables are the same since in the latter case, the $1/M_r$ factor is present in the time derivatives of both $\mathbf{u}$ and $U$.
\end{enumerate}
We thereby redefine our state and flux vectors as:
\begin{gather*}
    \mathbf{u} := \begin{bmatrix} \rho & \rho u & \rho v & \rho w & \rho e^t \end{bmatrix}^T, \ \mathbf{f_1} := \begin{bmatrix} \rho u & \rho u^2 + p/M_r^2 & \rho u v & \rho u w &(\rho e^t + p)u \end{bmatrix}^T, \\
    \mathbf{f_2} := \begin{bmatrix} \rho v & \rho u v & \rho v^2 + p/M_r^2 & \rho v w & (\rho e^t + p)v \end{bmatrix}^T, \
    \mathbf{f_3} := \begin{bmatrix} \rho w & \rho u w & \rho v w & \rho w^2 + p/M_r^2 & (\rho e^t + p)w \end{bmatrix}^T,
\end{gather*}
with $e^t := e + M_r^2 k$. \\ 
\indent The expressions of the entropy $U$ and its fluxes $(F_i)_{1 \leq i \leq 3}$ are unchanged, but since the vector of conserved variables $\mathbf{u}$ now contains a $M_r$ factor in the total energy component, the non-dimensional entropy variables we will work with are given by:
\begin{equation}
    \mathbf{v} = \begin{bmatrix}
            \frac{\gamma-s}{\gamma-1} - M_r^2\frac{\rho k}{p} & M_r^2 \frac{\rho}{p} \bm{u}^T & -\frac{\rho}{p} \end{bmatrix}^T.
\end{equation}
The potential function $\mathcal{F}$ is unchanged, and the temporal Jacobian is given by
\begin{gather*}
    H =  \begin{bmatrix}
    \rho & \rho u & \rho v & \rho w & \rho e^t  \\
      &  \rho u^2 + \frac{p}{M_r^2} & \rho u v & \rho u w & \big(\rho e^t + p \big) u  \\
     &  &  \rho v^2 + \frac{p}{M_r^2} & \rho v w & \big(\rho e^t + p \big) v \\
     &  &  & \rho w^2 + \frac{p}{M_r^2} & \big(\rho e^t + p \big) w \\
    sym & &  &  & \rho (e^t)^2 + p \big( \frac{p}{(\gamma-1)\rho} + M_r^2 (u^2 + v^2 + w^2)\big)
    \end{bmatrix}.
\end{gather*}
\indent While interesting in its own right, the mathematical structure of the incompressible and acoustic systems is not of concern in the present work. Things would be different if we were looking at constructing ES discretizations of the compressible system which reduce to structure-preserving discretizations of the incompressible and acoustic equations. Here we are simply revisiting, in the context of Tadmor's ES schemes, well-documented issues of compressible schemes in the low-Mach regime. The reference Mach number $M_r$ is kept strictly positive so that the entropy structure remains well-defined (one thing to remain careful with is the \textit{strict} convexity of $U$, which the definition of the entropy variables hinges upon \cite{Gouasmi_2, Gouasmi_3}).

\section{Discrete Analysis and Flux-Preconditioning}
\label{LMES:sec:flux_precond}
\indent Consider a general finite-volume discretization of (\ref{eq:PDE_base}). In a given cell $\Omega_i$ of volume $V_i$, we have:
\begin{equation}\label{eq:FVM}
    \frac{d \mathbf{u}_i}{d t} \ + \ \frac{1}{V_i}\int_{\delta \Omega_i} \mathbf{f^{*}} dS = 0,
\end{equation}
where $\mathbf{f^*} = \mathbf{f^{*}}(\mathbf{u}_i, \mathbf{u}_j, \mathbf{n})$ denotes the numerical flux across the cell trace $\delta \Omega_i$ ($\mathbf{u}_j$ is the neighboring cell state value, $\mathbf{n}$ is the normal vector). A standard choice for $\mathbf{f}^{*}$ is the Roe flux \cite{Roe, Pike}:
\begin{equation}\label{eq:flux_upwind}
    \mathbf{f^{*}}(\mathbf{u}_L, \mathbf{u}_R, \mathbf{n}) = \frac{1}{2}(\mathbf{f}(\mathbf{u}_L) + \mathbf{f}(\mathbf{u}_R)) - \frac{1}{2}|A|(\mathbf{u}_R - \mathbf{u}_L), \  |A| = R | \Lambda| R^{-1}.
\end{equation}
where the flux Jacobian $A$ is evaluated using the so-called Roe-averages ($A = A(\mathbf{u}_L, \mathbf{u}_R)$). \\
\indent In the low-Mach number regime the accuracy of such a scheme typically deteriorates as the Mach number goes to zero. Turkel \cite{Turkel0, Turkel1} explained that it is because the dissipation matrix $|A|$ contains terms which prevent the set of discrete equations solving the compressible system to converge to a set of discrete equations for the incompressible system in the low-Mach limit. To illustrate, Turkel considers \cite{Turkel0} the simple case of a 2-by-2 hyperbolic system with the following Jacobian matrix:
\begin{equation}\label{LMES:eq:2b2}
    A = \begin{bmatrix}
            u & a/M_r \\
            a/M_r & u
        \end{bmatrix}, \ R =  \begin{bmatrix}
            1 & 1 \\
            1 & -1 
        \end{bmatrix}, \ \Lambda = \begin{bmatrix} u+a/M_r & 0 \\ 0 & u-a/M_r \end{bmatrix}. 
\end{equation}
In the subsonic regime, $|u+a/M_r| = u+a/M_r$ and $|u-a/M_r| = -(u-a/M_r)$. This change of sign leads to a dissipation matrix that does not possess the same scaling behavior as the original Jacobian. The reader can easily verify that:
\begin{equation*}
    |A| = \begin{bmatrix} a/M_r & u \\ u & a/M_r \end{bmatrix} = \begin{bmatrix} \mathcal{O}(1/M_r) & \mathcal{O}(1) \\ \mathcal{O}(1) & \mathcal{O}(1/M_r) \end{bmatrix} \ \neq \ \begin{bmatrix} \mathcal{O}(1) & \mathcal{O}(1/M_r) \\ \mathcal{O}(1/M_r) & \mathcal{O}(1) \end{bmatrix}.
\end{equation*}
This difference in scaling behavior is the root cause of the accuracy degradation issues. The dissipation term is an important component of this flux (stability) hence it cannot be discarded because the scheme would be less robust (see appendix A). \\
\indent \textit{Flux-preconditioning} is one way to compromise between stability and correct low-Mach behavior. It consists in replacing the dissipation matrix $|A|$ with $P^{-1}|P A|$ where $P$ is an invertible preconditioning matrix. The preconditioned numerical flux now writes:
\begin{equation}
    \mathbf{f^{*}}(\mathbf{u}_L, \mathbf{u}_R, \mathbf{n}) = \frac{1}{2}(\mathbf{f}(\mathbf{u}_L) + \mathbf{f}(\mathbf{u}_R)) - \frac{1}{2}P^{-1}|PA|(\mathbf{u}_R - \mathbf{u}_L).
\end{equation}
$P$ should correct the asympotic behavior of the dissipation term in the low-Mach regime and only be active in this regime ($P \rightarrow I$ as $M_r \rightarrow 1$). \\
\indent The design of $P$ is not straightforward, even though it is clear that the acoustic eigenspace of the dissipation matrix should be targeted. The analysis can be significantly simplified by using similarity transformations, which amount to considering the compressible Euler equations in a alternative set of variables $\mathbf{z}$. Define:
\begin{equation}\label{eq:similarity}
     A_{\mathbf{z}} := Q^{-1} A Q, \ Q := \bigg( \frac{\partial \mathbf{u}}{\partial \mathbf{z}}\bigg).    
\end{equation}
First, a preconditioning matrix $P_{\mathbf{z}}$ is sought so that $P_{\mathbf{z}}^{-1}|P_{\mathbf{z}}A_{\mathbf{z}}|$ has appropriate Mach number scalings. The preconditioning matrix $P$ in terms of the conservative variables $\mathbf{u}$ is then derived from the similarity relation $P = Q P_{\mathbf{z}} Q^{-1}$. Indeed, one has:
\begin{equation*}
    P^{-1} |P A| = Q P_\mathbf{z}^{-1} Q^{-1} | Q P_{\mathbf{z}} Q^{-1} Q A_{\mathbf{z}} Q^{-1} | = Q \ \big( P_\mathbf{z}^{-1}  (| P_{\mathbf{z}}  A_{\mathbf{z}}|) \big) \ Q^{-1}
\end{equation*}
As a matter of course, this strategy is efficient only if $A_{\mathbf{z}}$ has a simpler structure than $A$. With the \textit{differential entropy variables}\footnote{These variables are referred to as the ``entropy variables" in the literature \cite{Turkel1994, Barsukow}. The naming ``differential entropy variables" is introduced to distinguish them from the entropy variables $\mathbf{v}$  ES schemes are centered around.} defined by:
\begin{equation}\label{eq:diff_entropy_var}
    d\mathbf{z} = (dp/(\rho a M_r), \ du, \ dv, \ dw, \ dp - a^2 d\rho), 
\end{equation}
the similarity matrix $Q$ writes:
\begin{equation}
    Q = \begin{bmatrix}
                                                   M_r \rho / a &                    0 &                  0 &                     0 & -1/a^2  \\
                                                M_r \rho u / a &                \rho &                  0 &                     0 & -u/a^2  \\
                                                M_r \rho v / a &                    0 &              \rho &                     0 & -v/a^2  \\
                                                M_r \rho w / a &                   0 &                   0 &                \rho & -w/a^2 \\
    \rho k M_r^3/a + \rho a M_r /(\gamma - 1) & M_r^2 \rho u & M_r^2 \rho v & M_r^2 \rho w & - M_r^2 k / a^2
    \end{bmatrix}.
\end{equation}
The mapped Jacobian has the elegant structure:
\begin{equation}\label{LMES:eq:Az}
    A_{\mathbf{z}} = \begin{bmatrix}
        u_n     & n_1 a/M_r & n_2 a/M_r & n_3 a/M_r &   0 \\
        n_1 a/M_r &     u_n &       0 &       0 &   0 \\
        n_2 a/M_r &       0 &     u_n &       0 &   0 \\
        n_3 a/M_r &       0 &       0 &     u_n &   0 \\
        0       &       0 &       0 &       0 & u_n
    \end{bmatrix}.
\end{equation}
Its eigenstructure $A_{\mathbf{z}} = R_{\mathbf{z}} \Lambda R_{\mathbf{z}}^{-1}$ is given by:
\begin{equation}
    R_{\mathbf{z}} = \begin{bmatrix}
             0 &    0 &    0 & 1 & 1 \\
             0 & -n_3 &  n_2 & n_1 & n_1 \\
           n_3 &    0 & -n_1 & n_2 & n_2 \\
          -n_2 &  n_1 &    0 & n_3 & n_3 \\
          -n_1 & -n_2 & -n_3 & 0 &   0
    \end{bmatrix}
   , \ 
    \Lambda = diag( u_{n}, \ u_{n}, \ u_{n}, \ u_{n}, \ u_{n} + a/M_r, \ u_{n} - a/M_r] ).
\end{equation}
The 2-by-2 hyperbolic system (\ref{LMES:eq:2b2}) of Turkel in \cite{Turkel0} is a specific case of system (\ref{LMES:eq:Az}). For this system, Turkel \textit{et al.} \cite{Turkel1994, Turkel0} established the following necessary condition on $P_{\mathbf{z}}$ for convergence in the low-Mach limit:
\begin{equation}\label{LMES:Turkel_cond}
    P_{\mathbf{z}}^{-1}|P_{\mathbf{z}} A_{\mathbf{z}}| = 
    \begin{bmatrix}
            \mathcal{O}(1/M_r^2) & \mathcal{O}(1/M_r) & \mathcal{O}(1/M_r) & \mathcal{O}(1/M_r) &             0 \\
            \mathcal{O}(1/M_r)   &   \mathcal{O}(1) &   \mathcal{O}(1) &   \mathcal{O}(1) &             0 \\
            \mathcal{O}(1/M_r)   &   \mathcal{O}(1) &   \mathcal{O}(1) &   \mathcal{O}(1) &             0 \\
            \mathcal{O}(1/M_r)   &   \mathcal{O}(1) &   \mathcal{O}(1) &   \mathcal{O}(1) &             0 \\
                             0 &                0 &                0 &                0 & \mathcal{O}(1)
    \end{bmatrix}.
\end{equation}
They also showed that this is achieved with the Turkel preconditioning matrix: 
\begin{equation}\label{LMES:Turkel_mat}
    P_{\mathbf{z}} = \begin{bmatrix}
        p^2 & 0 & 0 & 0 & 0 \\
          0 & 1 & 0 & 0 & 0 \\
          0 & 0 & 1 & 0 & 0 \\
          0 & 0 & 0 & 1 & 0 \\
          0 & 0 & 0 & 0 & 1
    \end{bmatrix},
\end{equation}
where $p = \min(\max(M_r, M_{cut}), 1)$. The parameter $p$ is defined in such a way that $P_{\mathbf{z}}$ is always invertible (the cut-off Mach number $M_{cut}$ prevents $p \rightarrow 0$ and $P_{\mathbf{z}}$ singular) and $P_{\mathbf{z}}$ approaches the identity matrix when $M_r \rightarrow 1$. We have $P_{\mathbf{z}}A_{\mathbf{z}} = R_{p\mathbf{z}} \Lambda_p R_{p\mathbf{z}}^{-1}$ with
\begin{gather*}
    R_{p\mathbf{z}} = \begin{bmatrix}
     \begin{bmatrix}
             0 &    0 &    0 \\
             0 & -n_3 &  n_2 \\
           n_3 &    0 & -n_1 \\
          -n_2 &  n_1 &    0 \\
          -n_1 & -n_2 & -n_3
    \end{bmatrix}
    \frac{1}{(K_1 - K_2)} \begin{bmatrix}
             K_1 & K_2 \\
             n_1 & n_1 \\
             n_2 & n_2 \\
             n_3 & n_3 \\
               0 &   0
    \end{bmatrix}
    \end{bmatrix}
   , \ 
    \Lambda_p = diag( [u_{n}, \ u_{n}, \ u_{n}, \ u_{n}, \ u_{np} + a_p, \ u_{np} - a_p] ), \\
    u_{np} = \frac{1}{2}u_n(p^2+1), \ a_p = (u_{np}^2 + p^2((a/M_r)^2 - u_n^2))^{1/2}, \ K_1 =  (u_{np} - u_n + a_p) M_r / a, \ 
    K_2 = (u_{np} - u_n - a_p) M_r / a.
\end{gather*}
In the subsonic regime, $u_n^2 < (a /M_r)^2  \implies u_{np} < a_p  \implies  |u_{np} - a_p| = a_p - u_{np}$. The preconditioned dissipation matrix writes (we assume $u_n > 0$ throughout this paper, with no loss of generality):
\begin{gather*}
    P_{\mathbf{z}}^{-1} | P_{\mathbf{z}} A_{\mathbf{z}}| = \begin{bmatrix}
            C_0 &                     n_1 C_1 &                     n_2 C_1 &                     n_3 C_1 &   0 \\
        n_1 C_2 & n_1^2 C_3 + (1 - n_1^2) u_n &                 n_1 n_2 C_4 &                 n_1 n_3 C_4 &   0 \\
        n_2 C_2 &                 n_2 n_1 C_4 & n_2^2 C_3 + (1 - n_2^2) u_n &                 n_2 n_3 C_4 &   0 \\
        n_3 C_2 &                 n_3 n_1 C_4 &                 n_3 n_2 C_4 & n_1^3 C_3 + (1 - n_3^2) u_n &   0 \\
              0 &                           0 &                           0 &                           0 & u_n
    \end{bmatrix}, \\
    C_0 = (a_p^2 + u_{np}^2 - u_n u_{np}) / (a_p p^2), \ C_1 = M_r u_{np} (a_p + u_n - u_{np}) (a_p - u_n + u_{np}) / (a a_p p^2), \\ C_2 = (a u_{np}) / (M_r a_p), \ C_3 = a_p + u_{np}(u_n - u_{np})/a_p, \\ C_4 = (a_p - u_{np}) (a_p - u_n + u_{np}) / a_p,
\end{gather*}
and meets condition (\ref{LMES:Turkel_cond}). \\
\indent A different perspective (which we will revisit later in this work) is provided in the work of Guillard \& Viozat \cite{Guillard1} who observed that in the incompressible regime, pressure fluctuations in space typically scale as $M_r^2$. By applying the process described in section \ref{LMES:sec:lowMach} to the discrete equations, they were able to rigorously demonstrate that certain terms in the dissipation matrix of the upwind flux can lead to pressure fluctuations in space which scale as $M_r$ instead. They show that with Turkel's preconditioner (\ref{LMES:Turkel_mat}), the proper scaling of pressure fluctuations is recovered. \\
\indent The second preconditioning matrix we consider was recently introduced by Miczek \textit{et al.} \cite{Miczek_T, Miczek} for unsteady calculations. It writes:
\begin{equation}\label{LMES:Miczek_mat}
    P_{\mathbf{z}} = \begin{bmatrix}
               1 & n_1 p & n_2 p & n_3 p & 0 \\
          -n_1 p &     1 &     0 &     0 & 0 \\
          -n_2 p &     0 &     1 &     0 & 0 \\
          -n_3 p &     0 &     0 &     1 & 0 \\
               0 &     0 &     0 &     0 & 1
    \end{bmatrix}.
\end{equation}
with $p = 1- 1/\delta , \ \delta = \min(\max(M_r, M_{cut}), 1)$. This time we have $P_{\mathbf{z}}A_{\mathbf{z}} = R_{p\mathbf{z}} \Lambda_p R_{p\mathbf{z}}^{-1}$ with
\begin{gather*}
    R_{p\mathbf{z}} = \begin{bmatrix}
    \begin{bmatrix}
             0 &    0 &    0 \\
             0 & -n_3 &  n_2 \\
           n_3 &    0 & -n_1 \\
          -n_2 &  n_1 &    0 \\
          -n_1 & -n_2 & -n_3
    \end{bmatrix}
    \frac{1}{(K_1 - K_2)} \begin{bmatrix}
             K_1 & K_2 \\
             n_1 & n_1 \\
             n_2 & n_2 \\
             n_3 & n_3 \\
               0 &   0
    \end{bmatrix}
    \end{bmatrix}, \\
    \Lambda_p = diag( u_{n}, \ u_{n}, \ u_{n}, \ u_{n}, \ u_{n} + a_p, \ u_{n} - a_p] ), \\
    a_p = \sqrt{(p^2+1) a^2/M_r^2 - p^2 u_n^2}, \\ K_1 =  (a + M_r p u_n) / (M_r a_p - a p), \ 
    K_2 = -(a + M_r p u_n) / (M_r a_p + a p).
\end{gather*}
It is easily shown that $u_n < a/M_r \implies u_n^2 < a_p^2 \implies |u_n - a_p| = a_p - u_n$, therefore the preconditioned dissipation matrix writes:
\begin{gather*}
    P_{\mathbf{z}}^{-1} | P_{\mathbf{z}} A_{\mathbf{z}}| = \begin{bmatrix}
            C_0 &                     n_1 C_1 &                     n_2 C_1 &                     n_3 C_1 &   0 \\
        n_1 C_2 & n_1^2 C_3 + (1 - n_1^2) u_n &                 n_1 n_2 C_4 &                 n_1 n_3 C_4 &   0 \\
        n_2 C_2 &                 n_2 n_1 C_4 & n_2^2 C_3 + (1 - n_2^2) u_n &                 n_2 n_3 C_4 &   0 \\
        n_3 C_2 &                 n_3 n_1 C_4 &                 n_3 n_2 C_4 & n_1^3 C_3 + (1 - n_3^2) u_n &   0 \\
              0 &                           0 &                           0 &                           0 & u_n
    \end{bmatrix}, \\
    C_0 = (a_p + u_n)/(p^2 + 1) + 2 u_n(K_2 - p) / ((p^2 + 1)(K_1 - K_2)), \\
    C_1 = (p((K_2-K_1) a_p + (K_1+K_2) u_n) - 2 K_1 K_2 u_n) / ((p^2 + 1)(K_1 - K_2)), \\
    C_2 = (2 u_n + p ((K_1-K_2) a_p + (K_1+K_2) u_n)) / ((p^2 + 1) (K_1 - K_2)), \\ 
    C_3 = ((K_1-K_2) a_p - (K_1+K_2) u_n - 2 K_1 K_2 p u_n) / ((p^2 + 1) (K_1 - K_2)), \\
    C_4 = - u_n + ((K_1-K_2) a_p - (K_1+K_2) u_n - 2 K_1 K_2 p u_n) / ((p^2 + 1)(K_1 - K_2)).
\end{gather*}
It satisfies Turkel's necessary condition (\ref{LMES:Turkel_cond}) as we have
\begin{equation}\label{LMES:Miczek_cond}
    P_{\mathbf{z}}^{-1} | P_{\mathbf{z}} A_{\mathbf{z}}| = \begin{bmatrix}
                \mathcal{O}(1) & \mathcal{O}(1/M_r) & \mathcal{O}(1/M_r) & \mathcal{O}(1/M_r) &             0 \\
            \mathcal{O}(1/M_r)   &   \mathcal{O}(1) &   \mathcal{O}(1) &   \mathcal{O}(1) &             0 \\
            \mathcal{O}(1/M_r)   &   \mathcal{O}(1) &   \mathcal{O}(1) &   \mathcal{O}(1) &             0 \\
            \mathcal{O}(1/M_r)   &   \mathcal{O}(1) &   \mathcal{O}(1) &   \mathcal{O}(1) &             0 \\
                             0 &                0 &                0 &                0 & \mathcal{O}(1)
    \end{bmatrix}.
\end{equation}
This flux-preconditioning matrix was designed to meet the more stringent condition (\ref{LMES:Miczek_cond}) that $P_{\mathbf{z}}^{-1}|P_{\mathbf{z}}A_{\mathbf{z}}|$ has the same Mach number scalings as $A_{\mathbf{z}}$. It is argued \cite{Miczek_T, Miczek, Barsukow} that meeting condition (\ref{LMES:Miczek_cond}) (which implies Turkel's), improves the accuracy of the scheme in both the incompressible and acoustic low-Mach limits. It was recently shown by Bruel \textit{et al.} \cite{Bruel} that while flux-preconditioning with the Turkel matrix improves the accuracy in the incompressible limit, it also leads to a numerical scheme which overly dissipates acoustic waves (more than the standard Roe flux would). \\
\indent Several other preconditioning matrices have been proposed in the literature \cite{Weiss, Lee, Merkle} with the acceleration of steady state calculations as the primary focus. We do not cover them in this work.

\section{Flux-Preconditioning and Entropy-Stability}
\label{LMES:sec:ES_precond}
\subsection{Preliminaries}
\begin{definition}[Tadmor \cite{ES_Tadmor_1987}]
The semi-discrete finite-volume scheme (\ref{eq:FVM}) is called Entropy Conservative (EC) if it \textit{implies}\footnote{an EC/ES discretization will solve the same number of discrete equations as a standard one. The difference with standard discretizations is that the discrete equations imply an additional (physically meaningful) one.} a finite-volume discretization of the entropy equation, that is:
\begin{equation}\label{eq:FVM_entropy}
    \frac{d}{dt}U(\mathbf{u}_{i}) \ + \ \frac{1}{V_i} \int_{\delta \Omega_i} F^{*} dS \ = \ 0, 
\end{equation}
where $F^{*} = F^{*}(\mathbf{u}_L, \mathbf{u}_R, \mathbf{n})$ is a consistent entropy numerical flux. If the scheme (\ref{eq:FVM}) implies instead the inequality: 
\begin{equation}\label{eq:FVM_entropy_ineq}
   \frac{d}{dt}U(\mathbf{u}_{i}) \  + \ \frac{1}{V_i} \int_{\delta \Omega_i} F^{*} dS \ < \ 0,
\end{equation}
it is called Entropy Stable (ES).
\end{definition}
As stated in the introduction, there are several different ways to construct ES schemes. Godunov-type schemes in particular are undoubtedly the most popular ones at the moment. The present work is solely concerned to ES schemes built from Tadmor's ground work \cite{ES_Tadmor_1987}. A study of the behavior of Godunov-type schemes in the low-Mach regime can be found in Guillard \& Murrone \cite{Guillard3}. \\
\indent At first-order, the main difference between conventional finite-volume schemes and EC/ES schemes lies in the choice of the numerical flux $\mathbf{f^{*}}$. The following two theorems outline their construction:
\begin{theorem}[Tadmor \cite{ES_Tadmor_1987}] \label{th:Tadmor_EC}
    The finite-volume scheme (\ref{eq:FVM}) is EC \textit{if and only if} the interface flux $\mathbf{f}^{*}$ satisfies the interface condition:
\begin{equation}\label{eq:ECcond0}
    [\mathbf{v}] \cdot \mathbf{f}^{*}= [\mathcal{F}],
\end{equation}
where $\mathcal{F}$ is the potential function defined by (\ref{eq:potential_def}). One such flux $\mathbf{f}^{*}$ is called Entropy-Conservative (EC) and its corresponding entropy flux $F^{*} = F^{*} (\mathbf{u}_L, \mathbf{u}_R, \mathbf{n})$ is \textit{explicitly given} by:
\begin{equation}\label{eq:EC_Eflux}
    F^{*} \ = \ \overline{\mathbf{v}} \cdot \mathbf{f}^{*} \ - \ \overline{\mathcal{F}}, 
\end{equation}
where the bar notation denotes the arithmetic average.
\end{theorem}
\begin{theorem}[Tadmor \cite{ES_Tadmor_1987}]\label{theorem:ES}
    The finite-volume scheme (\ref{eq:FVM}) is ES \textit{if and only if} the interface flux $\mathbf{f}^{*}$ satisfies the interface condition:
\begin{equation}\label{eq:EScond0}
    [\mathbf{v}] \cdot \mathbf{f}^{*} <[\mathcal{F}],
\end{equation}
where $\mathcal{F}$ is the potential function defined by (\ref{eq:potential_def}). This condition is met by fluxes of the form
    \begin{equation}\label{eq:ES_flux}
    \mathbf{f}^{*} = \mathbf{f}_{EC}^{*} - \frac{1}{2}D[\mathbf{v}].
    \end{equation}
    where $f_{EC}^{*}$ is an EC flux (denote $F_{EC}^{*}$ the associated entropy flux), and $D$ is a positive definite dissipation matrix.
    \begin{equation}\label{eq:FVM_ES}
    \frac{d U(\mathbf{u}_i)}{d t}  \ + \ \frac{1}{V_i}\int_{\delta \Omega_i} F^{*} dS \ = \ - \frac{1}{V_i} \mathcal{E}_i, \ \mathcal{E}_i = \int_{\delta \Omega_i} \mathcal{E}  dS.
    \end{equation}
    The interface entropy flux $F^{*}$ is given by:
    \begin{equation}\label{eq:ES_Eflux}
        F^{*} = F^{*}_{EC} - \overline{\mathbf{v}} \cdot D[\mathbf{v}].
    \end{equation}
    The local entropy production at the interface is given by:
    \begin{equation}\label{eq:Eprod}
    \mathcal{E} \ = \ \frac{1}{4}[\mathbf{v}] \cdot D [\mathbf{v}].
    \end{equation}
\end{theorem}

\subsection{Entropy Conservative Fluxes}
\label{LMES:sec:EC}
\hspace*{0.1 cm} As we have seen in section 3, flux-preconditioning only affects the dissipative component of the standard upwind flux because the central flux does not introduce terms that introduce inappropriate Mach number scalings. What about  EC fluxes? Unless the PDE is scalar, the entropy conservation condition (\ref{eq:ECcond0}) does not uniquely determine $\mathbf{f}_{EC}^{*}$. The first EC flux was introduced by Tadmor \cite{ES_Tadmor_1987}:
\begin{equation}\label{eq:EC_Tadmor0}
    \mathbf{f_{EC}^{*}} = \int_{0}^{1} \mathbf{f}(\mathbf{v}(\xi)) \ d\xi, \ \mathbf{v}(\xi) := \mathbf{v}_{L} + \xi (\mathbf{v}_R - \mathbf{v}_L).
\end{equation}
It is clear that this flux has the same $M_{r}$ scaling as the central flux. The EC flux (\ref{eq:EC_Tadmor0}) is not used in practice because it lacks a closed form (evaluating it would require using a numerical quadrature, which would introduce approximation errors). Tadmor subsequently introduced a variant of $(\ref{eq:EC_Tadmor0})$ which does not require quadrature, but remains too computationally intensive. Our endeavors will eventually bring us back to Tadmor's second flux (section 7.4). \\
\indent Using algebraic manipulations analogous to that of \cite{Roe}, Roe \cite{ES_Roe1, ES_Roe2} proposed a simple, closed-form EC flux for the Euler equations that is more popular than the previous two. In non-dimensional variables, this flux writes $\mathbf{f_{EC}^{*}} = [f_1^{*}, f_2^{*}, f_3^{*}, f_4^{*}, f_5^{*}]$ with:
\begin{align}\label{eq:EC_Roe}
    f_1 =& \  (\overline{m_2}n_1 + \overline{m_3}n_2 + \overline{m_4}n_3 ) m_5^{ln}, \nonumber \\
    f_2 =& \ \frac{n_1}{M_r^2} \frac{\overline{m_5}}{\overline{m_1}} + \frac{\overline{m_2}}{\overline{m_1}}f_1, \nonumber \\
    f_3 =& \ \frac{n_2}{M_r^2} \frac{\overline{m_5}}{\overline{m_1}} + \frac{\overline{m_3}}{\overline{m_1}}f_1, \\
    f_4 =& \ \frac{n_3}{M_r^2} \frac{\overline{m_5}}{\overline{m_1}} + \frac{\overline{m_4}}{\overline{m_1}}f_1, \nonumber \\
    f_5 =& \ \frac{1}{2} \frac{\gamma + 1}{\gamma - 1} \frac{f_1}{\overline{m_1} m_1^{ln}} + \frac{M_r^2}{2\overline{m_1}} \big( \overline{m_2} f_2 + \overline{m_3}f_3 + \overline{m_4} f_4 \big), \nonumber
\end{align}
with the algebraic variables $(m_1, \ m_2, \ m_3, \ m_4, \ m_5) = (\sqrt{\frac{\rho}{p}}, \ \sqrt{\frac{\rho}{p}}u, \ \sqrt{\frac{\rho}{p}}v,  \ \sqrt{\frac{\rho}{p}}w,  \sqrt{\rho p})$. Logarithmic averages \cite{ES_Roe1, ES_Ismail} in $m_1$ and $m_5$ are denoted by $m_1^{\ln}$ and $m_5^{ln}$, respectively. This flux has the same Mach number scaling as the central flux. Using a simpler set of algebraic variables, Chandrasekhar \cite{ES_Chandra} developed another EC flux given by:
\begin{align}\label{eq:EC_Chandra}
    f^{*}_1 =& \  \rho^{ln} u_n, \nonumber \\
    f^{*}_2 =& \ \frac{n_1}{M_r^2}\frac{\overline{\rho}}{\overline{\rho/p}} + \overline{u} f_{1}, \nonumber \\
    f^{*}_3 =& \ \frac{n_2}{M_r^2}\frac{\overline{\rho}}{\overline{\rho/p}} + \overline{v} f_{1}, \\
    f^{*}_4 =& \ \frac{n_3}{M_r^2}\frac{\overline{\rho}}{\overline{\rho/p}} + \overline{w} f_{1}, \nonumber \\
    f^{*}_5 =& \ \bigg(\frac{1}{(\gamma-1)(\rho/p)^{ln}} - M_r^2 \overline{k}\bigg) f_{1} + M_r^2 \big( \overline{u} f_2 + \overline{v} f_3 + \overline{w} f_4\big). \nonumber 
\end{align}
\indent While fluxes (\ref{eq:EC_Tadmor0}), (\ref{eq:EC_Roe}) and (\ref{eq:EC_Chandra}) all have the correct scaling, we refrain from stating that all EC fluxes have the correct scaling. Expanding condition (\ref{eq:ECcond0}) gives:
\begin{gather*}
  \bigg[\frac{\gamma - s}{\gamma - 1} - M_r^2 \frac{\rho}{p} k \bigg] f_1 \ + \ M_r^2 \bigg( \bigg[\frac{\rho u}{p} \bigg] f_2 \ + \ \bigg[\frac{\rho v}{p}\bigg] f_3 \ + \ \bigg[\frac{\rho w}{p}\bigg] f_4 \bigg) \ - \ \bigg[\frac{\rho}{p}\bigg] f_5 = [\rho u_n].
\end{gather*}
In the same way that this condition does not fully determine $\mathbf{f}_{EC}^{*}$, we cannot use it to impose its scaling. \\ \\
\indent As stated in section 3, Flux-Preconditioning was introduced in part because completely discarding the dissipation component of (\ref{eq:flux_upwind}) is not viable since the central flux alone lacks stability. For ES fluxes (\ref{eq:ES_flux}) and ES schemes in general, the situation is different. Discarding the dissipation component of an ES flux can be an option depending on the temporal discretization of (\ref{eq:FVM}). 
\begin{theorem}[Tadmor \cite{ES_Tadmor_2003} ]\label{th:Tadmor_BE}
Consider the fully-discrete finite-volume scheme:
\begin{equation}\label{eq:FVM_BE}
    \frac{1}{\Delta t} \big( \mathbf{u}_{i}^{n+1} - \mathbf{u}_{i}^{n} \big) \ + \ \frac{1}{V_i} \int_{\delta \Omega_i} \mathbf{f}^{*,n+1} dS \ = \ 0, 
\end{equation}
obtained by applying Backward Euler in time to (\ref{eq:FVM}), with $\mathbf{f}_{}^{*}$ an EC flux satisfying \ref{eq:ECcond0} (denote $F^{*}$ the associated entropy flux) and the $n$ and $n+1$ superscripts referring to discrete time instants $t^{n}$ and $t^{n+1}$, respectively. The scheme (\ref{eq:FVM_BE}) implies a fully discrete version of the entropy inequality (\ref{eq:entropy_ineq})
\begin{equation}\label{eq:FVM_entropy_EC_BE}
    \frac{1}{\Delta t} \big( U(\mathbf{u}_{i}^{n+1})  - U(\mathbf{u}_{i}^{n}) \big) \ + \ \frac{1}{V_i} \int_{\delta \Omega_i} F^{*,n+1} dS \ = \ - \frac{1}{\Delta t}\mathcal{E}_i^{BE},    
\end{equation}
With:
\begin{gather}\label{eq:entropy_prod_BE}
    \mathcal{E}_i^{BE} \ = \ \int_{-\frac{1}{2}}^{\frac{1}{2}} \bigg( \frac{1}{2} - \xi \bigg) \  \bigg( [\mathbf{v}]^{n+\frac{1}{2}} \cdot H^{n+\frac{1}{2}} [\mathbf{v}]^{n+\frac{1}{2}} \bigg) \ d\xi \ > 0, \  [\mathbf{v}]^{n+\frac{1}{2}} \ := \ \mathbf{v}_i^{n+1} - \mathbf{v}_i^{n}, \\
   \ H^{n+\frac{1}{2}} := H(\mathbf{v}^{n+\frac{1}{2}}), \ 
   \mathbf{v}^{n+\frac{1}{2}} (\xi) \ := \ \frac{1}{2} \big( \mathbf{v}_i^{n} + \mathbf{v}_i^{n+1} \big) \ + \ \xi [\mathbf{v}]^{n+\frac{1}{2}},\nonumber
\end{gather}
and $H$ defined by (\ref{eq:H_def}). A similar result holds if an ES flux (\ref{eq:ES_flux}) is used instead:
\begin{equation}\label{eq:FVM_entropy_ES_BE}
    \frac{1}{\Delta t} \big( U(\mathbf{u}_{i}^{n+1})  - U(\mathbf{u}_{i}^{n}) \big) \ + \ \frac{1}{V_i} \int_{\delta \Omega_i} F^{*,n+1} dS \ = \ - \frac{1}{V_i} \mathcal{E}_i^{n+1} \ - \ \frac{1}{\Delta t}\mathcal{E}_i^{BE},
\end{equation}
where $\mathcal{E}_i^{n+1}$ is the discrete entropy production in space (\ref{eq:Eprod}) at $\mathbf{u}^{n+1}$ in cell $i$.
\end{theorem}
In section 5, we compared four different fully-discrete ES schemes (\ref{eq:FVM_BE}), three using an ES flux, and one using an EC flux (\ref{eq:EC_Chandra}) alone. The best results are obtained with the latter choice (the central flux leads to unstable results \cite{Miczek_T, Miczek, Gouasmi_Thesis}). Analytical arguments regarding the lack of entropy stability of the central flux can be found in \cite{ES_Tadmor_1987}. \\ 
\indent For the Forward Euler in time, Tadmor established \cite{ES_Tadmor_2003} unconditional lack of entropy-stability with an EC flux in space. If an ES flux is used in space, the entropy-stability of the fully-discrete scheme depends on whether the entropy produced in space outweighs the entropy lost in time (this balance can only be evaluated after the next state $\mathbf{u}^{n+1}$). This configuration is not of interest to us.

\subsection{Entropy-Stable Dissipation and Preconditioning}
As noted in Barth \cite{ES_Barth} (section 2.4), the standard upwind dissipation operator $R |\Lambda| R^{-1}[\mathbf{u}]$ is not ES but the current standard ES dissipation operator is largely inspired by it. The following theorem shows that an upwind-type ES dissipation operator can be obtained by recasting the standard dissipation operator in terms of the entropy variables ($[\mathbf{u}] \ \xleftarrow \ H [\mathbf{v}]$):
\begin{equation}\label{eq:ES_diss_base}
    \frac{1}{2}D[\mathbf{v}] = \frac{1}{2}R |\Lambda| R^{-1} H [\mathbf{v}].
\end{equation}
\begin{theorem}[Barth \cite{ES_Barth}]\label{th:Barth}
     Let $A$ be a diagonalizable matrix ($A = R \Lambda R^{-1}$) and let $H$ be a symmetric positive definite matrix such that $AH$ is symmetric. Then there exists a symmetric positive definite and block diagonal matrix $T$ such that:
     \begin{enumerate}
         \item $RT$ is an eigenvector matrix of $A$, i.e. $A = (RT) \Lambda (RT)^{-1}$.
         \item $H = (RT) (RT)^{T}$ which implies $AH = (RT) \Lambda (RT)^{T}$.
     \end{enumerate}
The matrix $T$ can be inferred from $T^2 = R^{-1} H R^{-T}$.
\end{theorem}
\noindent Since $RT$ is a eigenvector matrix of $A$, the above theorem can be expressed in a simpler way as follows:
\begin{corollary}
Let $A$ be a diagonalizable matrix and let $H$ be a symmetric positive definite matrix such that $AH$ is symmetric. Then there exist an eigenvector matrix $R$ such that $A = R \Lambda R^{-1}$ and $H = RR^T$.
\end{corollary}\label{Barth_cor}
The above result was first established by Merriam \cite{ES_Merriam} for the compressible Euler equations. Barth generalized Merriam's finding later on \cite{ES_Barth}. Using corollary \ref{Barth_cor}, equation (\ref{eq:ES_diss_base}) becomes:
\begin{equation}\label{eq:ES_diss}
    \frac{1}{2}D[\mathbf{v}] = \frac{1}{2}R |\Lambda| R^T [\mathbf{v}].
\end{equation}
The classic upwind dissipation term $R|\Lambda|R^{-1}[\mathbf{u}]$ and its entropy-stable variant $R|\Lambda|R^{T}[\mathbf{v}]$ are thus not different in essence. For infinitesimal variations $(d\mathbf{u}, d\mathbf{v})$, they are equal as we have:
\begin{equation*}
 R |\Lambda| R^{-1} d\mathbf{u} = |A| d\mathbf{u} =  |A| (H d\mathbf{v}) = (R |\Lambda|R^{-1}) (R R^{T} d\mathbf{v}) = R |\Lambda| R^T d\mathbf{v}
\end{equation*}
From this relation, it is fair to assume that both dissipation operators will have the same scaling hence the same accuracy issues in the low-Mach limit (this is observed in practice - see next section). Furthermore, we can now introduce the candidate flux-preconditioned ES numerical flux as follows:
\begin{equation}\label{LMES:ES_flux_P}
    \mathbf{f^{*}} \ = \ \mathbf{f_{EC}^{*}} \ - \ \frac{1}{2}P^{-1}|PA|H [\mathbf{v}].
\end{equation}
\indent The compatibility of flux-preconditioning with entropy stability now boils down to a linear algebra problem: 
\begin{equation}\label{FPES}
\mbox{\textit{Under which conditions on the invertible matrix $P$ is $D_{P} = P^{-1}|PA|H$ positive definite?}} 
\end{equation}
\indent If $P = I$, positive definiteness follows from the eigenscaling theorem because $H$ is symmetric positive definite and symmetrizes $A$ from the right. Writing $H = R R^T$ as before is not helpful unless the eigenvectors of $|PA|$ are related to the eigenvalues of $|A|$ in a convenient way. If $H$ symmetrizes $PA$ from the right, then $|PA|H$ is symmetric positive definite but it is not clear if this matrix would remain positive definite upon multiplication on the left by $P^{-1}$. In addition, the condition that $H$ symmetrizes $PA$ might be too stringent to work with (for the compressible Euler equations in 3 dimensions, $A$ and $H$ are full matrices). \\
\indent A key result which enabled us to move forward is that \textit{the positive definiteness and symmetry properties of a matrix can be established using congruence transforms}. Since $H$ symmetrizes $A$, $HP^T$ symmetrizes $PA$ and we can rewrite $D_P$ as:
\begin{equation}\label{LMES:eq:Pcong}
    D_{P} = P^{-1}|PA|H = P^{-1}|PA|H P^T P^{-T} = P^{-1} \ (|PA|H P^T) \ (P^{-1})^T
\end{equation}
Equation (\ref{LMES:eq:Pcong}) shows that $D_{P}$ is positive definite if and only if $|PA|HP^T$ is positive definite. For the eigenscaling theorem (\ref{th:Barth}) to apply, we need $H P^T$ to be symmetric positive definite. At this stage, this still appears as a complicated a condition to work with ($P$ invertible, $H$ full). \\
\indent In section \ref{LMES:sec:flux_precond}, we recalled that preconditioners are typically developed for a mapped system first. Let $A_{\mathbf{z}} = Q^{-1} A Q$ and $P_{\mathbf{z}}$ be the associated preconditioner, then $P^{-1}|PA| = Q P_{\mathbf{z}}^{-1}|P_{\mathbf{z}}A_{\mathbf{z}}|Q^{-1}$. From there, we note that since $H$ symmetrizes $A$ from the right, then $H_{\mathbf{z}} = Q^{-1} H Q^{-T}$ symmetrizes $A_{\mathbf{z}}$ from the right as we have:
\begin{equation*}
    AH \ = \ (Q A_{\mathbf{z}} Q^{-1}) \ (Q H_{\mathbf{z}} Q^{T}) \ = \ Q \ (AH_{\mathbf{z}}) \ Q^{T}.
\end{equation*}
We can then further decompose $D_{P}$ as:
\begin{equation}\label{LMES:eq:Pcong_map}
    D_{P} = P^{-1}|PA|H = Q P_{\mathbf{z}}^{-1}|P_{\mathbf{z}}A_{\mathbf{z}}| H_{\mathbf{z}} Q^{T} = (Q P_{\mathbf{z}}^{-1}) \ |P_{\mathbf{z}}A_{\mathbf{z}}| H_{\mathbf{z}} P_{\mathbf{z}}^T \ (Q P_{\mathbf{z}}^{-1})^T
\end{equation}
Equation (\ref{LMES:eq:Pcong_map}) shows that $D_{P}$ is positive definite if and only if $|P_{\mathbf{z}} A_{\mathbf{z}}| H_{\mathbf{z}} P_{\mathbf{z}}^T$ is positive definite. Since $H_\mathbf{z} P_{\mathbf{z}}^T$ symmetrizes $P_{\mathbf{z}}A_{\mathbf{z}}$ from the right, then the eigenscaling theorem (\ref{th:Barth}) applies if $H_\mathbf{z} P_{\mathbf{z}}^T$ is symmetric positive definite. \\
\indent With the differential entropy variables $d\mathbf{z} = (dp/(\rho a M_r), \ du, \ dv, \ dw, \ dp - a^2 d\rho)$ introduced in section 3, the Jacobian $A_{\mathbf{z}}$ (equation (\ref{LMES:eq:Az})) is symmetric and from its structure, we can assume that $P_\mathbf{z}$ will have the general form:
\begin{equation*}
P_{\mathbf{z}} = \begin{bmatrix}
        P_{5\times5} & O_{5\times1} \\
        O_{1\times5} & 1 
    \end{bmatrix}.
\end{equation*}
Remarkably, the matrix $H_{\mathbf{z}}$ has a very simple structure:
\begin{equation*}
    H_{\mathbf{z}} = \frac{a^2}{\gamma \rho M_r^2} \begin{bmatrix}
        1 & 0 & 0 & 0 &                    0 \\
        0 & 1 & 0 & 0 &                    0 \\
        0 & 0 & 1 & 0 &                    0 \\
        0 & 0 & 0 & 1 &                    0 \\
        0 & 0 & 0 & 0 & M_r^2 a^2 \rho^2(\gamma - 1)
    \end{bmatrix}.
\end{equation*}
It is easy to show that $H_{\mathbf{z}}$ commutes with $P_{\mathbf{z}}$, $A_{\mathbf{z}}$ and with $|P_{\mathbf{z}}A_{\mathbf{z}}|$. This allows us to ultimately rewrite $D_P$ as:
\begin{equation}\label{LMES:eq:Pcong_map_final}
    D_{P} \ = \ (Q P_{\mathbf{z}}^{-1}) \ |P_{\mathbf{z}}A_{\mathbf{z}}| H_{\mathbf{z}} P_{\mathbf{z}}^T \ (Q P_{\mathbf{z}}^{-1})^T \ = \ (Q H_{\mathbf{z}}^{1/2} P_{\mathbf{z}}^{-1}) \ |P_{\mathbf{z}}A_{\mathbf{z}}| P_{\mathbf{z}}^T \ (Q H_{\mathbf{z}}^{1/2}P_{\mathbf{z}}^{-1})^T.
\end{equation}
and prove the following: 
\begin{theorem}[Flux-Preconditioning and Entropy-Stability]
     For the compressible Euler system (\ref{LMES:EulerM1}), the preconditioned numerical flux (\ref{LMES:ES_flux_P}) is ES if and only if the matrix $|P_{\mathbf{z}} A_{\mathbf{z}}| P_{\mathbf{z}}^{T}$, with $(A_{\mathbf{z}}, P_{\mathbf{z}})$ defined by equations (\ref{eq:similarity}),(\ref{eq:diff_entropy_var}) and (\ref{LMES:eq:Az}), is positive definite. This is achieved under the sufficient but not necessary condition that $P_{\mathbf{z}}$ is symmetric positive definite.  
\end{theorem}
\indent $P_{\mathbf{z}}$ symmetric positive definite is a sufficient condition because of the eigenscaling theorem (\ref{th:Barth}). Indeed, $A_{\mathbf{z}}$ symmetric implies that $(P_{\mathbf{z}}A_{\mathbf{z}})P_{\mathbf{z}}^T$ is as well. In other words, $P_{\mathbf{z}}^{T}$ symmetrizes $P_{\mathbf{z}}A_{\mathbf{z}}$ from the right. Turkel's matrix (\ref{LMES:Turkel_mat}) qualifies. This condition is not necessary, as Miczek's $P_{\mathbf{z}}$ is not symmetric yet leads to an ES flux. We found the latter result while trying to answer a follow-up question to (\ref{FPES}): \\
\begin{equation}\label{FPES_1}
\mbox{\textit{Can we find $P_{\mathbf{z}}$ such that $P_{\mathbf{z}} A_{\mathbf{z}}| P_{\mathbf{z}}^{T}$ positive definite and $P_{\mathbf{z}}^{-1}|P_{\mathbf{z}}A_{\mathbf{z}}|$ scales like $A_{\mathbf{z}}$ with respect to $M_r$?}} \vspace{0.2cm}
\end{equation}
We have not managed to find a symmetric positive definite matrix $P_{\mathbf{z}}$ which satisfies the scaling requirements. To simplify the analysis, let's consider the scenario where the flow and the interface normal are aligned with the x-direction. This brings us back to the Turkel's 2-by-2 system (\ref{LMES:eq:2b2}). Problem (\ref{FPES_1}) simplifies to:
\begin{equation*}
    \mbox{\textit{Can we find $p, p_1$ and $p_2$ such that
    $P_{\mathbf{z}}^{-1} |P_{\mathbf{z}} A_{\mathbf{z}}| = \begin{bmatrix} 
            \mathcal{O}(1) & \mathcal{O}(1/M_r) \\
            \mathcal{O}(1/M_r) & \mathcal{O}(1) \end{bmatrix} $ and $
    P_{\mathbf{z}} = \begin{bmatrix}
            p_1 & p \\
            p   & p_2
        \end{bmatrix}$ is positive definite?}}
\end{equation*}
We have $P_{\mathbf{z}}A_{\mathbf{z}} = R_p \Lambda_p R_p^{-1}$ with:
\begin{gather*}
    \Lambda_p = diag([0.5(u_p+a_p), \ 0.5(u_p-a_p)]), \ R_p = \begin{bmatrix}
        r_1 & r_2 \\
        1 & 1
    \end{bmatrix}, \\
    u_p = (p_1+p_2)u + 2 a p/M_r, \ a_p = \sqrt{u_p^2 + 4 det(P)(a^2/M_r^2 - u^2)}, \\
    r_1 = (M_r a_p - a p_2(p_1-p_2)/p) / (2 a p_2 + 2 M_r p u) + (p_1 - p_2)/(2 p), \\
    r_2 = (- M_r a_p - a p_2(p_1-p_2)/p) / (2 a p_2 + 2 M_r p u) + (p_1 - p_2)/(2 p).
\end{gather*}
$det(P) = p_1 p_2 - p^2 > 0$ and $trace(P) = p_1 + p_2 > 0$ impose $p_1$ and $p_2$ to be positive. $det(P)>0$ and $a^2/M_r^2 - u^2 > 0$ in the subsonic regime, therefore $u_p < a_p \ \implies \ |u_p - a_p| = - (u_p - a_p)$ and we have:
\begin{gather*}
    P_{\mathbf{z}}^{-1}|P_{\mathbf{z}}A_{\mathbf{z}}| = \begin{bmatrix}
            a_{11} & a_{12} \\ a_{12} & a_{22} 
        \end{bmatrix} \\ a_{11} = \frac{1}{a_p} 
            \big(u^2(p_1-p_2) + 2 a^2 p_2/M_r^2 + 2 a p u/M_r\big), \
            a_{12} = \frac{1}{a_p} \big(2 p u^2 + a(p_1+p_2)u/M_r\big), \\
            a_{21} = a_{12}, \ a_{22} =  \frac{1}{a_p} \big(-u^2(p_1-p_2) + 2 a^2 p_1/M_r^2 + 2 a p u/M_r\big).
\end{gather*}
Looking at the expression of $a_p$, we see that in the limit $M_r \rightarrow 0$, $a_p$ can scale either as $p/M_r$, $\sqrt{p_1 p_2}/M_r$, $p_1$ or $p_2$.
\begin{itemize}
    \item If $a_p \approx p/M_r$: $a_{11} \approx u^2(p_1-p_2)M_r/p + 2 a^2 p_2/(pM_r) + 2au = \mathcal{O}(1)$ requires $p_2/p$ to scale as $M_r$ at most. Likewise, $p_1/p$ must scale as $M_r$ at most for $a_{22}$ to be $\mathcal{O}(1)$. But then $a_{12} \approx 2M_ru^2 + a(p_1/p + p_2/p)u = \mathcal{O}(M_r)$ does not scale as $1/M_r$. 
    \item If $a_p \approx p_1$: the second term in $a_{22}$ scales as $1/M_r^2$.
    \item If $a_p \approx p_2$: the second term in $a_{11}$ scales as $1/M_r^2$.
    \item If $a_p \approx \sqrt{p_1 p_2}/M_r$: Denote $X = \sqrt{p_1/p_2}$. Then $a_{11} \approx u^2(X-1/X)M_r + 2 a^2/(XM_r) + 2 a p u/(\sqrt{p_1 p_2}) = \mathcal{O}(1)$ imposes that $X$ scales as $1/M_r$. But then the second term in $a_{22}$ scales at $1/M_r^2$ instead of $1$.
\end{itemize} 
In each case, it seems\footnote{we recognize that the above scaling arguments are not of the utmost rigor} that the Mach number scaling requirements cannot be met. \\
\indent Miczek's flux-preconditioner can be found by seeking $P_{\mathbf{z}}$ in the form: 
\begin{equation*}
    P_{\mathbf{z}} = \begin{bmatrix}
    1 & p \\ -p & 1
    \end{bmatrix}.
\end{equation*}
$P_{\mathbf{z}}$ is not symmetric but it is positive definite for any $p$ since its symmetric part is the identity matrix. We have $ P_{\mathbf{z}}A_{\mathbf{z}} = R_p \Lambda_p R_p^{-1}$ with:
\begin{gather*}
   \Lambda_p = diag([u + a_p, \ u - a_p]), \ R_p = \begin{bmatrix}
            (-M_r a_p + a p)/(a - M_r p u) & ( M_r a_p + a p)/(a - M_r p u) \\
            1 & 1
        \end{bmatrix}, \\
        a_p = \sqrt{u^2 + det(P)(a^2/M_r^2-u^2)} > u.
\end{gather*}
We have:
\begin{equation*}
    P_{\mathbf{z}}^{-1}|P_{\mathbf{z}} A_{\mathbf{z}}| = \frac{1}{a_p}
        \begin{bmatrix}
            a^2/M_r^2 & p u^2  + a(M_ru - ap)/M_r^2 \\
            - p u^2  + a(M_r u - a p)/M_r^2 & a^2/M_r^2
        \end{bmatrix}.
\end{equation*}
For the first term to be $\mathcal{O}(1)$ we need $a_p = \mathcal{O}(1/M_r^2)$ which imposes $p = \mathcal{O}(1/M_r)$. The scaling of $A_{\mathbf{z}}$ is completely recovered. In the subsonic regime, Miczek set $p = 1 - 1/M_r$  so that in the limit $M_r \rightarrow 1$, $P_{\mathbf{z}} \rightarrow I$. \\
\indent Finally, we have that $|P_{\mathbf{z}}A_{\mathbf{z}}|P_{\mathbf{z}}^{T}$ is symmetric positive definite since its symmetric part has a determinant $a^2(a^2 - M_r^2 a^2)det(P)^2/(M_r^4 a_p^2)$ and a trace $2a^2det(P)/(M_r^2 a_p)$ that are both positive. In section 6, this entropy stability result will be established in a more elegant way for the full system. 

\section{Numerical experiments}
\label{LMES:sec:Num}

\indent In this section, we examine four different first-order ES schemes (\ref{eq:FVM_BE}) (Theorem 3.3) in two simple flow configurations representative of the incompressible and acoustic limits. In section \ref{LMES:sec:Gresho}, we consider the Gresho vortex. In section \ref{LMES:sec:Sound}, we consider a right-moving sound wave in one dimension. Periodic boundary conditions are set in both problems. A CFL of 1 is used for Backward Euler time integration. \\
\indent In space, we use the classic ES Roe flux \cite{ES_Roe1} (the entropy fix of Ismail \& Roe \cite{ES_Ismail} is not needed as we are not dealing with shock configurations), the ES Turkel flux, the ES Miczek flux, with the EC flux of Chandrasekhar \cite{ES_Chandra} as the base (the same results were observed with the EC flux of Roe (\ref{eq:EC_Roe}) \cite{ES_Roe1, ES_Roe2}). We also consider the EC flux of Chandrasekhar alone.  \\
\indent The calculations are made using a code which solves the discrete equations in dimensional form using a standard Newton-GMRES method \cite{Knoll}. The numerical fluxes in dimensional form are obtained by setting $M_r = 1$ and computing $p$ locally using the Mach number associated with the average state. We use hat notation to denote the dimensional flow variables.

\subsection{Gresho Vortex}\label{LMES:sec:Gresho}
\begin{figure}[!htbp]
    \centering
    \includegraphics[scale = 0.5]{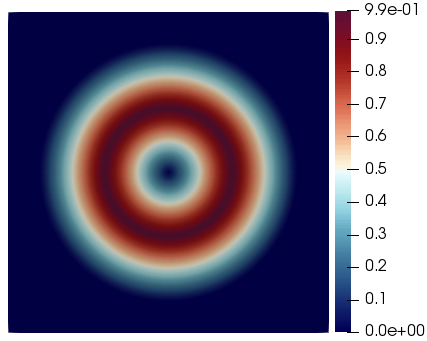}
    \caption{Gresho Vortex: Initial Mach number distribution $M/M_r$ for $M_r = 3 \times 10^{-3}$.}
    \label{fig:gresho_IC}
\end{figure}

\begin{figure}[h!]
    \centering
    \subfigure[EC Roe $ - M_r = 3 \times 10^{-1}$ ]{\includegraphics[scale = 0.43]{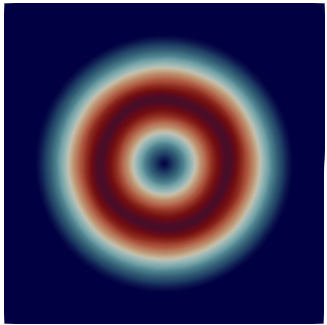}}
    \subfigure[EC Roe $ - M_r = 3 \times 10^{-2}$ ]{\includegraphics[scale = 0.43]{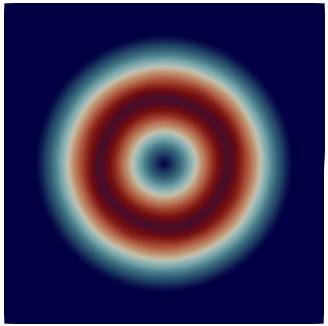}}
    \subfigure[EC Roe $ - M_r = 3 \times 10^{-3}$ ]{\includegraphics[scale = 0.43]{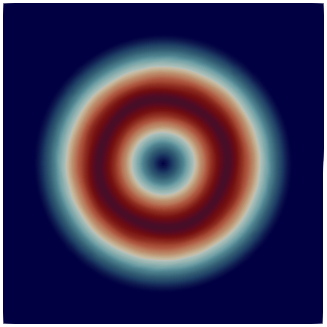}}
    \subfigure[ES Miczek $ - M_r = 3 \times 10^{-1}$ ]{\includegraphics[scale = 0.43]{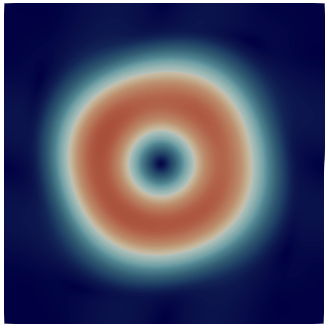}}
    \subfigure[ES Miczek $ - M_r = 3 \times 10^{-2}$ ]{\includegraphics[scale = 0.43]{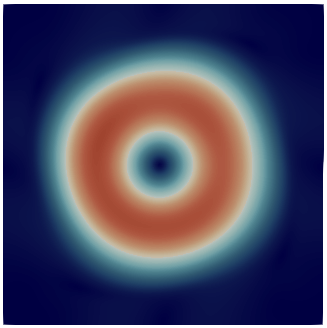}}
    \subfigure[ES Miczek $ - M_r = 3 \times 10^{-3}$ ]{\includegraphics[scale = 0.43]{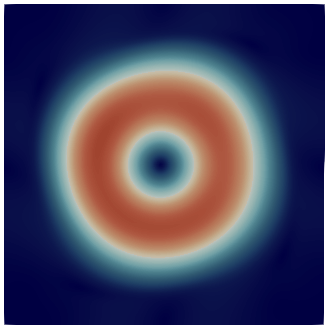}}
    \subfigure[ES Turkel $ - M_r = 3 \times 10^{-1}$ ]{\includegraphics[scale = 0.43]{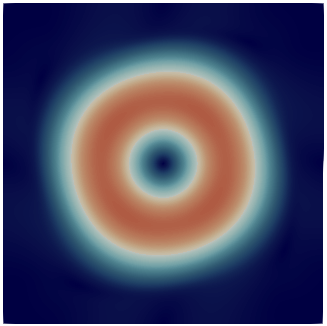}}
    \subfigure[ES Turkel $ - M_r = 3 \times 10^{-2}$ ]{\includegraphics[scale = 0.43]{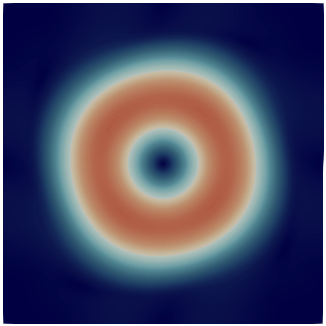}}
    \subfigure[ES Turkel $ - M_r = 3 \times 10^{-3}$ ]{\includegraphics[scale = 0.43]{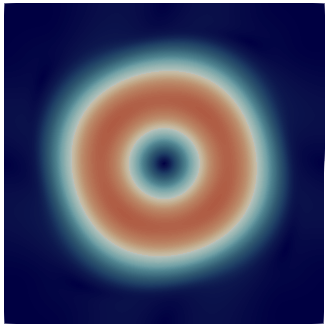}}
    \subfigure[ES Roe $ - M_r = 3 \times 10^{-1}$ ]{\includegraphics[scale = 0.43]{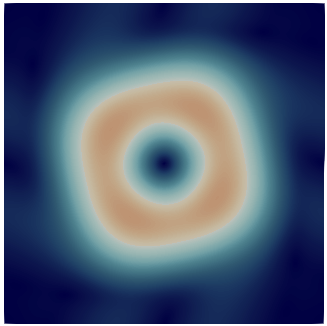}}
    \subfigure[ES Roe $ - M_r = 3 \times 10^{-2}$ ]{\includegraphics[scale = 0.43]{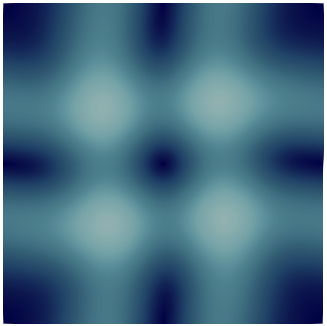}}
    \subfigure[ES Roe $ - M_r = 3 \times 10^{-3}$ ]{\includegraphics[scale = 0.43]{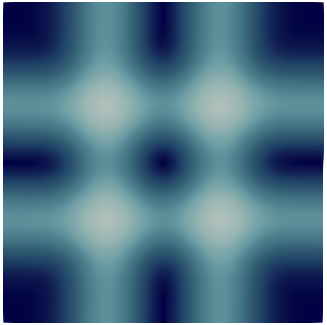}}
    \caption{Gresho Vortex: $M/M_r$ profiles at $t = 1$. Same legend as figure \ref{fig:gresho_IC}-(a).}
    \label{fig:gresho_all}
\end{figure}

\begin{figure}[htbp!]
    \centering
    \includegraphics[scale = 0.8]{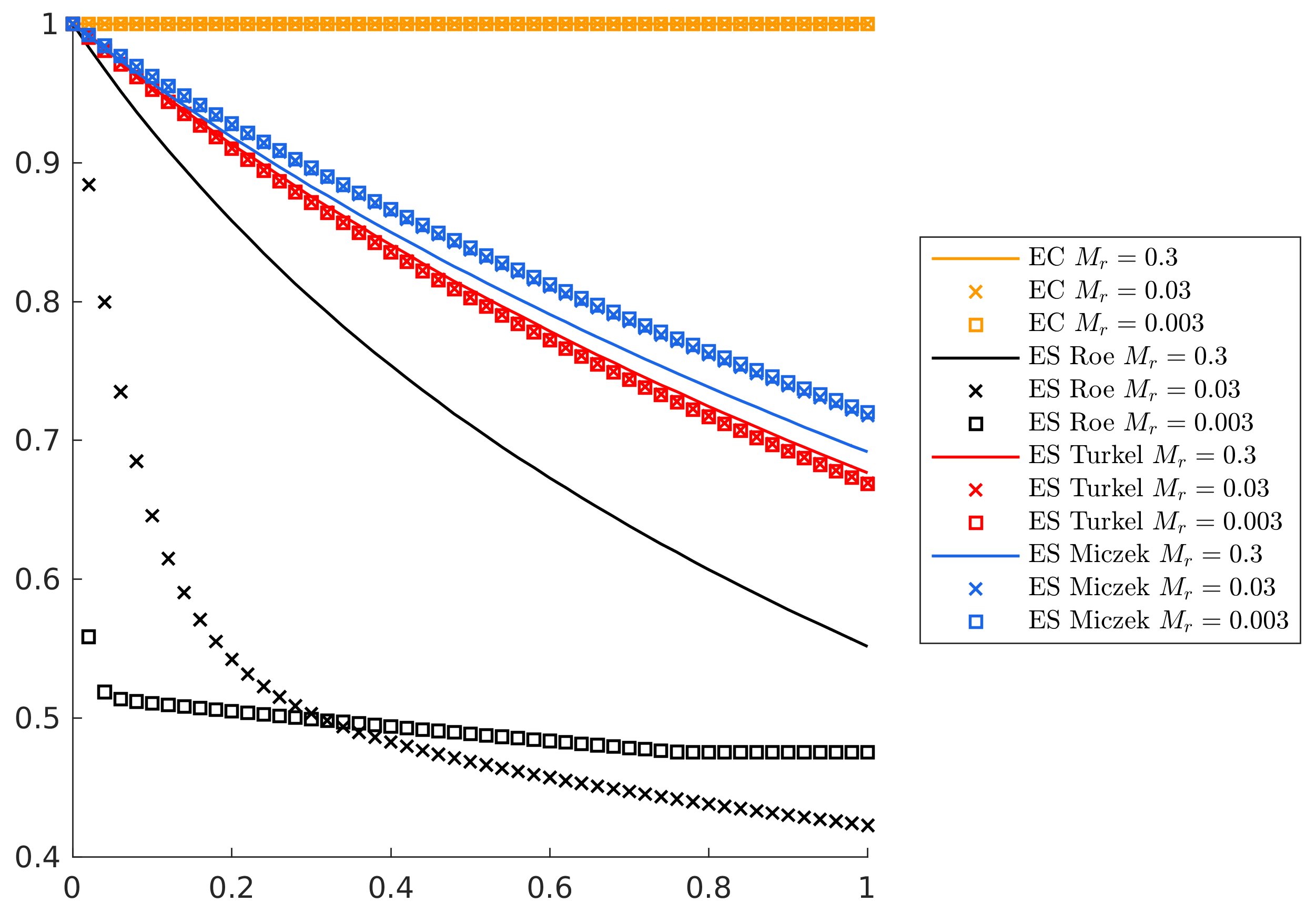}
    \caption{Gresho vortex: Total kinetic energy $k/k_0$ evolution over time for different ES fluxes at different Mach numbers. }
    \label{fig:Gresho_Ek}
\end{figure}

\begin{figure}[htbp!]
    \centering
    \includegraphics[scale = 0.16]{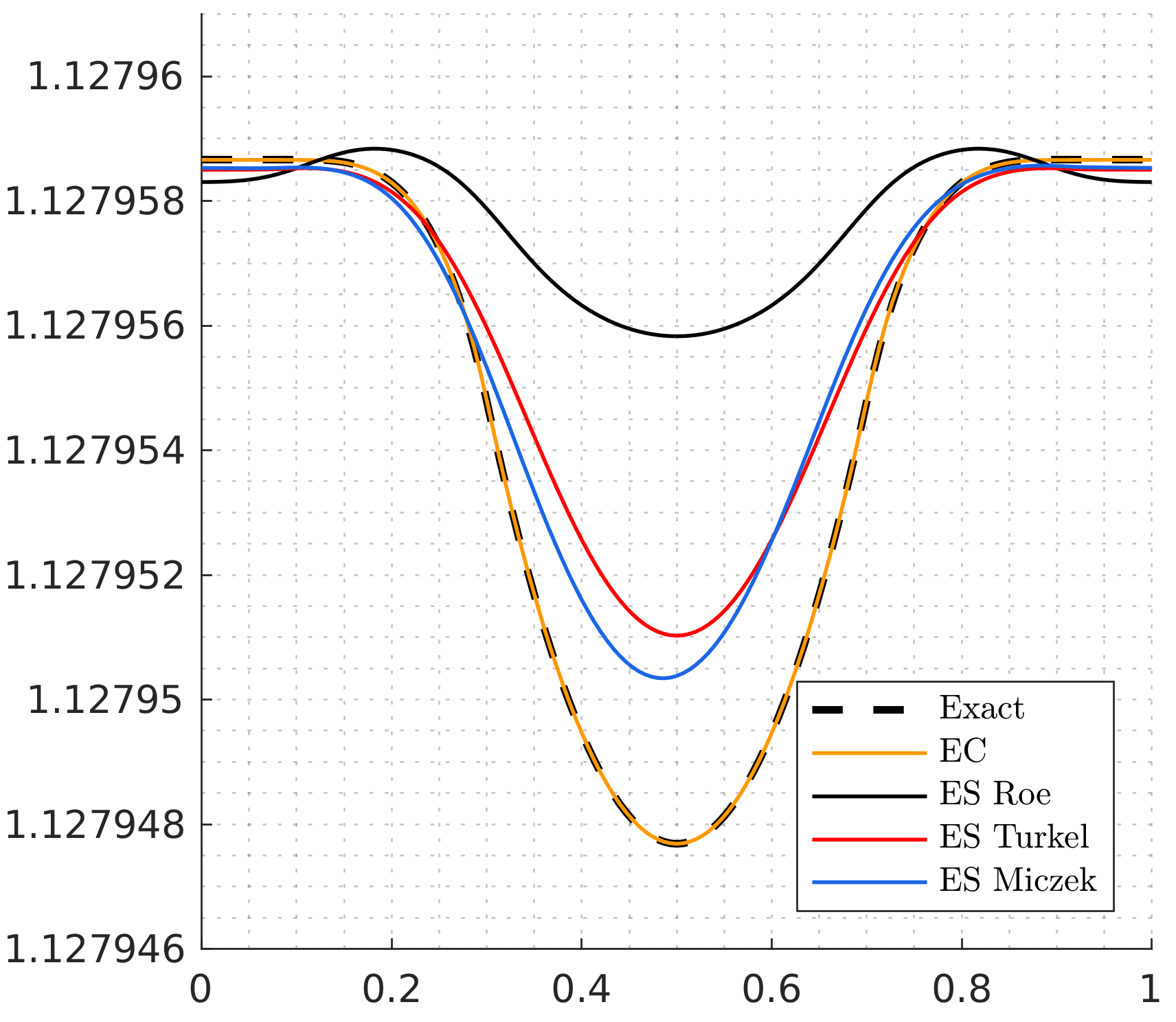}
    \caption{Gresho vortex: Centerline pressure profile $\hat{p}(x,0.5)$ after one rotation at $M_r = 3 \times 10^{-3}$. }
    \label{fig:Gresho_pressure}
\end{figure}

\begin{figure}[htbp!]
    \centering
    \subfigure[$t = 0.04$]{\includegraphics[scale = 0.15]{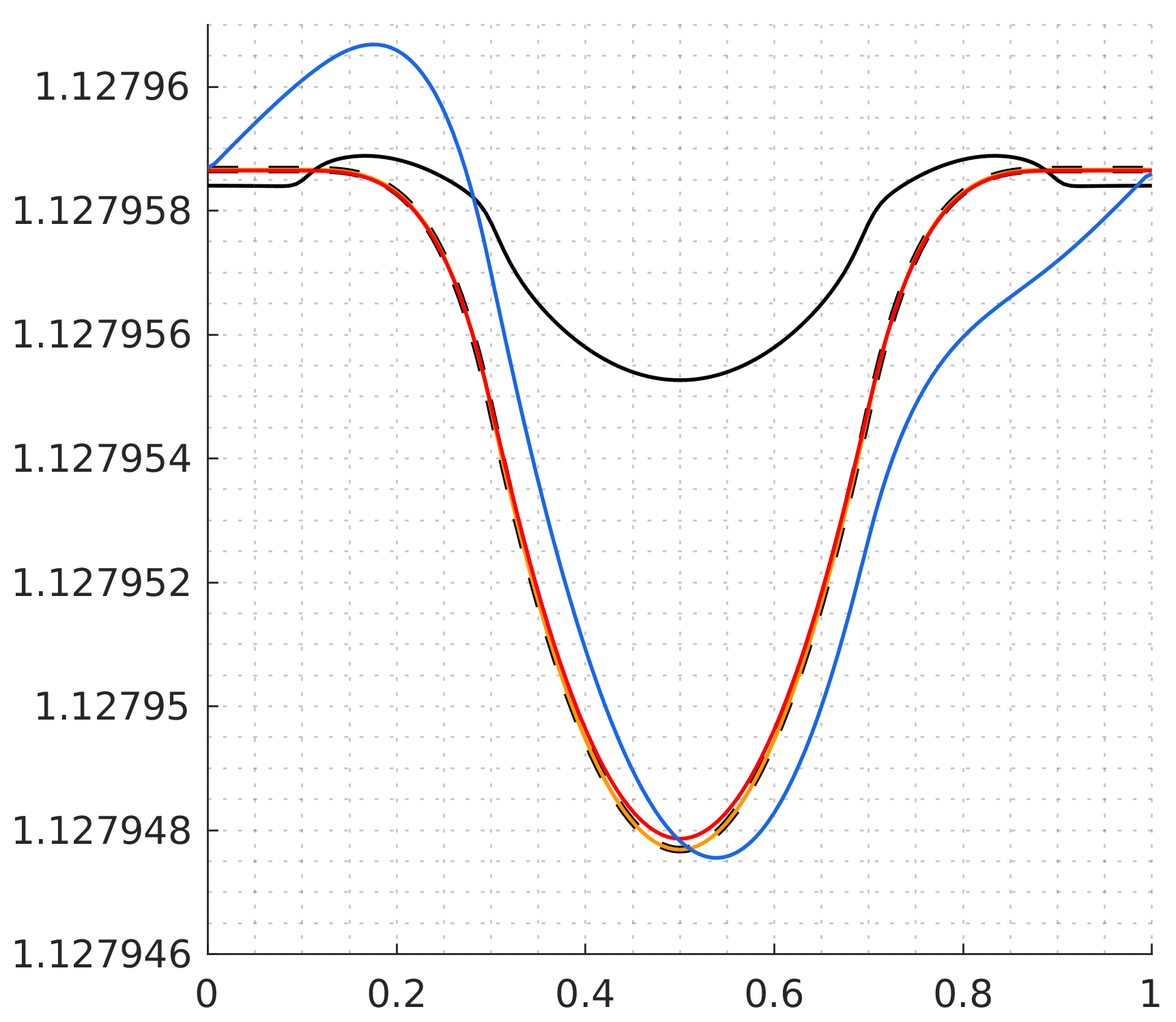}}
    \subfigure[$t = 0.08$]{\includegraphics[scale = 0.15]{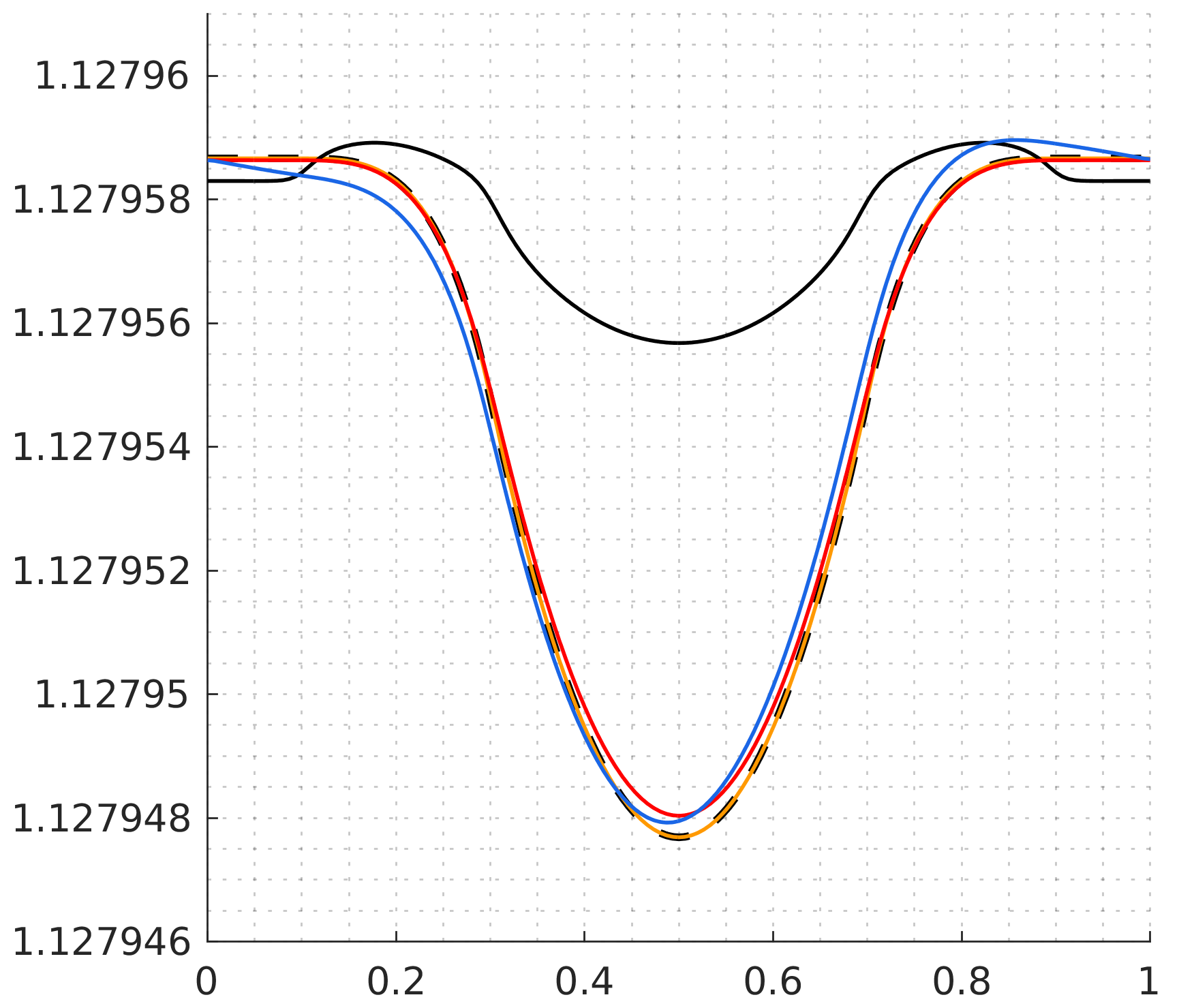}}
    \caption{Gresho Vortex: Centerline pressure profiles at early instants highlighting the spurious transient observed with the ES Miczek flux (blue). $M_r = 3 \times 10^{-3}$. Same legend as figure \ref{fig:Gresho_pressure}.}
    \label{fig:Gresho_pressure_early}
\end{figure}

\hspace*{0.1 cm}The Gresho vortex \cite{Gresho, Liska, Miczek} is a steady-state solution of the incompressible Euler equations in two dimensions ($\Omega = [0, 1] \times [0,1]$). Let $R$ be the radius of the vortex and $r$ be the radial coordinate. Density is constant $\hat{\rho} = \rho_r$. The velocity field is given by:
\begin{equation}
    \hat{\bm{u}} =  \hat{u}_{\phi} \bm{e}_{\phi}, \ \hat{u}_{\phi} = u_r \left\{
                \begin{array}{ll}
                  r/R, \ \ \ \ \ \ 0 \leq r < R \\
                  2 - r/R, \ R \leq r \leq 2R \\
                  0, \ \ \ \ \ \ \ \ \ \ 2R \leq r
                \end{array}
              \right.
\end{equation}
$u_{\phi}$ denotes the tangential velocity, and the mapping between cartesian $(\bm{u_x}, \bm{u_y})$ and radial coordinates is defined by:
\begin{equation*}
    r = \sqrt{x^2 + y^2}, \ \bm{e}_{\phi} = -sin(\phi) \bm{u_x} + cos(\phi) \bm{u_y} = -\bigg( \frac{y}{r} \bigg)  \bm{u_x} + \bigg( \frac{x}{r} \bigg) \bm{u_y}.
\end{equation*}
The reference time scale is set as the vortex period $t_r = 2 \pi R / u_{\phi}(R) = 2 \pi R / u_r$. The pressure $\hat{p}$ must provide the centripetal force:
\begin{align*}
\hat{p} =& \ p_r + \int_{0}^{r} \rho_r \frac{u_{\phi}^2(\overline{r})}{\overline{r}} \ d\overline{r} \
        = \ p_r + \rho_r u_r^2
    \left\{
                \begin{array}{ll}
                  (r/R)^2/2
                  , \ \ \ \ \ \ \ \ \ \ \ \ \ \ \ \ \ \ \ \ \ \ \ \ \ \ \ \ \ \ \ \ \ \ \ \ 0 \leq r < R \\
                  (r/R)^2/2 + 4 (1 - (r/R) + \ln(r/R)), \ R \leq r \leq 2R \\
                  -2 + 4 \ln 2, \ \ \ \ \ \ \ \ \ \ \ \ \ \ \ \ \  \ \ \ \ \ \ \ \ \ \ \ \ \ \ \ \ 2R \leq r
                \end{array}
              \right.
\end{align*}
$p_r$ is a strictly positive constant. The reference Mach number $M_r$ for this setup is defined as the one at $r = R$:
\begin{equation}\label{LMES:p_to_u}
    M_r = \frac{u_r}{\sqrt{\gamma (p_r/\rho_r + u_r^2/2)}} \ \iff \ p_r = \rho_r u_r^2 \bigg(\frac{1}{\gamma M_r^{2}} - \frac{1}{2}\bigg).
\end{equation}
This relation shows how the reference quantities $(\rho_r, \ u_r, \ p_r)$ relate to the reference Mach number $M_r$.  We take $R = 0.2$, $\rho_r = 1.0$ and $u_r = 2 \pi R M_r \implies t_r = 1/M_r$. $p_r$ is determined by equation (\ref{LMES:p_to_u}). Spatial fluctuations in $p$ are of order $M_r^2$ while spatial fluctuations in $\bm{u}$ are of order $1$. \\
\indent As in Miczek \textit{et al.} \cite{Miczek}, we fix the grid ($150 \times 150$ cells) and run the ES schemes at different $M_r \in \{ 3 \times 10^{-1}, \ 3 \times 10^{-2}, \ 3 \times 10^{-3} \}$ for one vortex revolution, that is until $t = 1$. Figure \ref{fig:gresho_IC} shows the initial solution for $M_r = 3 \times 10^{-3}$. 
Figure \ref{fig:gresho_all} shows snapshots of the solution with each scheme at different Mach numbers, and provides a clear illustration of the accuracy degradation issues in the low-Mach regime (ES Roe). The other three schemes do not show a visible dependency on the reference Mach number. The best results are obtained with the EC flux. The difference between the ES Turkel and ES Miczek fluxes is not clearly visible from these plots. \\
\indent Figure \ref{fig:Gresho_Ek} shows a normalized kinetic energy evolution for all four schemes at different Mach numbers. For the ES Roe flux, we see that the rate at which the kinetic energy decays increases with the Mach number. We can also see that the normalized kinetic energy for $M_r = 3 \times 10^{-2}$ becomes bigger than for $M_r = 3 \times 10^{-3}$. This was slightly visible in figure \ref{fig:gresho_all}. The Gresho vortex is a \textit{stationary} solution, hence it is not surprising that numerical solution would eventually reach a steady state (we could then see the faster kinetic energy decay as the scheme converging to a wrong solution faster as $M_r \xrightarrow{} 0$). For the ES Miczek and ES Turkel fluxes, the kinetic energy decay appears to be independent of the Mach number. In each case, we see that the $M_r = 3 \times 10^{-1}$ curve is not matching exactly with the $M_r = \{3 \times 10^{-2}, \ 3 \times 10^{-3} \}$ ones. We believe that this is because the $M_r = 3 \times 10^{-1}$ configuration does not completely fall into the incompressible regime. Figure \ref{fig:Gresho_Ek} overall suggests that the ES Miczek flux performs better than the ES Turkel flux. This is also supported by figures \ref{fig:Gresho_Sound_S}(a)-(c)-(e) which show that the ES Turkel flux produces more entropy than the ES Miczek flux. \\
\indent Figure \ref{fig:Gresho_pressure} shows the pressure distribution along the centerline $y = 0.5$, after one revolution at $M_r = 3 \times 10^{-3}$. The solution with the ES Miczek flux is clearly not in phase with the exact solution. The same anomaly is observed at different Mach numbers. Figure \ref{fig:Gresho_pressure_early} suggests that this anomaly is the consequence of a spurious transient in the early stages of the vortex rotation. We found that the duration of this transient decreases with the Mach number.

\subsection{Acoustic wave}\label{LMES:sec:Sound}
\begin{figure}[htbp!]
    \centering
    \subfigure[$M_r = 10^{-2}$]{\includegraphics[scale = 0.15]{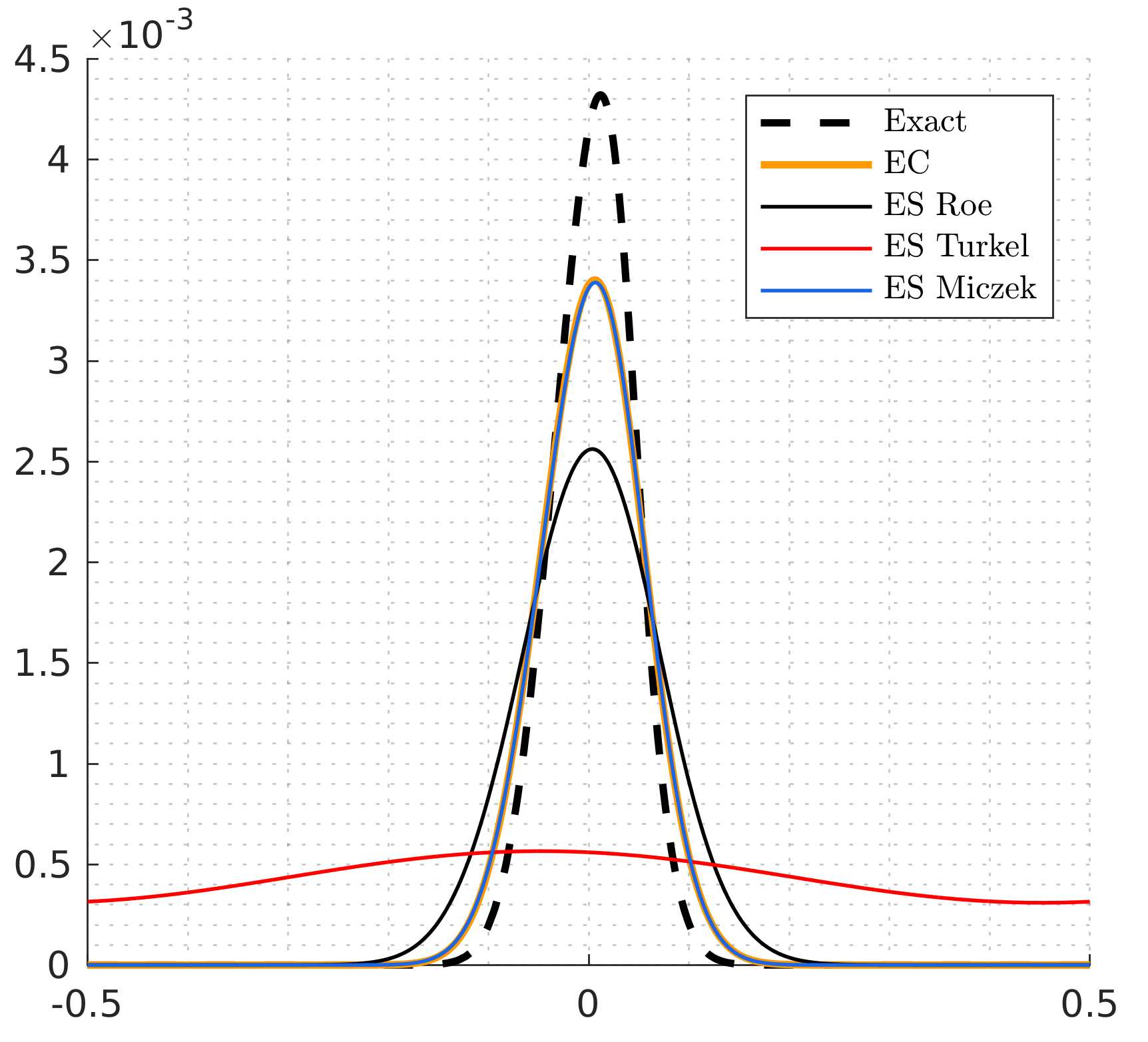}}
    \subfigure[$M_r = 10^{-3}$]{\includegraphics[scale = 0.15]{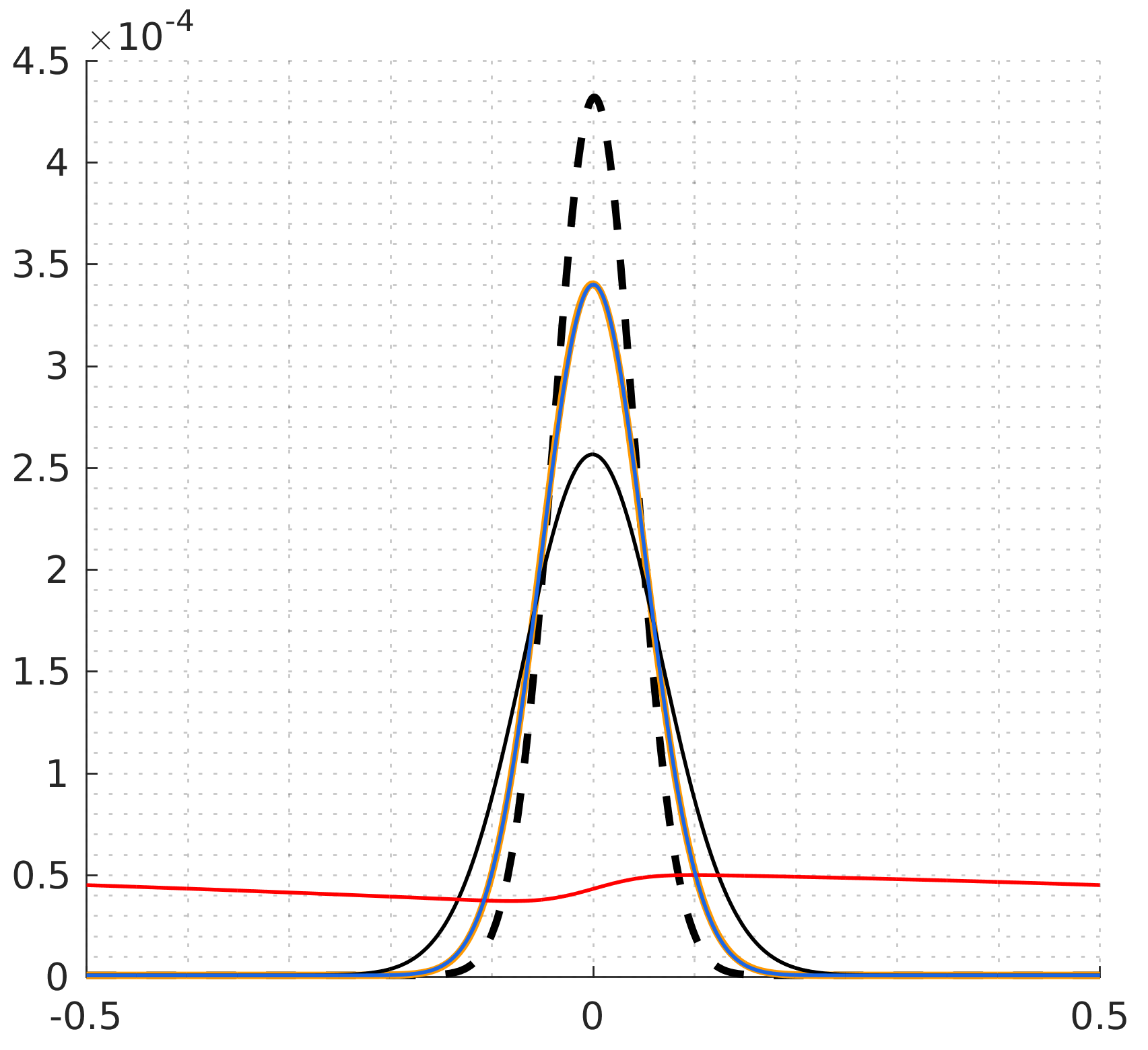}}
    \subfigure[$M_r = 10^{-4}$]{\includegraphics[scale = 0.15]{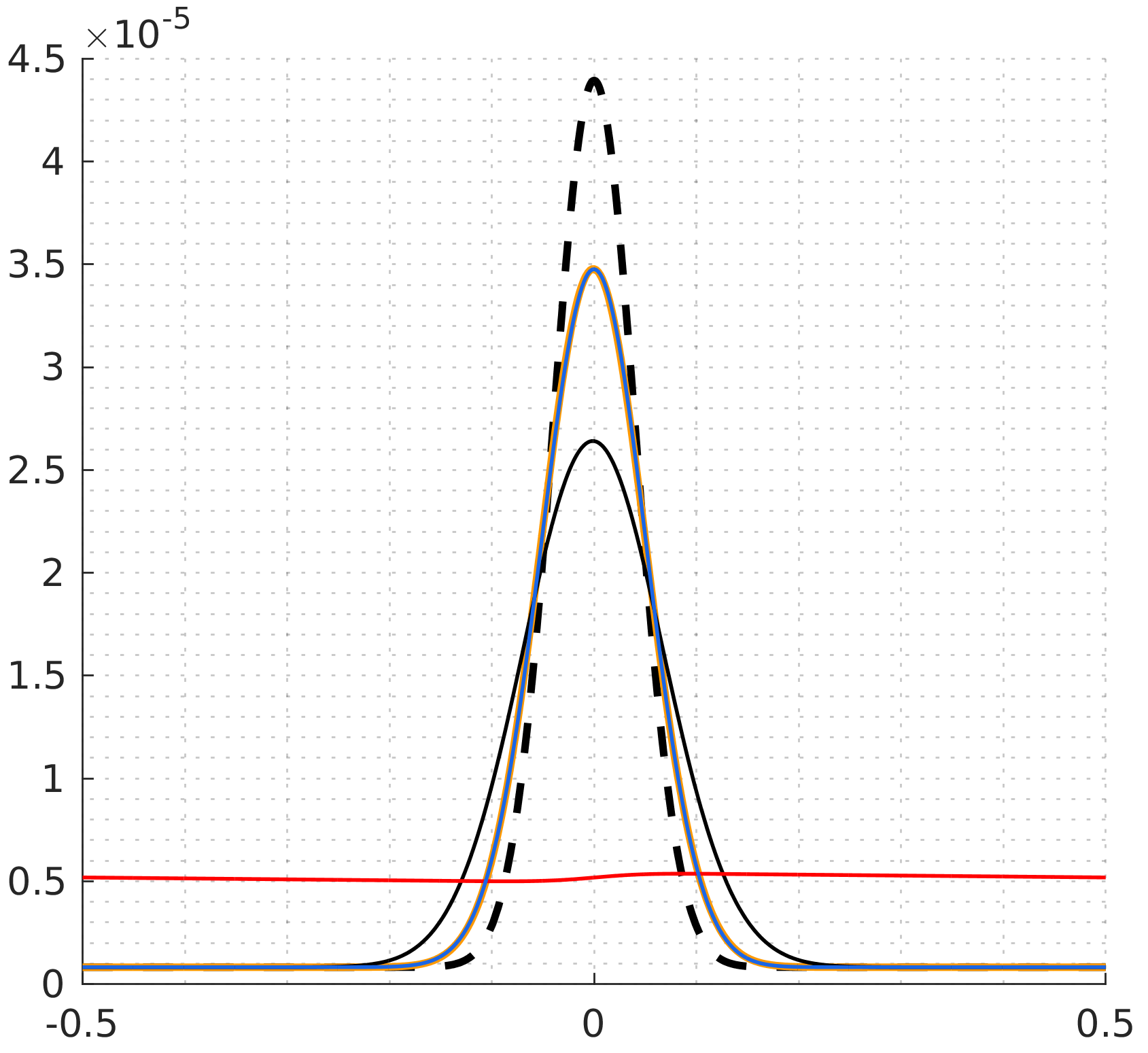}}
    \caption{Sound wave: Pressure profiles at $t = 1$ for different Mach numbers.}
    \label{fig:Sound_P_snaps}
\end{figure}

\begin{figure}[htbp!]
    \centering
    \subfigure[Gresho Vortex - $M_r = 3 \times 10^{-1}$]{\includegraphics[scale = 0.7]{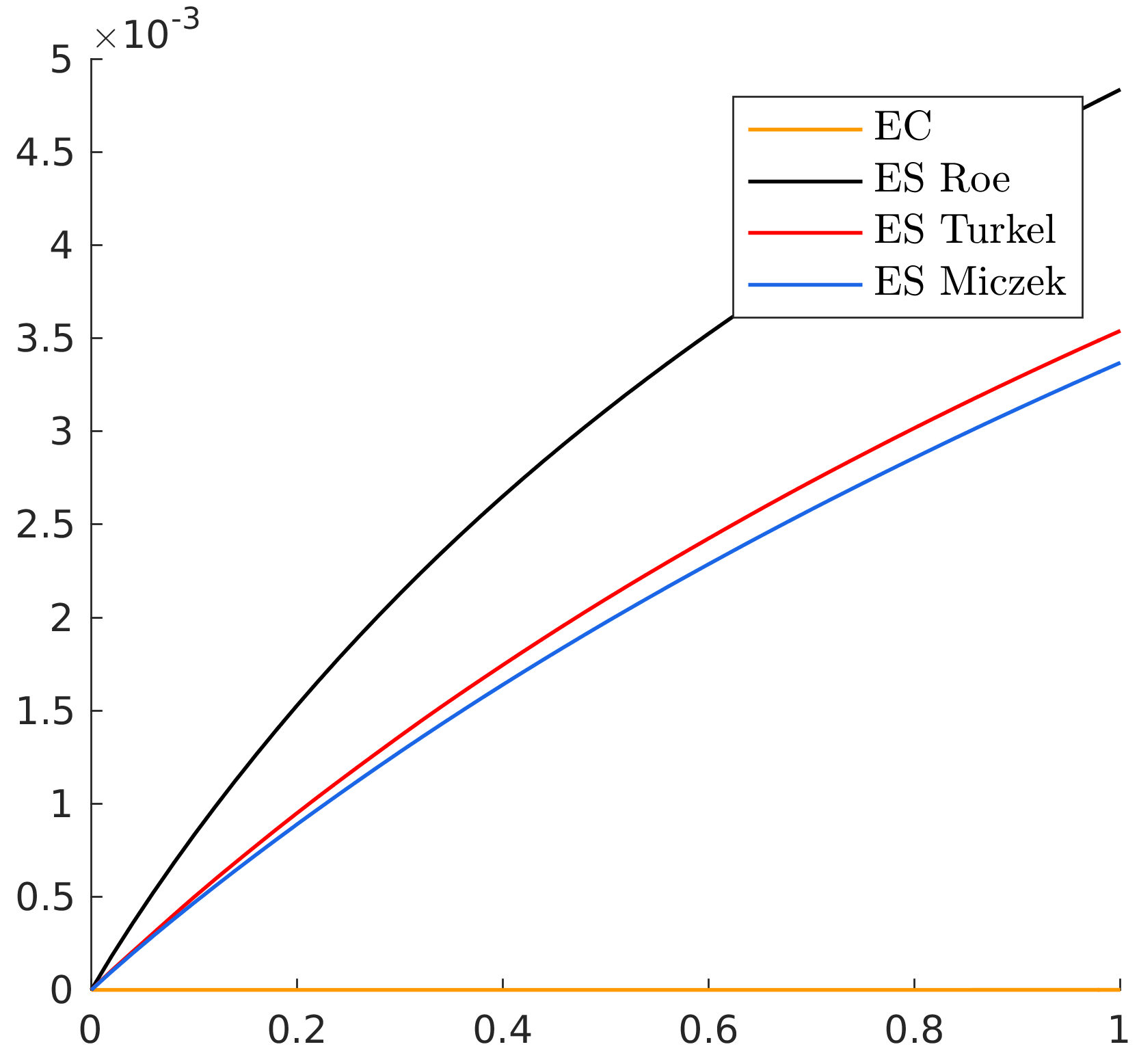}}
    \subfigure[Sound wave - $M_r = 10^{-2}$]{\includegraphics[scale = 0.127]{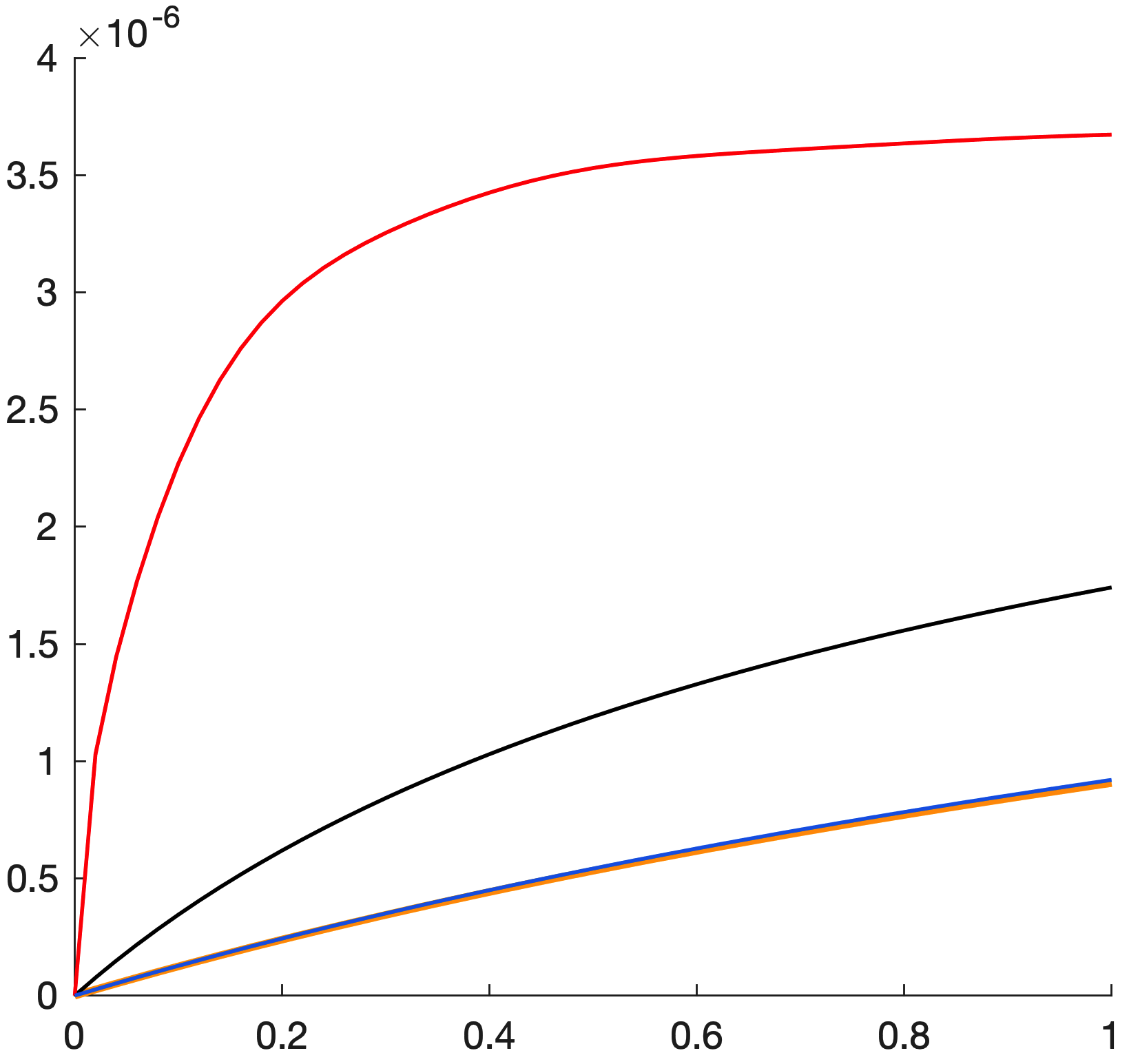}}
    \subfigure[Gresho Vortex - $M_r = 3 \times 10^{-2}$]{\includegraphics[scale = 0.7]{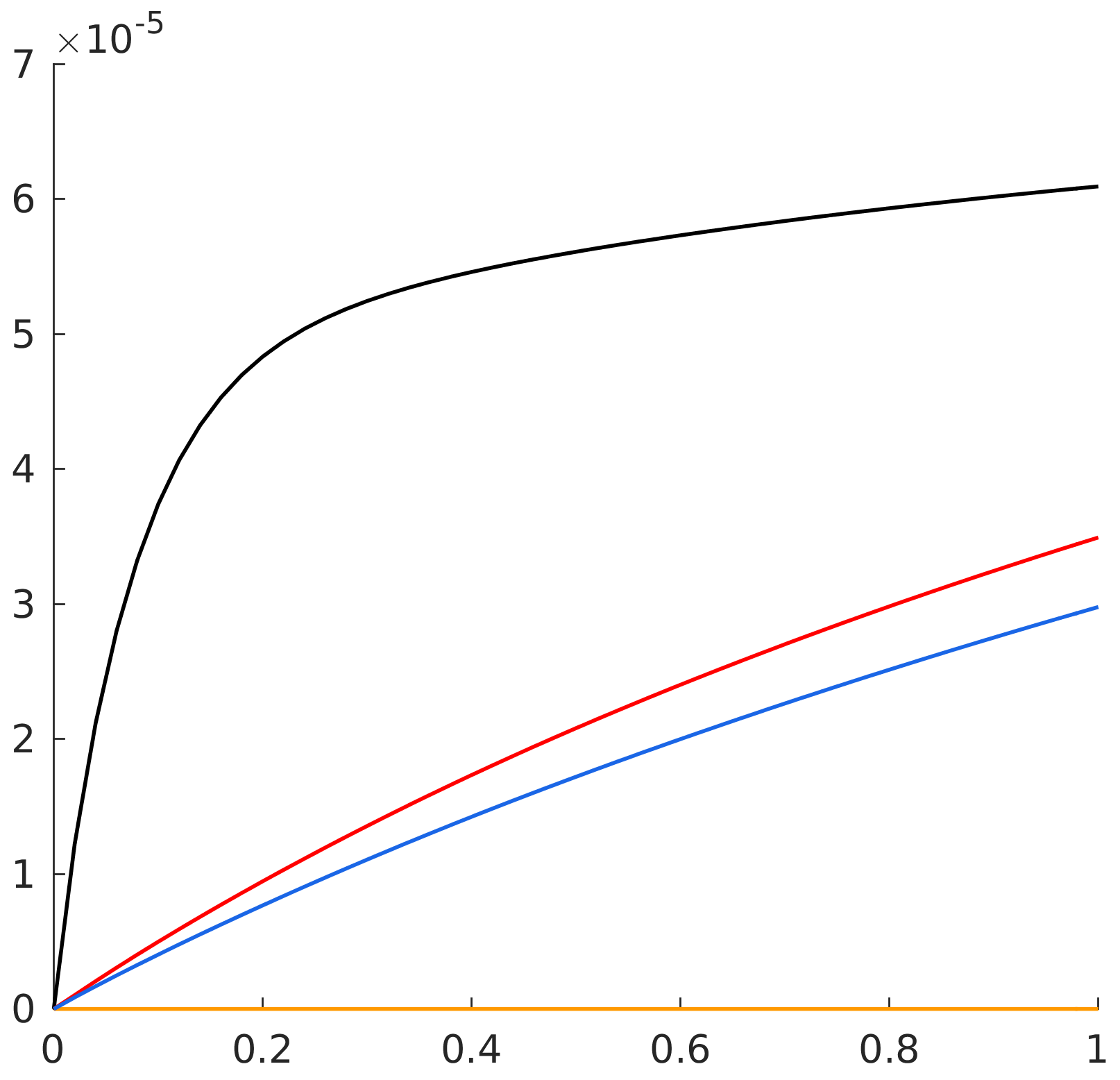}}
    \subfigure[Sound wave - $M_r = 10^{-3}$]{\includegraphics[scale = 0.7]{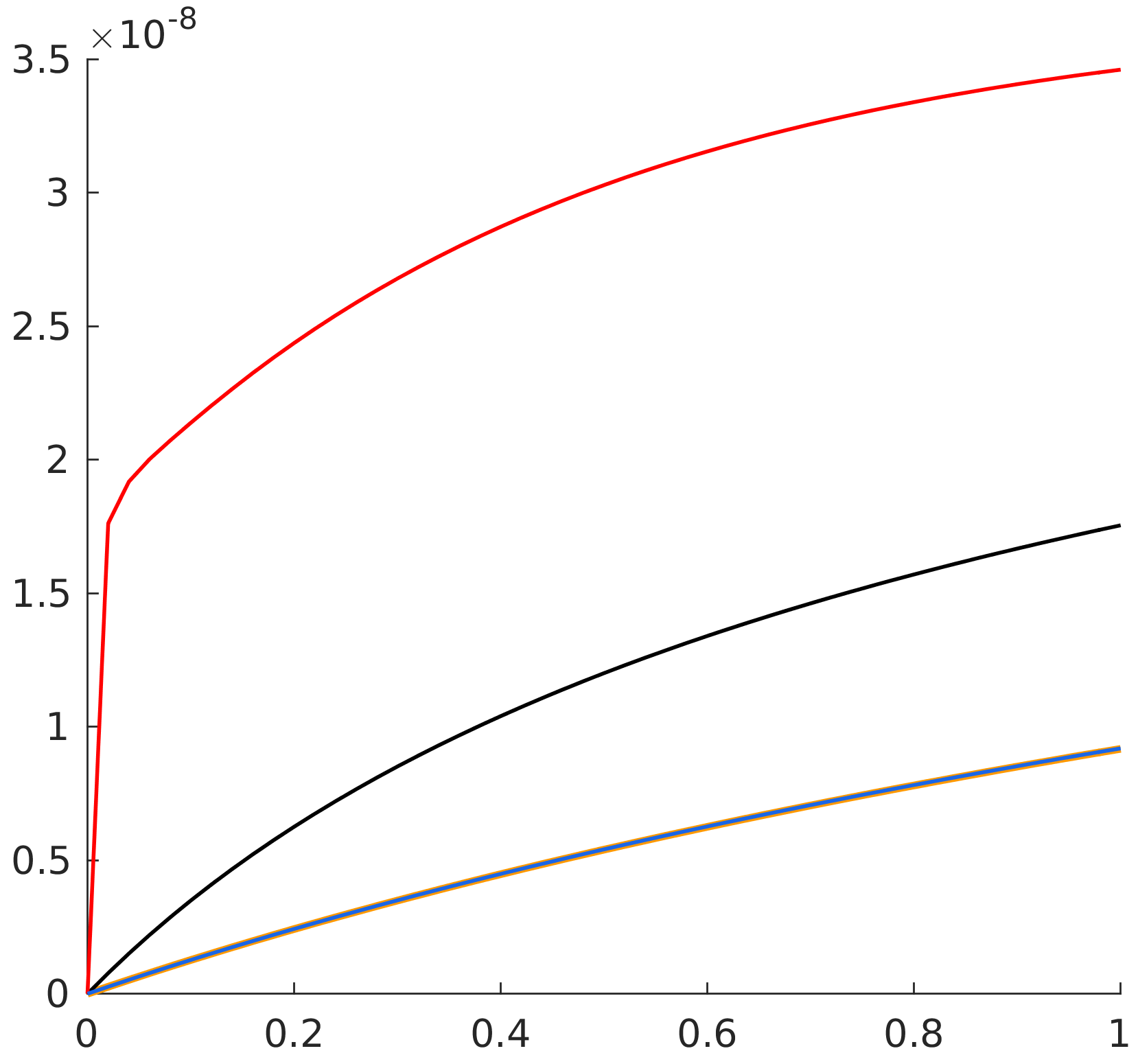}}
    \subfigure[Gresho Vortex - $M_r = 3 \times 10^{-3}$]{\includegraphics[scale = 0.7 ]{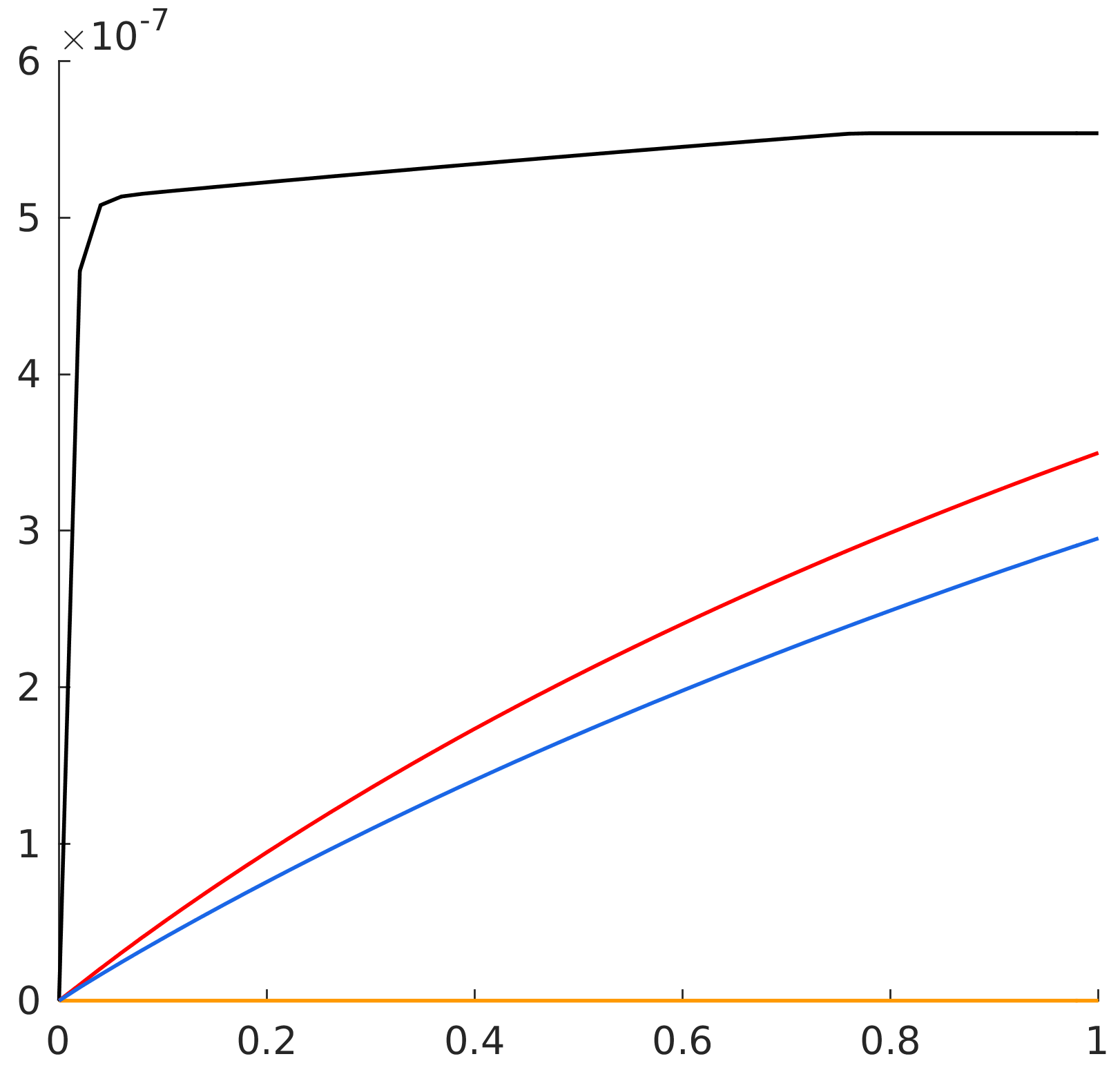}}
    \subfigure[Sound wave - $M_r = 10^{-4}$]{\includegraphics[scale = 0.7]{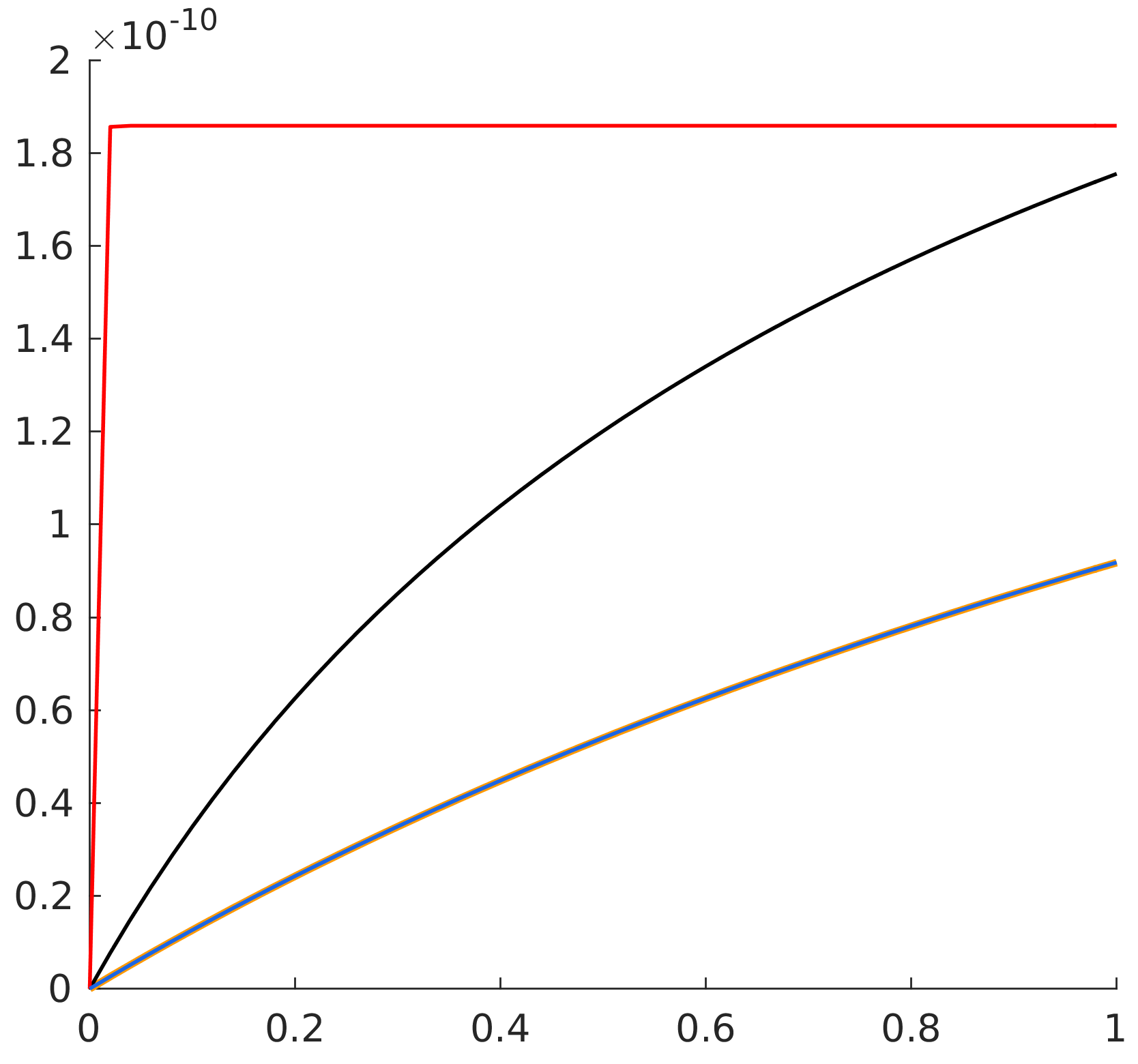}}
    \caption{Total entropy $(\rho s) - (\rho s)_0$ over time for all fluxes for the Gresho Vortex (left) and the sound wave (right).}
    \label{fig:Gresho_Sound_S}
\end{figure}

\begin{figure}[htbp!]
    \centering
    \includegraphics[scale = 0.8]{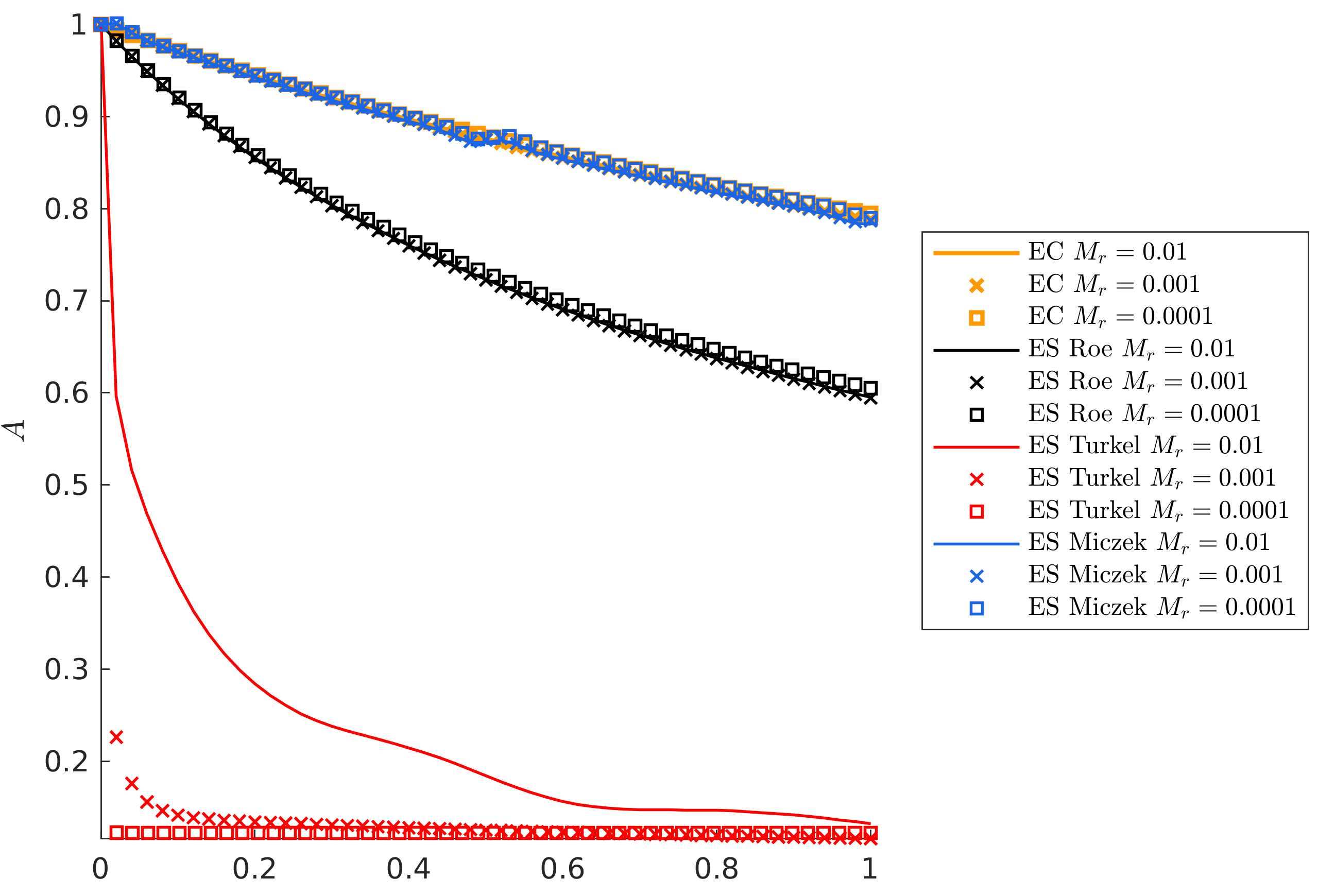}
    \caption{Sound wave: Normalized amplitude evolution for all fluxes at different Mach numbers.}
    \label{fig:Sound_A}
\end{figure}

\begin{figure}[htbp!]
    \centering
    \subfigure[$t = 0.1$]{\includegraphics[scale = 0.14]{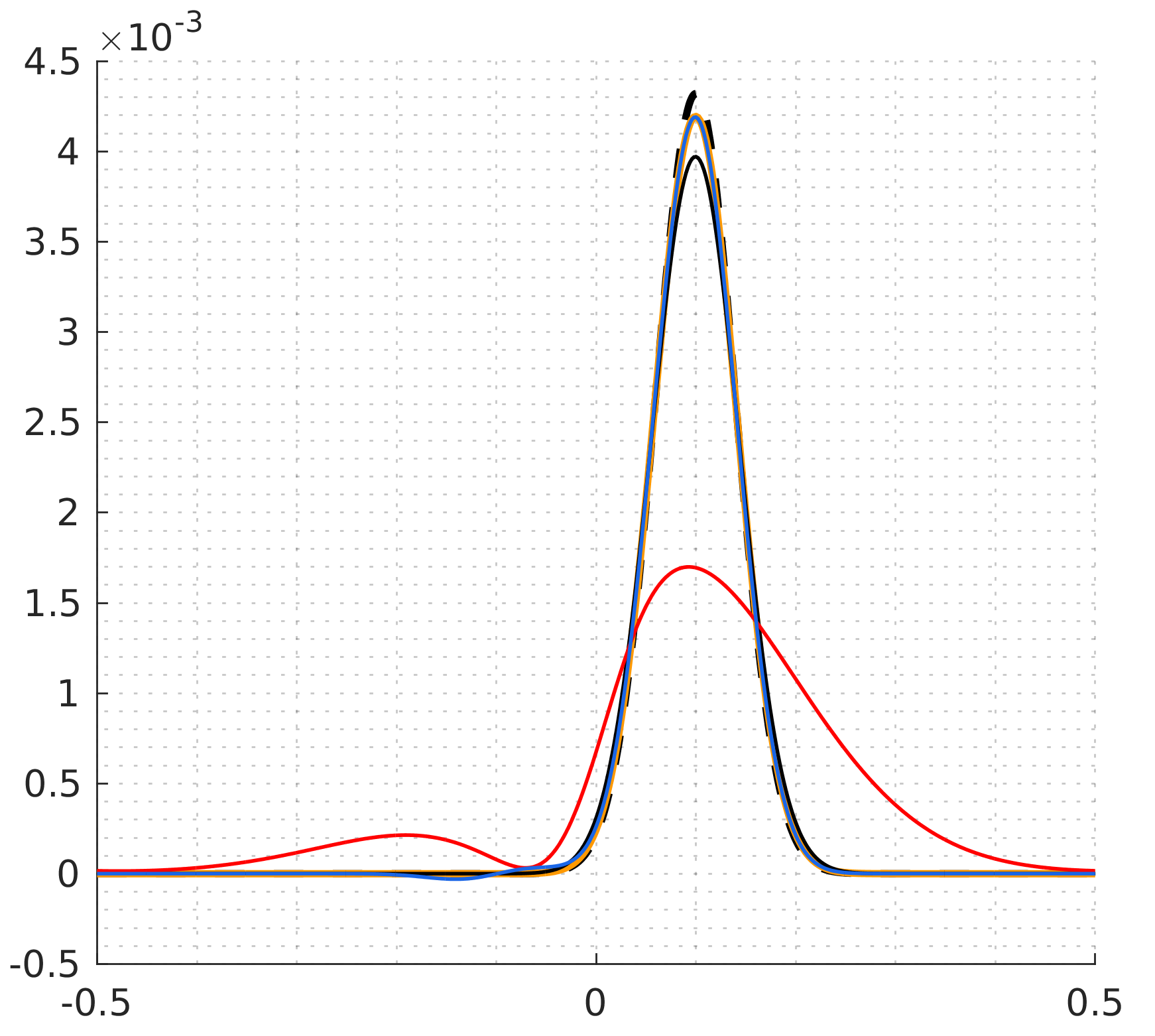}}
    \subfigure[$t = 0.1$ (zoom)]{\includegraphics[scale = 0.14]{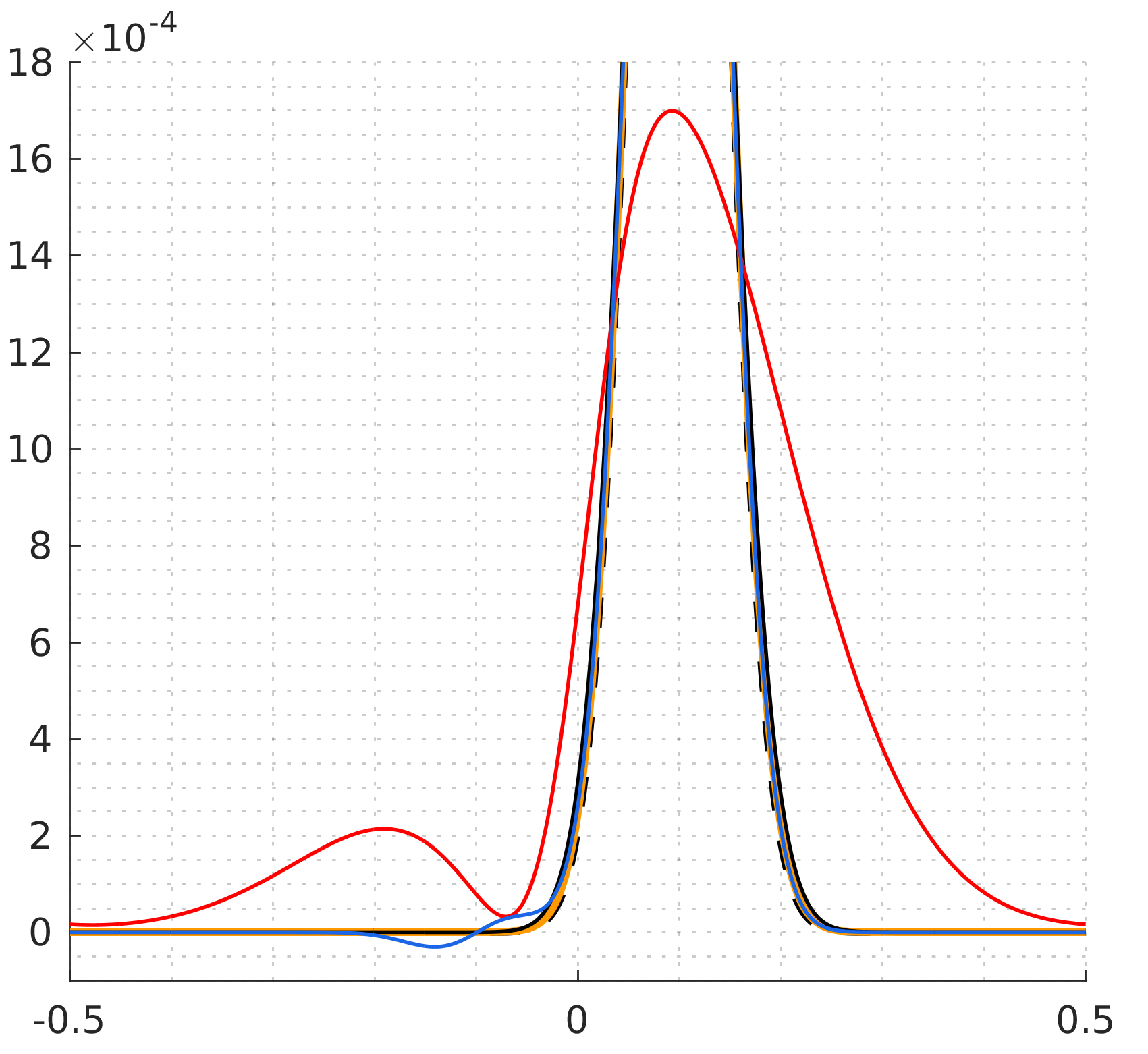}}
%
    \subfigure[$t = 0.2$ (zoom)]{\includegraphics[scale = 0.14]{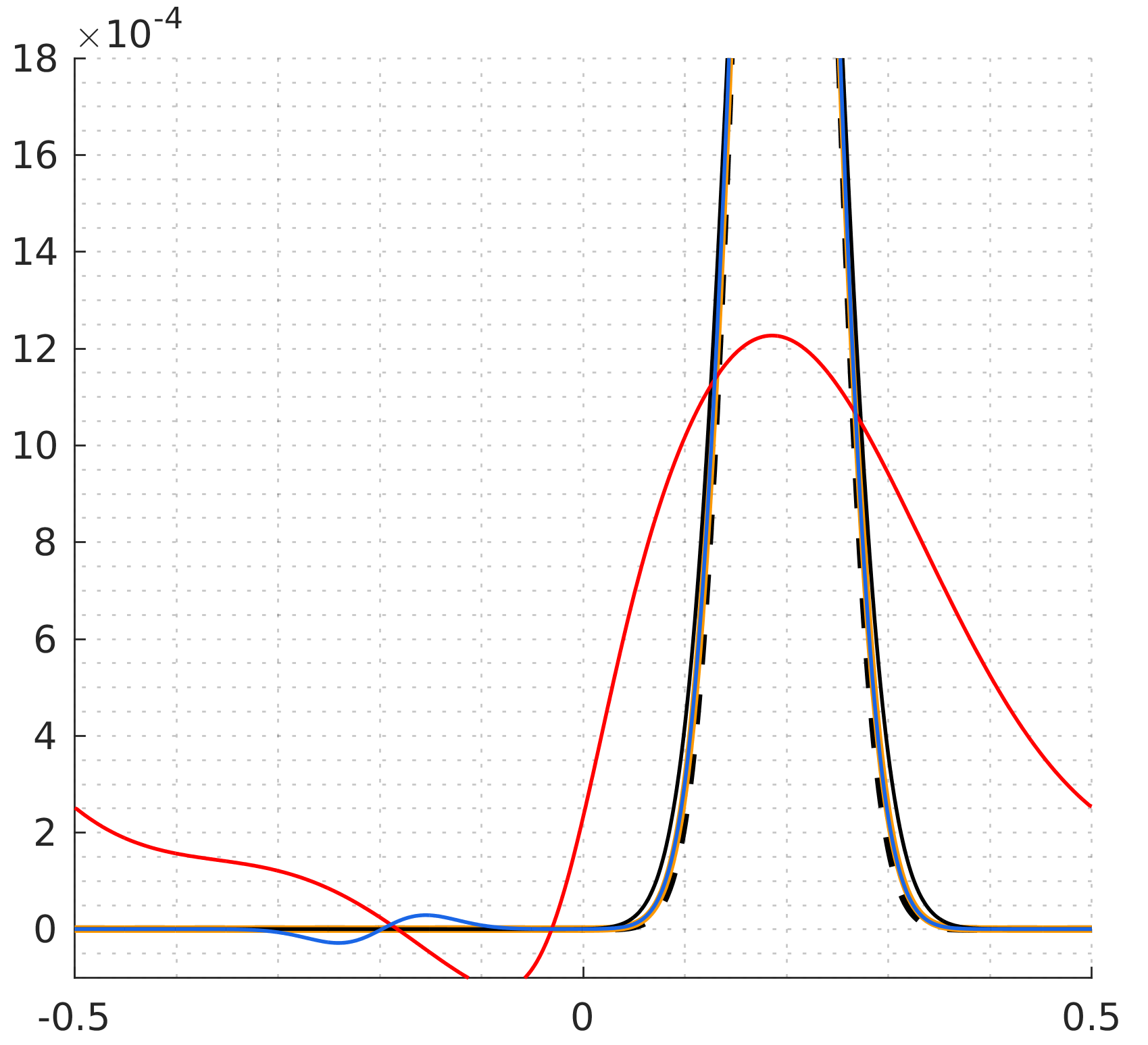}}
    \subfigure[$t = 0.3$ (zoom)]{\includegraphics[scale = 0.14]{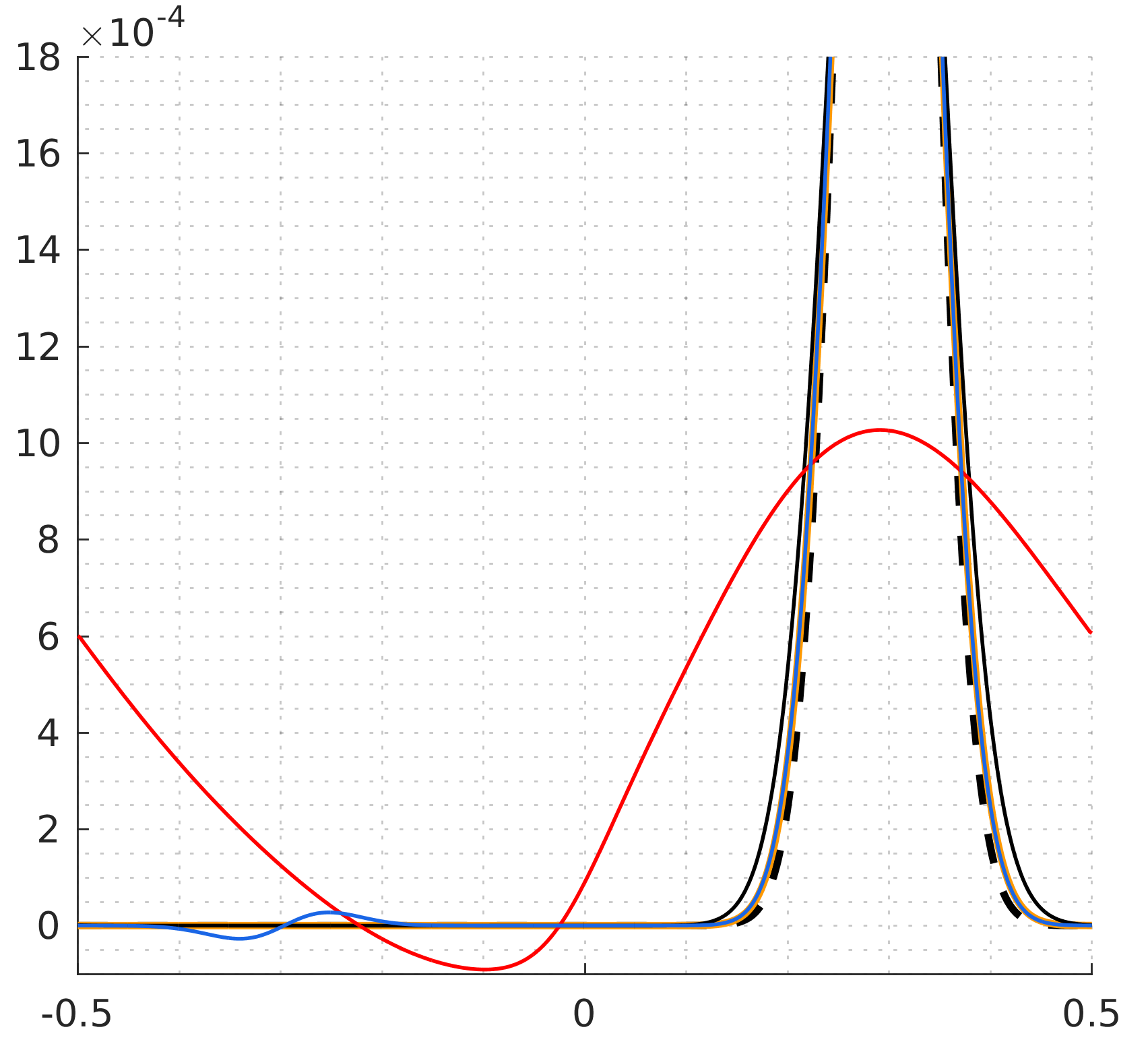}}
    \caption{Sound wave: Pressure profiles showing that the spurious wave (blue) reported in figure \ref{fig:Gresho_pressure_early} is an acoustic wave propagating at a speed of one. Same legend as figure \ref{fig:Sound_P_snaps}(a).}
    \label{fig:Sound_snap_MT_2}
\end{figure}

\indent One way to set up a right-moving acoustic wave is to consider, as in Bruel \textit{et al.} \cite{Bruel}, a free stream $(\rho_{\infty}, u_{\infty}, a_{\infty})$ and introduce fluctuations such that the Riemann invariants associated with the left moving acoustic wave and the entropy wave are constant throughout the domain. We set $\Omega = [-0.5, \ 0.5]$ and perturn density as follows:
\begin{equation*}
    \hat{\rho}(x,0) = \rho_{\infty} \big( 1 + M_r \psi(x) \big), 
\end{equation*}
where $\psi(x) = \exp(-\alpha x^2)$ defines a gaussian pulse centered at the center of the domain. We set $\alpha = \ln(10^3)/0.15^2$ so that $\psi(x) < 10^{-3}$ for $|x| < 0.15$. The flow is isentropic, hence $\hat{p}(x,0) = \hat{\rho}(x,0)^{\gamma} \implies \hat{a} = \sqrt{\gamma} (\hat{\rho}(x,0))^{\frac{\gamma-1}{2}}$. The corresponding velocity perturbation must satisfy
\begin{equation*}
    \big(\hat{u}(x,0) - u_{\infty} \big) - \frac{2 \big(\hat{a}(x,0) - a_{\infty} \big)}{\gamma - 1} = 0 \ \implies \  \hat{u}(x,0) = u_{\infty} + \frac{2\big( \hat{a}(x,0) - a_{\infty} \big)}{\gamma - 1}.
\end{equation*}
$\hat{a}(x,0)$ and $\hat{p}(x,0)$ are imposed by the density. If the reference Mach number $M_r$ is small enough, we can write:
\begin{equation}
    \hat{u}(x,0) = u_r + \frac{2a_r}{\gamma - 1} \bigg( \big( 1 + M_r \psi \big)^{\frac{\gamma-1}{2}} - 1 \bigg) = u_{\infty} + a_{\infty} M_r \psi + \mathcal{O}(M_r^2). 
\end{equation}
We set $u_{\infty}= 0$ and $a_{\infty} = 1$, so that the speed of propagation of the acoustic wave is roughly one. The reference time scale $t_r = 1$ is the time it takes for the acoustic wave to do one period. We have $\rho_r = \rho_{\infty}$, $u_r = M_r$ and $a_r = a_{\infty}$. Hence, spatial fluctuations in $(\rho, p)$ are of order $M_r$, while fluctuations in $u$ are of order $1$. \\
\indent We tested the four schemes on a grid of $500$ cells for $M_r \ \in \ \{10^{-2}, \ 10^{-3}, \ 10^{-4} \}$. 
Figures \ref{fig:Sound_P_snaps}(a)-(c) show the numerical solution at $t = 1$ for different Mach numbers. The reference solution\footnote{An exact solution can be calculated using the method of characteristics, which requires a nonlinear solver \cite{Bruel}. The solution for this problem is simple enough for a fine numerical solution to be trusted.} is obtained using a 4-th order TecNO scheme \cite{ES_Fjordholm} on a grid of $1000$ cells with a 4-th order Runge-Kutta time integration and a CFL of 0.5.  We see that the ES Roe flux, ES Miczek flux and EC flux lead to a self-similar numerical solution. We can see that the acoustic wave is almost completely gone with the ES Turkel flux. This is in agreement with the analysis and results of Bruel \textit{et al.} \cite{Bruel} for the barotropic Euler equations. The ES Miczek flux does not have this problem. Furthermore, it seems to perform just as well as the EC flux. This is also supported by figures \ref{fig:Gresho_Sound_S}(b)-(d)-(f). \\ 
\indent Figure \ref{fig:Sound_A} shows the temporal evolution of a normalized sound wave amplitude $A$ defined as:
\begin{equation}\label{LMES:A}
    A(t) = \frac{\max_{x \in \Omega} p(x,t)}{\max_{x \in \Omega} p(x,0)}.
\end{equation}
We can see that the rate at which the ES Turkel flux damps the sound wave increases as the Mach number decreases, while all other fluxes show a self-similar behavior. For the ES Miczek flux, we notice slight perturbations in $A$ which seem to occur around $t = \{0, \ 0.5, \ 1.0 \}$. Figure \ref{fig:Sound_snap_MT_2} suggests that these perturbations are caused by a spurious left-moving acoustic wave, created at $(x,t) = (0,0)$, that meets the right-moving acoustic wave when it reaches the periodic boundary and when it reaches the center of the domain at the end of the period.

\subsection{Summary}\label{LMES:sec:summary1}
\hspace*{0.1 cm} At this point, we have demonstrated, both analytically and numerically, that Flux-Preconditioning is compatible with Entropy-Stability. Numerical results are overall consistent with previous studies:
\begin{enumerate}[label=\textbf{(S.\arabic*)}]
    \item The ES Roe flux suffers from the same accuracy issues in the incompressible low-Mach limit as those previously reported with the classic Roe flux \cite{Miczek_T, Miczek, Barsukow}. This does not come as a surprise considering that the dissipation operators are not fundamentally different. These issues were not observed with the sound wave, in agreement with \cite{Bruel}. The temporal variation of the wave amplitude (\ref{LMES:A}) is independent of the Mach number.  \label{S1}
     \item The ES Turkel flux has a more consistent behavior in the incompressible limit, but at the price of damping acoustic waves harder as the Mach number decreases \cite{Bruel}. \label{S2}
    \item The ES Miczek flux performs well in both limits if we ignore the spurious transient in the Gresho vortex and the spurious left-moving acoustic wave, none of which were reported in \cite{Miczek_T, Miczek, Barsukow}. These are hard to notice on contour plots.  \label{S3}
\end{enumerate}
The EC flux performs the best in both cases. This confirms that for standard ES schemes, it is the dissipation component of (\ref{eq:ES_flux}) that causes the accuracy issues. This also suggests that the simplest fix in the context of (implicit) ES schemes could be to simply discard the dissipation part of the ES flux. This will be investigated in future work for space-time high-order \cite{ES_Diosady} discretizations in complex mixed flow configurations. We expect the stiffness issues associated with both the low-Mach regime and high-order implicit discretization to add a significant layer of complexity to the analysis. \\
\indent What follows in the remaining two sections is entirely motivated by the authors' goal to better understand the local behavior of ES schemes. The errors observed with the ES Miczek flux are intriguing. For the sound wave in particular, the spurious left-moving acoustic wave is reminiscent of anomalies the authors studied previously \cite{Gouasmi_0, Gouasmi_2} (none of which could be tied to the violation of an entropy inequality). For all these reasons, we decided to delve into these issues, with an emphasis on the physical quantity ES schemes have an actual handle on.

\section{The Accuracy Degradation from an Entropy Production Perspective}
\label{LMES:sec:Entropy_analysis}
\indent In the incompressible limit, Guillard \& Viozat \cite{Guillard1} showed that pressure fluctuations in space are of order $M^2$, i.e. we can write $p(\mathbf{x},t)  =  p_0(t)  +  M^2 p_2(\mathbf{x},t)$. Assuming constant density $\rho = \rho_0$, we can write:
\begin{align*}
        \rho s =& \ \rho_0 \bigg( \ln \big(p_0 + M_r^2 p_2 + \mathcal{O}(M_r^3)\big) - \gamma \ln (\rho_0) \bigg) 
          =  \rho_0 \bigg( s_0 + \ln\big(1 + M_r^2 (p_2/p_0) + \mathcal{O}(M_r^3) \big) \bigg) 
          = \rho_0 s_0 + M_r^2 \rho_ 0 (p_2/p_0) + \mathcal{O}(M_r^3).
\end{align*}
with $s_0 = \ln p_0 - \gamma \ln \rho_0$. Therefore, we state:
\begin{enumerate}[label=\textbf{(E.1)}]
    \item In the incompressible limit, entropy $\rho s$ fluctuations in space should be of order $M_r^2$. \label{E1}
\end{enumerate}
Similarly \cite{Guillard2}:
\begin{enumerate}[label=\textbf{(E.2)}]
    \item In the acoustic limit, entropy $\rho s$ fluctuations in space should be of order $M_r$. \label{E2}
\end{enumerate}
\indent In the incompressible limit, there is the additional requirement that kinetic energy should be conserved. To precisely and rigorously 
relate discrete changes in kinetic energy to discrete changes in entropy is not straightforward, if at all possible. Let's assume periodic boundary conditions so that discrete conservation of total energy implies that it remains constant globally. We can write:
\begin{gather*}
        \Delta (\rho e + M_r^2 \rho k) = 0 \ \iff \  \Delta (\rho k) = -\frac{1}{M_r^2}\Delta (\rho e) = -\frac{1}{(\gamma - 1)M_r^2} \Delta (p) = -\frac{1}{(\gamma - 1)M_r^2} \Delta \bigg(\exp\bigg( \frac{\rho s - \gamma \rho \ln (\rho)}{\rho}\bigg)
        \bigg),
\end{gather*}
where $\Delta$ refers to the global change, that is the sum $\sum_{i}\Delta_i$ of local changes in each cell $i$. Assuming constant density, this relation simplifies to:
\begin{equation}\label{LMES:KES}
    \Delta (\rho k) = \frac{-1}{\exp(\rho)(\gamma - 1)M_r^2} \Delta \big(\exp(\rho s)\big).
\end{equation}
Equation (\ref{LMES:KES}) relates the global change in kinetic energy $\rho k$ to the global change in the \textit{exponential of} the entropy, which ES schemes do not explicitly control. It is certainly tempting to say that since the exponential function is monotonically increasing, $\Delta (\rho s) > 0 \implies \Delta \big( \exp(\rho s) \big) > 0$. This statement is true locally, but ES schemes are not explicitly designed to achieve $\Delta_i (\rho s) > 0$, because equation (\ref{eq:FVM_ES}) also features flux contributions (in other words, one could have $\Delta_i (\rho s) < 0$). We therefore refrain from making hasty interpretations.

\subsection{Entropy Production Breakdowns (EPBs)}\label{sec:entropy_prod}
\begin{figure}[htbp!]
    \centering
    \subfigure[$\hat{\mathcal{E}}_{u_n, \tau_{xy}}$]{\includegraphics[scale = 0.38]{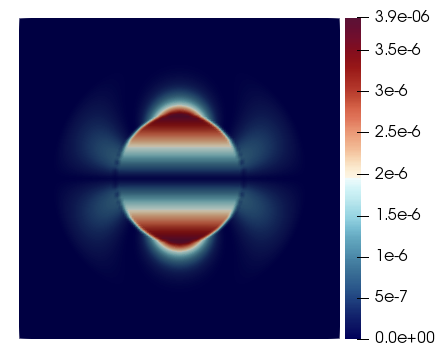}}
    \subfigure[$\hat{\mathcal{E}}_{u_n, \tau_{yx}}$]{\includegraphics[scale = 0.38]{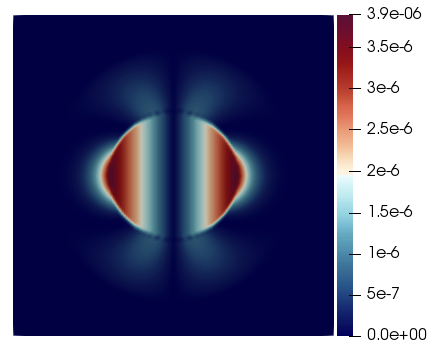}}
    \subfigure[$\hat{\mathcal{E}}_{u_n, s}$]{\includegraphics[scale = 0.38]{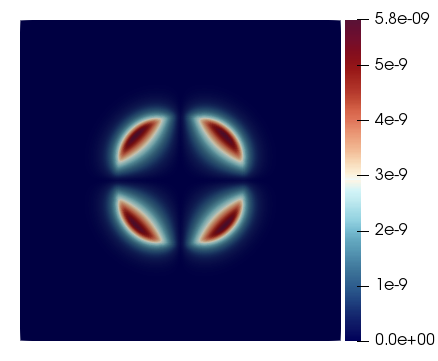}}
    \caption{Gresho vortex: Entropy production fields associated with the convective modes at $t = 0$ and $M_r = 3 \times 10^{-2}$. These are common to all ES fluxes. }
    \label{fig:Gresho_Roe_dS_un}
\end{figure}

\begin{figure}[htbp!]
    \centering
    \subfigure[$\hat{\mathcal{E}}_{u_n+a}$]{\includegraphics[scale = 0.4]{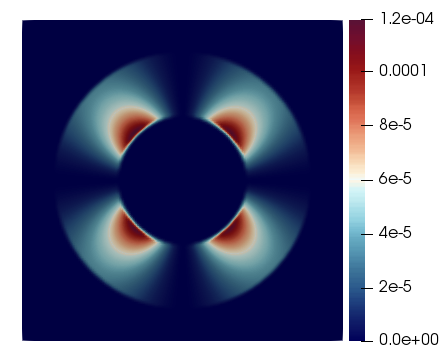}}
    \subfigure[$\hat{\mathcal{E}}_{u_n-a}$]{\includegraphics[scale = 0.4]{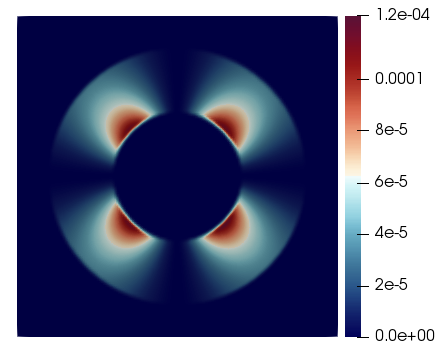}}
    \caption{Gresho vortex: Acoustic Entropy production fields at $t = 0$ of the ES Roe flux at $M_r = 3 \times 10^{-2}$.}
    \label{fig:Gresho_Roe_dS}
\end{figure}

\begin{figure}[htbp!]
    \centering
    \subfigure[$\hat{\mathcal{E}}_{u_n+a}$]{\includegraphics[scale = 0.53]{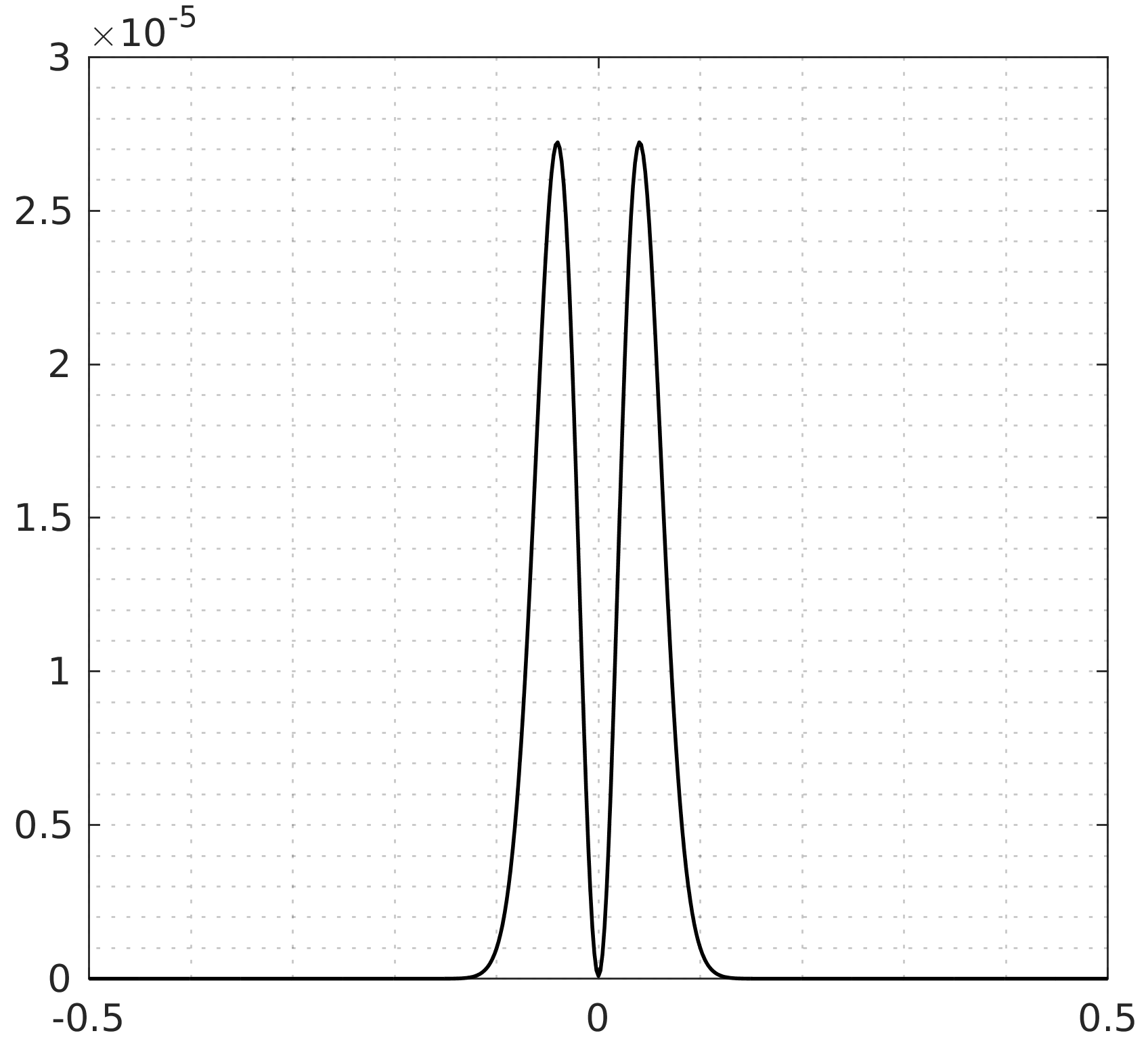}}
    \subfigure[$\hat{\mathcal{E}}_{u_n-a}$]{\includegraphics[scale = 0.53]{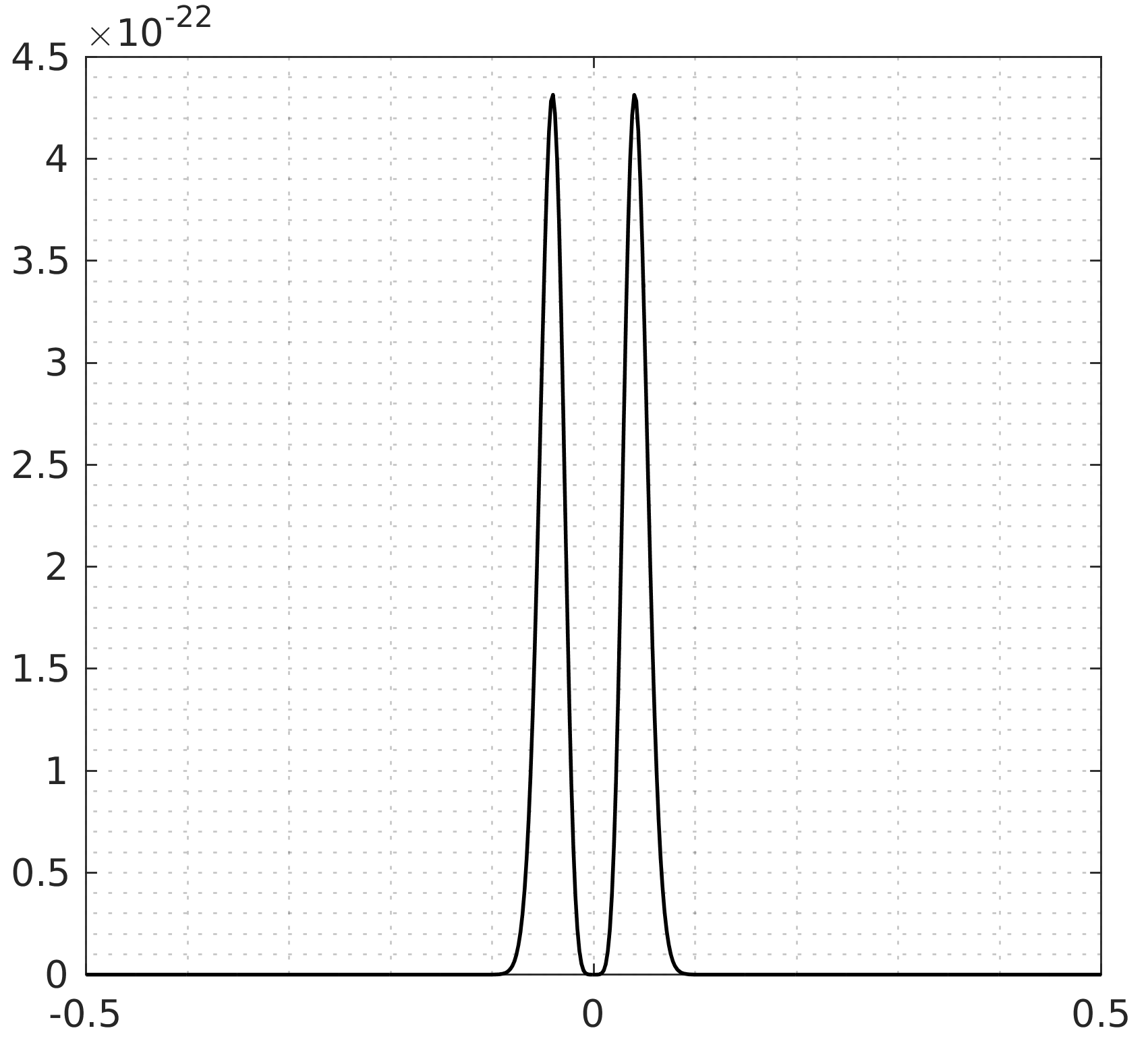}}
    \subfigure[$\hat{\mathcal{E}}_{u_n,s}$]{\includegraphics[scale = 0.53]{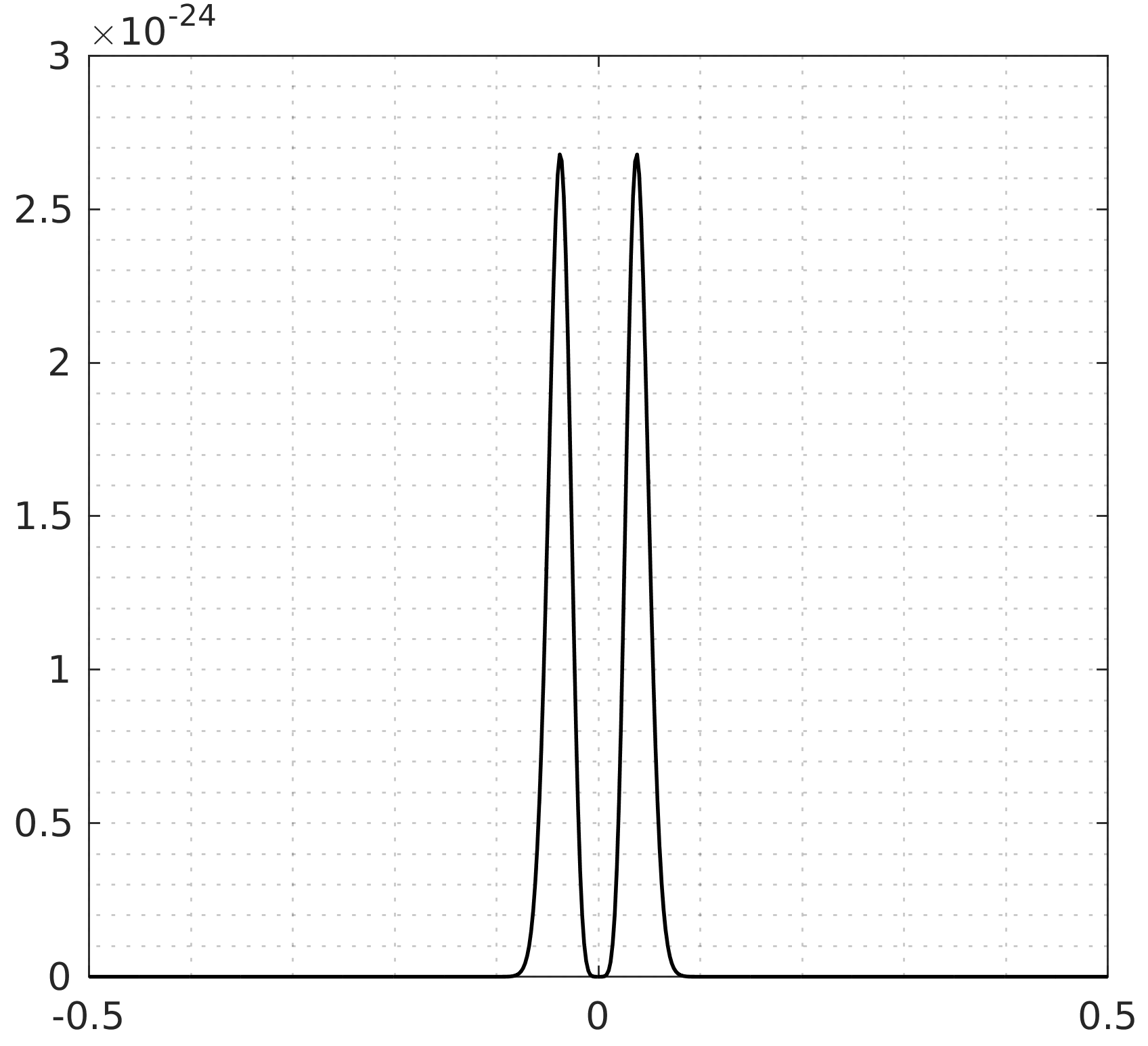}}
    \caption{Sound Wave: Entropy production fields at different Mach numbers at $t = 0$ and $M_r = 10^{-2}$ for the ES Roe flux.}
    \label{fig:Sound_Roe_dS}
\end{figure}

\begin{figure}[htbp!]
    \centering
    \subfigure[Gresho Vortex - $M_r = 3 \times 10^{-1}$]{\includegraphics[scale = 0.62]{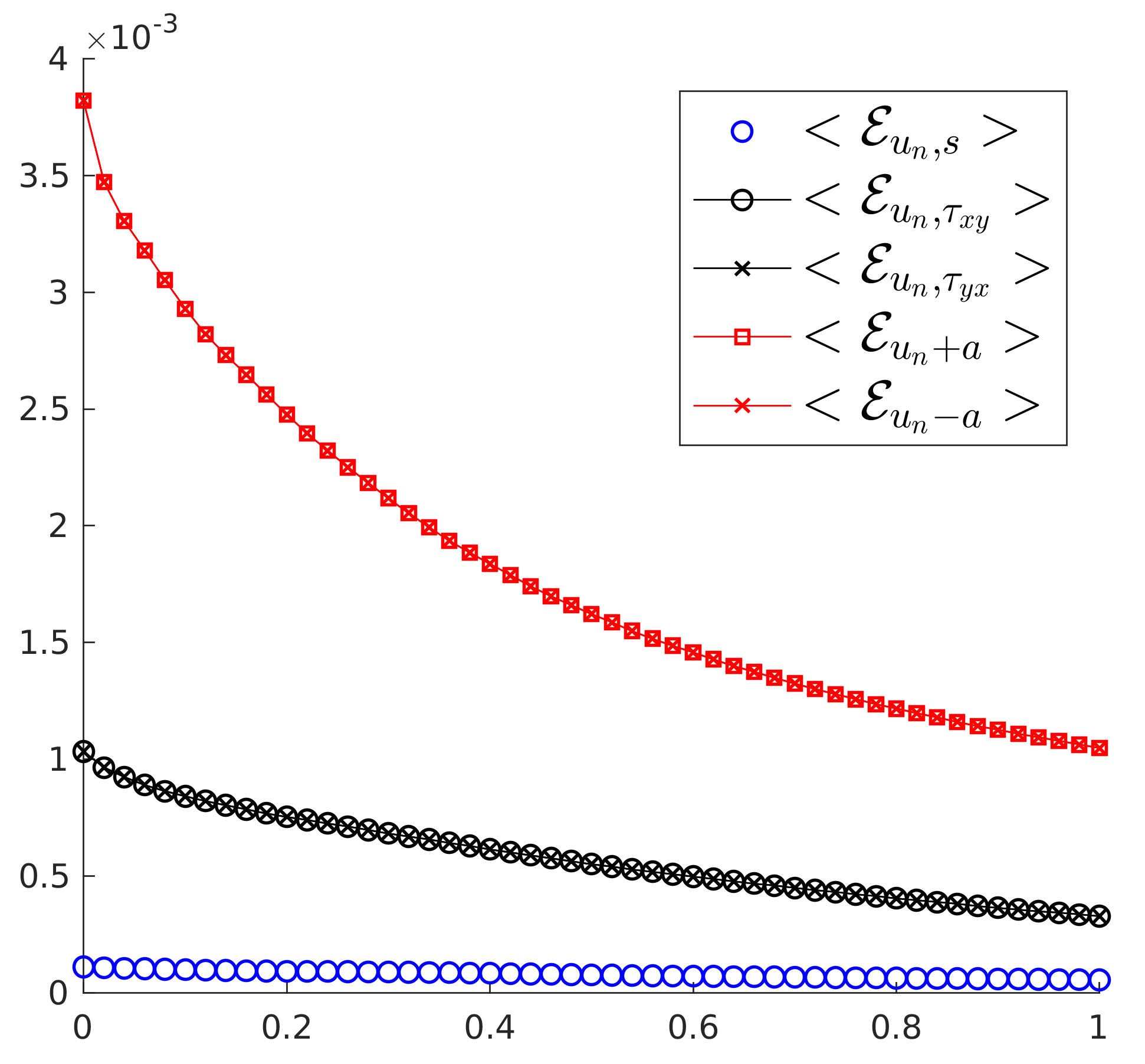}}
    \subfigure[Sound Wave - $M_r = 10^{-2}$]{\includegraphics[scale = 0.113]{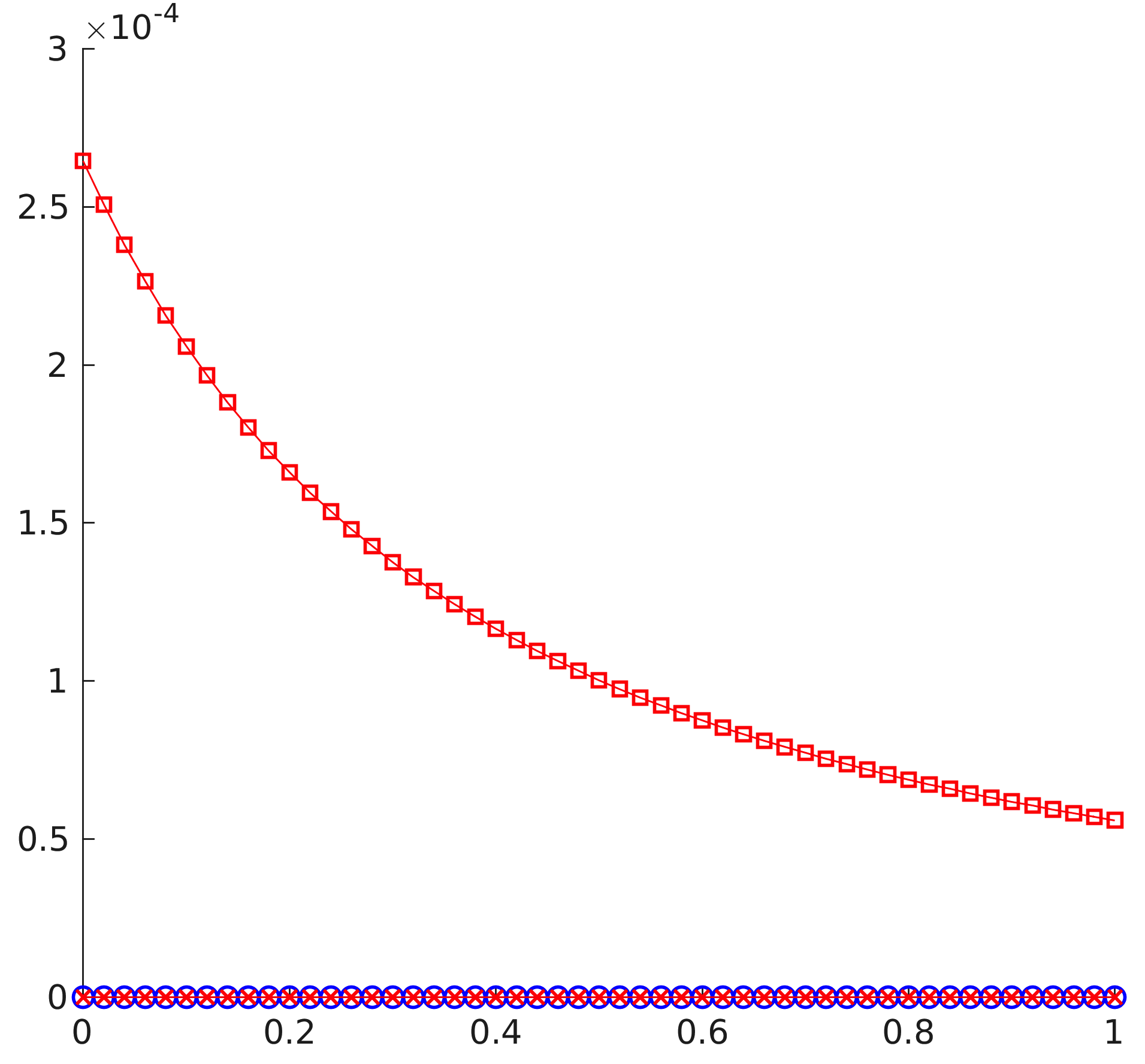}}
   \subfigure[Gresho Vortex - $M_r = 3 \times 10^{-2}$]{\includegraphics[scale = 0.62]{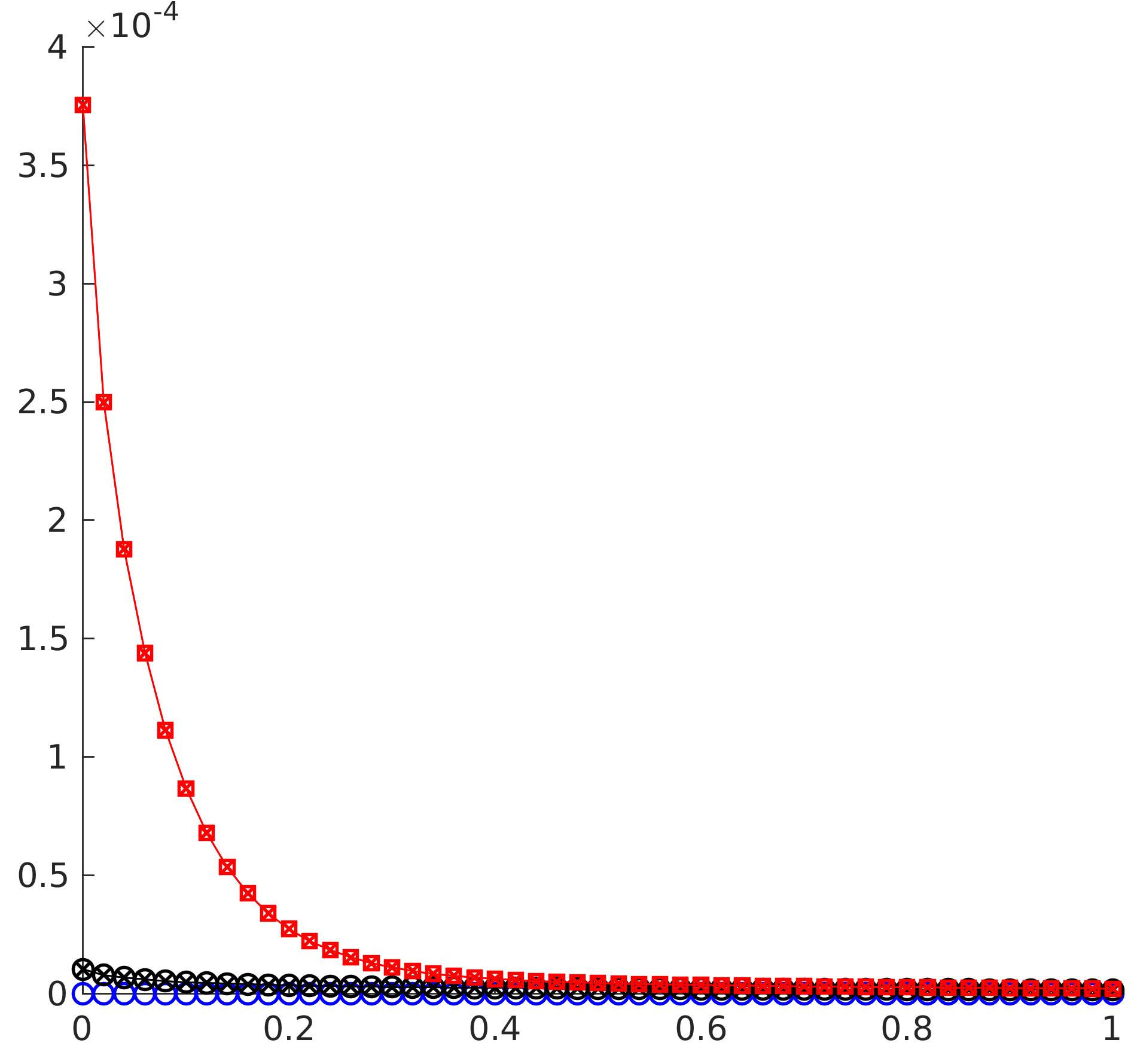}}
    \subfigure[Sound Wave - $M_r = 10^{-3}$]{\includegraphics[scale = 0.62]{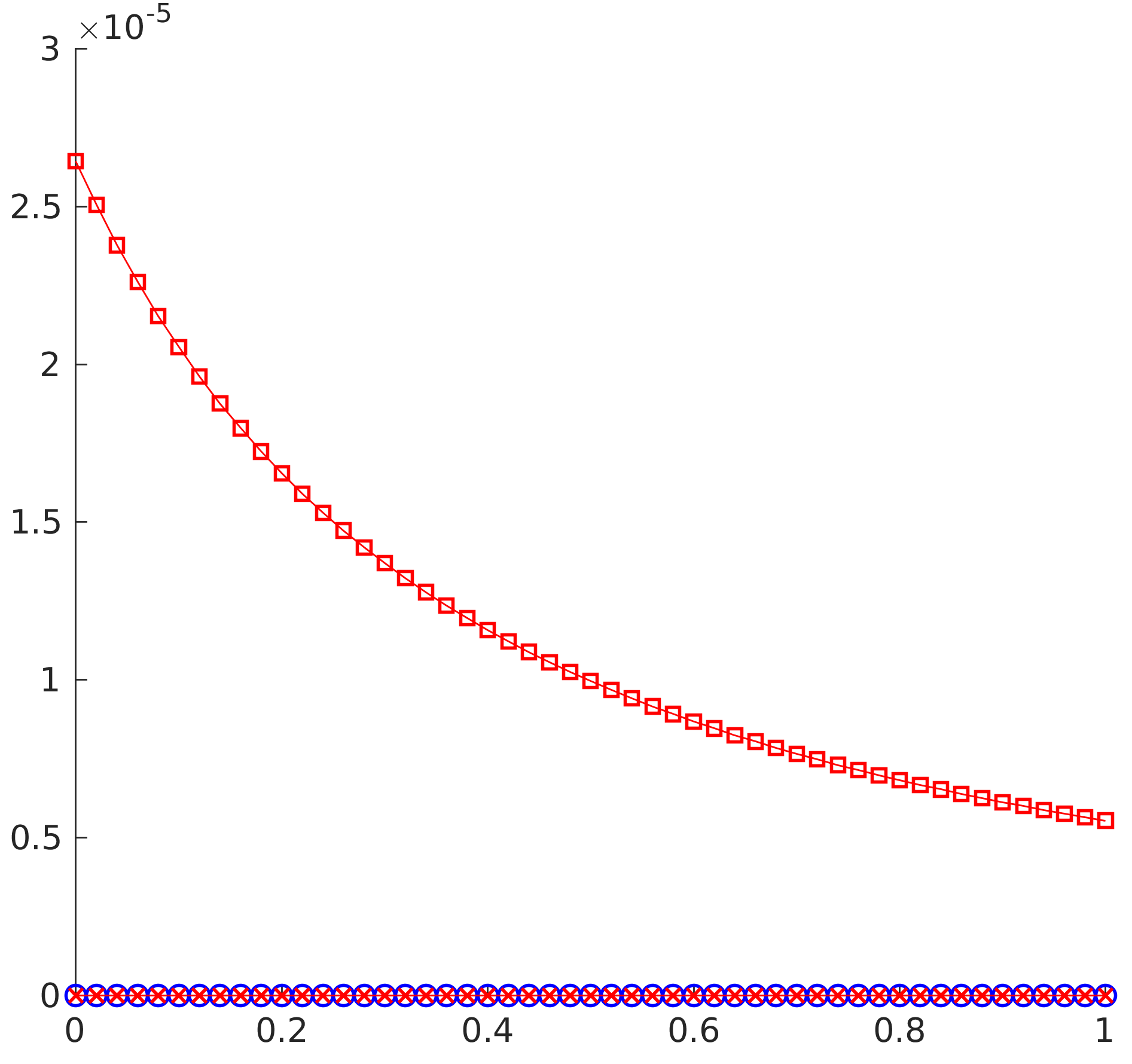}}
    \subfigure[Gresho Vortex - $M_r = 3 \times 10^{-3}$]{\includegraphics[scale = 0.62]{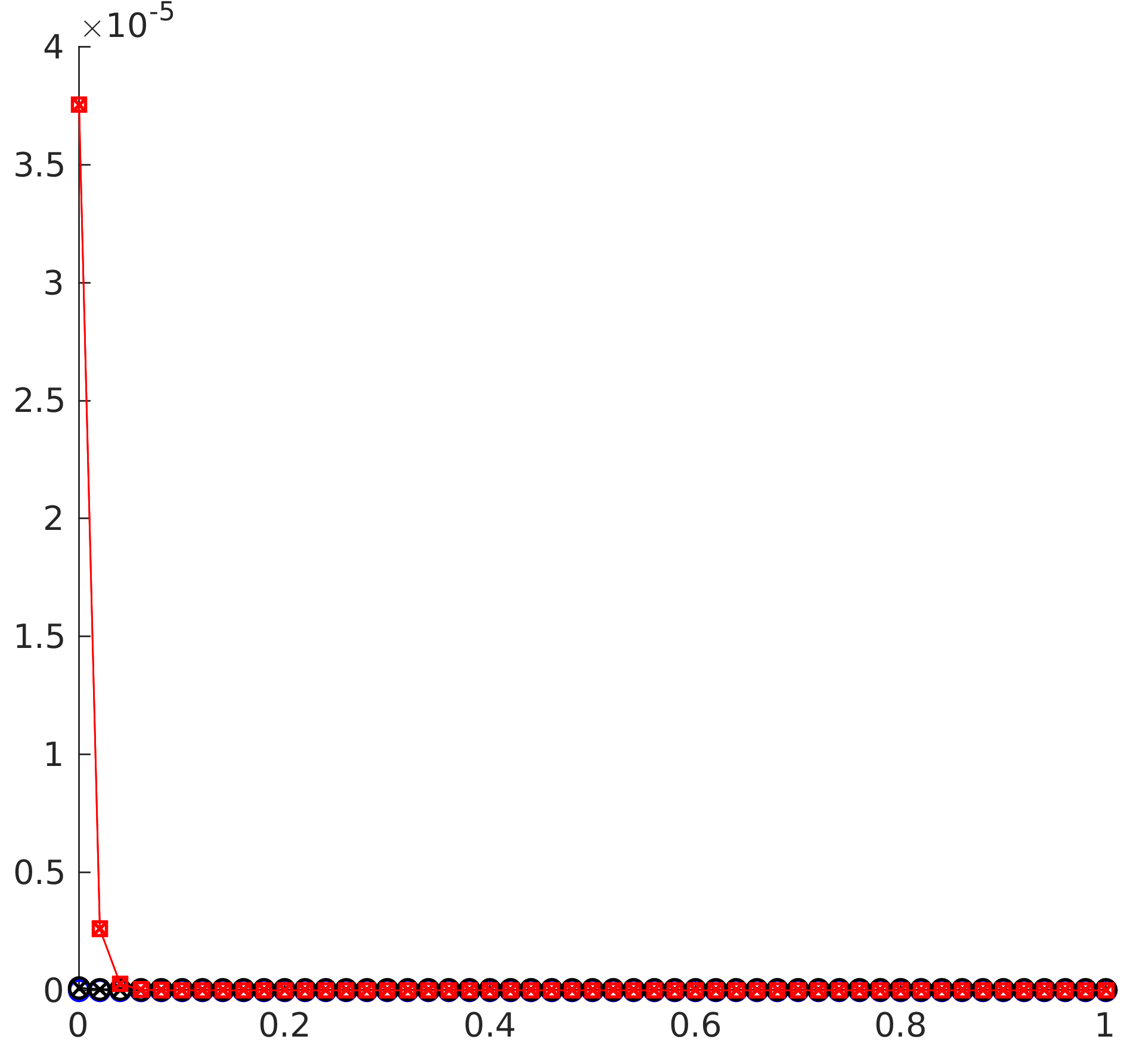}}
    \subfigure[Sound Wave - $M_r = 10^{-4}$]{\includegraphics[scale = 0.62]{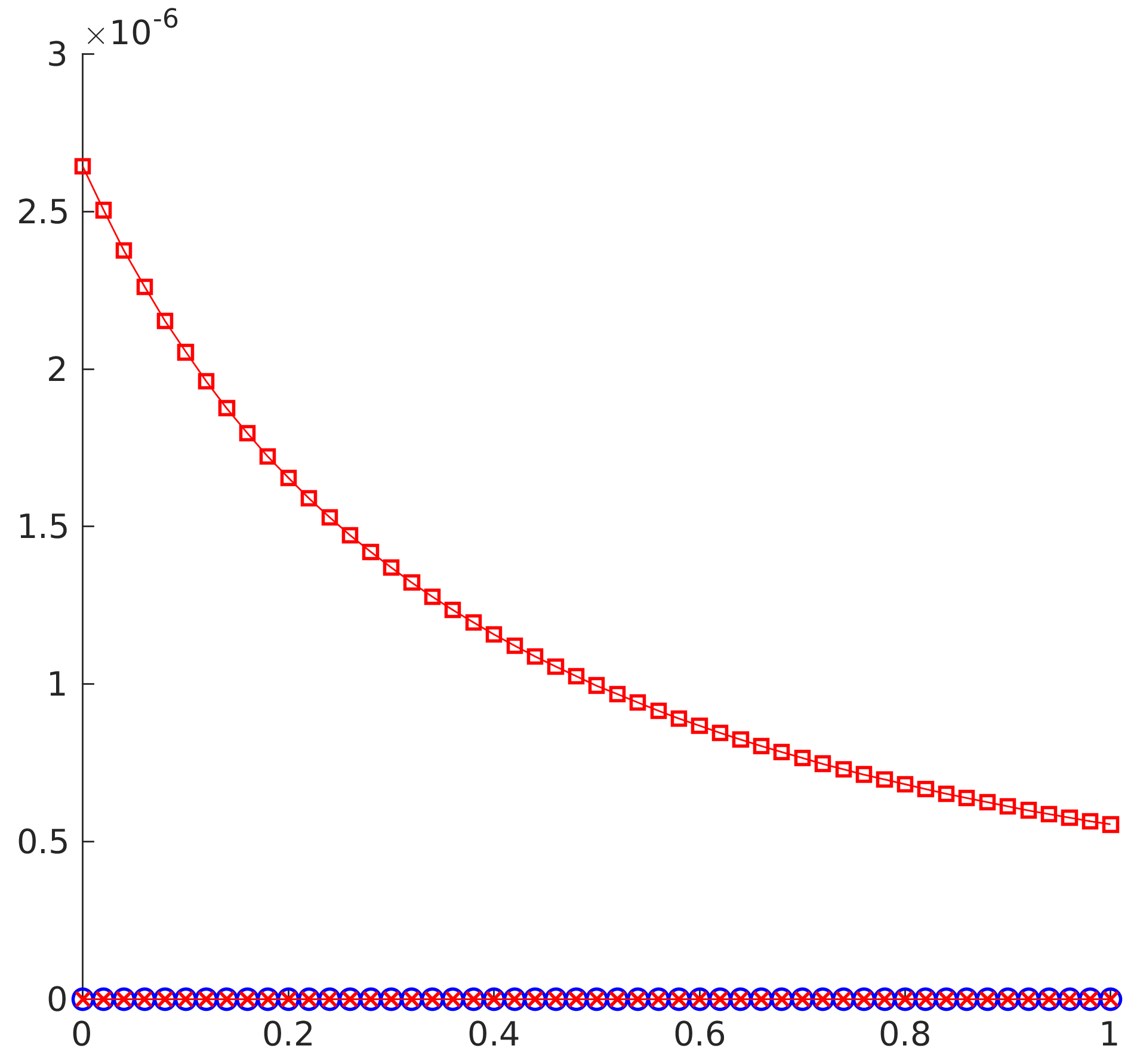}}
    \caption{ Integral of each entropy production field with time at different Mach numbers for the ES Roe flux on the Gresho Vortex (left) and the sound wave (right).}
    \label{fig:Roe_global_dS}
\end{figure}

\indent The most remarkable feature of ES schemes is the relation that holds at the semi-discrete level for entropy in each cell (\ref{eq:FVM_ES}), which we rewrite here:
\begin{equation*}
    \frac{d U(\mathbf{u}_i)}{d t} \ + \ \frac{1}{V_i}\int_{\delta \Omega_i} F^{*} dS \ = \ - \frac{1}{V_i}\mathcal{E}_i, \ \mathcal{E}_i = \int_{\delta \Omega_i} \mathcal{E}  dS.
\end{equation*}
The cell valued field $\mathcal{E}_i$ tells us how much entropy is produced in space in response to the jumps in entropy variables across interfaces. It can therefore give us an idea of the magnitude of the entropy fluctuations the scheme creates. To this end, we proceed to derive a more detailed expression for $\mathcal{E}$. Ignoring the $1/4$ factor in (\ref{eq:Eprod}), we have:
\begin{equation*}
    \mathcal{E} \ = \ [\mathbf{v}]^T D[\mathbf{v}] \ = \ [\mathbf{v}]^T R |\Lambda|  R^T [\mathbf{v}].
\end{equation*}
Now let $\mathbf{r}_1, \ \dots, \mathbf{r}_N$ denote the columns of the eigenvector matrix $R$ where $N$ is the number of eigenvalues (possibly repeated). We have $R = [\mathbf{r}_1, \ \dots \ ,\mathbf{r}_N]$, $|\Lambda| = diag(|\lambda_1|, \ \dots \, |\lambda_N|)$. Define $\bm{\mu}^T = [\mu_1, \ \dots, \ \mu_N] = [\mathbf{v}]^T R$. We have:
\begin{equation*}
    \mathcal{E} \ = \ \bm{\mu}^T |\Lambda| \bm{\mu} \ = \ \sum_{k=1}^N |\lambda_k| \mu_k^2,
\end{equation*}
 and we can see how the total entropy production breaks down into the positive contributions associated with each eigenvector or "mode" $\mathbf{r_i}$. This decomposition is inspired by how Roe \& Pike \cite{Pike} rewrote the Roe flux:
\begin{equation} \label{LMES:Pike}
    R |\Lambda| R^{-1} [\mathbf{u}] \ = \ \sum_{k=1}^{N} |\lambda_k| \alpha_k \mathbf{r_k}, \ \bm{\alpha} = R^{-1}[\mathbf{u}].  
\end{equation}
It is also inspired by the family of closed-form EC fluxes Tadmor proposed in \cite{ES_Tadmor_2003} (we discuss them in section 7). \\
\indent The vector $\bm{\alpha}$ in equation (\ref{LMES:Pike}) is known as a vector of \textit{wave strengths}. We can interpret $\bm{\mu} = R^{T}[\mathbf{v}]$ in our decomposition as a vector of wave strengths as well, as for infinitesimal variations we have:
\begin{equation*}
    d\mathbf{u} = H d\mathbf{v} = R R^T d\mathbf{v} \implies R^{-1} d\mathbf{u} = R^{T} d\mathbf{v}.
\end{equation*}
\indent For the compressible Euler system (\ref{LMES:EulerM1}), $R = [\mathbf{r}_{u_n,1}, \ \mathbf{r}_{u_n,2}, \ \mathbf{r}_{u_n,3}, \ \mathbf{r}_{u_n+a}, \ \mathbf{r}_{u_n-a}]$ ($N = 5$) and we have:
\begin{equation}\label{LMES:dS_euler}
    \mathcal{E} = |u_n| (\mu_{u_n,1}^2 + \mu_{u_n,2}^2 + \mu_{u_n,3}^2) + |u_n + (a/M_r)| \mu_{u_n+a}^2 + |u_n - (a/M_r)| \mu_{u_n-a}^2.
\end{equation}
It breaks down into entropy production due to convective modes (first three terms, which we gather into $\mathcal{E}_{u_n}$) and entropy production due to acoustic modes (remaining two terms, which we denote $\mathcal{E}_{u_n +a}$ and $\mathcal{E}_{u_n-a}$ respectively). We expect the latter to be the key in understanding the low-Mach problems. The entropy production field $\hat{\mathcal{E}}$ that the code solving the dimensional system (\ref{eq:Euler}) computes is related to $\mathcal{E}$ by the simple relation:
\begin{equation}
    \hat{\mathcal{E}} = (\rho_r u_r) \times \mathcal{E}.
\end{equation}
Given that in both test problems the reference Mach number $M_r$ is adjusted by changing the the reference velocity $u_r$ only, we use $ \hat{\mathcal{E}} = M_r \times \mathcal{E} $ instead. We also define the global quantity (sum over all cells):
\begin{equation}\label{LMES:total_dS}
    \langle \mathcal{E} \rangle = \sum_{i}  \mathcal{E}_i,
\end{equation}
which we will subsequently use to visualize the global influence of each entropy production field in (\ref{LMES:dS_euler}) on the total entropy production in space. \\ 
\indent We now proceed to develop the expressions of $\mathcal{E}_{u_n}, \mathcal{E}_{u_n-a}$ and $\mathcal{E}_{u_n+a}$. We will first treat the convective modes (entropy \& shear waves) since they are common to all three ES fluxes. We then treat the acoustic modes. Note that these dissipation matrices are evaluated at averaged states (we simply take arithmetic averages for all ES fluxes). For smooth flow configurations, we do not expect the choice of average values to have a impact. \\ \\
\indent \textbf{Convective modes.} The scaled eigenvectors associated with $\lambda = u_n$ are given by:
\begin{gather*}
    \mathbf{r}_{u_n,1} = K_{q} (n_1 K_0 \mathbf{r_0} + (a/M_r) \mathbf{r_{s1}}), \ \mathbf{r}_{u_n,2} = K_{q} (n_2 K_0 \mathbf{r_0} + (a/M_r) \mathbf{r_{s2}}), \ \mathbf{r}_{u_n,3} = K_{q} (n_3 K_0 \mathbf{r_0} + (a/M_r) \mathbf{r_{s3}}), \\
    \mathbf{r_0} = 
        \begin{bmatrix}
            1 \\ u \\ v \\ w \\ M_r^2 k
        \end{bmatrix},
    \mathbf{r_{s1}} = 
        \begin{bmatrix}
            0 \\ 0 \\ n_3 \\ -n_2 \\ M_r^2(n_3 v - n_2 w)
        \end{bmatrix}, 
    \mathbf{r_{s2}} = 
        \begin{bmatrix}
            0 \\ -n_3 \\ 0 \\ n_1 \\ M_r^2(n_1 w - n_3 u)
        \end{bmatrix}, \
    \mathbf{r_{s3}} = 
        \begin{bmatrix}
            0 \\  n_2 \\ -n_1 \\ 0 \\ M_r^2(n_2 u - n_1 v)
        \end{bmatrix}, \\ K_q = (\rho/\gamma)^{1/2}, \ K_0 = (\gamma-1)^{1/2}.
\end{gather*}
$\mathbf{r_0}$ is an entropy wave. Let $\mu_0 = \mathbf{r_0}^T [\mathbf{v}]$ be the corresponding wave strength, we can show that:
\begin{align*}
    \mu_0 =& \ \frac{-[s]}{\gamma-1} - \frac{M_r^2}{4} \bigg[\frac{\rho}{p}\bigg] \bigg( [u]^2 + [v]^2 + [w]^2 \bigg).
\end{align*}
$\mathbf{r_{s1}}, \mathbf{r_{s2}}$ and $ \mathbf{r_{s3}}$ are shear waves (they satisfy $ n_1 \mathbf{r_{s1}} + n_2 \mathbf{r_{s2}} + n_3 \mathbf{r_{s3}} = 0$). The corresponding wave strengths $\mu_{s1} = \mathbf{r_{s1}}^T[\mathbf{v}], \ \mu_{s2} = \mathbf{r_{s2}}^T[\mathbf{v}]$ and $\mu_{s3} = \mathbf{r_{s3}}^T[\mathbf{v}]$ are given by:
\begin{gather*}
    \mu_{s1} = M_r^2 \overline{\bigg( \frac{\rho}{p} \bigg)} [\mathcal{V}_1], \ \mu_{s2} = M_r^2 \overline{\bigg( \frac{\rho}{p} \bigg)} [\mathcal{V}_2], \ \mu_{s3} = M_r^2 \overline{\bigg( \frac{\rho}{p} \bigg)} [\mathcal{V}_3], \\
     \mathcal{V}_1 = n_3 v - n_2 w, \ \mathcal{V}_3 = n_1 w - n_3 u, \  \mathcal{V}_3 = n_2 u - n_1 v.
\end{gather*}
We now have:
\begin{align*}
    \mathcal{E}_{u_n}  =& \ |u_n| K_q^2 \bigg( \ (n_1 K_0 \mu_0 + (a/M_r) \mu_{s1})^2 \ + \ (n_2 K_0 \mu_0 + (a/M_r) \mu_{s2})^2 \ + \ (n_3 K_0 \mu_0 + (a/M_r) \mu_{s3})^2 \ \bigg) \\
    =& \ |u_n| K_q^2 \bigg( \ K_0^2 \mu_0^2 \ + \ (a/M_r)^2 (\mu_{s1}^2 + \mu_{s2}^2 + \mu_{s3}^2) \ \bigg) \\
    =& \ |u_n| \bigg(\  \frac{(\gamma-1)}{\gamma}\rho \mu_0^2 \ + \ \frac{\rho a^2}{\gamma M_r^2} (\mu_{s1}^2 + \mu_{s2}^2 + \mu_{s3}^2) \bigg).
\end{align*}
Injecting the discrete wave strength expressions, we get:
\begin{equation}\label{eq:dS_q}
    \mathcal{E}_{u_n} = |u_n| \bigg(\  \frac{\gamma - 1}{\gamma}\rho \mu_0^2 \ + \ \alpha \rho M_r^2 ([\mathcal{V}_1]^2 + [\mathcal{V}_2]^2 + [\mathcal{V}_3]^2) \bigg), \ \alpha =  \frac{a^2}{\gamma} \overline{\bigg( \frac{\rho}{p} \bigg)}^2,
\end{equation}
which we rewrite as the sum of a contribution of an entropy wave contribution $\mathcal{E}_{u_n, s}$ (first term) and a contribution due to shear waves $\mathcal{E}_{u_n, \tau}$ (remaining three terms).
\begin{equation*}
    \mathcal{E}_{u_n} = \mathcal{E}_{u_n, s} + \mathcal{E}_{u_n, \tau}.
\end{equation*}
For cartesian grids aligned with the $(x,y,z)$ coordinate system, the normal vectors are along the basis vectors and the entropy production due to shear $\mathcal{E}_{u_n, \tau}$ can be broken down into 6 terms:
\begin{itemize}
    \item Along $x$, shear in y ($\mathcal{E}_{u_n, \tau_{xy}} = |u_n| \mu_3^2$) and shear in z ($\mathcal{E}_{u_n, \tau_{xz}} = |u_n| \mu_2^2$).
    \item Along $y$, shear in z ($\mathcal{E}_{u_n, \tau_{yz}} = |u_n| \mu_1^2$) and shear in x ($\mathcal{E}_{u_n, \tau_{yx}} = |u_n| \mu_3^2$).
    \item Along $z$, shear in y ($\mathcal{E}_{u_n, \tau_{zy}} = |u_n| \mu_1^2$) and shear in x ($\mathcal{E}_{u_n, \tau_{zx}} = |u_n| \mu_2^2$).
\end{itemize}
This gives:
\begin{equation}\label{LMES:dS_shear}
    \mathcal{E}_{u_n, \tau} \ = \  \mathcal{E}_{u_n, \tau_{xy}} + \mathcal{E}_{u_n, \tau_{xz}} + \mathcal{E}_{u_n, \tau_{yx}} + \mathcal{E}_{u_n, \tau_{yz}} + \mathcal{E}_{u_n, \tau_{zx}} + \mathcal{E}_{u_n, \tau_{zy}}
\end{equation}
\vspace*{0.1cm} \\
\indent \textbf{Acoustic modes.} The scaled acoustic eigenvectors $\mathbf{r}_{4,5} = \mathbf{r}_{u_n \pm a}$ are given by :
\begin{gather*}
    \mathbf{r}_{u_n\pm a} = K_a \begin{bmatrix}
        1 \\ u \pm n_1 (a/M_r) \\ v \pm n_2 (a/M_r) \\ w \pm n_3 (a/M_r) \\ h + M_r^2 k \pm u_n a M_r 
    \end{bmatrix}, \ 
    K_a = \bigg(\frac{\rho}{2 \gamma}\bigg)^{1/2}.
\end{gather*}
The acoustic wave strengths $\mu_{u_n \pm a} =  \mathbf{r}_{u_n\pm a}  [\mathbf{v}]$ are given by:
\begin{equation*}
    \mu_{u_n \pm a} = K_a \bigg(\mu_0  \ - \ h \bigg[ \frac{\rho}{p}\bigg] \ \pm \ M_r a \overline{\bigg(\frac{\rho}{p}\bigg)} [ u_n ]  \bigg).
\end{equation*} 
The entropy production field due to acoustic modes therefore writes:
\begin{equation*}
    \mathcal{E}_{u_n \pm a} = |u_n \pm (a/M_r)| K_a^2\bigg(\mu_0 \ - \ h \bigg[ \frac{\rho}{p}\bigg] \ \pm \ M_r a \overline{\bigg(\frac{\rho}{p}\bigg)} [ u_n ]  \bigg)^2.
\end{equation*}
\indent \textbf{Summary.} Overall, the discrete entropy production field $\mathcal{E} = [\mathbf{v}]^TD[\mathbf{v}]$ can be decomposed as:
\begin{equation}\label{LMES:dS_upwind}
    \mathcal{E} \ = \ \mathcal{E}_{u_n, s} \ + \ \big(\mathcal{E}_{u_n, \tau_{xy}} + \mathcal{E}_{u_n, \tau_{xz}} + \mathcal{E}_{u_n, \tau_{yx}} + \mathcal{E}_{u_n, \tau_{yz}} + \mathcal{E}_{u_n, \tau_{zx}} + \mathcal{E}_{u_n, \tau_{zy}}\big) \
 + \ \mathcal{E}_{u_n+a} \ + \ \mathcal{E}_{u_n-a}
\end{equation}
\indent Each of these entropy production fields can be visualized. Figures \ref{fig:Gresho_Roe_dS_un} and \ref{fig:Gresho_Roe_dS} show them for the Gresho vortex at $t = 0$. This is, to the best of our knowledge, the first time that a concrete view on how an ES scheme produces entropy \textit{locally} is given. What is striking is that the acoustic entropy production fields are 2 to 5 orders of magnitudes bigger than the convective ones. Figures \ref{fig:Roe_global_dS}(a)-(b)-(e) show that the acoustic entropy production fields make for most of the entropy produced by the ES Roe flux. \\
\indent Similarly, figures \ref{fig:Sound_Roe_dS}(a)-(c) show the entropy production fields at $t = 0$ for the acoustic wave. The entropy production field associated with entropy waves (there is no shear in this one-dimensional setup) and the entropy production field associated with left-moving acoustic waves are negligible compared to the entropy production field associated with right-moving acoustic waves. This makes sense, and over time, this difference in magnitude is sustained as shown in figures \ref{fig:Roe_global_dS}(b)-(d)-(f). 

\subsubsection{EPBs of Preconditioned operators}
\begin{figure}[htbp!]
    \centering
    \subfigure[Gresho Vortex - $M_r = 3 \times 10^{-1}$]{\includegraphics[scale = 0.62]{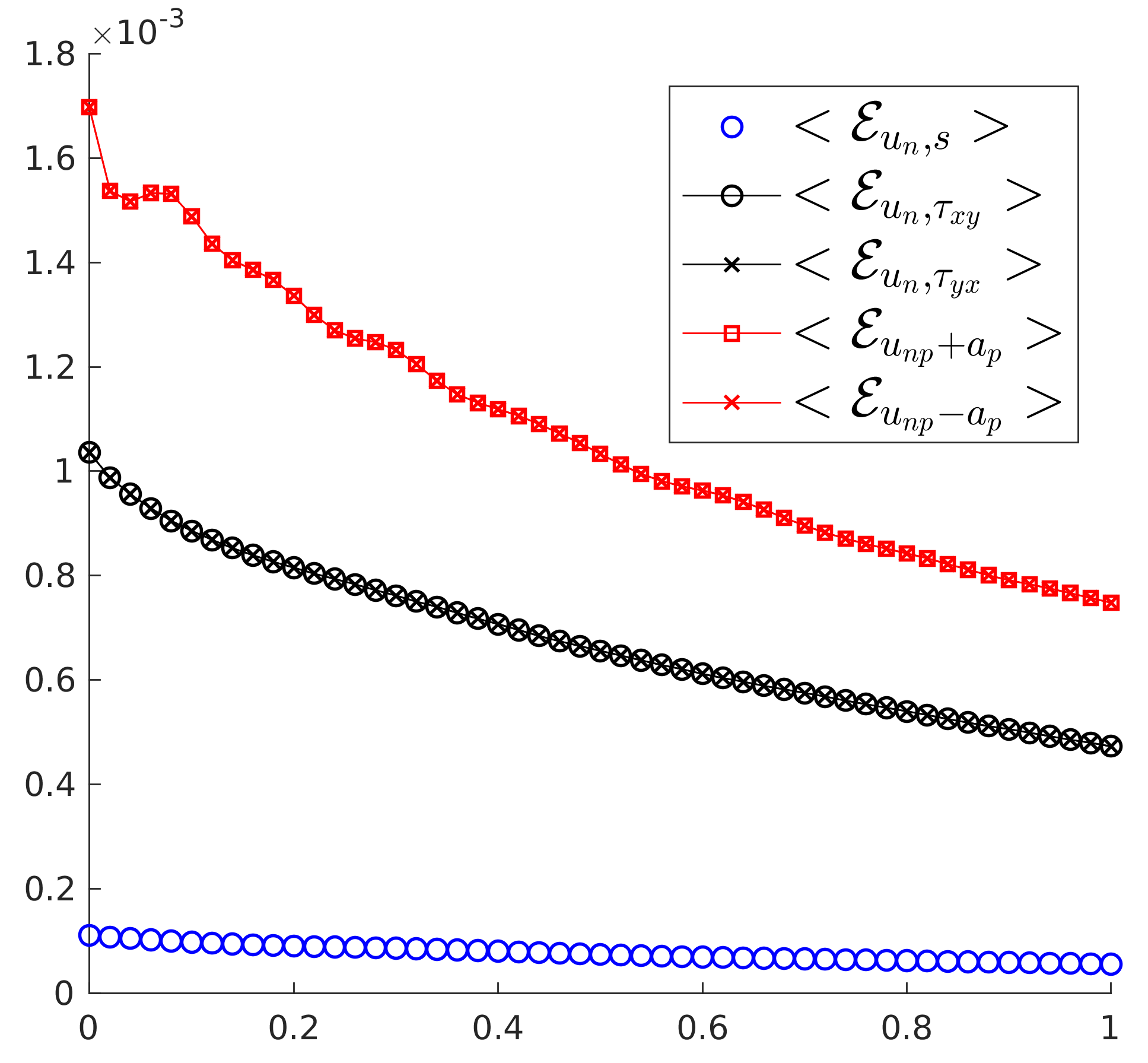}}
     \subfigure[Sound Wave - $M_r = 10^{-2}$]{\includegraphics[scale = 0.113]{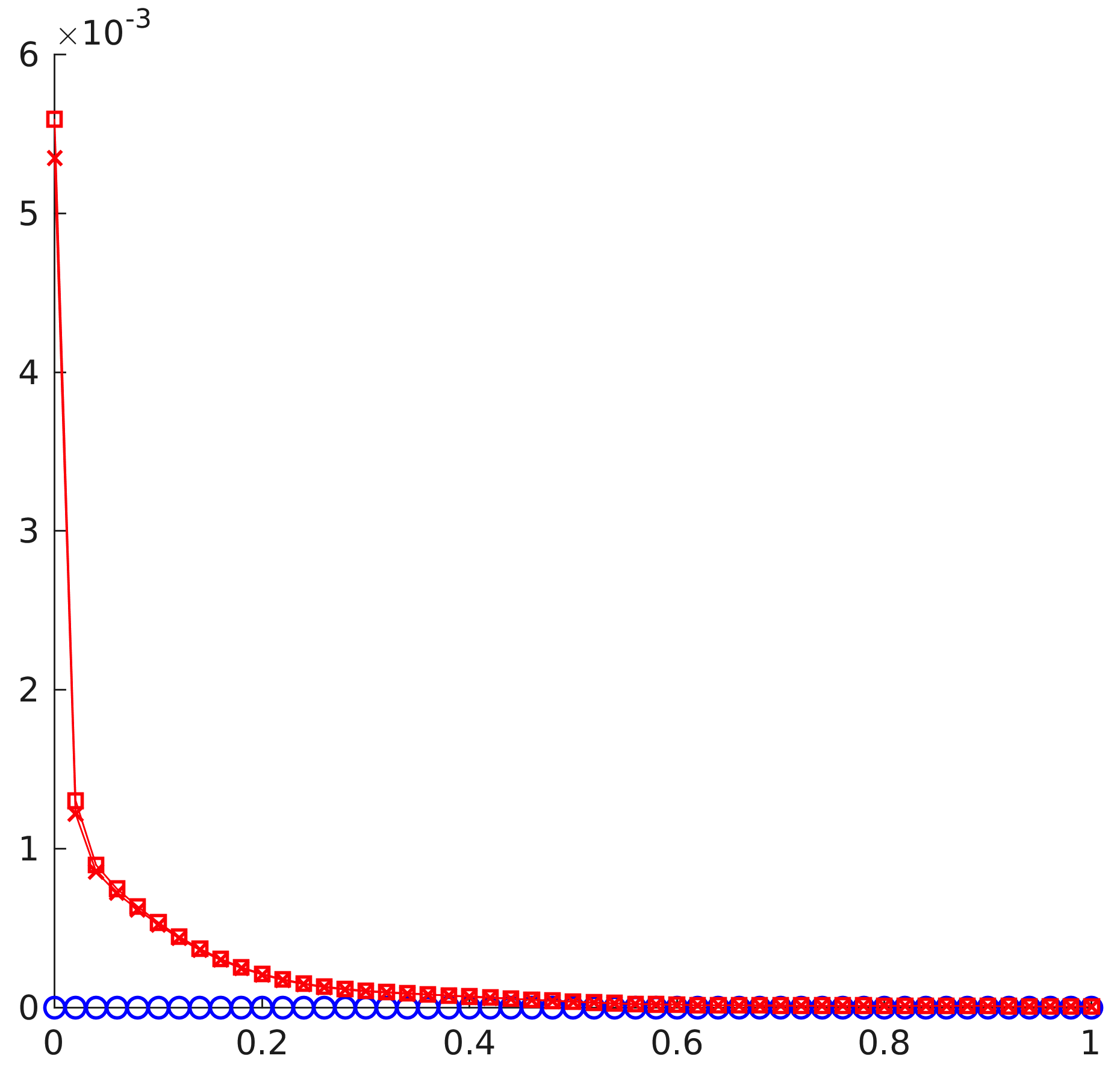}}
    \subfigure[Gresho Vortex - $M_r = 3 \times 10^{-2}$]{\includegraphics[scale = 0.62]{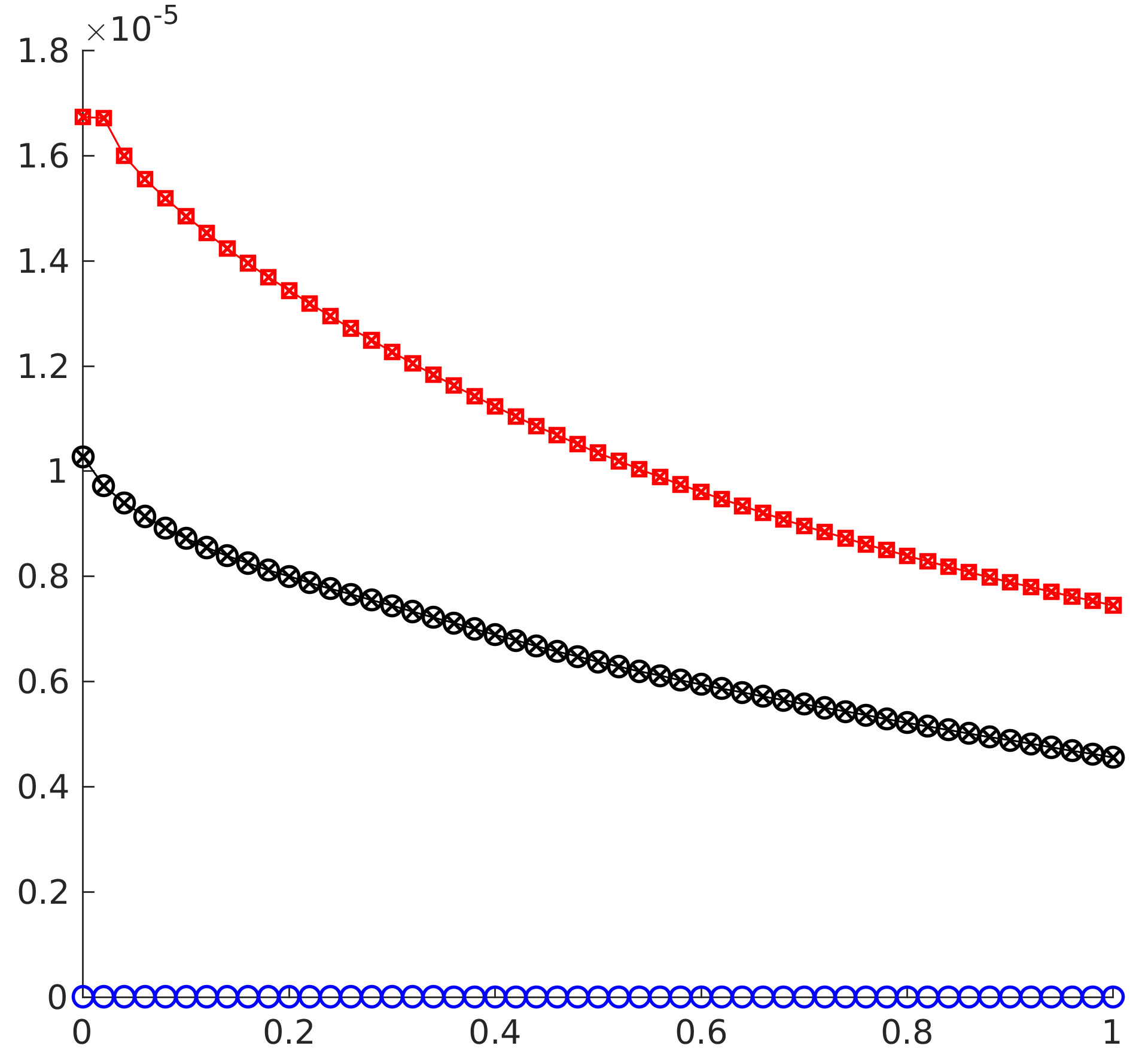}}
     \subfigure[Sound Wave - $M_r = 10^{-3}$]{\includegraphics[scale = 0.62]{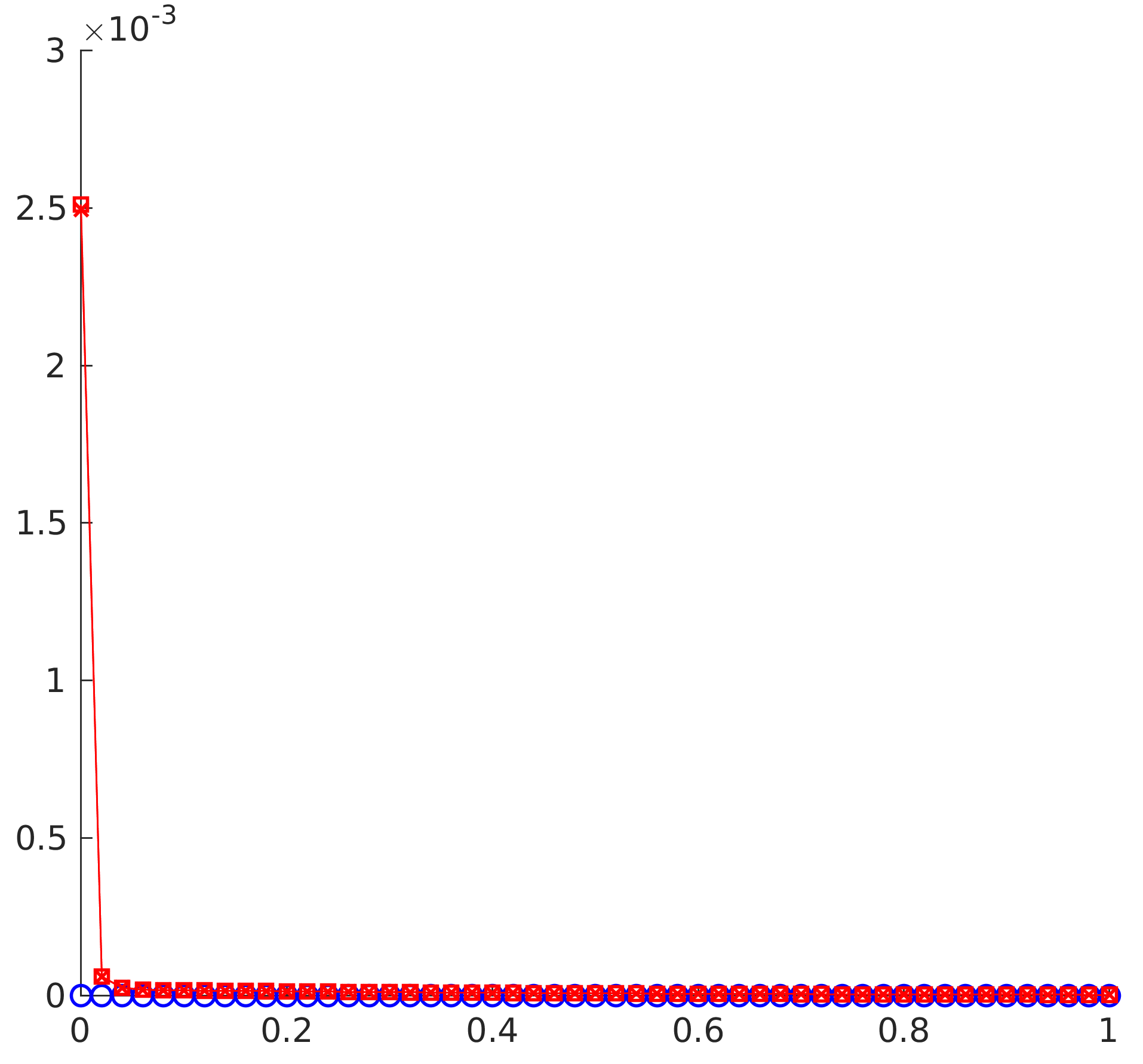}}
    \subfigure[Gresho Vortex - $M_r = 3 \times 10^{-3}$]{\includegraphics[scale = 0.62]{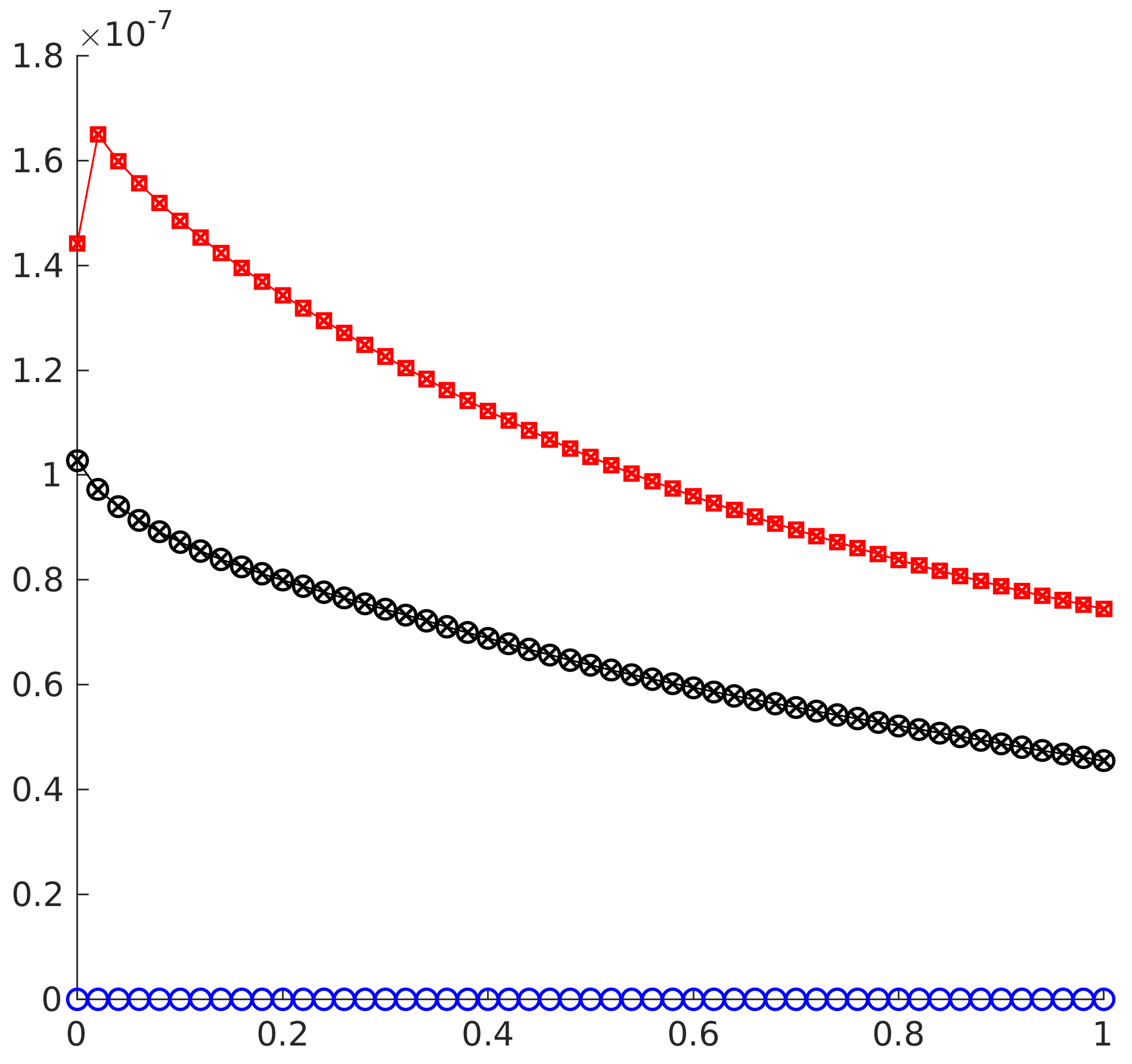}}
     \subfigure[Sound Wave - $M_r = 10^{-4}$]{\includegraphics[scale = 0.62]{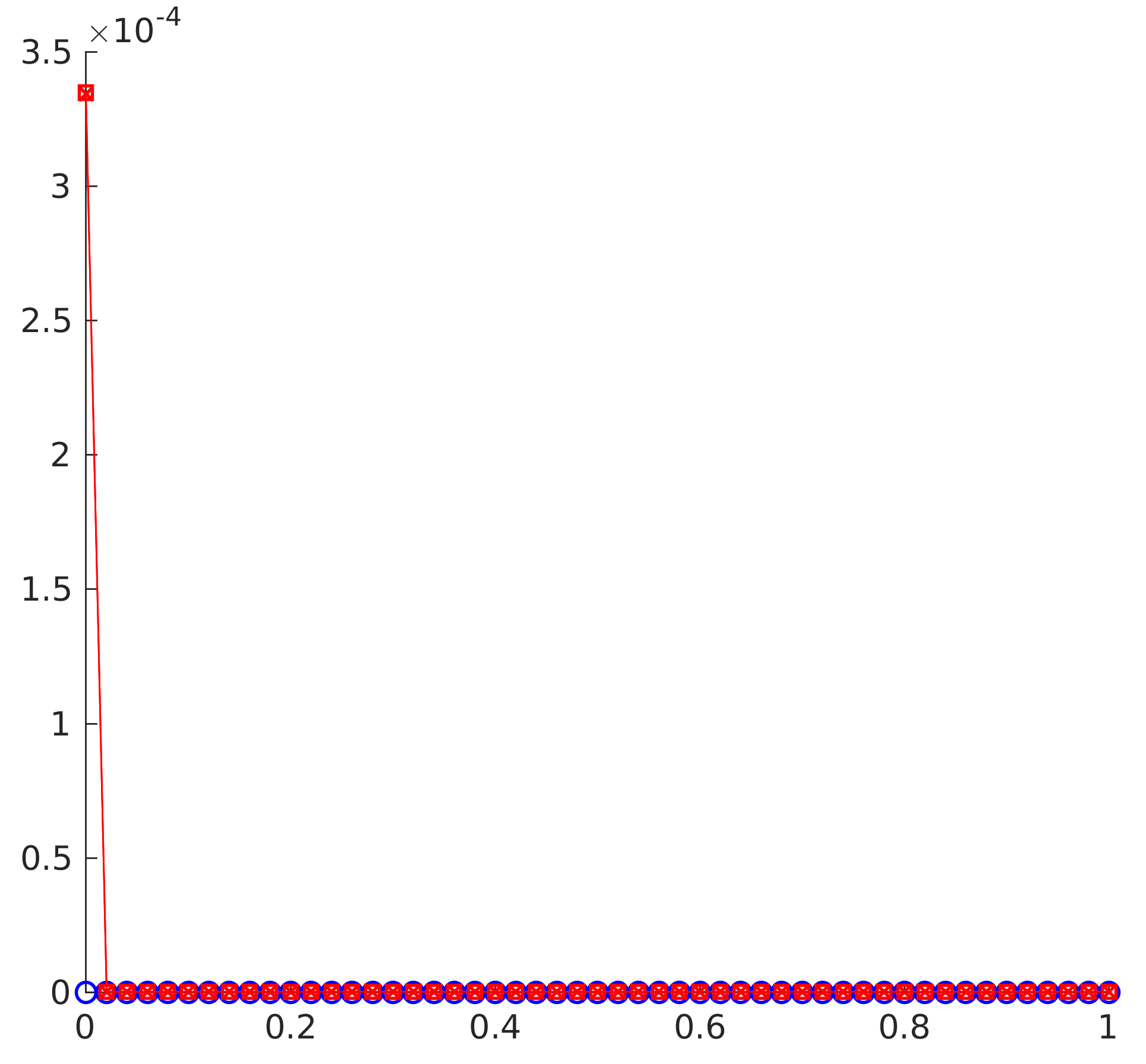}}
    \caption{Integral of each entropy production field with time at different Mach numbers for the ES Turkel flux on the Gresho Vortex (left) and the sound wave (right).}
    \label{fig:Tur_global_dS}
\end{figure}

\indent EPBs for the preconditioned operators $D_P[\mathbf{v}]$ of Turkel and Miczek can be obtained if we manage to express them in a symmetric form $R |\Lambda|R^{T}[\mathbf{v}]$. For this purpose, we can start from the congruence relation (\ref{LMES:eq:Pcong_map_final}) we established in section 4:
\begin{equation*}
    P^{-1}|PA|[\mathbf{v}] \ = \ (Q H_{\mathbf{z}}^{1/2} P_{\mathbf{z}}^{-1}) \ ( |P_{\mathbf{z}}A_{\mathbf{z}}| P_{\mathbf{z}}^T) \ (Q H_{\mathbf{z}}^{1/2} P_{\mathbf{z}}^{-1})^T.
\end{equation*}
We see that a symmetric form for $D_P$ can be inferred from one for $|P_{\mathbf{z}}A_{\mathbf{z}}| P_{\mathbf{z}}^T$ (and vice versa). A simple trick to proceed, which we picked up from Diosady \& Murman \cite{ES_Diosady}, consists in forcing the eigenscaling theorem (\ref{th:Barth}) by introducing the matrix $T^2_p$ defined by $P_{\mathbf{z}}^{T} = R_{p\mathbf{z}} T_p^2 R_{p\mathbf{z}}^T$. From there, we have:
\begin{equation}\label{LMES:eq:Pcong_map_scaled}
    |P_{\mathbf{z}}A_{\mathbf{z}}|P_{\mathbf{z}}^T = R_{p\mathbf{z}} |\Lambda_p|T_p^2 R_{p\mathbf{z}}^T \ \implies \  D_{P} = R_{p} (|\Lambda_p| T_p) R_{p}^T, \ R_{p} = Q H_{\mathbf{z}}^{1/2} P_{\mathbf{z}}^{-1} R_{p\mathbf{z}}.
\end{equation}
If $T_{p}$ is diagonal, an EPB can be introduced along the column vectors of $R_p$. Turkel's matrix qualifies with $T_p^2$ given by:
\begin{equation}\label{eq:T2_Turkel}
    T_p^2 = diag([1, \ 1, \ 1, \ K_2^2+p^2, \ K_1^2+p^2]),
\end{equation}
and $R_{p} = [\mathbf{r}_{u_n,1}, \ \mathbf{r}_{u_n,2},  \ \mathbf{r}_{u_n,3}, \ \mathbf{r}_{u_{np}+a_p}, \ \mathbf{r}_{u_{np}-a_p}  ]$ with:
\begin{equation}\label{eq:Rp_Turkel}
  \mathbf{r}_{u_{np}+a_p} = \sqrt{\frac{\rho}{\gamma}} \frac{K_1}{p^2(K_1 - K_2)} 
        \begin{bmatrix} 
            1 \\ u + n_1 (a / M_r) (p^2/K_1) \\ v + n_2 (a / M_r)(p^2/K_1) \\ w + n_3 (a / M_r)(p^2/K_1) \\ h^t + u_n a M_r (p^2/K_1)
        \end{bmatrix}, \ \mathbf{r}_{u_{np}-a_p} = \sqrt{\frac{\rho}{\gamma}} \frac{K_2}{p^2(K_1 - K_2)} 
        \begin{bmatrix} 
            1 \\ u + n_1 (a / M_r) (p^2/K_2) \\ v + n_2 (a / M_r)(p^2/K_2) \\ w + n_3 (a / M_r)(p^2/K_2) \\ h^t + u_n a M_r (p^2/K_2)
        \end{bmatrix}
\end{equation}
The corresponding entropy production fields differ from (\ref{LMES:dS_upwind}) in the acoustic part only. We have:
\begin{gather}\label{LMES:dS_Turkel}
    \mathcal{E}_{p} \ := \ [\mathbf{v}] \cdot D_P[\mathbf{v}] \ = \ [\mathbf{v}]^T R_{p} (\Lambda_p T_p^2) R_{p}^T [\mathbf{v}] \ = \ \mathcal{E}_{u_n, s} \ + \ \mathcal{E}_{u_n, \tau} \ + \ \mathcal{E}_{u_{np}+a_p} \ + \ \mathcal{E}_{u_{np}-{a_p}},
\end{gather}
with:
\begin{equation}\label{eq:E_acoustic_Turkel}
    \mathcal{E}_{u_{np}+a_p} = \mu_{u_{np}+a_p}^2 (K_2^2 + p^2)|u_{np} + a_p|, \ \mathcal{E}_{u_{np}-a_p} = \mu_{u_{np}-a_p}^2 (K_1^2 + p^2)|u_{np} - a_p|, 
\end{equation}
and:
\begin{align*}
    \mu_{u_{np} + a_p} =& \ \mathbf{r}_{u_{np}+a_p}^T [\mathbf{v}] = \sqrt{\frac{\rho}{\gamma}} \frac{K_1}{p^2(K_1 - K_2)}  \bigg( \mu_0 - h \bigg[\frac{\rho}{p}\bigg] + \frac{a M_r p^2}{K_1} \overline{\bigg(\frac{\rho}{p}\bigg)}[u_n]  \bigg), \\
    \mu_{u_{np} - a_p} =& \ \mathbf{r}_{u_{np}-a_p}^T [\mathbf{v}] = \sqrt{\frac{\rho}{\gamma}} \frac{K_2}{p^2(K_1 - K_2)}  \bigg( \mu_0 - h \bigg[\frac{\rho}{p}\bigg] + \frac{a M_r p^2}{K_2} \overline{\bigg(\frac{\rho}{p}\bigg)}[u_n]  \bigg).
\end{align*}
Figures \ref{fig:Gresho_Tur_dS}(a)-(b) show the modified acoustic entropy production fields for the Gresho vortex at $t = 0$. The two fields are of the same magnitude and they appear to be in some sort of symmetry. Figures \ref{fig:Tur_global_dS}(a)-(c)-(e) show that the total acoustic entropy production fields are of the same order as those associated with the shear waves over time. \\
\indent For the sound wave, the initial entropy production fields are shown in figures \ref{fig:Sound_Tur_dS}(a)-(b). The preconditioning leads to modified acoustic eigenvectors which can no longer be tied to right-moving and left-moving moving waves. The flow consists of a right-moving acoustic wave, yet we see that both entropy production fields $\mathcal{E}_{u_{np}\pm a_p}$ are active. Figures \ref{fig:Tur_global_dS}(b)-(d)-(f) show the overwhelming domination of both acoustic entropy production fields. \\

\begin{figure}[htbp!]
    \centering
    \subfigure[$M_r = 3 \times 10^{-1}$]{\includegraphics[scale = 0.7]{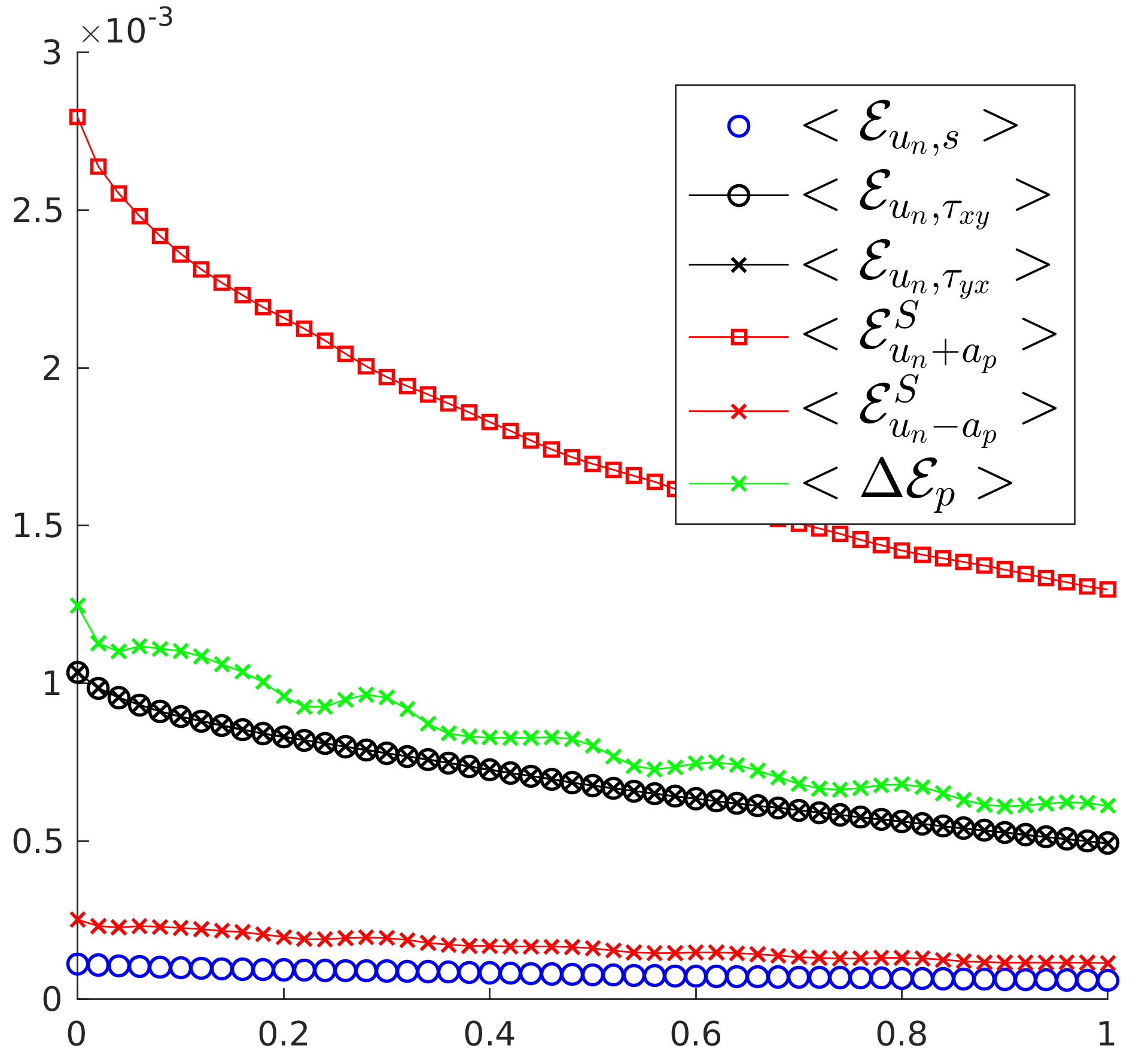}}
    \subfigure[$M_r = 3 \times 10^{-2}$]{\includegraphics[scale = 0.7]{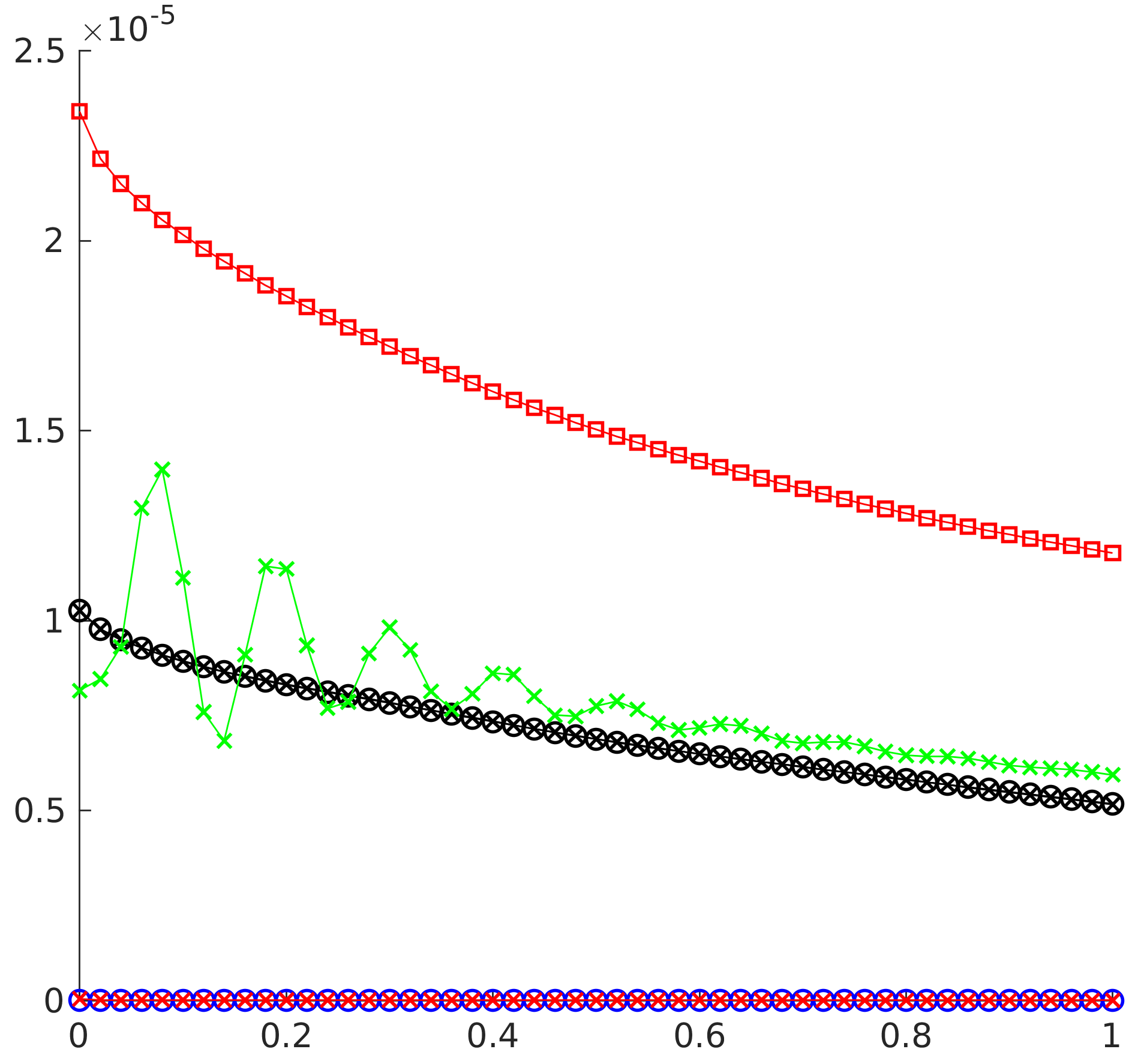}}
    \subfigure[$M_r = 3 \times 10^{-3}$]{\includegraphics[scale = 0.7]{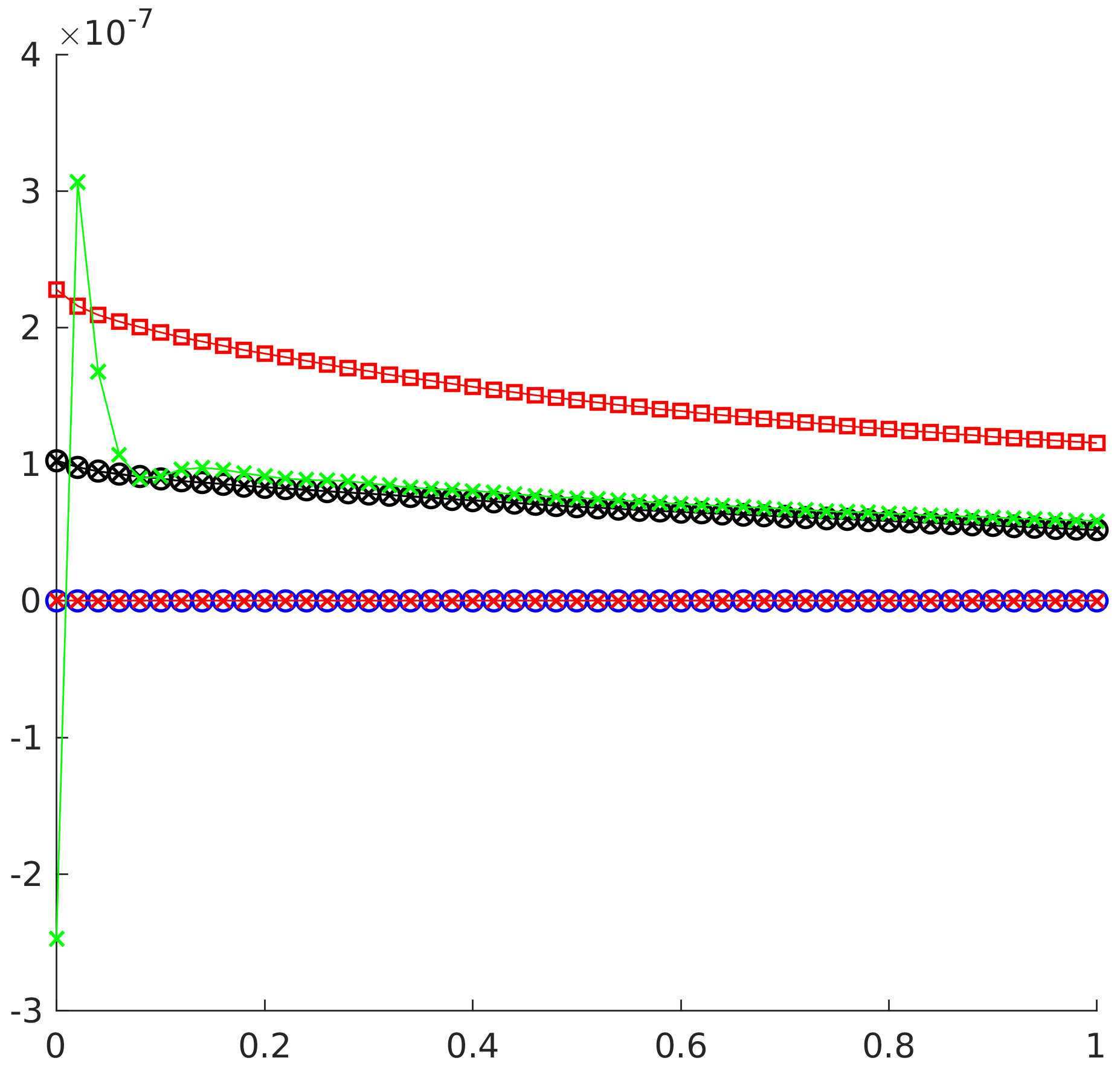}}
    \caption{Gresho Vortex: Integral of each entropy production field with time at different Mach numbers for the ES Miczek flux.}
    \label{fig:Gresho_Mic_global_dS}
\end{figure}

\begin{figure}[htbp!]
    \centering
    \subfigure[$t = 1$]{\includegraphics[scale = 0.84]{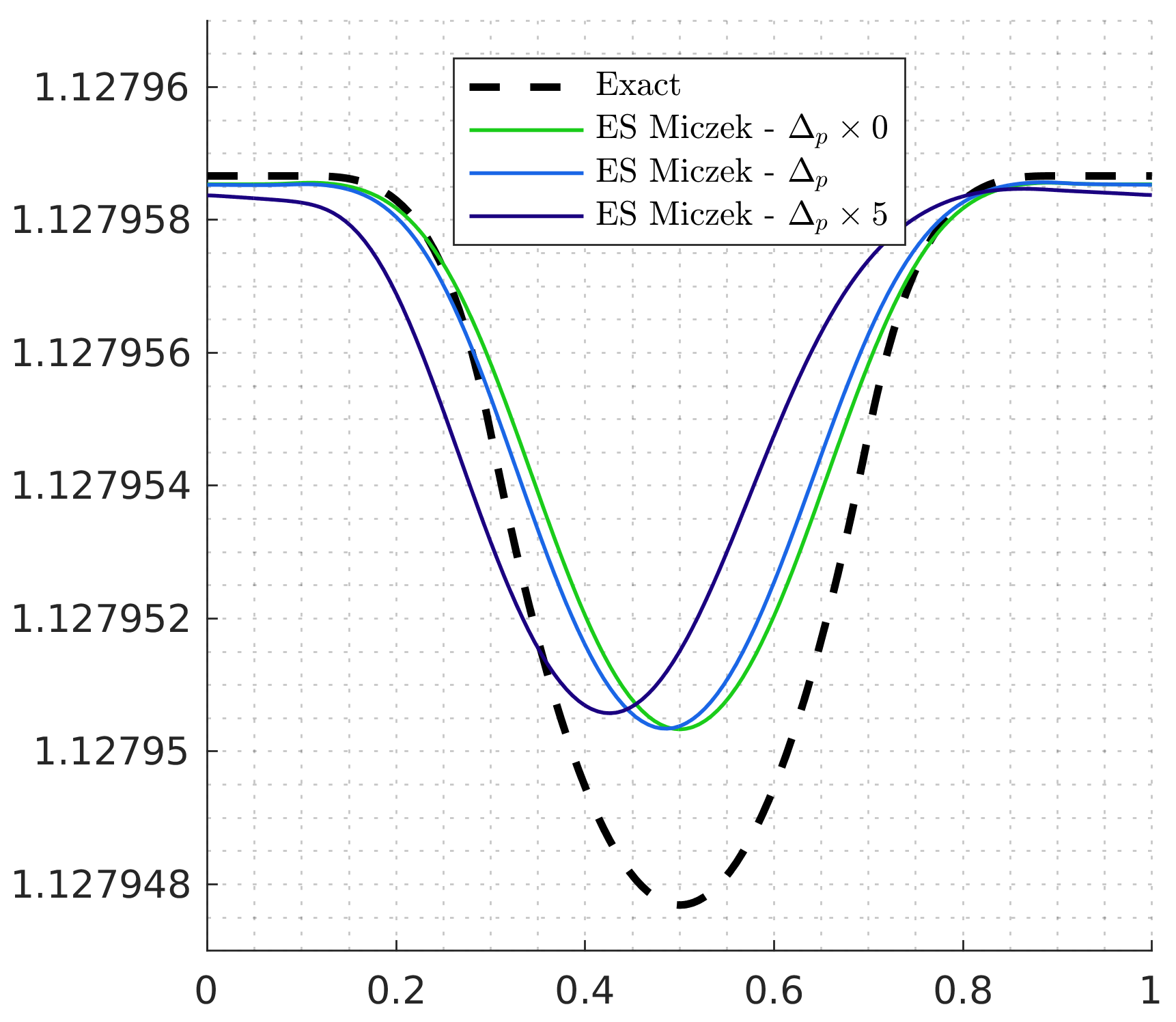}}
    \subfigure[$t = 0.04$]{\includegraphics[scale = 0.84]{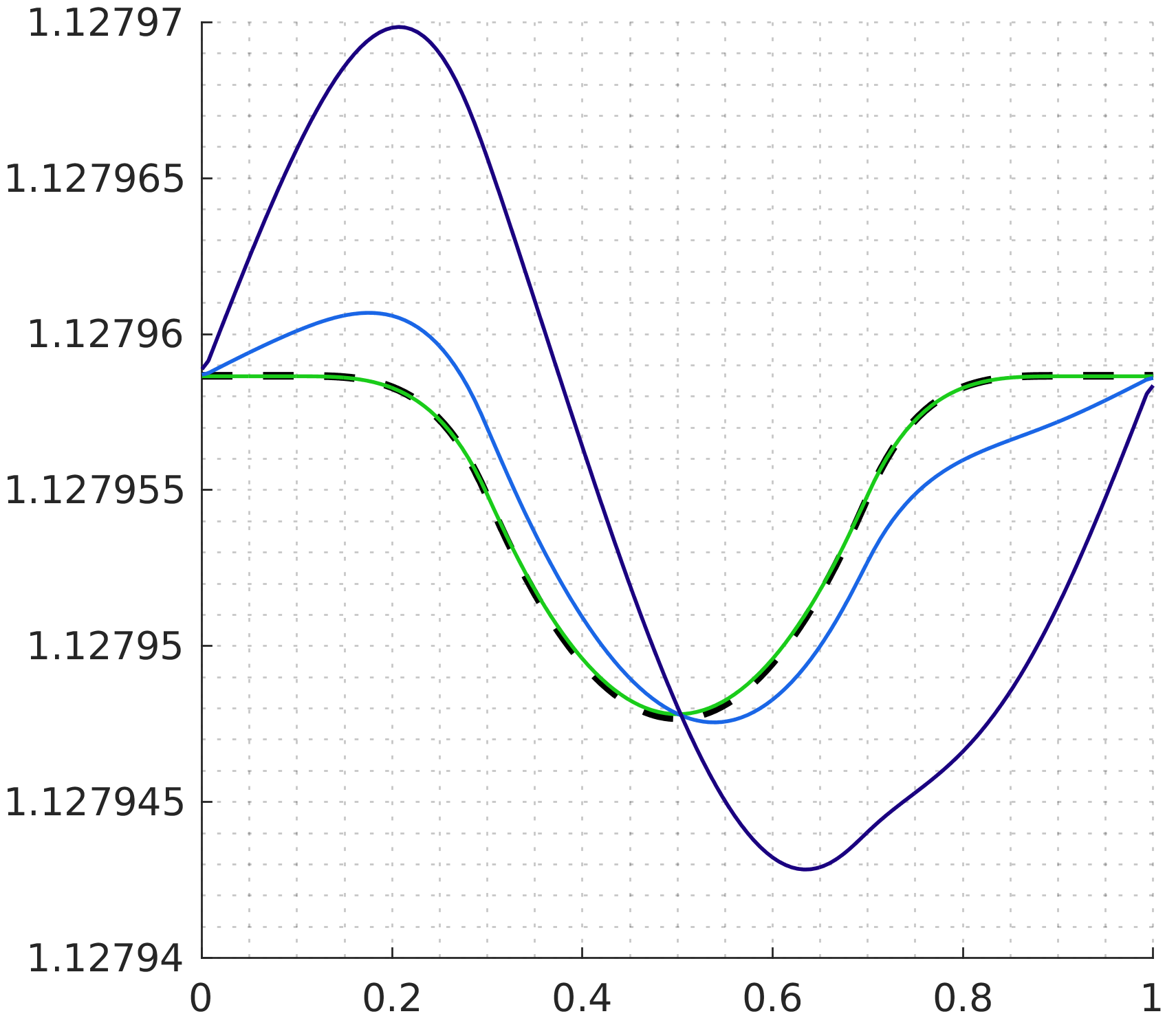}}
    \caption{Gresho Vortex: Pressure field at $M_r = 3 \times 10^{-3}$ for the Miczek flux when the skew-symmetric term is multiplied by a factor. The phase errors observed in figures \ref{fig:Gresho_pressure} and \ref{fig:Gresho_pressure_early} disappear if the skew-symmetric term is removed, and amplified if the skew-symmetric term is made bigger.}
    \label{fig:Gresho_pressure_Mic}
\end{figure}

\begin{figure}[htbp!]
    \centering
    \includegraphics[scale = 0.85]{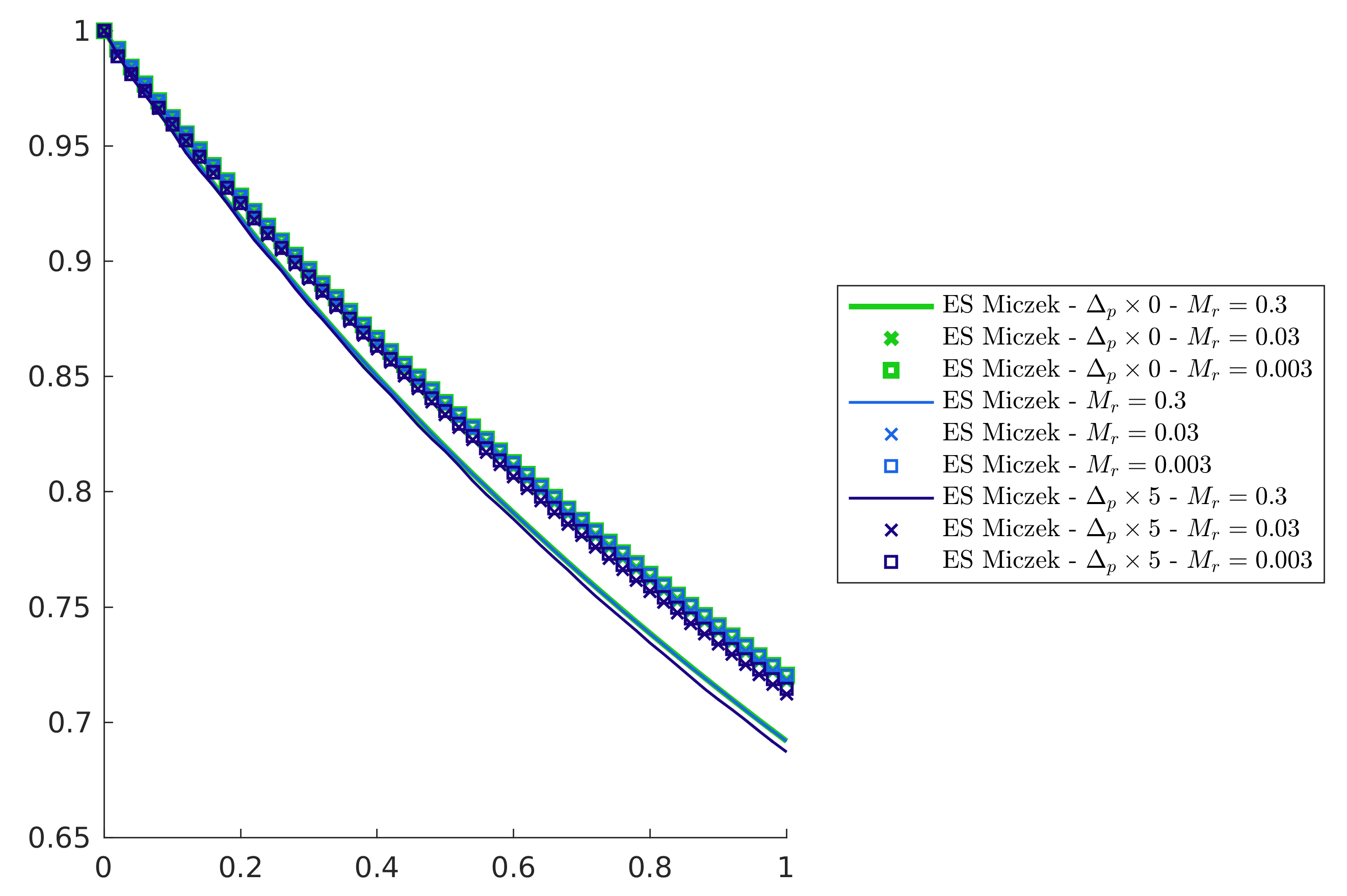}
    \caption{Gresho vortex: Total kinetic energy over time. The skew-symmetric matrix does not strongly impact the ability of the scheme to preserve the kinetic energy. }
    \label{fig:Gresho_Ek_Mic}
\end{figure}

\begin{figure}[htbp!]
    \centering
    \subfigure[Without $\Delta_p$]{\includegraphics[scale = 0.7]{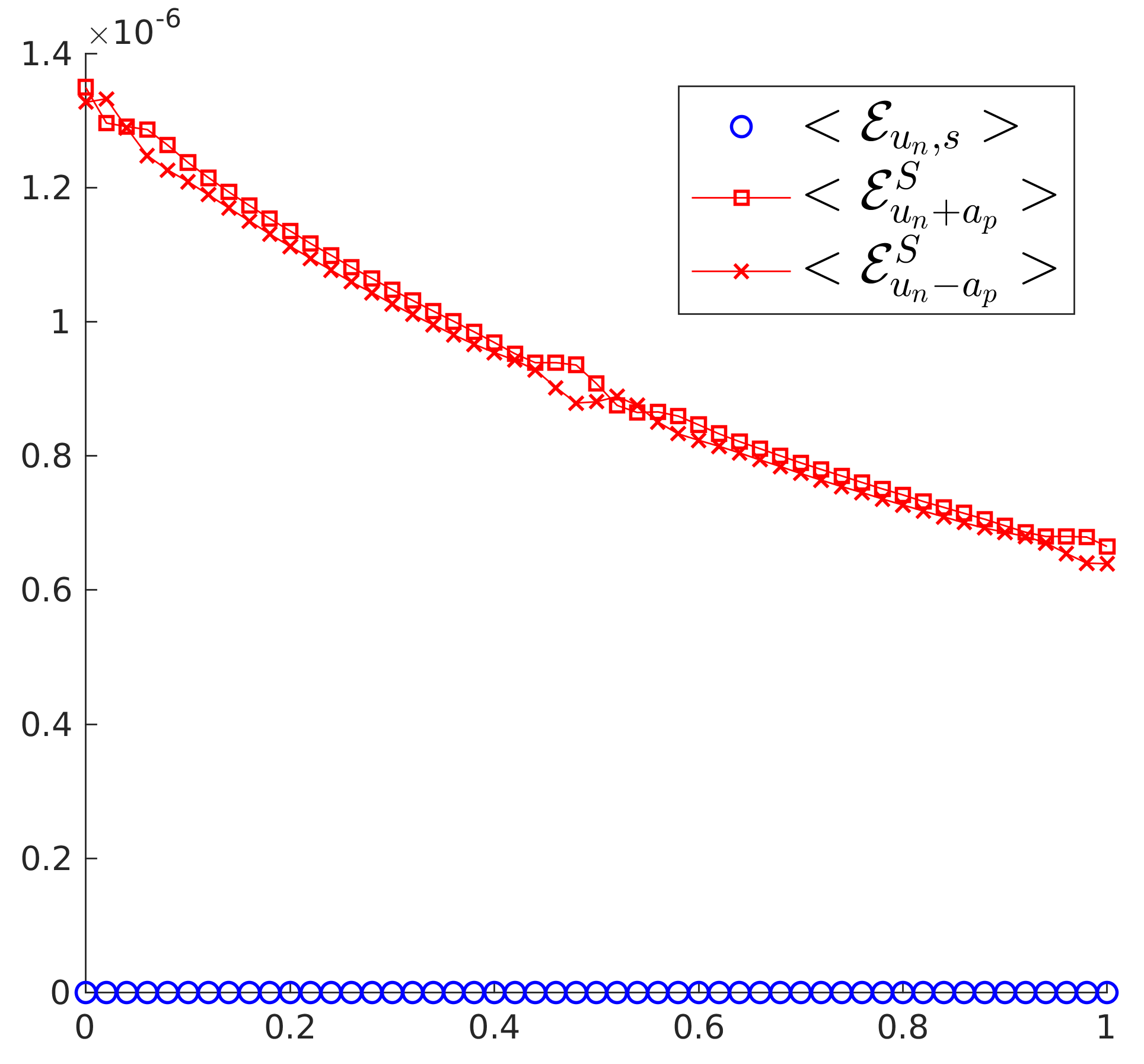}}
    \subfigure[With $\Delta_p$]{\includegraphics[scale = 0.7]{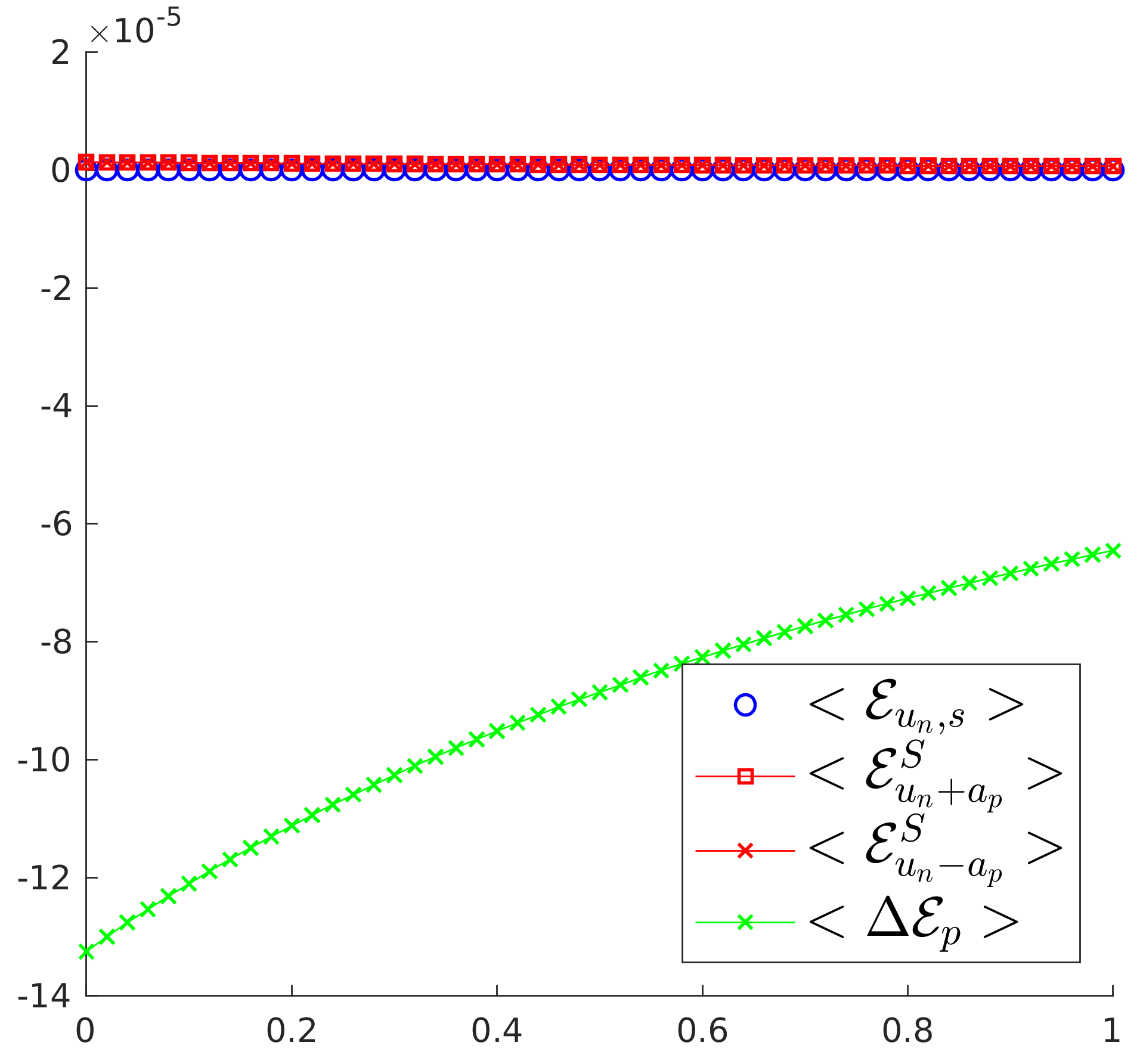}}
    \caption{Sound wave: Integral of entropy production fields for $M_r = 10^{-2}$ (same trends at lower $M_r$ values) for the ES Miczek flux, omitting the contribution of the skew-symmetric matrix (left) and including it (right).}
    \label{fig:Sound_Mic_global_dS}
\end{figure}

\begin{figure}[htbp!]
    \centering
    \subfigure[$t = 0.03$]{\includegraphics[scale = 0.8]{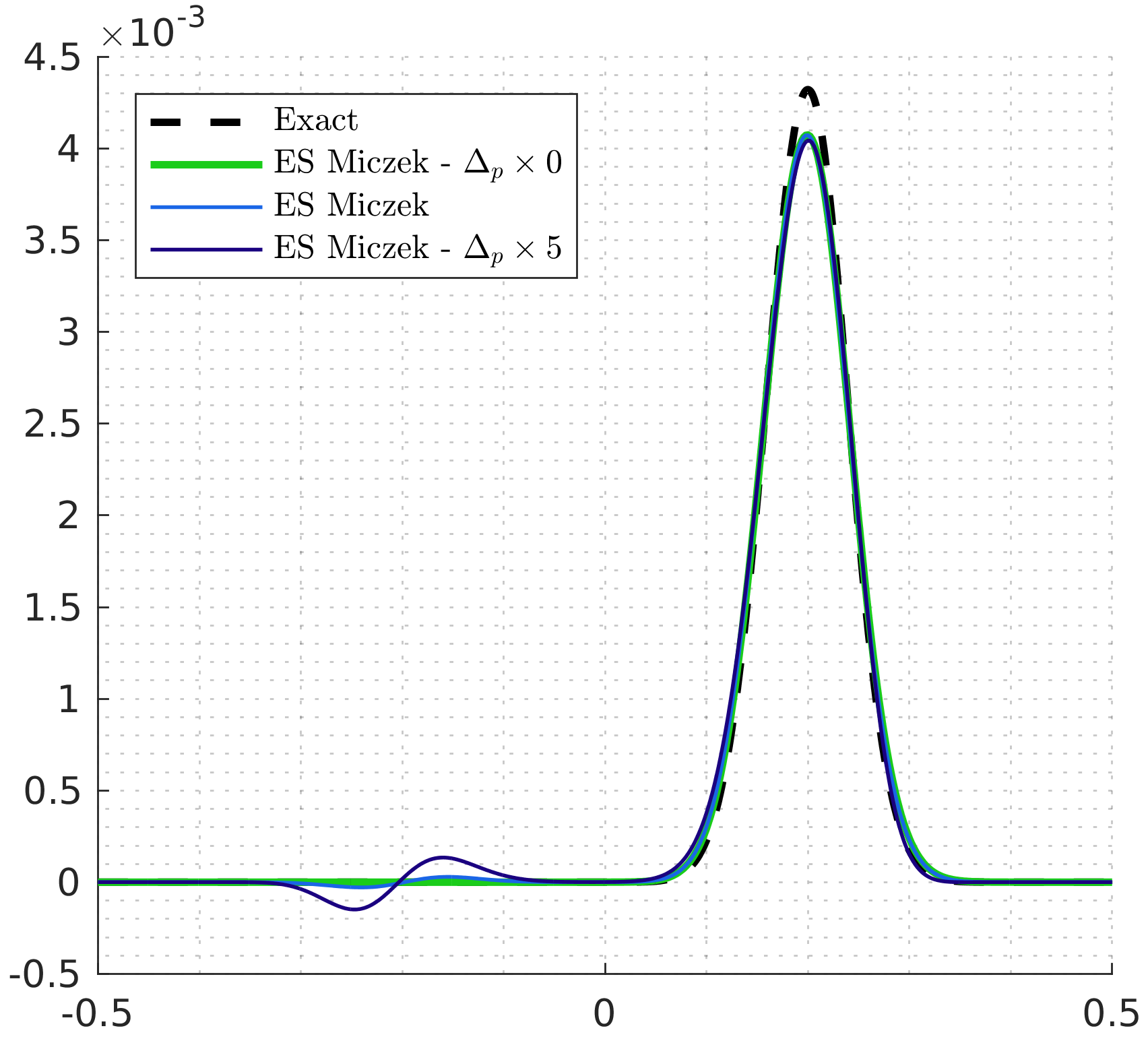}}
    \subfigure[$t = 0.03$, zoomed]{\includegraphics[scale = 0.8]{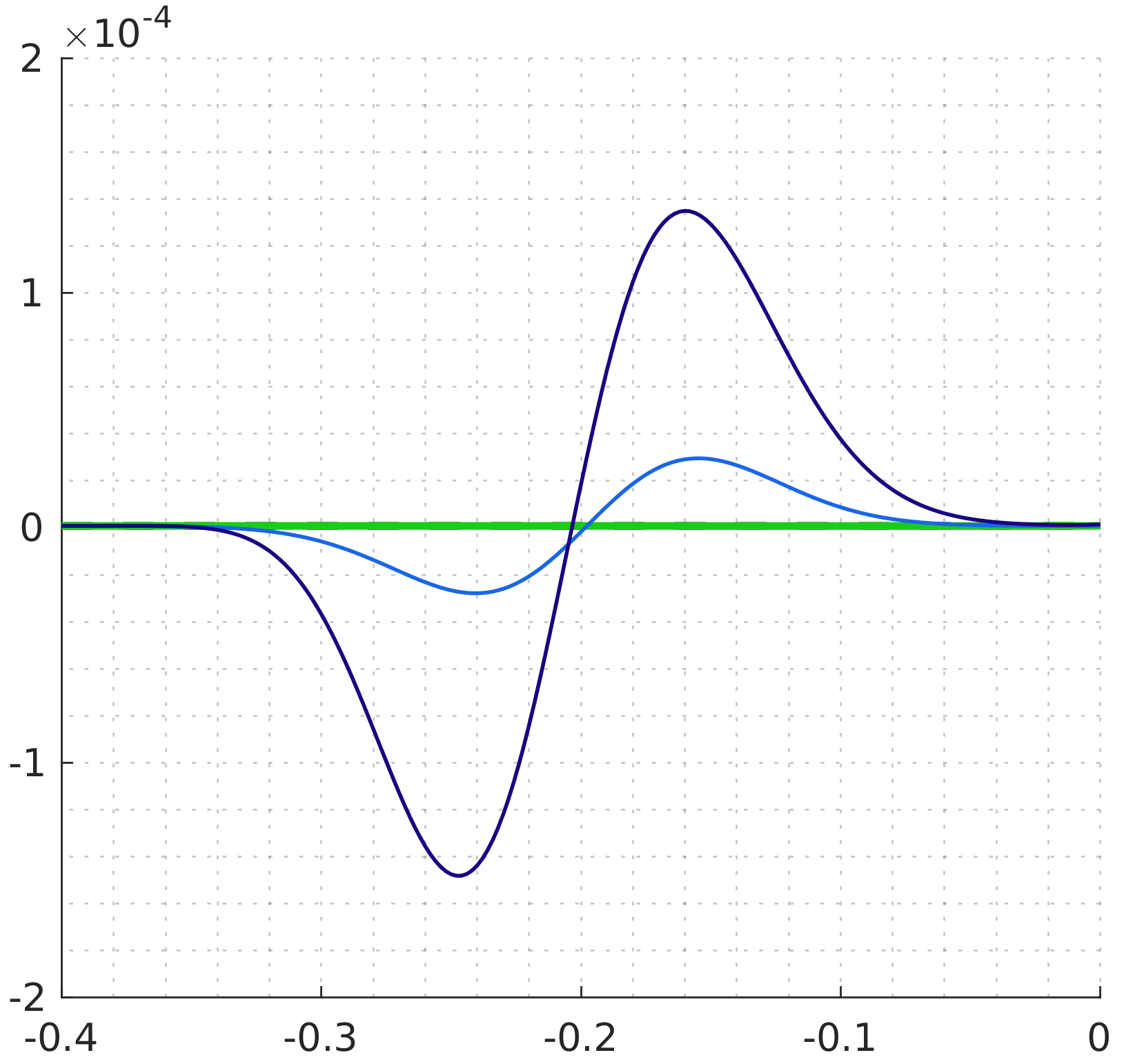}}
    \caption{Sound wave: Pressure snapshots showing that the spurious left-moving acoustic wave is due to the skew-symmetric term in the ES Miczek flux. $M_r = 10^{-2}$ }
    \label{fig:Sound_snaps_Mic}
\end{figure}

\indent For Miczek's matrix, we do not benefit from the eigenscaling theorem, but we do know that $T_p^2$ is positive definite. We have $R_p = [\mathbf{r}_{u_n,1}, \ \mathbf{r}_{u_n,2}, \ \mathbf{r}_{u_n,3}, \ \mathbf{r}_{u_n+a_p}, \ \mathbf{r}_{u_n-a_p} ]$ with:
\begin{gather*}
    \mathbf{r}_{u_n+a_p} = \sqrt{\frac{\rho}{\gamma}} \frac{1}{(p^2+1)(K_1-K_2)}
        \begin{bmatrix} 
            (K_1-p) \\
          u (K_1-p)  + (a/M_r) n_1 (K_1 p+1) \\
          v (K_1-p)  + (a/M_r) n_2 (K_1 p+1) \\  
          w (K_1-p)  + (a/M_r) n_3 (K_1 p+1) \\
          h^t(K_1-p) + u_n (a M_r) (K_1 p+1)
        \end{bmatrix}, \\
         \mathbf{r}_{u_n-a_p} = \sqrt{\frac{\rho}{\gamma}} \frac{1}{(p^2+1)(K_1-K_2)}
        \begin{bmatrix} 
            (K_2-p) \\
          u (K_2-p)  + (a/M_r) n_1 (K_2 p+1) \\
          v (K_2-p)  + (a/M_r) n_2 (K_2 p+1) \\  
          w (K_2-p)  + (a/M_r) n_3 (K_2 p+1) \\
          h^t(K_2-p) + u_n (a M_r) (K_2 p+1)
        \end{bmatrix},
\end{gather*}
and:
\begin{equation*}
    T_p^2 = \begin{bmatrix}
            1 & 0 & 0 & 0       & 0 \\
            0 & 1 & 0 & 0       & 0 \\
            0 & 0 & 1 & 0       & 0 \\
            0 & 0 & 0 & K_2^2+1 & (K_2-K_1)p - K_1 K_2 - 1 \\
            0 & 0 & 0 & (K_1-K_2)p - K_1 K_2 - 1 & K_1^2 +1
        \end{bmatrix}.
\end{equation*}
\vspace{0.1cm} \\
$T_p^2$ is not diagonal, but $D_P$ can be further reduced by observing that in the subsonic regime, the last 2-by-2 bloc of $|\Lambda_p|T_p^2$ can be decomposed into symmetric and skew-symmetric parts as follows
\begin{gather*}
    \big(|\Lambda_p|T_p^2\big)_{(4:5,4:5)}  \ = \  \begin{bmatrix}
    |u_n + a_p| (K_2^2+1) & 0 \\
    0 & |u_n - a_p| (K_1^2+1)
    \end{bmatrix}
    + 
    \begin{bmatrix}
    0 & -\delta_p \\
    \delta_p & 0
    \end{bmatrix}, \\ \delta_p = 2 p (p^2 + 1) (a^2 - M_r^2 u_n^2)/(M_r (a - M_r p u_n)).
\end{gather*}
Instead of a symmetric form, we now have $D_P = R_{p} (
|\Lambda_p| \overline{T}^2_p + \Delta_p) R_{p}^T$ with:
\begin{equation*}
    \overline{T}^2_p = diag([1, \ 1, \ 1, \ K_2^2+1, \ K_1^2+1]), \ 
    \Delta_p = \begin{bmatrix}
                0 & 0 & 0 & 0 & 0 \\
                0 & 0 & 0 & 0 & 0 \\
                0 & 0 & 0 & 0 & 0 \\
                0 & 0 & 0 & 0 & -\delta_p \\
                0 & 0 & 0 & \delta_p & 0
                \end{bmatrix}.
\end{equation*}
Since $\overline{T}^2_p$ is diagonal positive and $\Delta_p$ is skew-symmetric, it follows that $D_P$ is positive definite for the Miczek flux-preconditioner (hence the compatibility with entropy-stability). The EPB for Miczek's flux is more subtle than for Turkel's or Roe's because of the skew symmetric matrix $\Delta_p$. We could ignore $\Delta_p$ in the EPB, since it does not contribute to $\mathcal{E}_{p}$. However, it turns out that this matrix plays a key role in the anomalies observed in the previous section. \\ 
\indent Let $D^{A}_P[\mathbf{v}]$ denote the acoustic part of Miczek's dissipation operator $R_{p} (\Lambda_p \overline{T}_p^2 + \Delta_p) R_{p}^T [\mathbf{v}]$. We have:
\begin{equation}\label{LMES:acoustic_part}
    D^{A}_P[\mathbf{v}] = \begin{bmatrix} \mathbf{r}_{u_n+a_p} & \mathbf{r}_{u_n-a_p} \end{bmatrix}
    \begin{bmatrix} |u_n+a_p| (1+K_2^2) & -\delta_p \\
                    \delta_p            & |u_n-a_p| (1+K_1^2)\end{bmatrix}
    \begin{bmatrix} \mu_{u_n+a_p} \\ \mu_{u_n-a_p} \end{bmatrix},
\end{equation}
where $\mu_{u_n \pm a_p} = \ \mathbf{r}_{u_n\pm a_p}^T [\mathbf{v}]$ is given by:
\begin{gather*}
    \mu_{u_n + a_p} = \sqrt{\frac{\rho}{\gamma}} \frac{1}{(p^2+1)(K_1-K_2)}  \bigg( (K_1 - p) \bigg(\mu_0 - h \bigg[\frac{\rho}{p}\bigg]\bigg) + a M_r (K_1 p+1) \overline{\bigg(\frac{\rho }{p}\bigg)}[u_n] \bigg),  \\
    \mu_{u_n - a_p} = \sqrt{\frac{\rho}{\gamma}} \frac{1}{(p^2+1)(K_1-K_2)}  \bigg( (K_2 - p) \bigg(\mu_0 - h \bigg[\frac{\rho}{p}\bigg]\bigg) + a M_r (K_2 p+1) \overline{\bigg(\frac{\rho }{p}\bigg)}[u_n] \bigg).
\end{gather*}
Expanding (\ref{LMES:acoustic_part}), we have:
\begin{multline}\label{LMES:acoustic_part_skew}
    D^{A}_{P}[\mathbf{v}] = \mathbf{r}_{u_n+a_p} \bigg(\mu_{u_n+a_p} |u_n+a_p| (1 + K_2^2) - \delta_p \mu_{u_n-a_p}\bigg)  + \mathbf{r}_{u_n-a_p} \bigg(\mu_{u_n-a_p} |u_n-a_p| (1 + K_1^2) + \delta_p \mu_{u_n+a_p}\bigg).
\end{multline}
Multiplying on the left by $[\mathbf{v}]^T$ gives an acoustic entropy production field:
\begin{equation}\label{LMES:eq:dS_mic0}
    [\mathbf{v}]^T D^{A}_{P}[\mathbf{v}] = \mathcal{E}_{u_n+a_p} + \mathcal{E}_{u_n-a_p},
\end{equation}
where the fields $\mathcal{E}_{u_n \pm a_p}$ break down into contributions $\{\mathcal{E}^S_{u_n \pm a_p}, \ \mp \Delta\mathcal{E}^p \}$ from the symmetric and skew-symmetric parts of the dissipation operator:
\begin{gather}\label{LMES:eq:dS_mic}
    \mathcal{E}_{u_n+a_p} = \mathcal{E}_{u_n+a_p}^S - \Delta \mathcal{E}_p, \ \mathcal{E}_{u_n-a_p} = \mathcal{E}_{u_n-a_p}^S  +  \Delta \mathcal{E}_p, \\
    \mathcal{E}_{u_n+a_p}^S = \mu_{u_n+a_p}^2 |u_n+a_p| (1+K_2^2), \ \mathcal{E}_{u_n-a_p}^S = \mu_{u_n-a_p}^2 |u_n-a_p| (1+K_1^2), 
    \\ \Delta \mathcal{E}_p = \delta_p \mu_{u_n+a_p} \mu_{u_n-a_p}.
\end{gather}
$\mathcal{E}_{u_n+a_p}$ and $\mathcal{E}_{u_n-a_p}$ are no longer positive in principle but their addition is always positive.  Equations (\ref{LMES:acoustic_part_skew}) and (\ref{LMES:eq:dS_mic}) suggest that \textit{while $\Delta_p$ does not change the discrete entropy production produced at an interface, it effects how this amount is distributed locally among acoustic modes}. \\
\indent Figures \ref{fig:Gresho_Mic_dS}(a)-(b) show the modified acoustic entropy production fields for the Gresho vortex at $t = 0$. They resemble those of the ES Turkel flux. Figures \ref{fig:Gresho_Mic_dS}(c)-(e) show the contributions of the symmetric and skew-symmetric terms. The skew-symmetric component is not negligible. Figures \ref{fig:Gresho_Mic_global_dS}(a)-(c) show that the total acoustic entropy production fields are of the same order as those associated with the shear waves over time. We also see that the total contribution $\langle\Delta \mathcal{E}_p\rangle$ from the skew-symmetric matrix evolves in time like a damped oscillator, with a characteristic time that decreases with the Mach number. This suggests that the spurious transient causing the phase errors we observed earlier has something to do with the skew-symmetric matrix. A simple way to confirm this is to multiply the skew-symmetric term by a factor and see how it impacts the solution. This is illustrated in figures \ref{fig:Gresho_pressure_Mic}(a)-(b). Taking out the skew-symmetric indeed removes the transient and phase errors. Making the skew-symmetric term stronger amplifies them. What's more, figure \ref{fig:Gresho_Ek_Mic} shows that the skew-symmetric term does not have a visible impact on the ability of the scheme to conserve the kinetic energy of the system. \\
\indent For the sound wave, the initial entropy production fields are showed in figures \ref{fig:Sound_Mic_dS}(a)-(e). The contribution from the skew-symmetric part is two orders of magnitude bigger than the contribution from the symmetric part. This is why, for visibility, we show the integrated entropy production fields in two parts (figure \ref{fig:Sound_Mic_global_dS}). While the perturbations we observed in figure \ref{fig:Sound_A} appear in the symmetric parts $\mathcal{E}_{u_n \pm a_p}^{S}$ of the acoustic entropy production fields, it turns out from figures \ref{fig:Sound_snaps_Mic}(a)-(b) that it is the skew-symmetric term again that is causing the appearance of a spurious left-moving acoustic wave. These observations lead us to the following statement:
\begin{equation*}
    \mbox{\textbf{Conjecture:} \textit{The skew-symmetric matrix $\Delta_p$ in the Miczek ES flux causes entropy transfers among acoustic waves.}}
\end{equation*}
We examine this claim in more detail in section 7.

\subsection{The Discrete Low-Mach Regime Revisited}
\begin{figure}[ht!]
    \centering
    \subfigure[Gresho Vortex]{\includegraphics[scale = 0.8]{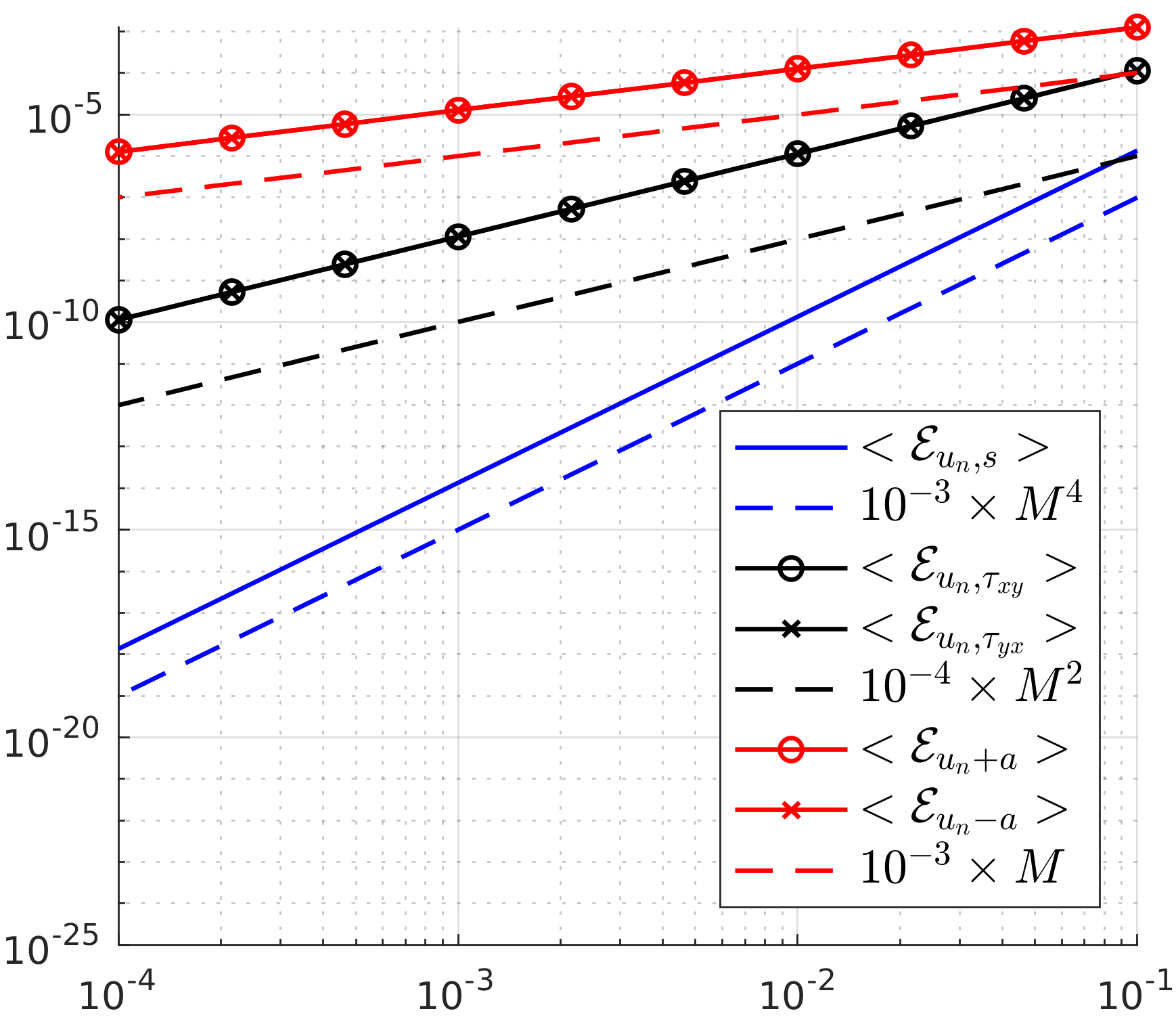}}
    \subfigure[Sound wave]{\includegraphics[scale = 0.8]{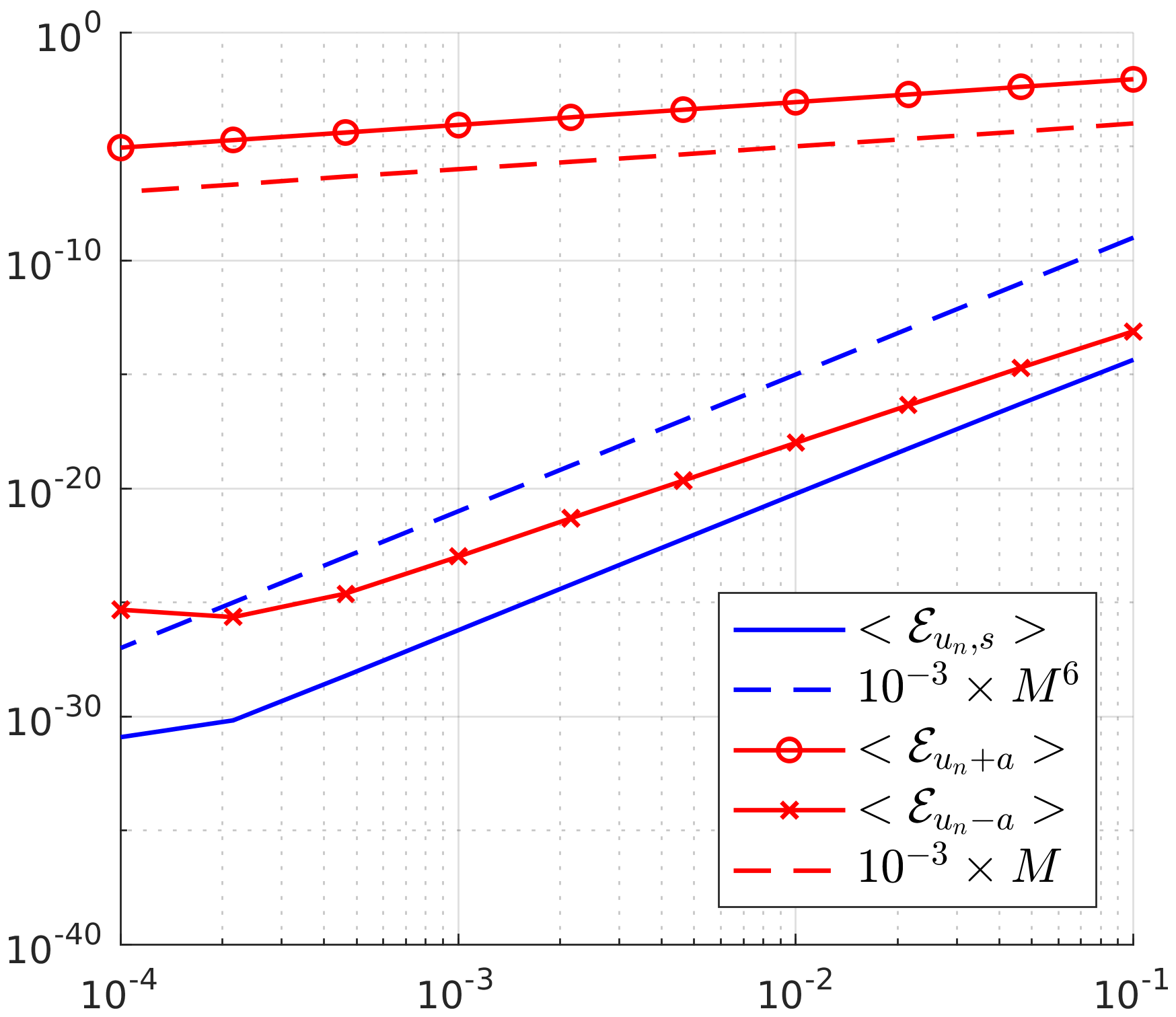}}
    \caption{Entropy production scalings - ES Roe flux}
    \label{fig:Roe_dS}
\end{figure}

\begin{figure}[ht!]
    \centering
    \subfigure[Gresho Vortex]{\includegraphics[scale = 0.14]{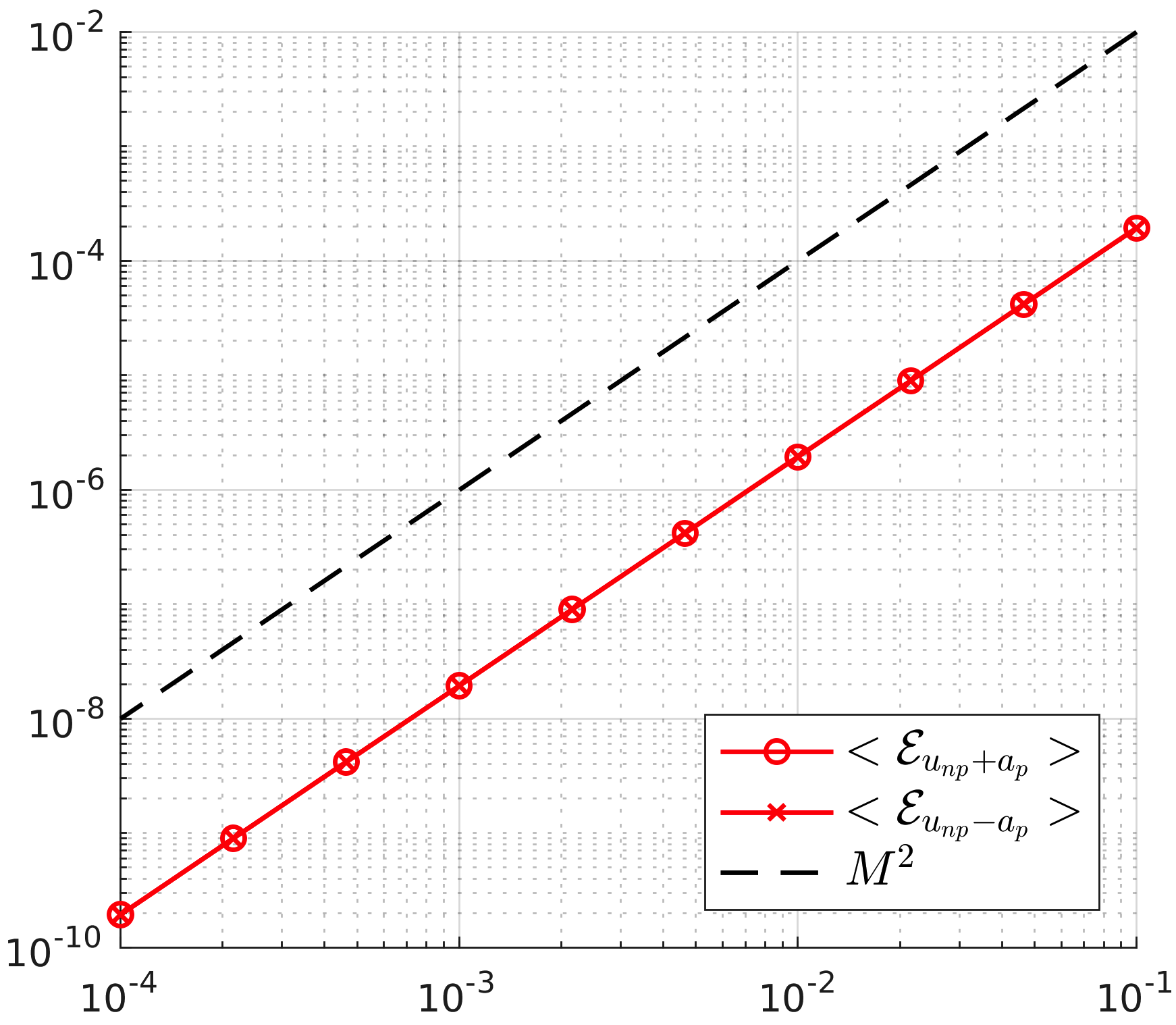}}
    \subfigure[Sound wave]{\includegraphics[scale = 0.77]{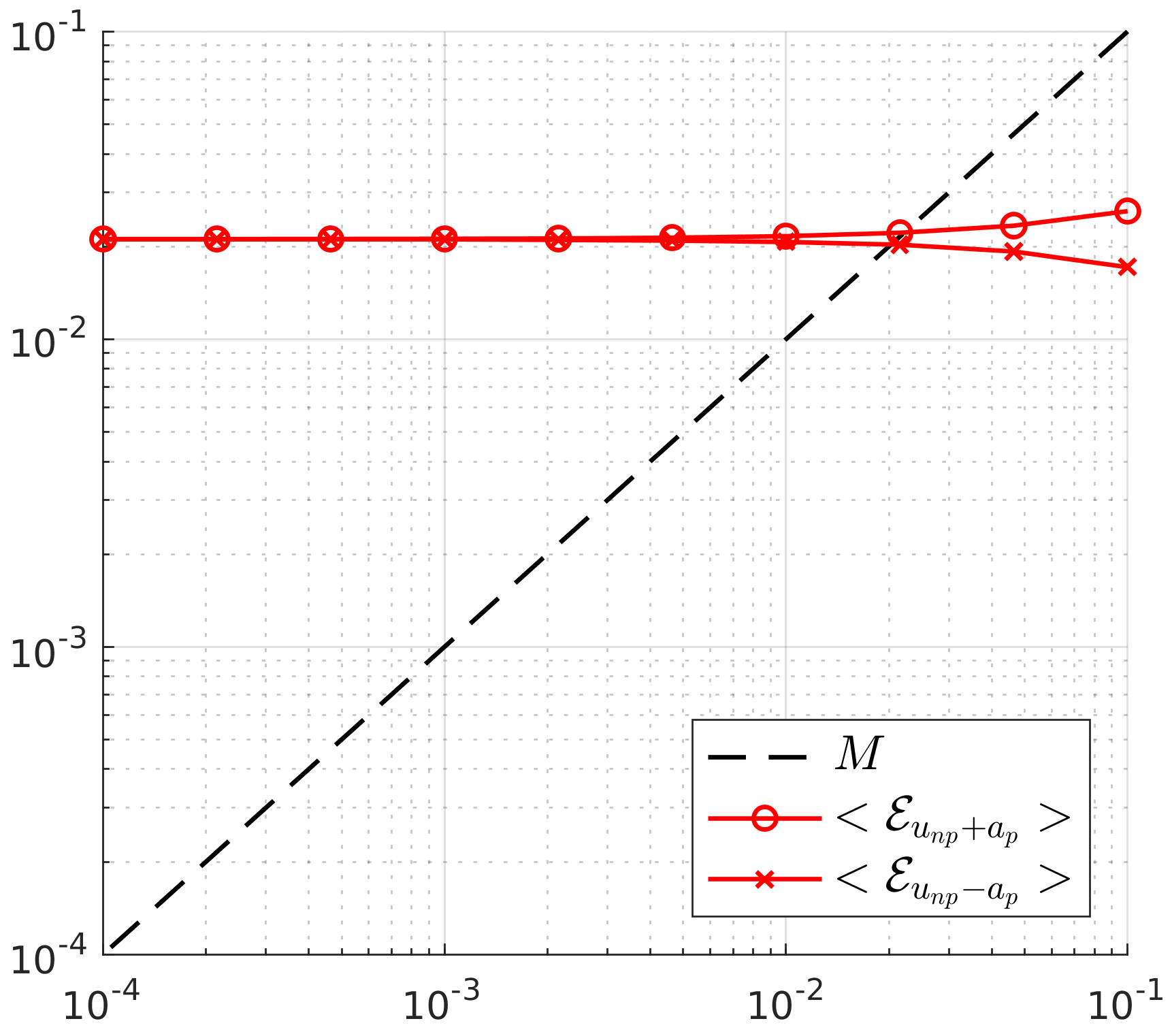}}
    \caption{Entropy production scalings - ES Turkel flux}
    \label{fig:Turkel_dS}
\end{figure}

\begin{figure}[htbp!]
    \centering
    \subfigure[Gresho Vortex]{\includegraphics[scale = 0.8]{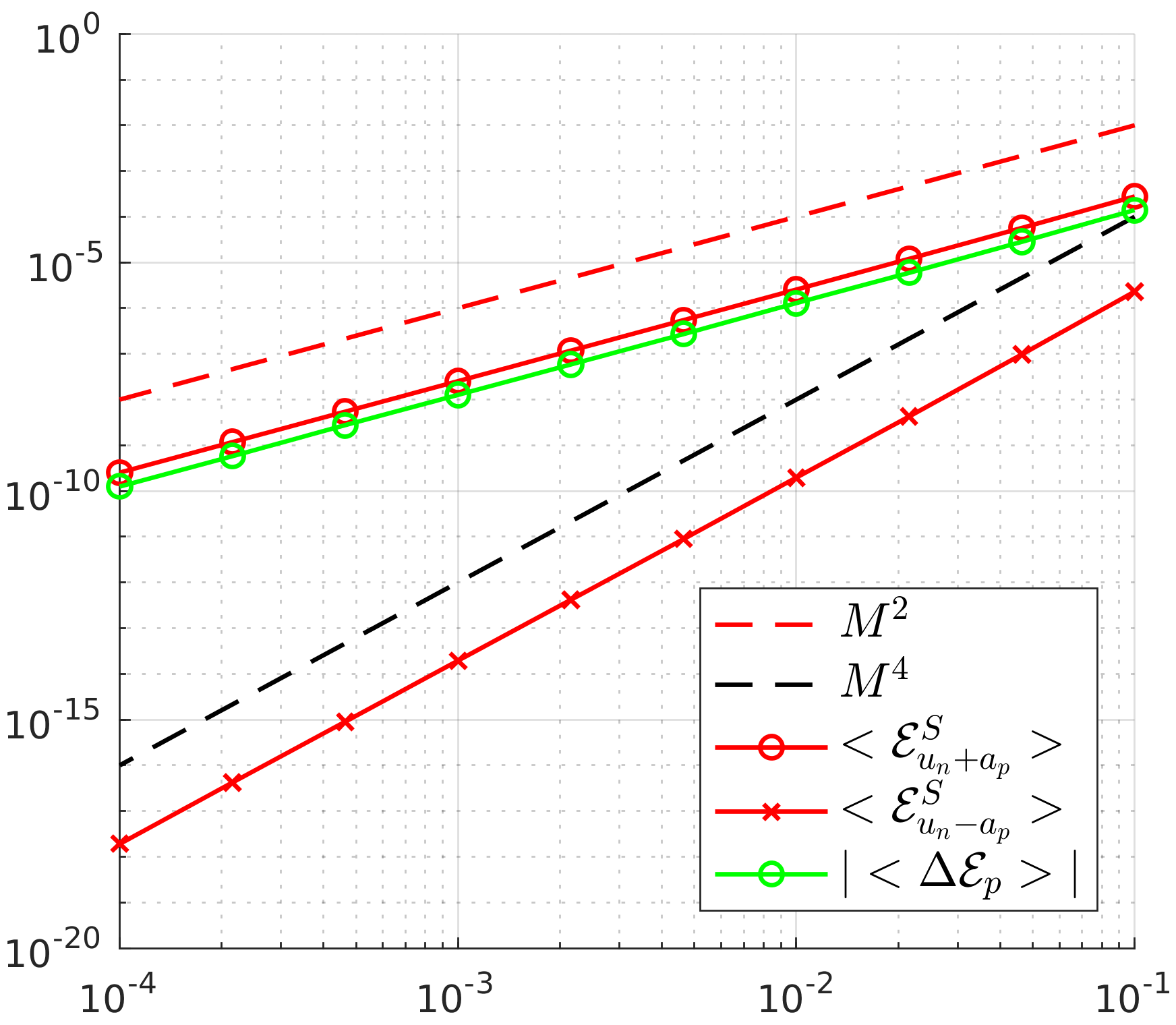}}
    \subfigure[Sound wave]{\includegraphics[scale = 0.8]{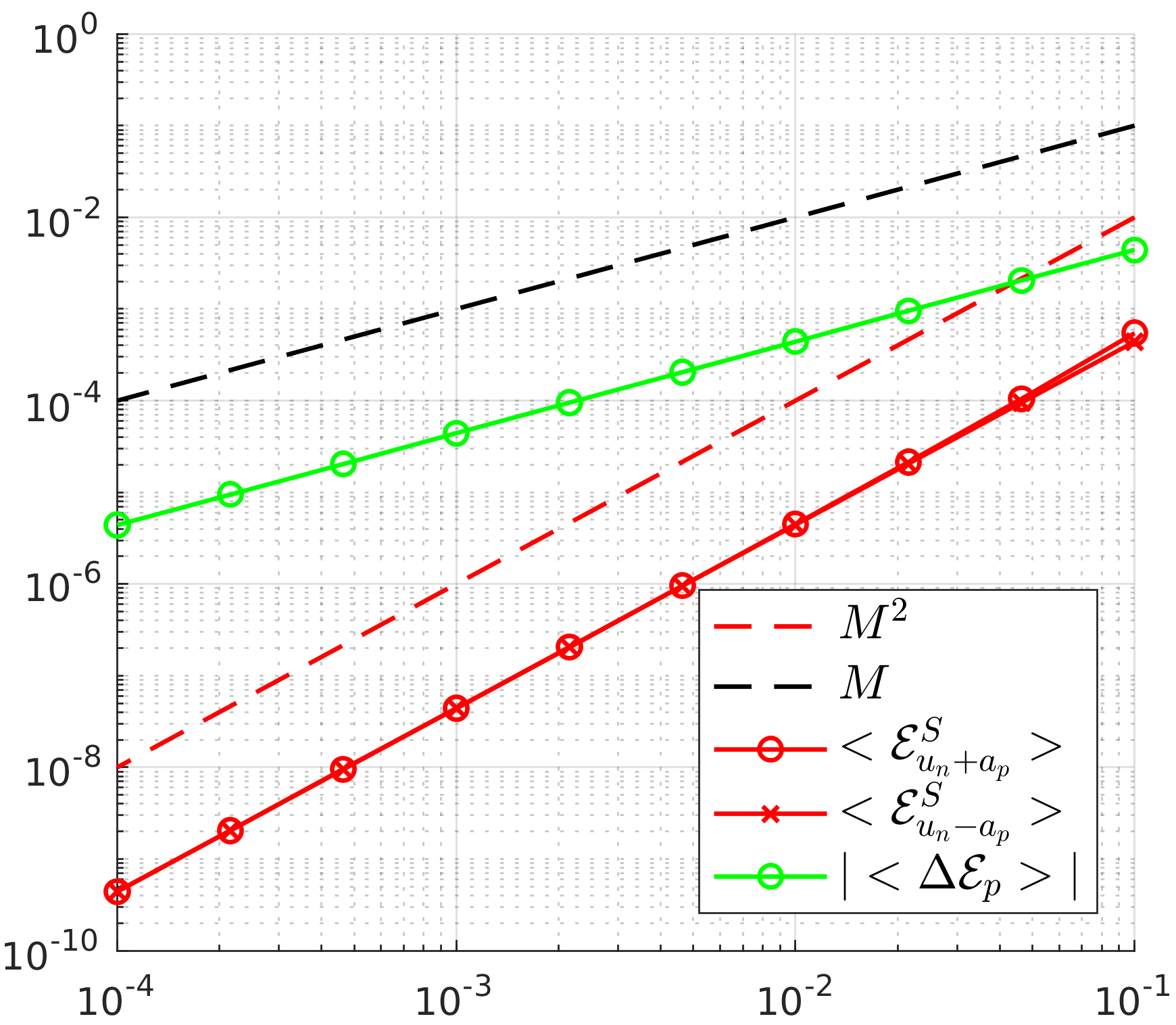}}
    \caption{Entropy production scalings - ES Miczek flux}
    \label{fig:Miczek_dS}
\end{figure}

\hspace*{0.1 cm} Using the analytical expressions we just derived, we can now determine how the entropy produced by each ES flux scales with respect to $M_r$ and establish whether \ref{E1} and \ref{E2} are satisfied. This effort, similar in spirit to the analysis of  Guillard \& Viozat \cite{Guillard1}, Guillard \& Nkonga \cite{Guillard2} and Bruel \textit{et al.} \cite{Bruel}, provides an explanation to \ref{S1}, \ref{S2} and \ref{S3} in terms of entropy production. \\
\indent To verify the scaling analysis, we computed, for each ES flux, the integrated (\ref{LMES:total_dS}) entropy production fields at $t = 0$ for the Gresho vortex and the sound wave at different reference Mach numbers. The scalings are shown in figures \ref{fig:Roe_dS}, \ref{fig:Turkel_dS} and \ref{fig:Miczek_dS}. \\ \\
\indent \textbf{Incompressible limit.} For the Gresho vortex, density is constant, pressure and velocity variations are of order $M_r^2$ and $1$, respectively. At the discrete level, this translates into:
\begin{multline}
    [\rho] = 0, [p] = \mathcal{O}(M_r^2), \ [u] = \mathcal{O}(1), [v] = \mathcal{O}(1) \ \implies \\ \ \ \ [s] = \mathcal{O}(M_r^2), \ \bigg[\frac{\rho k}{p}\bigg] = \mathcal{O}(1), \ \bigg[\frac{\rho u}{p}\bigg] = \mathcal{O}(1), \
    \bigg[\frac{\rho v}{p}\bigg] = \mathcal{O}(1), \
    \bigg[\frac{\rho}{p}\bigg] = \mathcal{O}(M_r^2).
\end{multline}
For the classic ES upwind dissipation, this gives:
\begin{align}\label{LMES:Gresho_ES}
    \mu_0 = \mathcal{O}(M_r^2) \ \ \mbox{and} \ \ \mu_{s1} = \mu_{s2} = 0, \ \mu_{s3} = \mathcal{O}(1) \ \implies& \ \mathcal{E}_{u_n, s} = \mathcal{O}(M_r^4), \  \mathcal{E}_{u_n, \tau} = \mathcal{O}(M_r^2). \\
    \mu_{u_n \pm a} = \mathcal{O}(M_r)  \ \implies& \ \mathcal{E}_{u_n \pm a} = \mathcal{O}(M_r).
\end{align} 
This implies that the overall discrete entropy production scales as $M_r$, that is one order of magnitude above what is expected. This explains the accuracy degradation observed. \\
\indent With the flux-preconditioner of Turkel, we have $p = \mathcal{O}(M_r)$, $a_p, u_{np} = \mathcal{O}(1)$ and $K_1, K_2, K_1 - K_2 = \mathcal{O}(M_r)$, therefore:
\begin{equation*}
    \mu_{u_{np}\pm a_p} = \mathcal{O}(1) \implies   \mathcal{E}_{u_{np}\pm a_p} = \mathcal{O}(M_r^2).
\end{equation*}
That is the correct scaling. Hence the consistent behavior. \\
\indent With the preconditioner of Miczek, we have $p = \mathcal{O}(1/M_r)$, $a_p = \mathcal{O}(1/M_r^2)$, $K_2 = \mathcal{O}(M_r)$ and $K_1 = \mathcal{O}(1/M_r)$ because its denominator writes:
\begin{align*}
    M_r a_p - a p =& \ M_r \sqrt{(p^2+1) a^2/M_r^2 - p^2 u_n^2} - a p \\
    =& \ a p \bigg(\sqrt{1 + 1/p^2 - M_r^2 u_n^2 / a^2} - 1 \bigg) = \mathcal{O}(M_r).
\end{align*}
Therefore:
\begin{align*}
    \mu_{u_n+a_p} = \mathcal{O}(M_r^2), \ \mu_{u_n-a_p} = \mathcal{O}(M_r^4) \ &\implies \ 
    \mathcal{E}_{u_n+a_p}^S = \mathcal{O}(M_r^2), \ \mathcal{E}_{u_n-a_p}^S = \mathcal{O}(M_r^4), \\
    \delta_p = \mathcal{O}(1/M_r^4) & \implies \Delta \mathcal{E}_p = \mathcal{O}(M_r^2).
\end{align*}
Here again, the discrete entropy production has the correct scaling. \\ \\ 
\indent \textbf{Acoustic limit.} For the sound wave configuration, density, velocity and pressure gradients are of orders $M_r$, $1$ and $M_r$, respectively. The specific entropy is constant. At the discrete level, this translates into:
\begin{equation*}
    [\rho] = \mathcal{O}(M_r), [p] = \mathcal{O}(M_r), \ [u] = \mathcal{O}(1), \ [s] = 0  \implies \ \bigg[\frac{\rho k}{p}\bigg] = \mathcal{O}(1), \ \bigg[\frac{\rho u}{p}\bigg] = \mathcal{O}(1), \ \bigg[\frac{\rho}{p}\bigg] = \mathcal{O}(M_r).
\end{equation*}
For the classic ES upwind dissipation, this gives:
\begin{align}\label{LMES:Sound_ES}
    \mu_0 = \mathcal{O}(M_r^3) \ \ \mbox{and} \ \ \mu_{s1} = \mu_{s2} = \mu_{s3} = 0 \ \implies& \ \mathcal{E}_{u_n, s} = \mathcal{O}(M_r^{6}), \ \mathcal{E}_{u_n, \tau} = 0.  \\
    \mu_{u_n \pm a} = \mathcal{O}(M_r)  \ \implies& \ \mathcal{E}_{u_n \pm a} = \mathcal{O}(M_r).
\end{align} 
This implies that the overall discrete entropy production scales as $M_r$, in agreement with \ref{E2}. \\
\indent With Turkel's preconditioner, we have:
\begin{equation*}
    \mu_{u_{np}\pm a_p} = \mathcal{O}(1/M_r) \implies   \mathcal{E}_{u_{np}\pm a_p} = \mathcal{O}(1),
\end{equation*}
meaning that entropy fluctuations will be one order of magnitude stronger than what is expected. This explains why sound waves are severely damped with this flux. \\
\indent With Miczek's preconditioner, we have:
\begin{equation*}
    \mu_{u_n+a_p} = \mathcal{O}(M_r^2), \ \mu_{u_n-a_p} = \mathcal{O}(M_r^3) \ \implies \ 
    \mathcal{E}_{u_n+a_p}^S = \mathcal{O}(M_r^2), \ \mathcal{E}_{u_n-a_p}^S = \mathcal{O}(M_r^2), \ \Delta \mathcal{E}_p = \mathcal{O}(M_r)
\end{equation*}
The discrete entropy production is one order of magnitude weaker than what is expected. This can explain why the sound wave is less damped than with the ES Roe flux.

\subsection{Connections with other low-Mach fixes}
\indent Several alternatives to Flux-Preconditioning have been proposed, some of which are discussed in Guillard \& Nkonga \cite{Guillard2}. They can be broken down into two categories. There are corrections based on the idea that the excessive dissipation in the low-Mach limit is due to the acoustic eigenvalues scaling as $\mathcal{O}(1/M_r)$ and therefore becoming infinitely large. Li \& Gu \cite{LiGu1, LiGu2} introduced an  all-speed  Roe-type scheme where the eigenvalues are modified as:
\begin{equation*}
    u_n \pm (a / M_r) \ \rightarrow \  u_n \pm f(M_r) (a / M_r),
\end{equation*}
and $f(M_r)$ is a correction introduced so that $f(M_r) (a / M_r)$ is bounded in the low-Mach limit. This type of correction does not impede entropy-stability. \\
\indent The second kind of correction \cite{Rieper, Dellacherie, Osswald} consists in modifying the jump terms in the normal velocity $[u_n]$ (see also Thornber \textit{et al.} \cite{Thornber1, Thornber2}). By and large, they multiply $[u_n]$ by a correction term of order $M_r$. These fixes are motivated in part by the work of Birken \& Meister \cite{Birken}, who showed that the flux-preconditioner of Turkel enforce a more stringent (by a factor $M_r$) CFL condition (a similar result was proved by Barsukow \textit{et al.} \cite{Barsukow} for the flux-preconditioner of Miczek). \\
\indent For the ES Roe flux, the acoustic part $D^{A}[\mathbf{v}]$ of the dissipation operator writes:
\begin{equation}\label{LMES:Roe_acoustic_part}
    D^{A}[\mathbf{v}] = \begin{bmatrix} \mathbf{r}_{u_n+a} & \mathbf{r}_{u_n-a} \end{bmatrix}
    \begin{bmatrix} |u_n+(a/M_r)|  & 0 \\
                    0           & |u_n-(a/M_r)| \end{bmatrix}
    \begin{bmatrix} \mu_{u_n+a} \\ \mu_{u_n-a} \end{bmatrix}
\end{equation}
where
\begin{equation*}
    \mu_{u_n \pm a} = K_a \bigg(\mu_0 - h \bigg[ \frac{\rho}{p}\bigg] \pm M_r a \overline{\bigg(\frac{\rho}{p}\bigg)} [ u_n ]  \bigg).
\end{equation*}
Let $\tilde{\mu}_{u_n \pm a}$ be the wave strength obtained after applying the correction $[u_n] \rightarrow M_r [u_n]$:
\begin{equation*}
    \tilde{\mu}_{u_n \pm a} = K_a \bigg(\mu_0 - h \bigg[ \frac{\rho}{p}\bigg] \pm M_r^2 a \overline{\bigg(\frac{\rho}{p}\bigg)} [ u_n ]  \bigg).
\end{equation*}
The resulting acoustic entropy production field $\tilde{\mathcal{E}}^{A}$ becomes:
\begin{equation*}
    \tilde{\mathcal{E}}^{A} = [\mathbf{v}]^T D^{A}[\mathbf{v}] = |u_n + (a/M_r)| \mu_{u_n + a} \tilde{\mu}_{u_n + a} \ + \ |u_n - (a/M_r)| \mu_{u_n - a} \tilde{\mu}_{u_n - a}.
\end{equation*}
It is not clear whether the resulting operator leads to an ES flux as the sign of $\tilde{\mathcal{E}}^{A}$ is not clear, but we have $\tilde{\mathcal{E}}^{A} = \mathcal{O}(M_r^2)$ in the incompressible limit, in agreement with \ref{E1}.

\section{Discussion}
\label{LMES:sec:Discussion}

\subsection{The origin of the skew-symmetric term}
\label{LMES:sec:skew_origin}
\hspace*{0.1 cm} Given that the Miczek flux-preconditioner was constructed so that $P^{-1}|PA|$ has the same scaling as $A$, the appearance of a skew-symmetric term in the scaled form of the Miczek ES dissipation operator could be explained by examining the acoustic entropy production field \textit{without upwinding}, that is with $\Lambda$ instead of $|\Lambda|$. We have:
\begin{equation}\label{LMES:eq:ES_A}
    \mathcal{E}_{u_n+a} = \mu_{u_n + a}^2 (u_n+(a/M_r)), \ \mathcal{E}_{u_n-a} = \mu_{u_n - a}^2 (u_n-(a/M_r)).
\end{equation}
If we assume $\mu_{u_n+a}^2 \approx \mu_{u_n-a}^2 \approx \mu_{u_n + a} \mu_{u_n-a}$, then we can write something similar to (\ref{LMES:eq:dS_mic}):
\begin{equation}\label{LMES:eq:ES_B}
    \mathcal{E}_{u_n+a} \approx \mathcal{E}^{S}_{u_n+a} + \Delta \mathcal{E}_a, \ \mathcal{E}_{u_n-a} \approx \mathcal{E}^{S}_{u_n-a} - \Delta \mathcal{E}_a,
\end{equation}
where:
\begin{equation*}
    \mathcal{E}^{S}_{u_n+a} = \mu_{u_n + a}^2 u_n, \ \Delta \mathcal{E}_a = \mu_{u_n + a} \mu_{u_n - a} (a / M_r), \ \mathcal{E}^{S}_{u_n-a} = \mu_{u_n - a}^2 u_n.
\end{equation*}
As a matter of course, the resulting dissipation operator is no longer guaranteed to be ES. The point is that the entropy production balance between acoustic fields described by equation (\ref{LMES:eq:ES_B}) might be what the skew-symmetric matrix $\Delta_p$ of the ES Miczek flux tries to reproduce. Whether recovering this balance is key in ensuring a good low-Mach behavior is a different story. The numerical results (section 5) advise against it, at least at first-order.
\subsection{A simple equivalent to the ES Miczek flux}
\begin{figure}[htbp!]
    \centering
    \subfigure[$t = 1$]{\includegraphics[scale = 0.15]{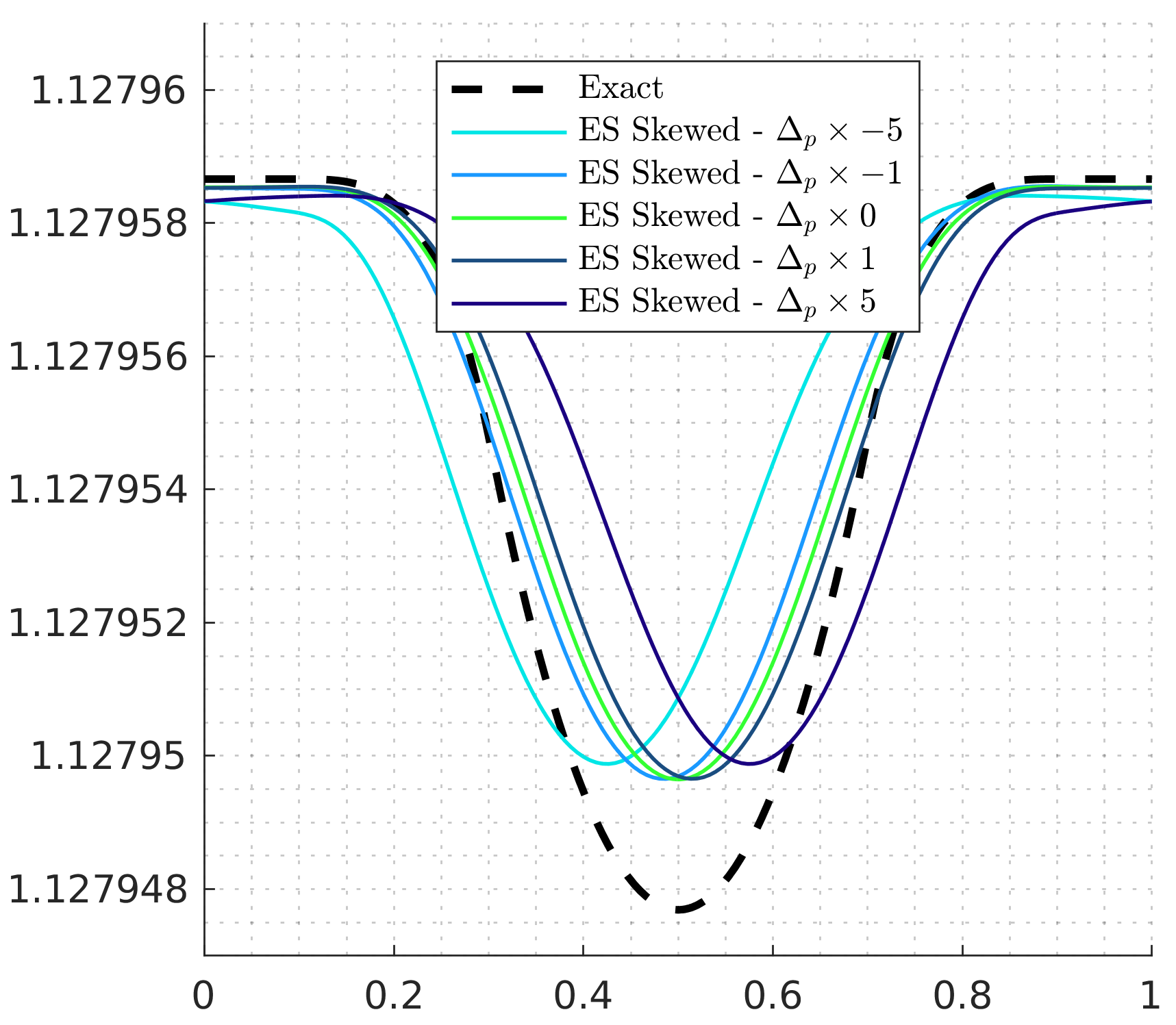}}
    \subfigure[$t = 0.04$]{\includegraphics[scale = 0.15]{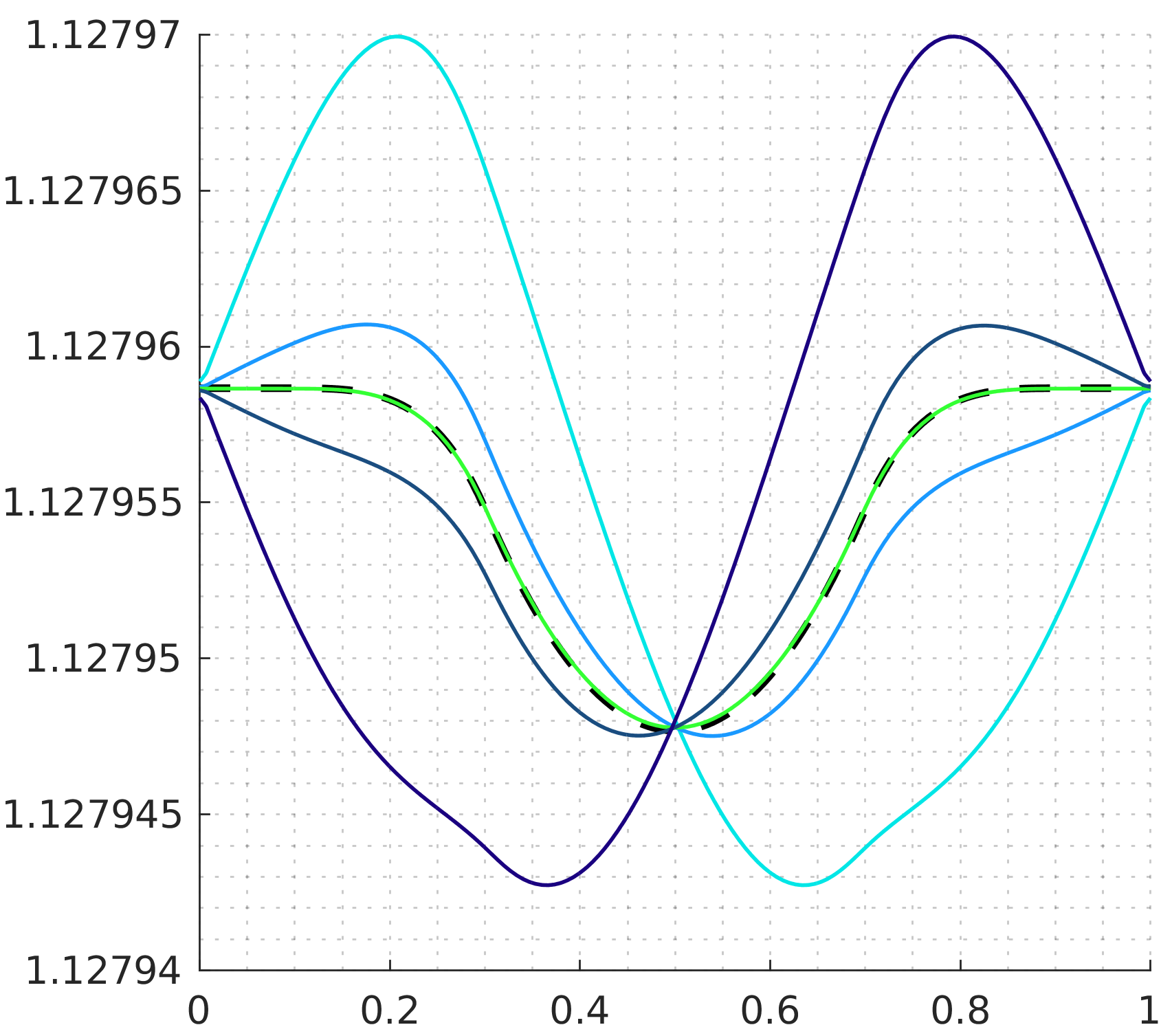}}
    \caption{Gresho Vortex: Pressure field at $M_r = 3 \times 10^{-3}$ for the skewed ES flux when the skew-symmetric term is multiplied by different factors. }
    \label{fig:Gresho_pressure_skew}
\end{figure}

\begin{figure}[htbp!]
    \centering
    \subfigure[$t = 0.03$]{\includegraphics[scale = 0.15]{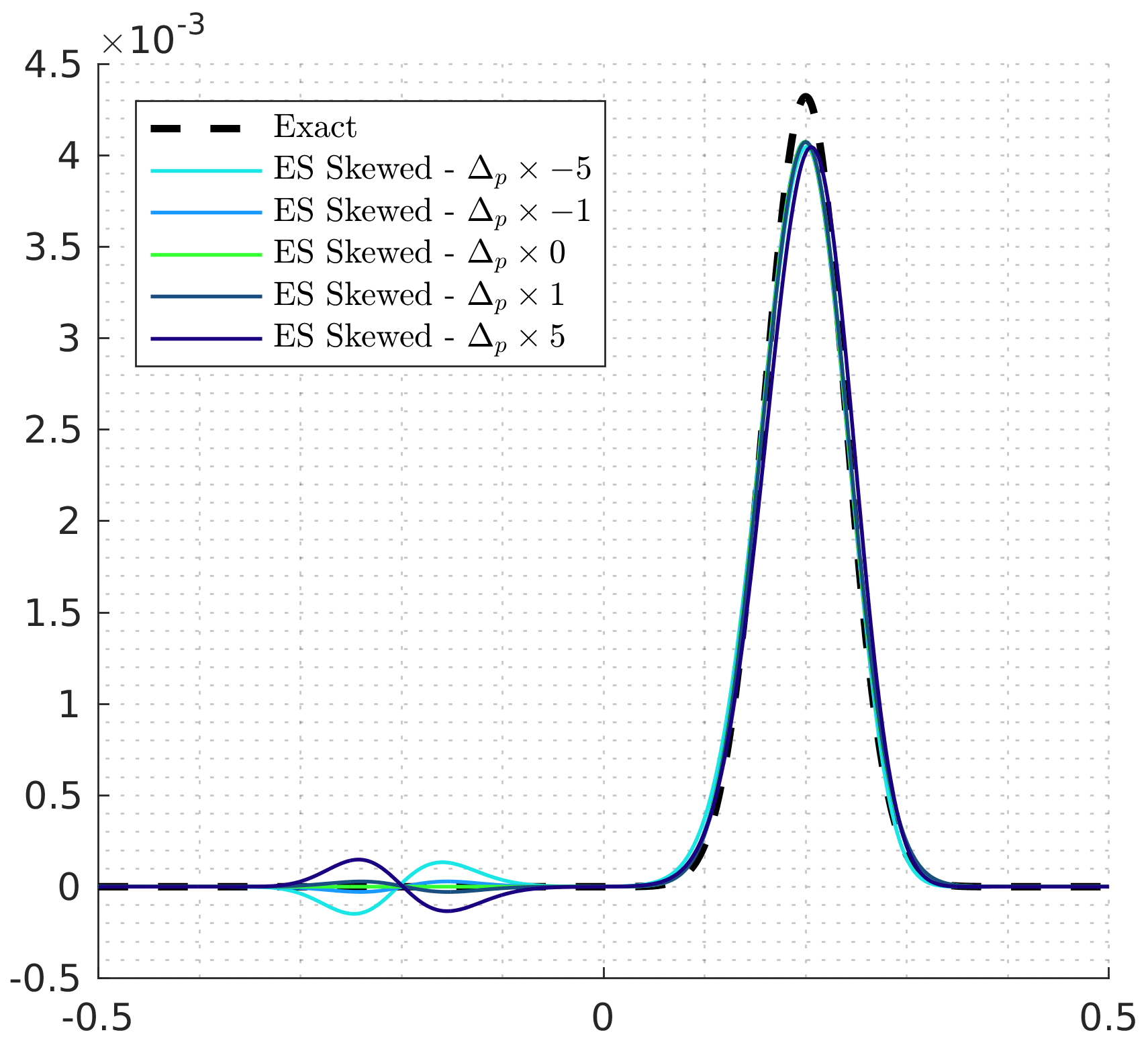}}
    \subfigure[$t = 0.03$, zoomed]{\includegraphics[scale = 0.15]{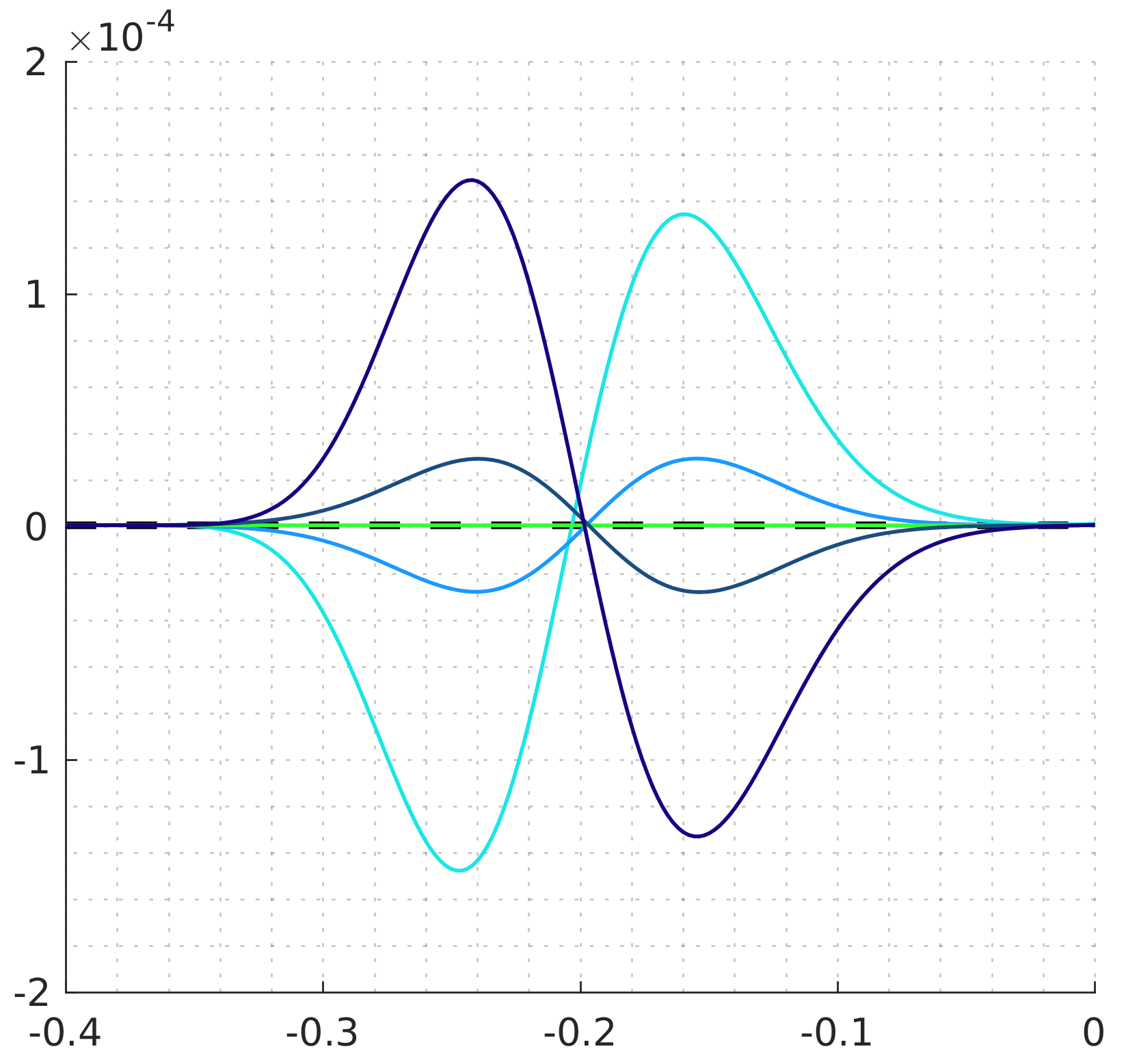}}
    \caption{Sound wave: Centerline pressure profiles with the skewed ES flux instead. $M_r = 10^{-2}$. }
    \label{fig:Sound_snaps_skew}
\end{figure}

\hspace*{0.1 cm} Consider the dissipation operator $ D_P = R (|\Lambda_p| + \Delta_p) R^T$ where $R$ is the scaled eigenvector matrix of the ES Roe flux (Section 6.1.) and:
\begin{gather}\label{LMES:dS_scheme}
   |\Lambda_p|  = \begin{bmatrix}
                    |u_n| &   0   &   0   &        0       &       0       \\
                      0   & |u_n| &   0   &        0       &       0       \\
                      0   &   0   & |u_n| &        0       &       0       \\
                      0   &   0   &   0   &  f_1 |u_n + (a/M_r)| &       0       \\
                      0   &   0   &   0   &        0       & f_2 |u_n - (a/M_r)|
                  \end{bmatrix}, \nonumber 
        \ \Delta_p = g
                  \begin{bmatrix}
                      0   &   0   &   0   &      0     &      0       \\
                      0   &   0   &   0   &      0     &      0       \\
                      0   &   0   &   0   &      0     &      0       \\
                      0   &   0   &   0   &      0     &  +\delta_p   \\
                      0   &   0   &   0   &  -\delta_p  &      0
                  \end{bmatrix}, 
\end{gather}
and $f_1, f_2$ and $g$ are functions of $M_r$. The eigenvectors $R$ are left untouched. Similarly to (\ref{LMES:eq:dS_mic}) and (\ref{LMES:eq:ES_B}), we have:
\begin{gather*}
    \mathcal{E}_{u_n-a} = \mathcal{E}_{u_n-a}^S - g \Delta \mathcal{E}_p, \  \mathcal{E}_{u_n+a} = \mathcal{E}_{u_n+a}^S +  g\Delta \mathcal{E}_p \\
   \mathcal{E}_{u_n-a}^S = \mu_{u_n-a}^2 f_2 |u_n-a|, \ \Delta \mathcal{E}_p = \delta_p \mu_{u_n-a} \mu_{u_n+a}, \ \mathcal{E}_{u_n + a}^S = \mu^2_{u_n+a} f_1 |u_n+a|).
\end{gather*}
This dissipation operator is ES as long as $f_1, f_2 \geq 0$, and we can emulate equation (\ref{LMES:eq:ES_B}) by taking $f_1 = |u_n| / |u_n + a/M_r|$, $f_2 = |u_n| / |u_n - a/M_r|$, $g = 1$ and $\delta_p = a/M_r$. This gives:
\begin{equation*}
    \mathcal{E}_{u_n-a}^S = \mu_{u_n-a}^2 |u_n|, \ \Delta \mathcal{E}_p = \mu_{u_n-a} \mu_{u_n+a} (a/M_r), \ \mathcal{E}_{u_n + a}^S = \mu^2_{u_n+a} |u_n|.
\end{equation*}
In the incompressible and acoustic limits, we have:
\begin{equation*}
    \mathcal{E}_{u_n-a}^S = \mathcal{O}(M_r^2), \ \Delta \mathcal{E}_p = \mathcal{O}(M_r), \ \mathcal{E}_{u_n + a}^S = \mathcal{O}(M_r^2),
\end{equation*}
which meets the low-Mach requirements \ref{E1} and \ref{E2}. This dissipation operator also meets Miczek's requirement that the dissipation matrix should have exactly the same Mach number scalings as $A$. Remarkably, we have:
\begin{equation*}
    R_{\mathbf{z}} (|\Lambda_p| + \Delta_p) R_{\mathbf{z}}^{-1} = \begin{bmatrix} 
            |u_n|   & -n_x a/M_r & -n_y a/M_r & -n_z a/M_r & 0 \\
            n_x a/M_r & |u_n|  & 0 & 0 & 0 \\
            n_y a/M_r & 0 & |u_n| & 0 & 0 \\
            n_z a/M_r & 0 & 0 & |u_n| & 0 \\
                0   & 0 & 0 &     0 & |u_n|
          \end{bmatrix}.
\end{equation*}
\indent To have this "skewed" ES dissipation operator return to Roe's as $M_r \rightarrow 1$, we can set $f_1 = M_r + (1-M_r)|u_n|/|u_n + (a/M_r)|$, $f_2 = M_r + (1-M_r)|u_n| / |u_n - (a/M_r)|$, and $g = 1-M_r$ for instance. Numerical results (figures  \ref{fig:Gresho_pressure_skew} and \ref{fig:Sound_snaps_skew}) show that this operator behaves just like the ES Miczek dissipation operator for both the Gresho vortex and the sound wave. The skewed ES dissipation operator (\ref{LMES:dS_scheme}) has not been introduced to compete with Miczek's flux or any of the aforementioned schemes (we remind the reader that the best results were observed with an EC flux in space). We introduced it to assess our intuition that skew-symmetric dissipation operators change the way entropy is locally produced. The skewed dissipation operator can be easily shown to induce a $\mathcal{O}(M_r^2)$ CFL condition, like the Turkel \cite{Birken} and Miczek \cite{Barsukow} dissipation operators (see appendix C in \cite{Gouasmi_Thesis}).

\subsection{Different ways of breaking down discrete entropy production}
\begin{figure}[htbp!]
    \centering
    \includegraphics[scale = 0.17]{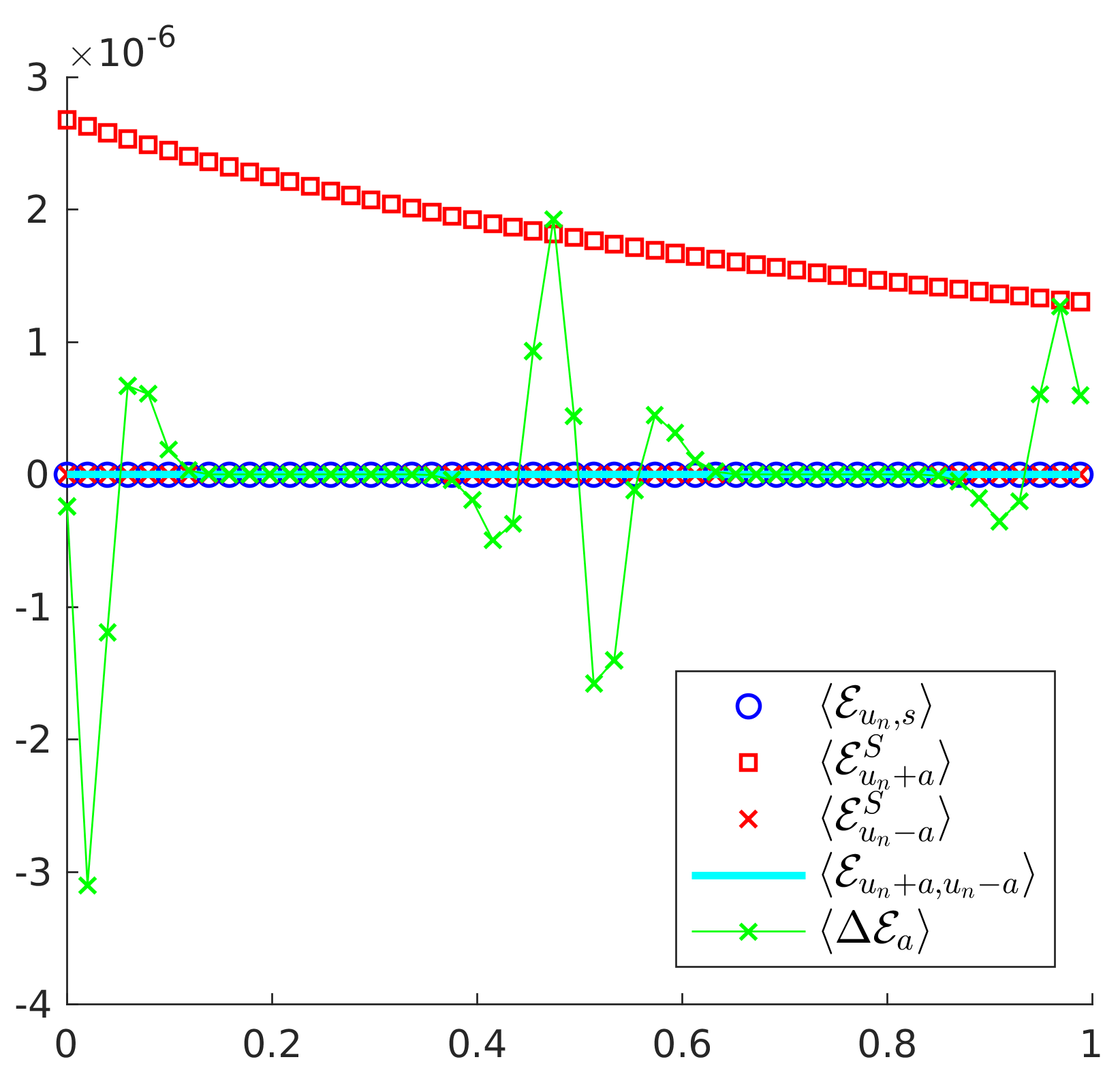}
    \caption{Sound wave: Integral of entropy production fields along the acoustic eigenvectors with time at $M_r = 10^{-2}$ for the ES Miczek flux.}
    \label{fig:Sound_Mic_global_dS2}
\end{figure}

\indent It is important to recognize that the EPB we introduced in section \ref{sec:entropy_prod} is not unique. Consider a dissipation operator consisting of two linearly independent modes $R = [\mathbf{r_1}, \ \mathbf{r_2} ]$. We have:
\begin{equation*}
    R |\Lambda| R^T [\mathbf{v}] = |\lambda_1| \mu_1 \mathbf{r_1} + |\lambda_2| \mu_2 \mathbf{r_2} \ \implies \ \mathcal{E} = [\mathbf{v}]^T R |\Lambda| R^T [\mathbf{v}] = \mathcal{E}_{1} + \mathcal{E}_2.
\end{equation*}
Let $\overline{R} = \{\mathbf{\bar{r}_1}, \ \mathbf{\bar{r}_2}\}$ be an alternative pair of linearly independent modes defined by the mapping:
\begin{equation*}
    \begin{cases}
    \mathbf{r_1} = l_{11} \mathbf{\bar{r}_1} + l_{12} \mathbf{\bar{r}_2}, \\
    \mathbf{r_2} = l_{21} \mathbf{\bar{r}_1} + l_{22} \mathbf{\bar{r}_2}.
    \end{cases}
    \ \implies \
    \begin{cases}
    \mu_1 = l_{11} \bar{\mu}_1 + l_{12} \bar{\mu}_2, \\
    \mu_2 = l_{21} \bar{\mu}_1 + l_{22} \bar{\mu}_2.
    \end{cases}
\end{equation*}
Then the dissipation operator can be expressed in terms of these modes:
\begin{equation*}
    R |\Lambda| R^T [\mathbf{v}] 
        =  \big( L_{11}\bar{\mu}_1 + L_{12} \bar{\mu}_2 \big) \mathbf{\bar{r}_1} + \big( L_{22}\bar{\mu}_2 + L_{21} \bar{\mu}_1 \big) \mathbf{\bar{r}_2},
\end{equation*}
with:
\begin{gather*}
    L_{11} = (|\lambda_1| l_{11}^2 + |\lambda_2| l_{21}^2) > 0, \ 
        L_{22} = (|\lambda_1| l_{12}^2 + |\lambda_2| l_{22}^2) > 0, \
        L_{12} = L_{21} = (|\lambda_1| l_{11}l_{12} + |\lambda_2| l_{21}l_{22}).
\end{gather*}
This gives the EPB:
\begin{equation}\label{eq:dS_alter}
    \mathcal{E} \ = \overline{\mathcal{E}}_1 + \overline{\mathcal{E}}_2, \ \overline{\mathcal{E}_1} = L_{11} \overline{\mu}_1^2 + L_{12} \overline{\mu}_1 \overline{\mu}_2, \ \overline{\mathcal{E}_2} = L_{22} \overline{\mu}_2^2 + L_{21} \overline{\mu}_1 \overline{\mu}_2
\end{equation}
Unlike the initial decomposition, each individual field $\overline{\mathcal{E}}_i$ is no longer guaranteed to be positive. This does not matter much given that the sign of $\mathcal{E}$ will not change. We also see that depending on the choice of modes, coupling terms within each field $\overline{\mathcal{E}}_i$ may appear. We can write:
\begin{equation}\label{eq:dS_alter2}
    \overline{\mathcal{E}}_k = \overline{\mathcal{E}}^S_k + \sum_{k' \neq k}  \overline{\mathcal{E}}_{kk'}, \ \overline{\mathcal{E}}^S_k = L_{kk}\bar{\mu}_k^2, \ \overline{\mathcal{E}}_{kk'} = L_{kk'}\bar{\mu}_k \bar{\mu}_{k'}.
\end{equation}
Note that the coupling terms do not cancel each other, i.e. $\overline{\mathcal{E}}_{kk'} + \overline{\mathcal{E}}_{k'k} = 0, \ k \neq k'$. In fact, they are equal $\overline{\mathcal{E}}_{kk'} = \overline{\mathcal{E}}_{k'k}$. \\
\indent If the dissipation operator has a skew-symmetric component:
\begin{equation*}
    R\Delta R^{T}[\mathbf{v}] = \delta (\mu_2 \mathbf{r_1} - \mu_1 \mathbf{r_2}) \implies [\mathbf{v}]^T R \Delta R^T [\mathbf{v}] = \Delta \mathcal{E} - \Delta\mathcal{E} = 0, \ \Delta \mathcal{E} = \delta \mu_{1}\mu_{2},
\end{equation*}
we can rewrite it in terms of $\{\mathbf{\bar{r}_1}, \ \mathbf{\bar{r}_2}\}$ instead. Remarkably, we have:
\begin{align*}
    R \Delta R^T [\mathbf{v}] \ = \ \delta (\mu_2 \mathbf{r_1} - \mu_1 \mathbf{r_2}) \ = \ \delta (l_{11}l_{22} - l_{12}l_{21}) \big( \bar{\mu}_2\mathbf{\bar{r}_1} - \bar{\mu}_1 \mathbf{\bar{r}_2} \big) \ = \ \overline{\delta}  \big( \bar{\mu}_2\mathbf{\bar{r}_1} - \bar{\mu}_1 \mathbf{\bar{r}_2} \big)
\end{align*}
The mapping is one-to-one hence $l_{11}l_{22} - l_{12}l_{21} \neq 0$. This shows that the skew-symmetric operator $R\Delta R^{T}[\mathbf{v}]$ (entropy transfer between modes $\{ \mathbf{r_1}, \ \mathbf{r_2} \}$) is equivalent to a skew-symmetric operator $\overline{R} \ \overline{\Delta} \ \overline{R}^T [\mathbf{v}]$ (entropy transfer between $\{ \mathbf{\overline{r}_1}, \ \mathbf{\overline{r}_2} \}$):
\begin{equation*}
    [\mathbf{v}]^T \overline{R} \ \overline{\Delta} \ \overline{R}^T [\mathbf{v}] = \overline{\Delta \mathcal{E}} - \overline{\Delta \mathcal{E}} = 0, \ \overline{\Delta \mathcal{E}} = \overline{\delta} \overline{\mu}_1 \overline{\mu}_2. 
\end{equation*}
\indent Using the above algebra, we can now introduce EPBs for the ES Turkel and ES Miczek fluxes in terms of the original acoustic eigenvectors instead of the modified ones. For Turkel's flux-preconditioner, we map from $\{\mathbf{r_1}, \ \mathbf{r_2}\} = \{\mathbf{r}_{u_{np}+a_p}, \ \mathbf{r}_{u_{np}-a_p}\}$ to $\{\mathbf{\overline{r}_1}, \ \mathbf{\overline{r}_2}\} = \{\mathbf{r}_{u_{n}+a}, \ \mathbf{r}_{u_{n}-a}\}$ using:
\begin{gather*}
    \mathbf{r}_{u_{np}+a_p} = \frac{1}{\sqrt{2}p^2(K_1 - K_2)} \bigg( (K_1 + p^2) \mathbf{r}_{u_n + a} + (K_1 - p^2) \mathbf{r}_{u_n - a} \bigg), \\ \mathbf{r}_{u_{np}-a_p} = \frac{1}{\sqrt{2}p^2(K_1 - K_2)} \bigg( (K_2 + p^2) \mathbf{r}_{u_n + a} + (K_2 - p^2) \mathbf{r}_{u_n - a} \bigg).
\end{gather*}
The new decomposition (\ref{eq:dS_alter})-(\ref{eq:dS_alter2}) (to contrast with (\ref{LMES:dS_Turkel})) writes:
\begin{equation}\label{eq:dS_Turkel_alter}
    \mathcal{E}_p \ = \ \big(\mathcal{E}^S_{u_{n}+a} + \mathcal{E}_{u_{n}+a,u_n-a}\big) + \big(\mathcal{E}^S_{u_{n}-a} + \mathcal{E}_{u_{n}-a, u_n+a}\big).
\end{equation}
Figures \ref{fig:Gresho_Tur_dS2} and \ref{fig:Sound_Tur_dS2} show the initial entropy production fields for the Gresho vortex and the sound wave using the decomposition (\ref{eq:dS_Turkel_alter}). These entropy production fields are more similar to those of the classic ES Roe flux (figures \ref{fig:Gresho_Roe_dS} and \ref{fig:Sound_Roe_dS}) than the ones along the modified acoustic eigenvectors (figures \ref{fig:Gresho_Tur_dS} and \ref{fig:Sound_Tur_dS}). We also see that the coupling term $\mathcal{E}_{u_n+a,u_n-a}$ can be negative. \\
\indent For Miczek's flux-preconditioner, modified and original acoustic eigenvectors satisfy:
\begin{align*}
    \mathbf{r}_{u_n+a_p} =& \ \frac{1}{\sqrt{2}(p^2+1)(K_1-K_2)}\bigg( \big(K_1(1 + p) + 1 - p\big) \mathbf{r}_{u_n+a} + \big(K_1(1-p) - (1+p)\big) \mathbf{r}_{u_n-a}\bigg), \\
    \mathbf{r}_{u_n-a_p} =& \ \frac{1}{\sqrt{2}(p^2+1)(K_1-K_2)}\bigg( \big(K_2(1 + p) + 1 - p\big) \mathbf{r}_{u_n+a} + \big(K_2(1-p) - (1+p)\big) \mathbf{r}_{u_n-a}\bigg).
\end{align*}
The new decomposition (\ref{eq:dS_alter})-(\ref{eq:dS_alter2}) writes:
\begin{equation}\label{LMES:eq:dS_mic_alter}
    \mathcal{E}_p \  = \  \big(\mathcal{E}^S_{u_{n}+a} \ + \ \mathcal{E}_{u_{n}+a,u_n-a} \ - \ \Delta \mathcal{E}_a \big) \ + \  \big(\mathcal{E}^S_{u_{n}-a} \ + \ \mathcal{E}_{u_{n}-a, u_n+a} \ + \ \Delta \mathcal{E}_a\big).
\end{equation}
Figures \ref{fig:Gresho_Mic_dS2} and \ref{fig:Sound_Mic_dS2} show the initial entropy production fields for the Gresho vortex and the sound wave using EPB (\ref{LMES:eq:dS_mic_alter}). The coupling term $\mathcal{E}_{u_n+a,u_n-a}$ is negligible. For the sound wave, figure \ref{fig:Sound_Mic_global_dS2} shows the integrated entropy production fields according to (\ref{LMES:eq:dS_mic_alter}). The culpability of the skew-symmetric part of the ES Miczek flux in the spurious transient is striking : the skew-symmetric contribution $\Delta \mathcal{E}_a$ kicks in around $t \in \{0, 0.5, 1\}$. This could not be seen with the previous EPB (figure \ref{fig:Sound_Mic_global_dS}).

\subsection{An Incomplete Picture}
\begin{figure}[htbp!]
    \centering
    \subfigure[$t = 1$]{\includegraphics[scale = 0.15]{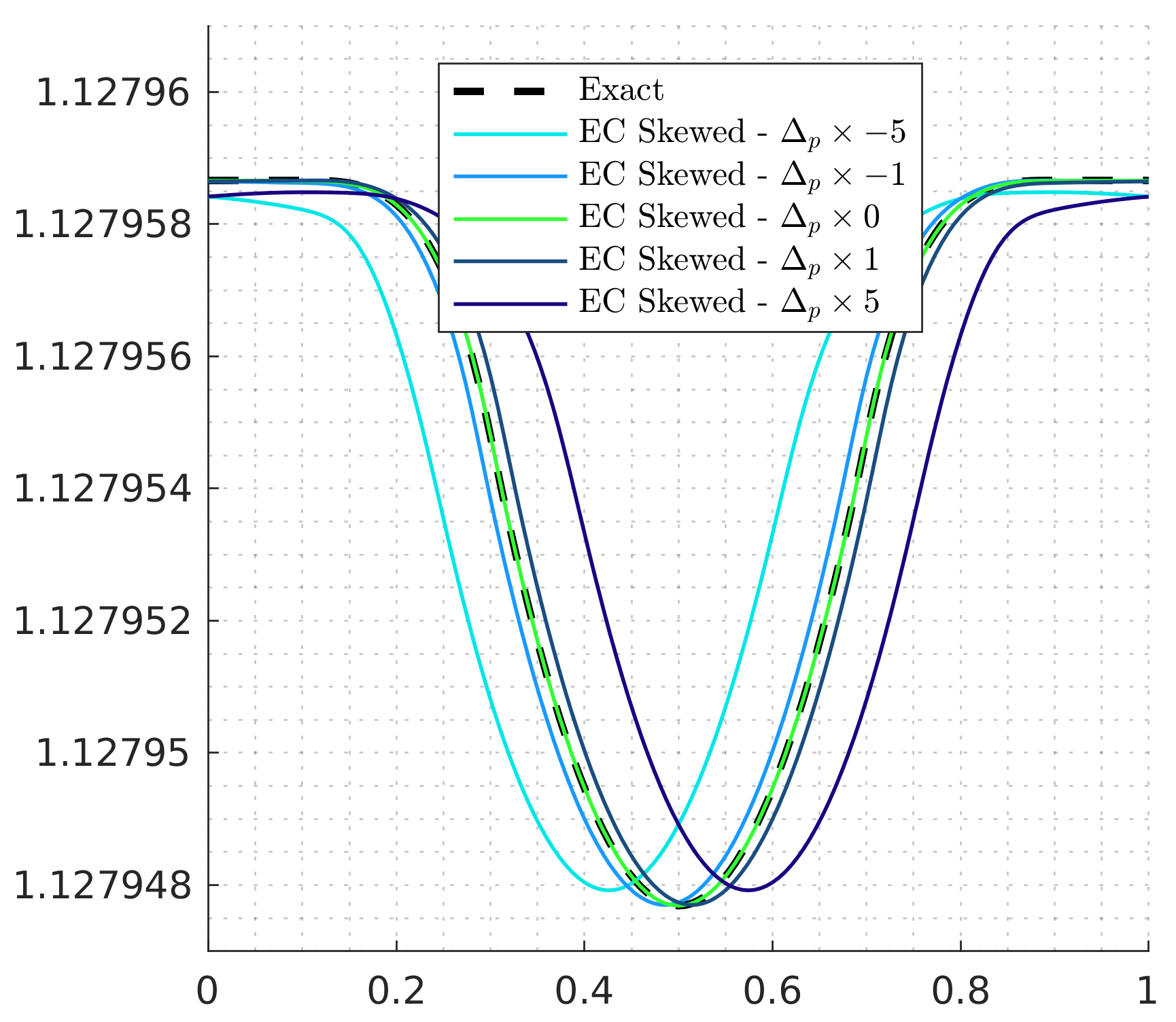}}
    \subfigure[$t = 0.04$]{\includegraphics[scale = 0.15]{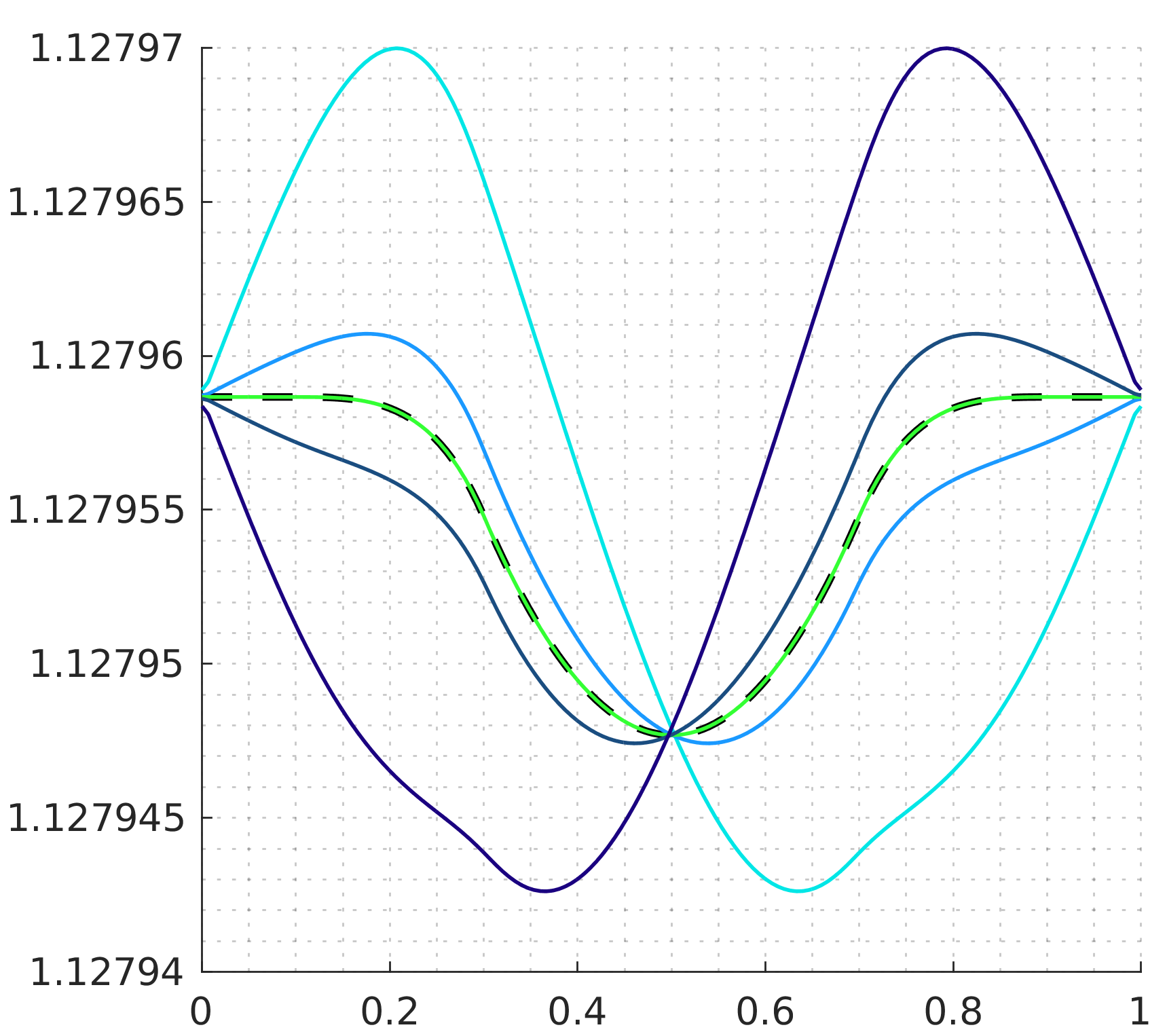}}
    \caption{Gresho Vortex: Centerline pressure profile with the skewed EC flux. $M_r = 3 \times 10^{-3}$.}
    \label{fig:Gresho_pressure_Mic_skewEC}
\end{figure}

\begin{figure}[htbp!]
    \centering
    \subfigure[$t = 0.03$]{\includegraphics[scale = 0.15]{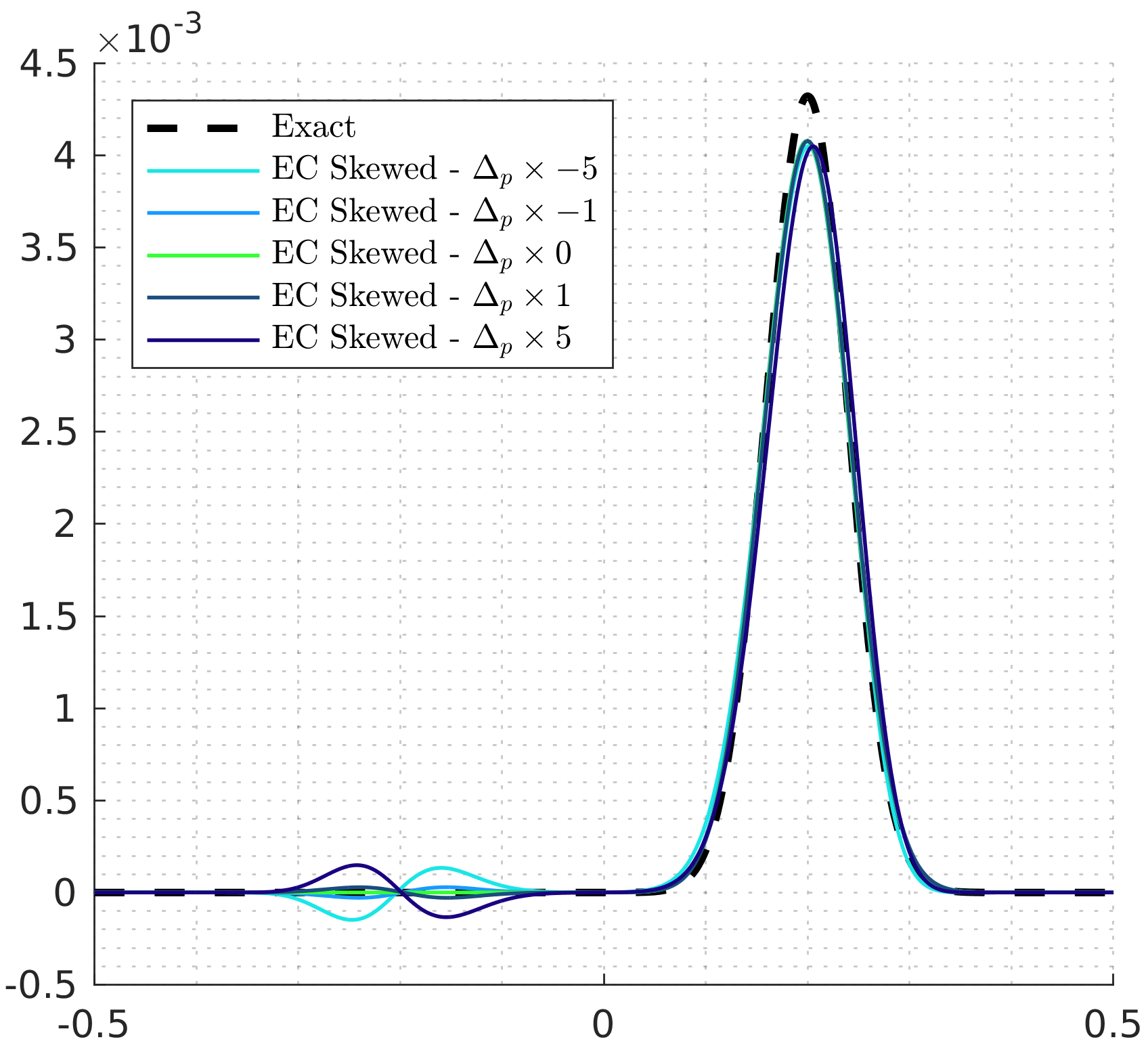}}
    \subfigure[$t = 0.03$, zoomed]{\includegraphics[scale = 0.15]{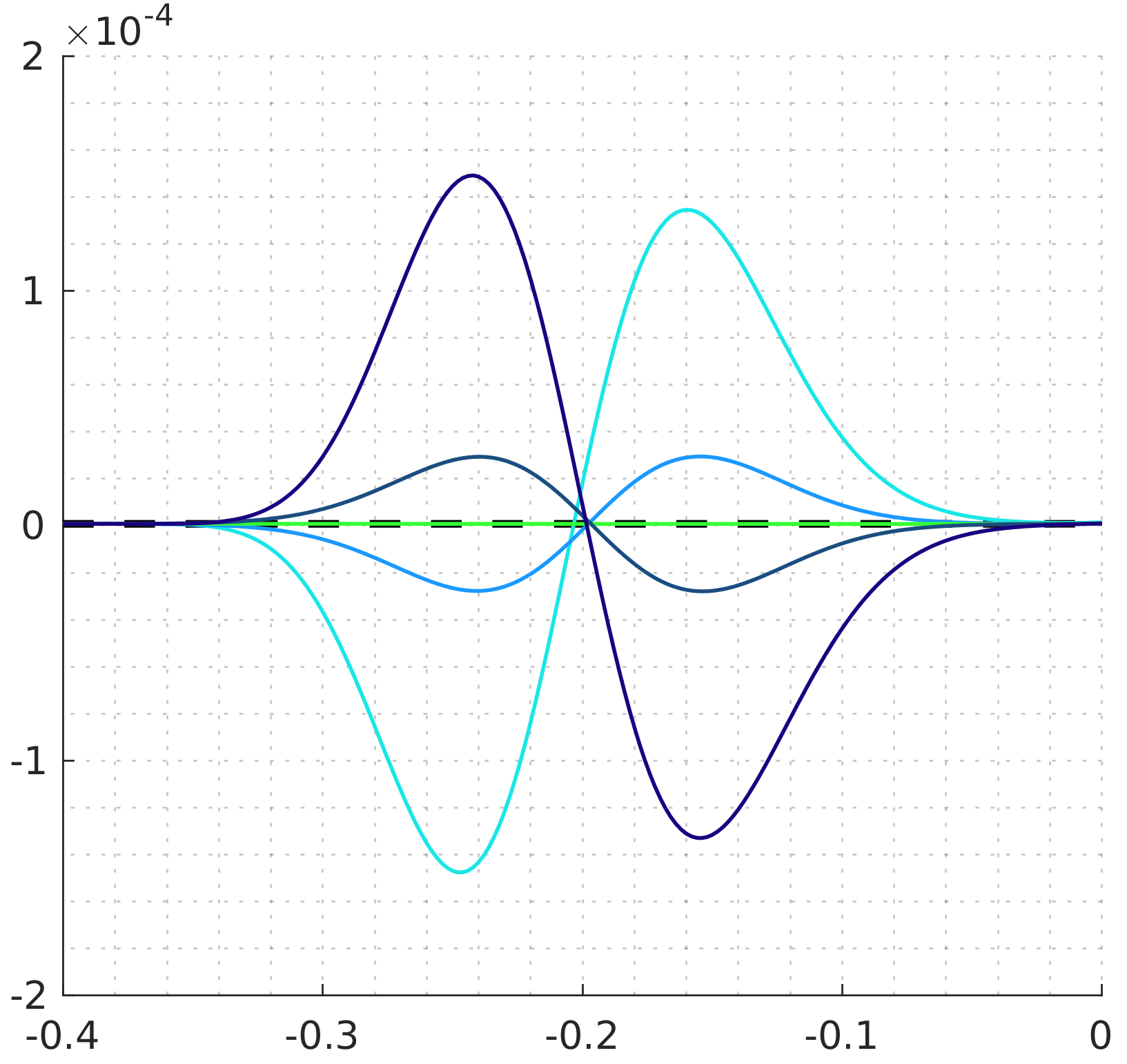}}
    \caption{Sound wave: Same pressure profiles as in figure \ref{fig:Sound_snaps_Mic} with the skewed EC flux instead. $M_r = 10^{-2}$.}
    \label{fig:Sound_snaps_skewEC}
\end{figure}

\indent In an unpublished manuscript, Roe \cite{ES_Roe2} made the following observation regarding EC fluxes:
\begin{corollary}[Roe \cite{ES_Roe2}]
Let $\mathbf{f}_{EC}^{*}$ an EC flux following theorem (\ref{th:Tadmor_EC}). Then for any skew-symmetric matrix $\Delta$, we have that the numerical flux $\mathbf{f^{*}}$ defined by:
\begin{equation}\label{eq:EC_flux_skewed}
    \mathbf{f^{*}} \ = \ \mathbf{f}_{EC}^{*} \ - \ \Delta [\mathbf{v}],
\end{equation}
is EC as well. 
\end{corollary}\label{Roe_EC}
\noindent This result is a simple consequence of $[\mathbf{v}]^{T} \Delta [\mathbf{v}] = 0$. While $\Delta$ does not contribute to the total discrete entropy production, it still has an impact on the discrete entropy equation (\ref{eq:FVM_entropy}) through the entropy flux $F^{*}$. It is easy to show that the discrete entropy flux writes:
\begin{equation*}
    F^{*} = \overline{\mathbf{v}} \cdot \big( \mathbf{f^{*}} - \Delta [\mathbf{v}] \big) - \overline{\mathcal{F}}.
\end{equation*}
The contribution from $\Delta$ is non-zero. We ran the skewed ES flux of section 7.2 \textit{without $R|\Lambda_p|R^{T}[\mathbf{v}]$} (this leaves a skewed EC flux (\ref{eq:EC_flux_skewed})) and observed the same anomalies (figures \ref{fig:Gresho_pressure_Mic_skewEC} and \ref{fig:Sound_snaps_skewEC}). This shows that the picture drawn by EPBs is not complete, and that perhaps we should take a few steps back and try to better understand discrete entropy conservation first. \\
\indent In section 4.2 we got a glimpse of how diverse EC fluxes can be. EC fluxes such as Roe's (\ref{eq:EC_Roe}) or Chandrasekhar's (\ref{eq:EC_Chandra}) are popular because of their algebraic simplicity but their local behavior is not easy to analyze, at least analytically. Given a dissipation operator $R|\Lambda|R^{T}[\mathbf{v}] = \sum_{k=1}^{N} \mathbf{r}_k |\Lambda_k| \mu_k$, one could look for an EC flux $\mathbf{f^*}$ of the form:
\begin{equation*}
    \mathbf{f^{*}} = \sum_{k=1}^{N} f_k \mathbf{r}_{k} \ \implies \ [\mathbf{v}] \cdot \mathbf{f^{*}} = \sum_{k=1}^{N} f_k \mu_{k}
\end{equation*}
Following theorem \ref{th:Tadmor_EC}, the implied sum above has to be equal to the jump $[\mathcal{F}]$. If $[\mathcal{F}]$ can be broken down as the sum of $N$ independent intermediate jumps $[\mathcal{F}]^{k}$, then one can solve for the $f_k^{*}$ and obtain the candidate EC flux:
\begin{equation}\label{eq:EC_Tadmor1}
    f_{k} \ = \ \frac{[\mathcal{F}]^{k}}{\mu_k} \ \implies \ \mathbf{f^{*}} \ = \ \sum_{k=1}^{N} \bigg(\frac{[\mathcal{F}]^{k}}{\mu_k} \bigg) \ \mathbf{r}_k.
\end{equation}
This candidate flux qualifies if it is consistent. This kind of \textit{modal} EC flux was already introduced by Tadmor \cite{ES_Tadmor_2003} about two decades ago. He set $[\mathcal{F}]^{k} = \mathcal{F}(\mathbf{v}^{*,k+1}) - \mathcal{F}(\mathbf{v}^{*,k})$ with an appropriately designed sequence $\big(\mathbf{v}^{*,k}\big)_{1\leq k \leq N+1}$ satisfying $\mathbf{v}^{*,1} = \mathbf{v}_L$, $\mathbf{v}^{*,N+1} = \mathbf{v}_R$. This modal representation might lead to useful insights into the local behavior of EC fluxes. One could for instance compare the magnitudes of the $f_k$ coefficients and use skew-symmetric interface operators $R \Delta R^{T}[\mathbf{v}]$ to alter the balance $ \ \sum_{k=1}^{N} \mu_k f_k = [\mathcal{F}] \ $ and see how it effects the discrete solution. \\
\indent When considering the anomalies associated with the ES Miczek flux in the second test problem, we talked about entropy transfer between (acoustic) \textit{waves}. It is important to recognize that for nonlinear flow configurations involving significant discontinuities and/or multi-dimensional physics, the notion of a wave become ambiguous. One can argue that the EPBs we introduced and worked with are inherently one-dimensional. \\
\indent Regarding the effect of the temporal discretization on discrete entropy dynamics, we could introduce a temporal EPB starting from $\mathcal{E}^{BE}$ (eq. (\ref{eq:entropy_prod_BE}) - theorem \ref{th:Tadmor_BE}) and using $H = RR^T$ (corollary \ref{Barth_cor}) but computing the fields accurately requires quadrature. In addition, they cannot be computed \textit{a priori}. Elements of discussion regarding the effect of temporal discretization can be found in \cite{Gouasmi_0, Gouasmi_1}.

\section{Conclusions}
\indent In this work, the behavior of ES schemes in the low-Mach regime was investigated. We showed that standard ES schemes suffer from the same accuracy degradation issues as standard upwind schemes (and for the same reasons). Using appropriate similarity and congruence transforms, we were able to define the extent to which the flux-preconditioning technique is compatible with entropy-stability. We introduced ES versions of the preconditioned upwind fluxes of Turkel and Miczek. Numerical results confirmed the analysis but also highlighted spurious transients with the flux-preconditioner of Miczek which were not reported until now. \\
\indent These unexpected anomalies, together with the recent work of Bruel \textit{et al.} \cite{Bruel} on the acoustic limit and the failure of the Turkel flux-preconditioner, led us to further investigate the matter. Leveraging Tadmor’s framework, we introduced discrete Entropy Production Breakdowns (EPBs) that allowed us to revisit the accuracy degradation issue in terms of entropy. In the same spirit as Guillard \& Viozat \cite{Guillard1} (incompressible limit) and Bruel \textit{et al.} \cite{Bruel} (acoustic limit), we showed that the accuracy degradation problems at the discrete level are caused by discrete entropy fluctuations that are inconsistent with those of the continuous system. \\
\indent An important outgrowth of the overall effort is the discovery that the spurious transients observed with the ES Miczek flux are caused by a skew-symmetric dissipation term which appeared when a scaled form $R |\Lambda|R^{T}[\mathbf{v}]$ of the preconditioned dissipation operator was sought. Analytical and numerical arguments suggest that this term induces entropy transfers between acoustic waves. While the role played by skew-symmetric terms and the scope of EPBs remain to be fully understood, we believe these findings shed new lights on the local behavior of EC/ES schemes and how to further improve them. These findings should also, hopefully, convince the reader that \textit{there is more to draw from EC/ES schemes than a global stability statement} such as inequality (\ref{eq:entropy_ineq}). \\ 
\indent Future work will continue the analysis in a more complex setting, including unstructured grids \cite{Bader}, high-order discretizations and mixed flow configurations \cite{Thornber2} of practical interest. The challenges associated with efficient time-integration and preconditioning (stiffness and steady-state convergence \cite{Turkel2005}) will bring yet another layer of difficulty to that effort.

\section*{Acknowledgments}
Ayoub Gouasmi and Karthik Duraisamy were funded by AFOSR through grant number FA9550-16-1-030 (Tech. monitor: Fariba Fahroo). Support from the NASA Space Technology Mission Directorate (STMD) through the Entry Systems Modeling (ESM) project and from the NASA High-End Computing (HEC) Program are gratefully acknowledged. \\ 
\indent Ayoub Gouasmi would like to thank Laslo Diosady, Philip Roe and Eitan Tadmor for productive conversations, and Eli Turkel for important clarifications regarding a proof in his work \cite{Turkel0} and the benefits of congruence transformations.

\appendix 

\section{Entropy Production Field Snapshots}
\label{app:EPF}
\begin{figure}[htbp!]
    \centering
    \subfigure[$\hat{\mathcal{E}}^S_{u_{n}+a}$]{\includegraphics[scale = 0.33]{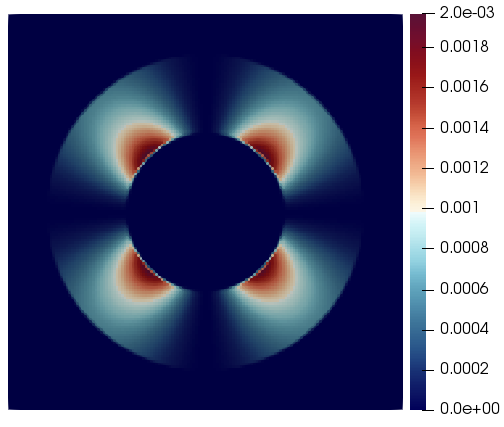}}
    \subfigure[$\hat{\mathcal{E}}^S_{u_{n}-a}$]{\includegraphics[scale = 0.33]{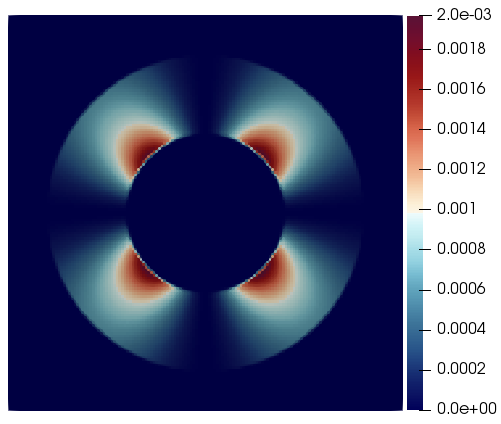}}
    \subfigure[$\hat{\mathcal{E}}_{u_{n}-a,u_n+a}$]{\includegraphics[scale = 0.33]{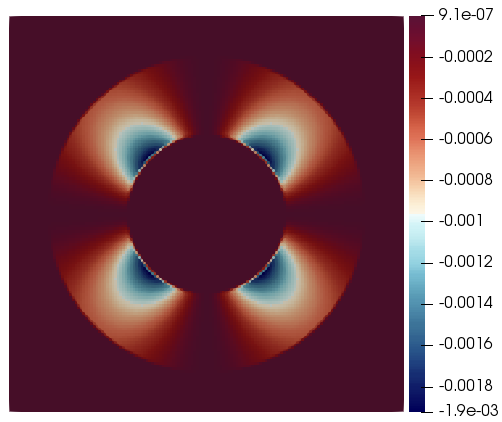}}
    \caption{Gresho Vortex: Entropy production fields along the original acoustic eigenvectors at $t = 0$ for the ES Turkel flux. $M_r = 3 \times 10^{-2}$.}
    \label{fig:Gresho_Tur_dS2}
\end{figure}
\begin{figure}[htbp!]
    \centering
    \subfigure[$\hat{\mathcal{E}}^S_{u_{n}+a}$]{\includegraphics[scale = 0.35]{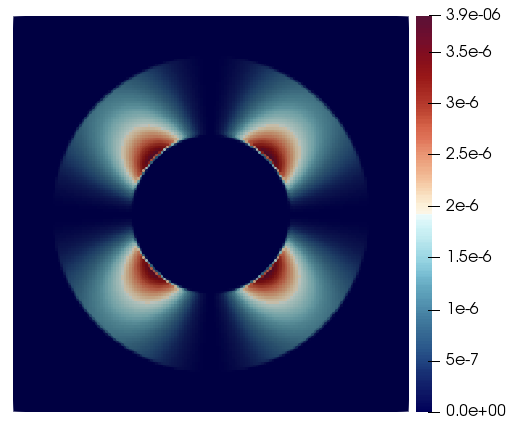}}
    \subfigure[$\hat{\mathcal{E}}^S_{u_{n}-a}$]{\includegraphics[scale = 0.35]{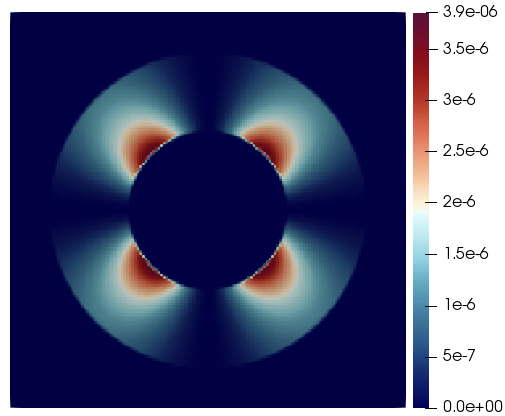}}
    \subfigure[$\hat{\mathcal{E}}_{u_{n}-a,u_n+a}$]{\includegraphics[scale = 0.35]{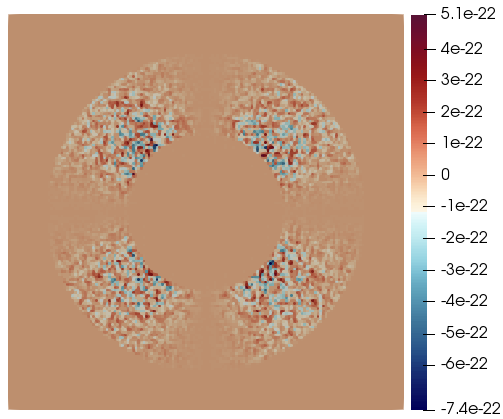}}
    \subfigure[$\Delta \hat{\mathcal{E}}_{a}$]{\includegraphics[scale = 0.35]{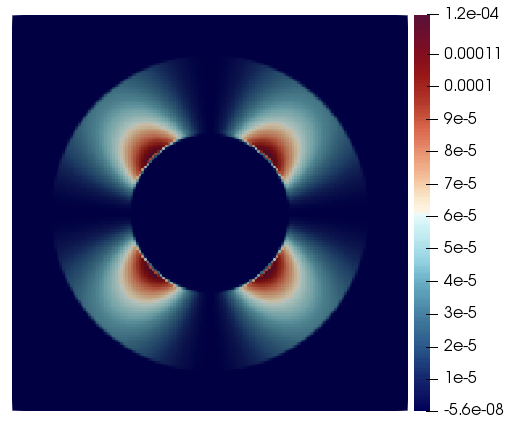}}
    \caption{Gresho Vortex: Entropy production fields along the original acoustic eigenvectors at $t = 0$ for the ES Miczek flux. $M_r = 3 \times 10^{-2}$.}
    \label{fig:Gresho_Mic_dS2}
\end{figure}
\begin{figure}[htbp!]
    \centering
    \subfigure[$\hat{\mathcal{E}}^S_{u_{n}+a}$]{\includegraphics[scale = 0.10]{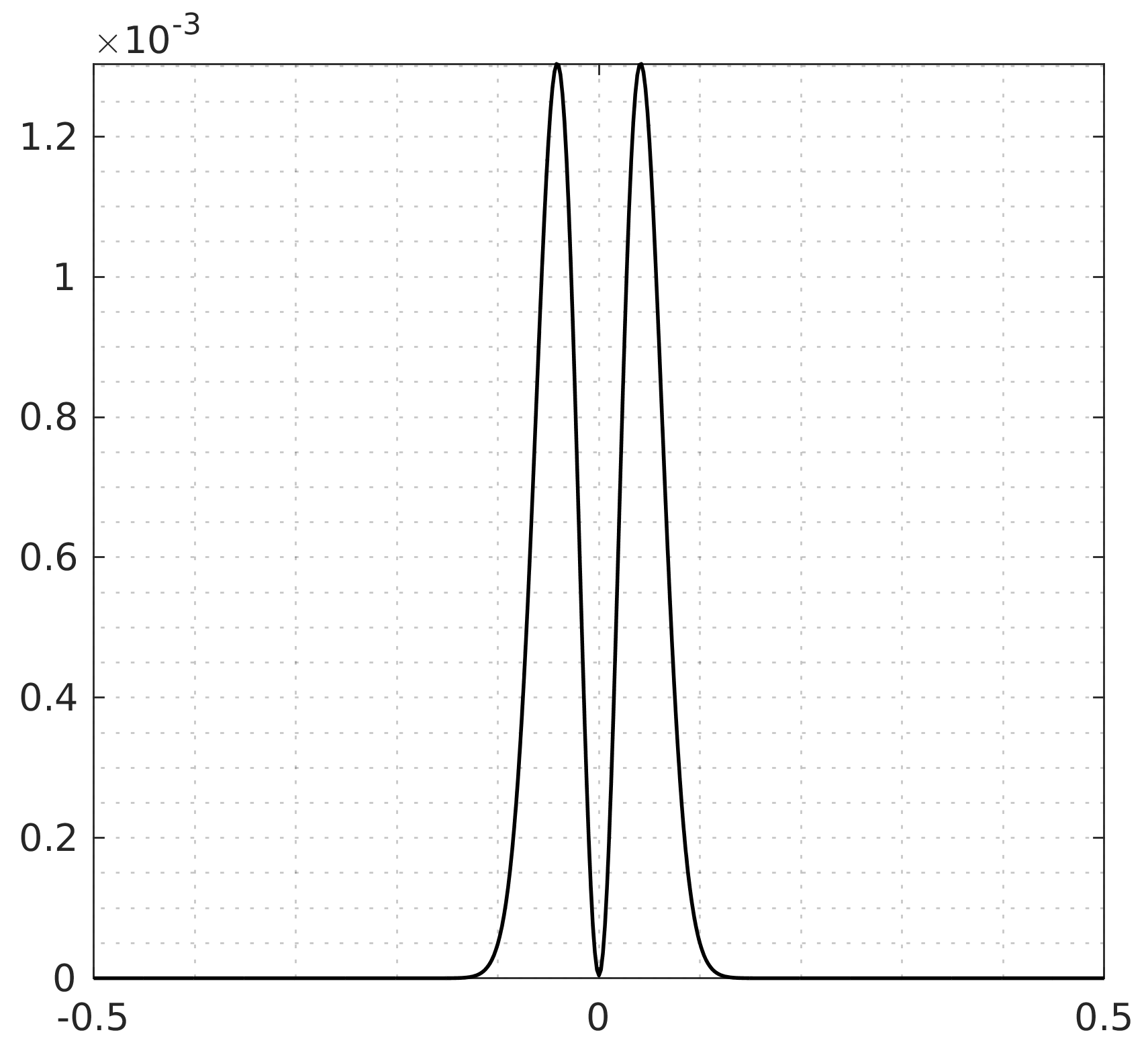}}
    \subfigure[$\hat{\mathcal{E}}^S_{u_{n}-a}$]{\includegraphics[scale = 0.10]{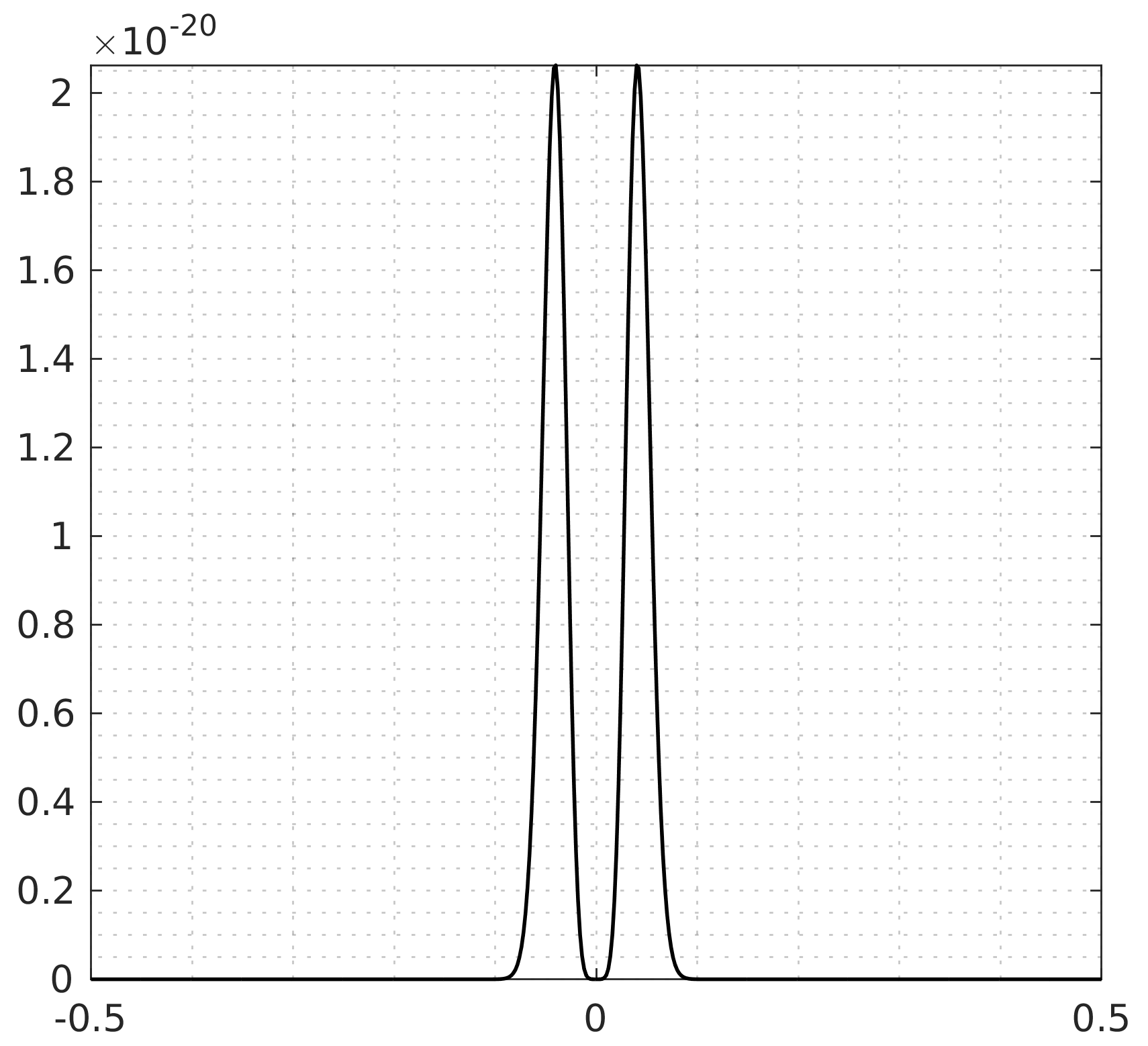}}
    \subfigure[$\hat{\mathcal{E}}_{u_{n}-a, u_n+a}$]{\includegraphics[scale = 0.10]{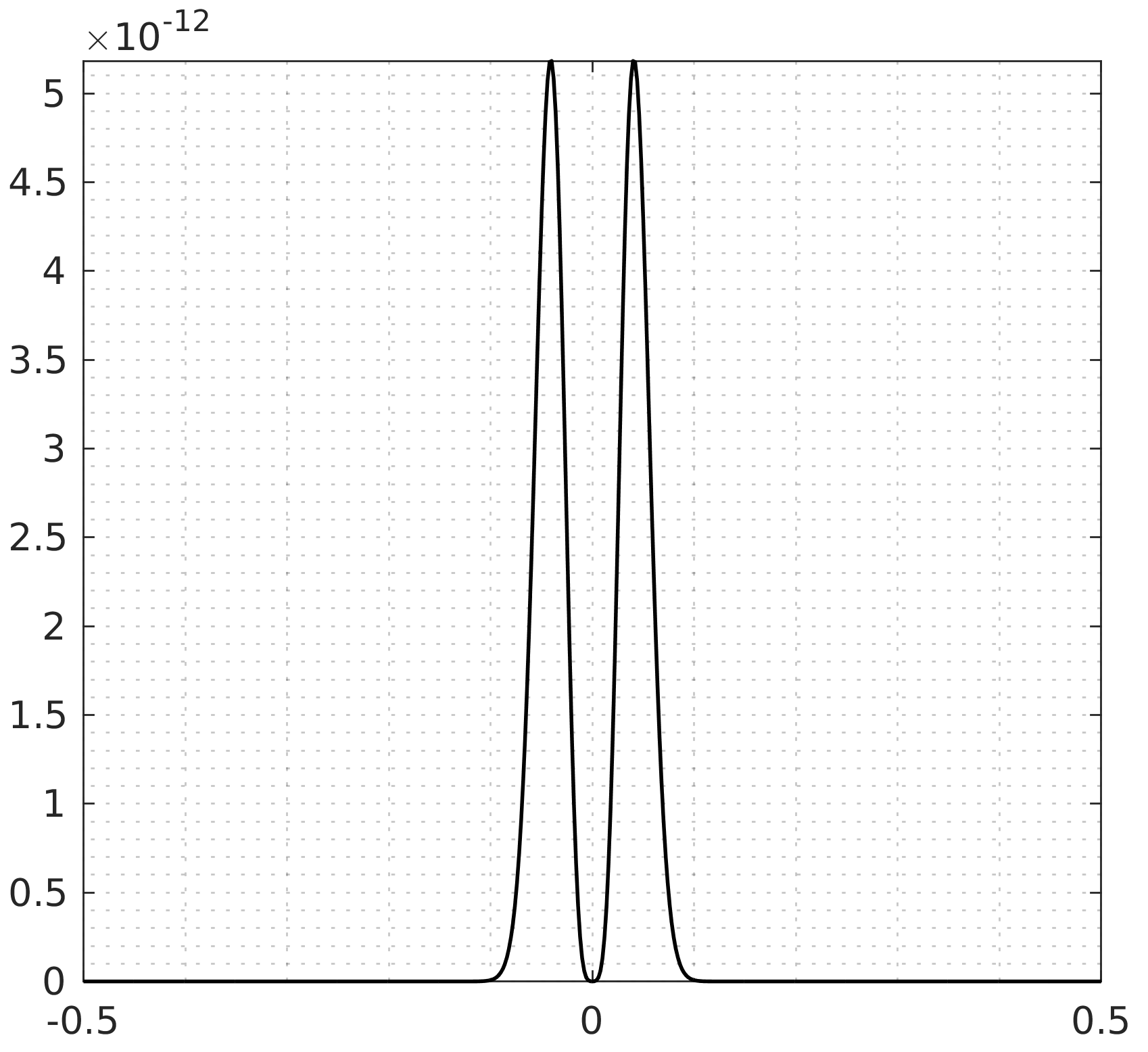}}
    \caption{Sound wave: Entropy production fields along the original acoustic eigenvectors at $t = 0$ for the ES Turkel flux $M_r = 10^{-2}$.}
    \label{fig:Sound_Tur_dS2}
\end{figure}
\begin{figure}[htbp!]
    \centering
    \subfigure[$\hat{\mathcal{E}}^S_{u_{n}+a}$]{\includegraphics[scale = 0.11]{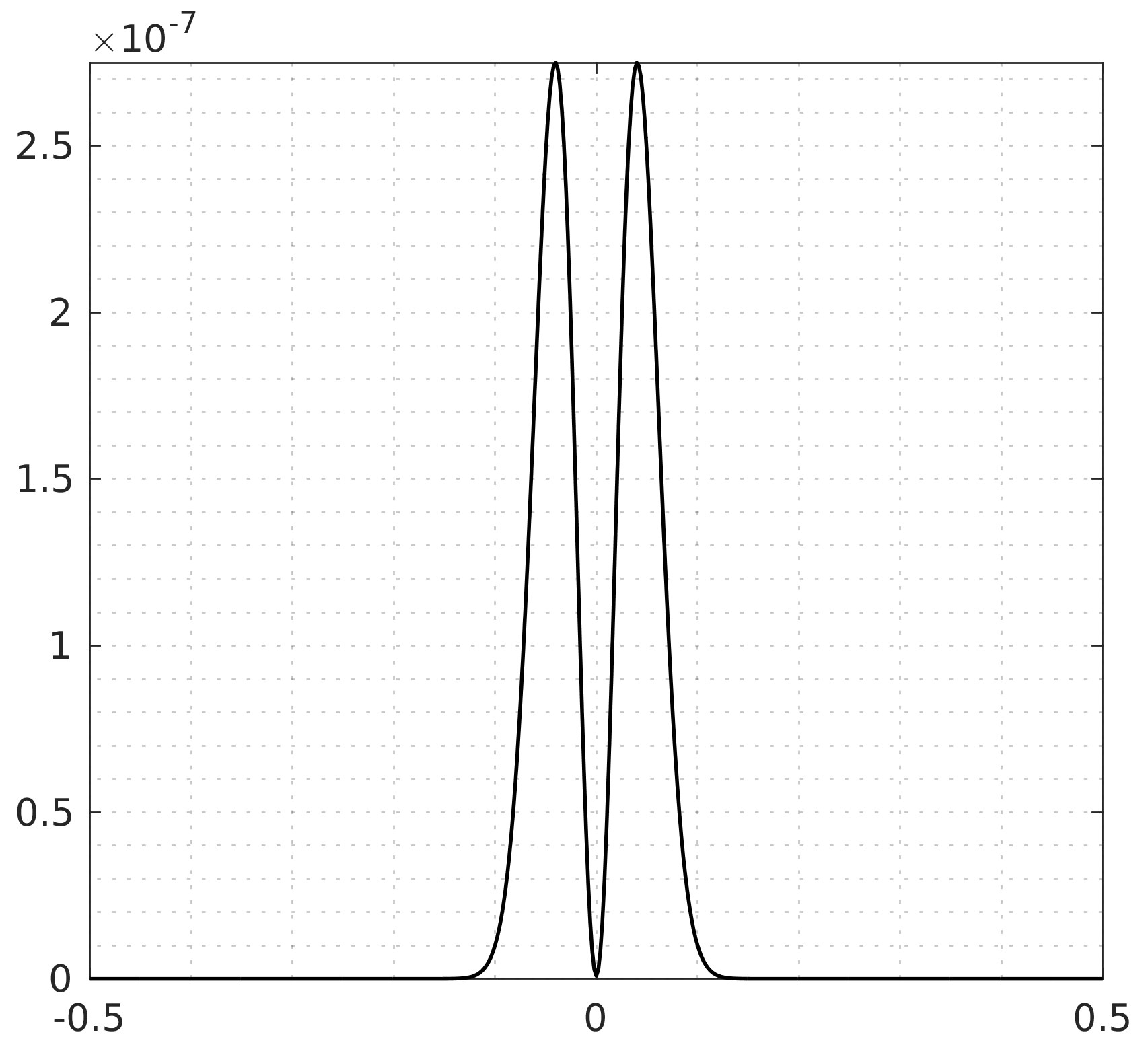}}
    \subfigure[$\hat{\mathcal{E}}^S_{u_{n}-a}$]{\includegraphics[scale = 0.11]{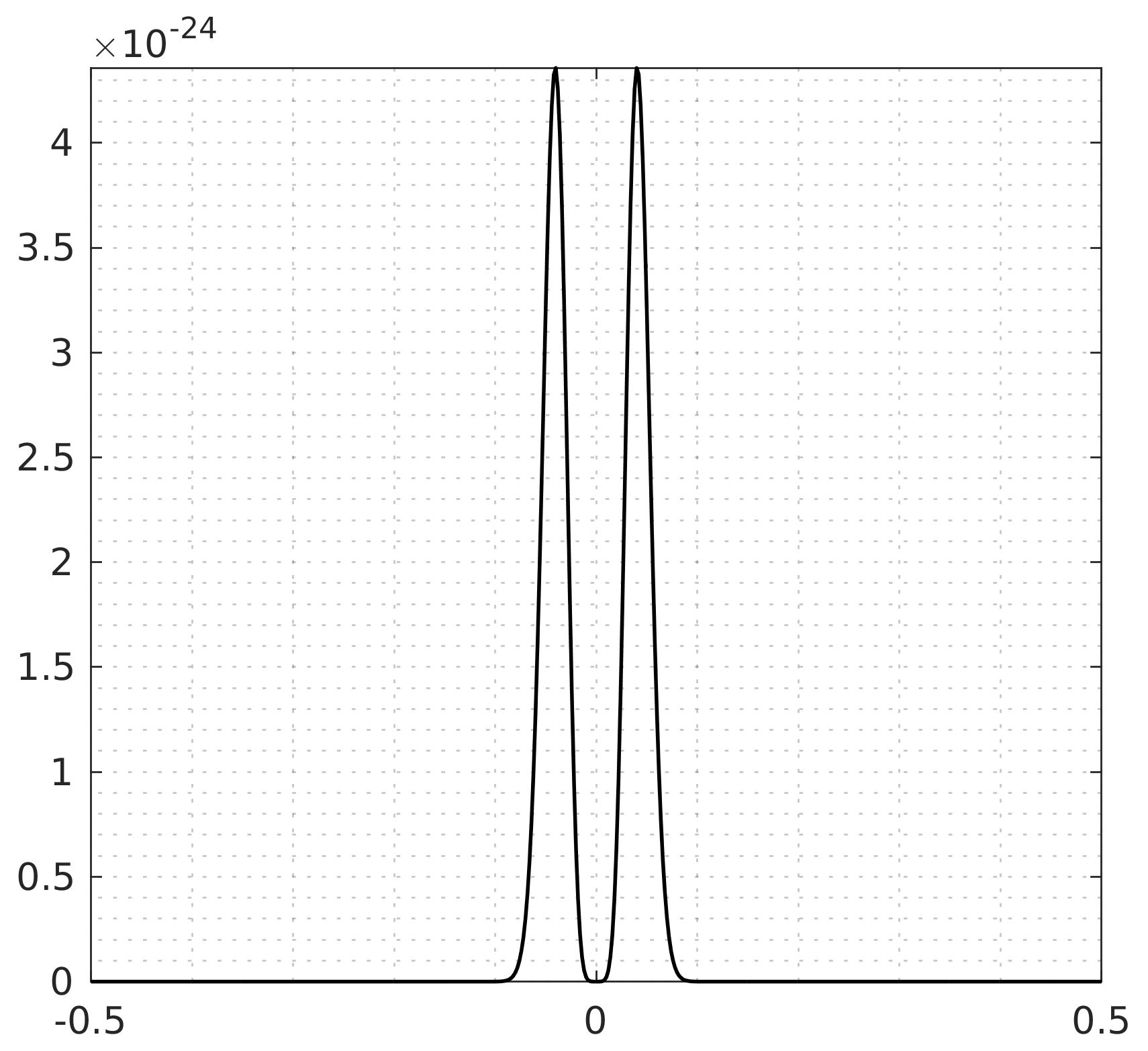}}
    \subfigure[$\hat{\mathcal{E}}_{u_{n}-a, u_n+a}$]{\includegraphics[scale = 0.11]{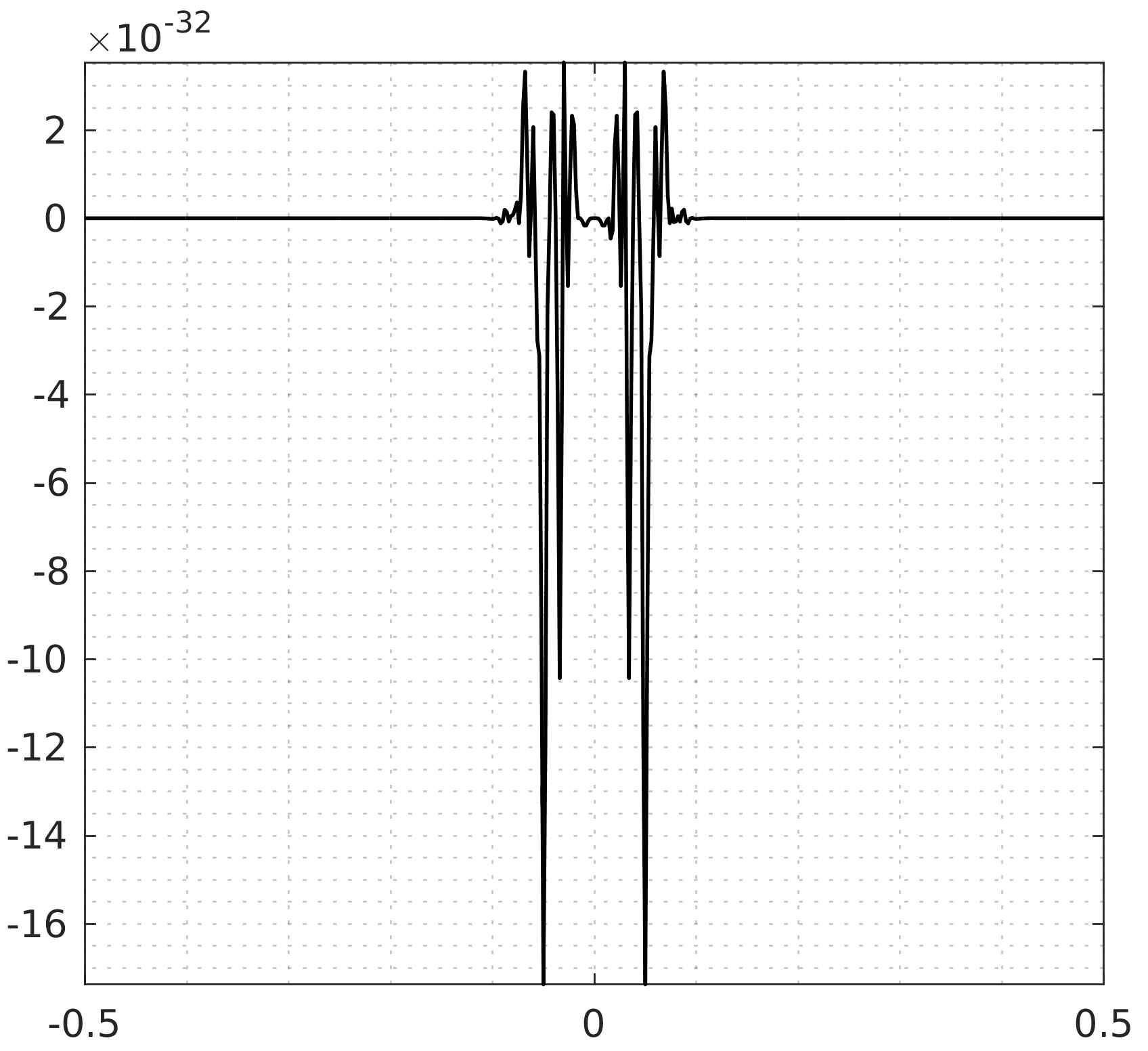}}
    \subfigure[$\Delta \hat{\mathcal{E}}_{a}$]{\includegraphics[scale = 0.11]{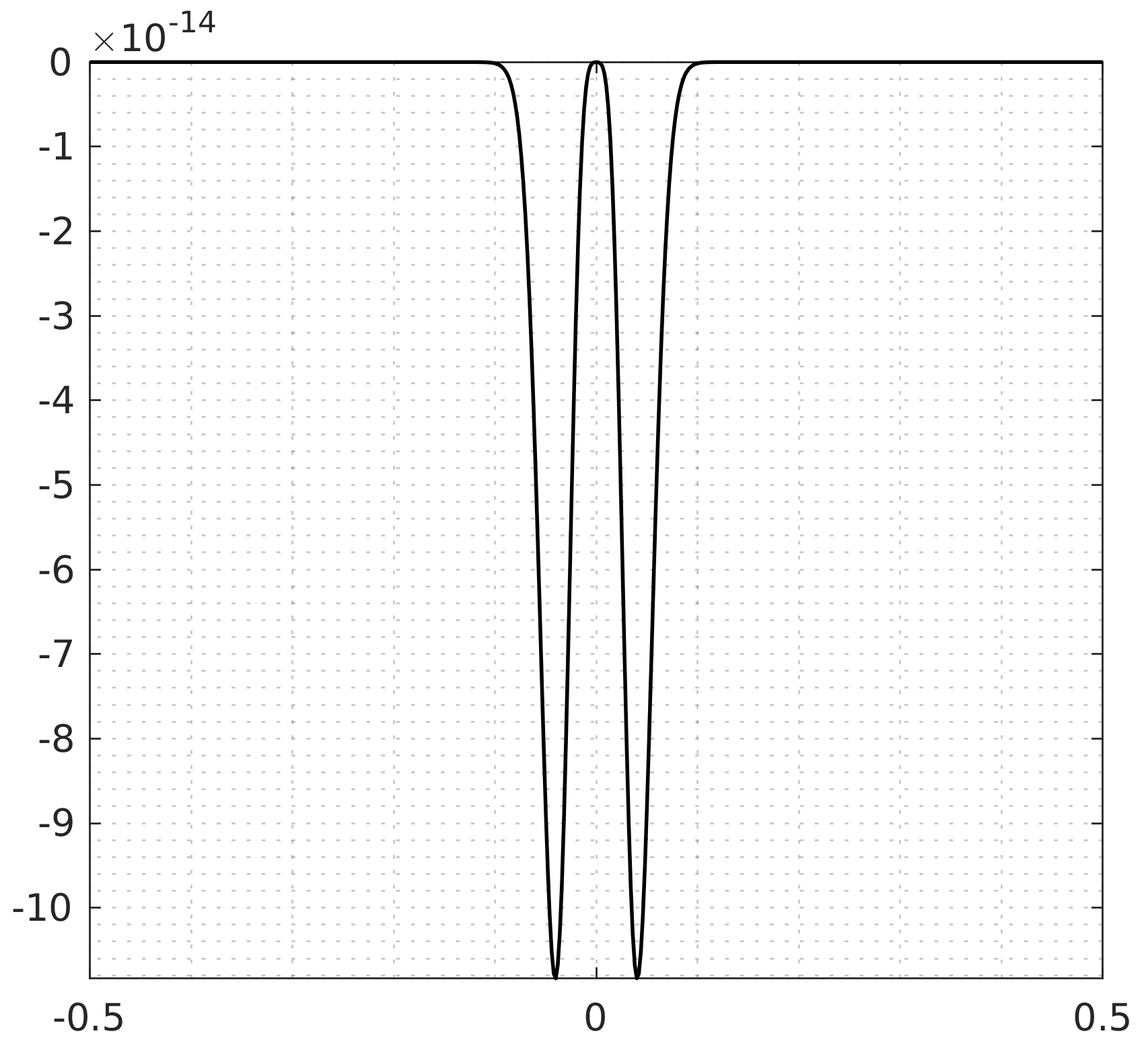}}
    \caption{Sound wave: Entropy production fields along the acoustic eigenvectors at $t = 0$ for the ES Miczek flux $M_r = 10^{-2}$.}
    \label{fig:Sound_Mic_dS2}
\end{figure}
\begin{figure}[htbp!]
    \centering
    \subfigure[$\hat{\mathcal{E}}_{u_{np}+a_p}$]{\includegraphics[scale = 0.5]{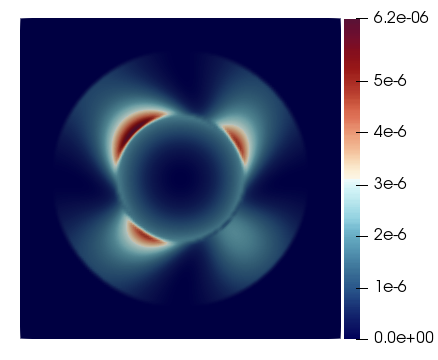}}
    \subfigure[$\hat{\mathcal{E}}_{u_{np}-a_p}$]{\includegraphics[scale = 0.5]{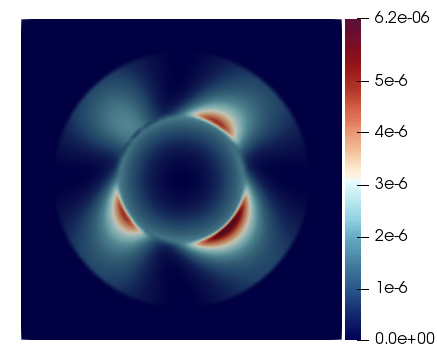}}
    \caption{Gresho Vortex: Entropy production fields at $t = 0$ for the ES Turkel flux $M_r = 3 \times 10^{-2}$.}
    \label{fig:Gresho_Tur_dS}
\end{figure}

\begin{figure}[htbp!]
    \centering
    \subfigure[$\hat{\mathcal{E}}_{u_{np}+a_p}$]{\includegraphics[scale = 0.7]{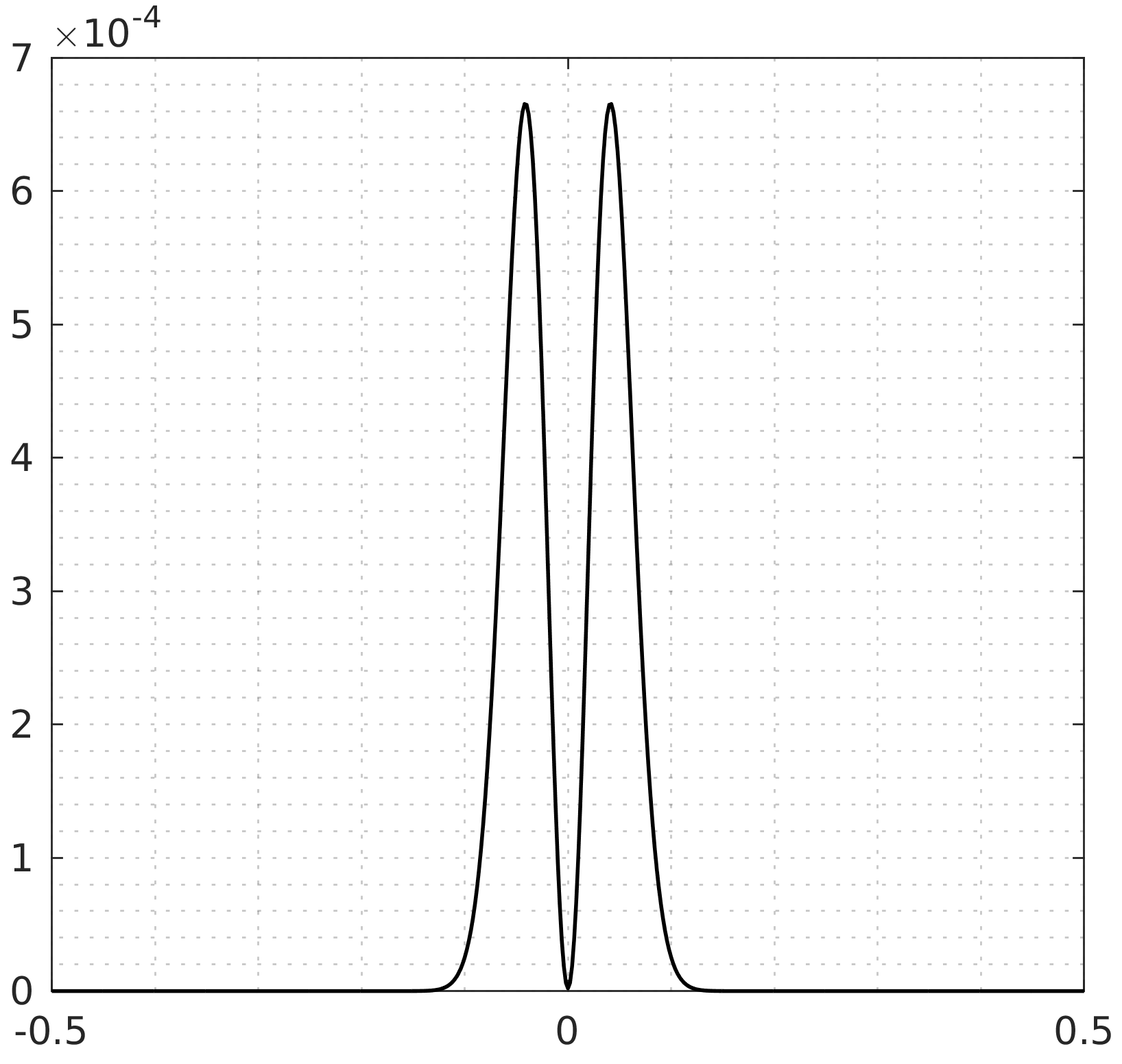}}
    \subfigure[$\hat{\mathcal{E}}_{u_{np}-a_p}$]{\includegraphics[scale = 0.7]{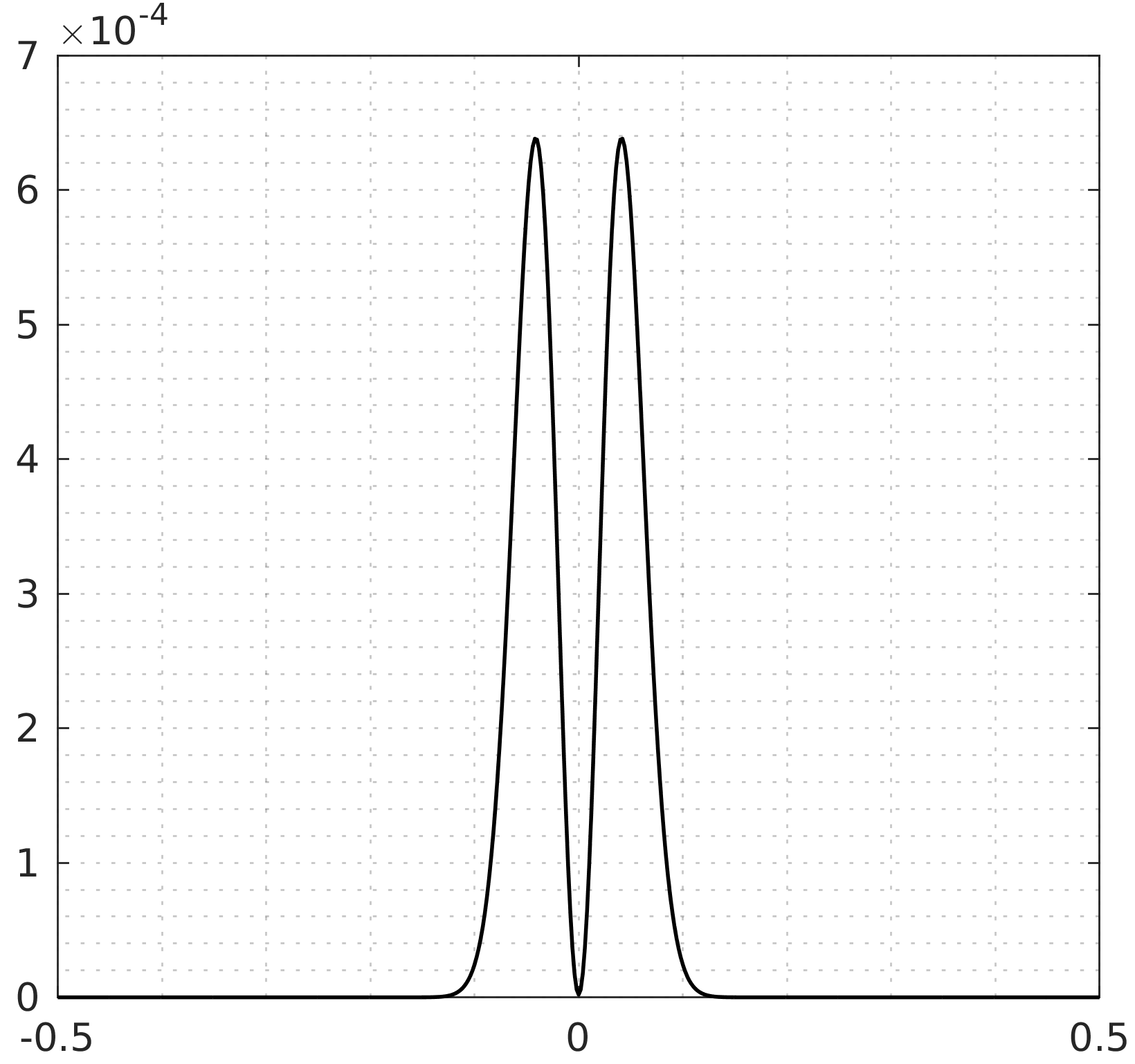}}
    \caption{Sound wave: Entropy production fields at $t = 0$ for the ES Turkel flux $M_r = 10^{-2}$.}
    \label{fig:Sound_Tur_dS}
\end{figure}

\begin{figure}[htbp!]
   \centering
    \subfigure[$\hat{\mathcal{E}}_{u_{np}+a_p} = \hat{\mathcal{E}}^S_{u_{np}+a_p} - \Delta\hat{\mathcal{E}}_{p}$]{\includegraphics[scale = 0.5]{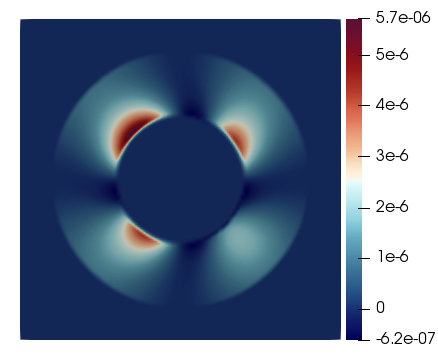}}
    \subfigure[$\hat{\mathcal{E}}_{u_{np}-a_p} = \hat{\mathcal{E}}^S_{u_{np}-a_p} + \Delta\hat{\mathcal{E}}_{p}$]{\includegraphics[scale = 0.5]{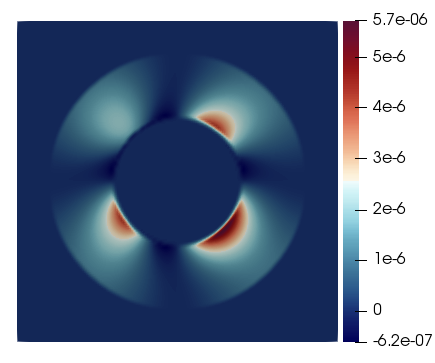}}
    \subfigure[$\hat{\mathcal{E}}^S_{u_{np}+a_p}$]{\includegraphics[scale = 0.5]{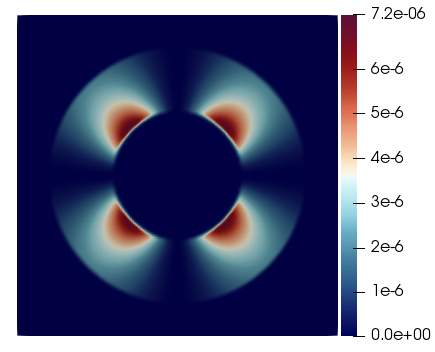}}
    \subfigure[$\hat{\mathcal{E}}^S_{u_{np}-a_p}$]{\includegraphics[scale = 0.5]{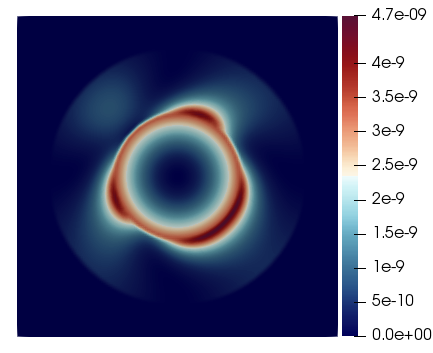}}
    \subfigure[$\Delta\hat{\mathcal{E}}_{p}$]{\includegraphics[scale = 0.5]{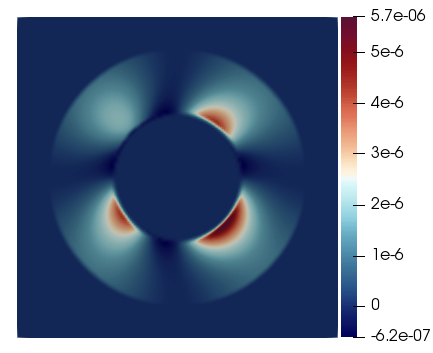}}
    \caption{Gresho Vortex: Entropy production fields at $t = 0$ for the ES Miczek flux $M_r = 3 \times 10^{-2}$.}
    \label{fig:Gresho_Mic_dS}
\end{figure}

\begin{figure}[htbp!]
    \centering
    \subfigure[$\hat{\mathcal{E}}_{u_{np}+a_p} = \hat{\mathcal{E}}^S_{u_{np}+a_p} - \Delta\hat{\mathcal{E}}_{p}$]{\includegraphics[scale = 0.7]{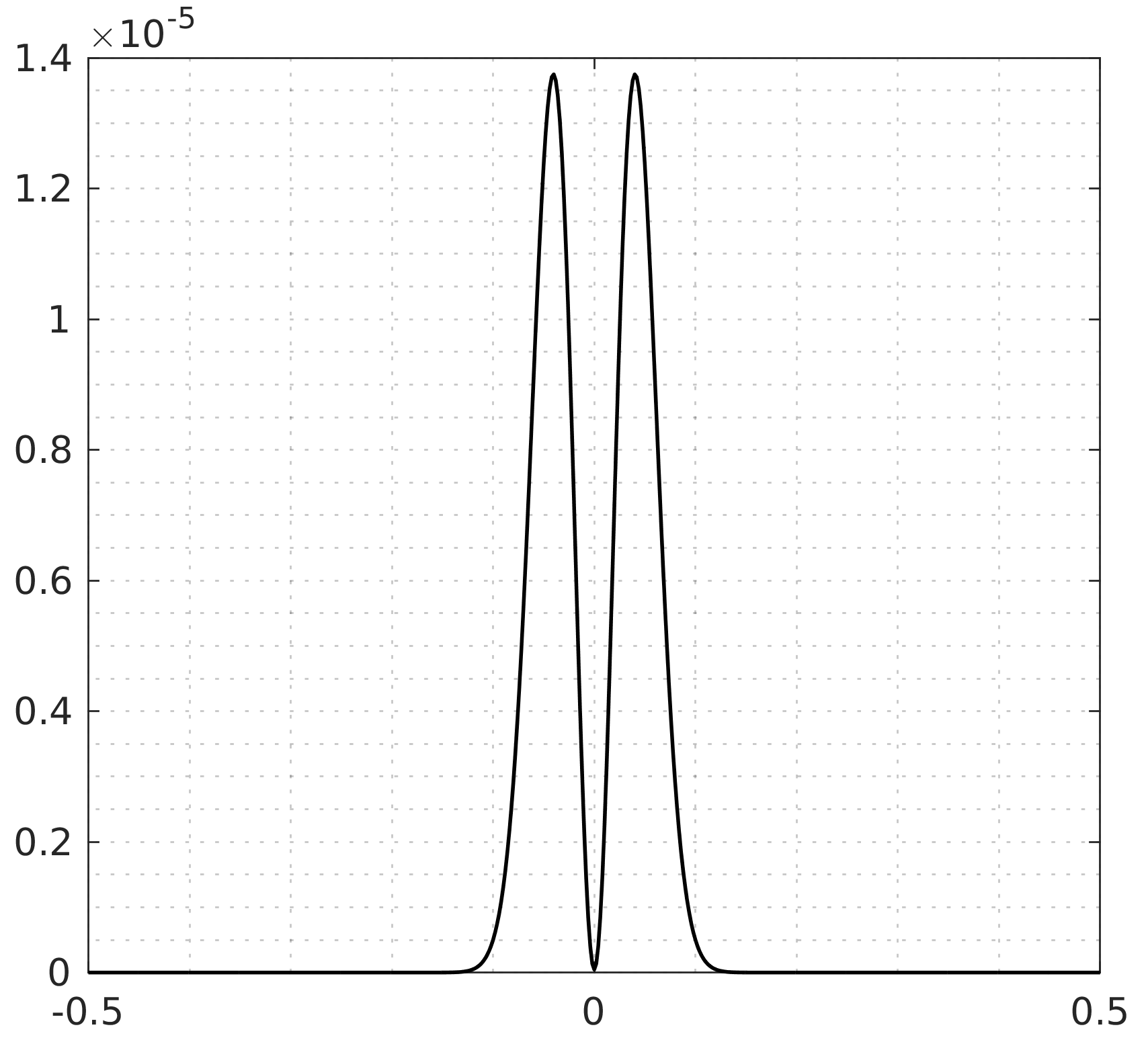}}
    \subfigure[$\hat{\mathcal{E}}_{u_{np}-a_p} = \hat{\mathcal{E}}^S_{u_{np}+a_p} 
    + \Delta\hat{\mathcal{E}}_{p}$]{\includegraphics[scale = 0.7]{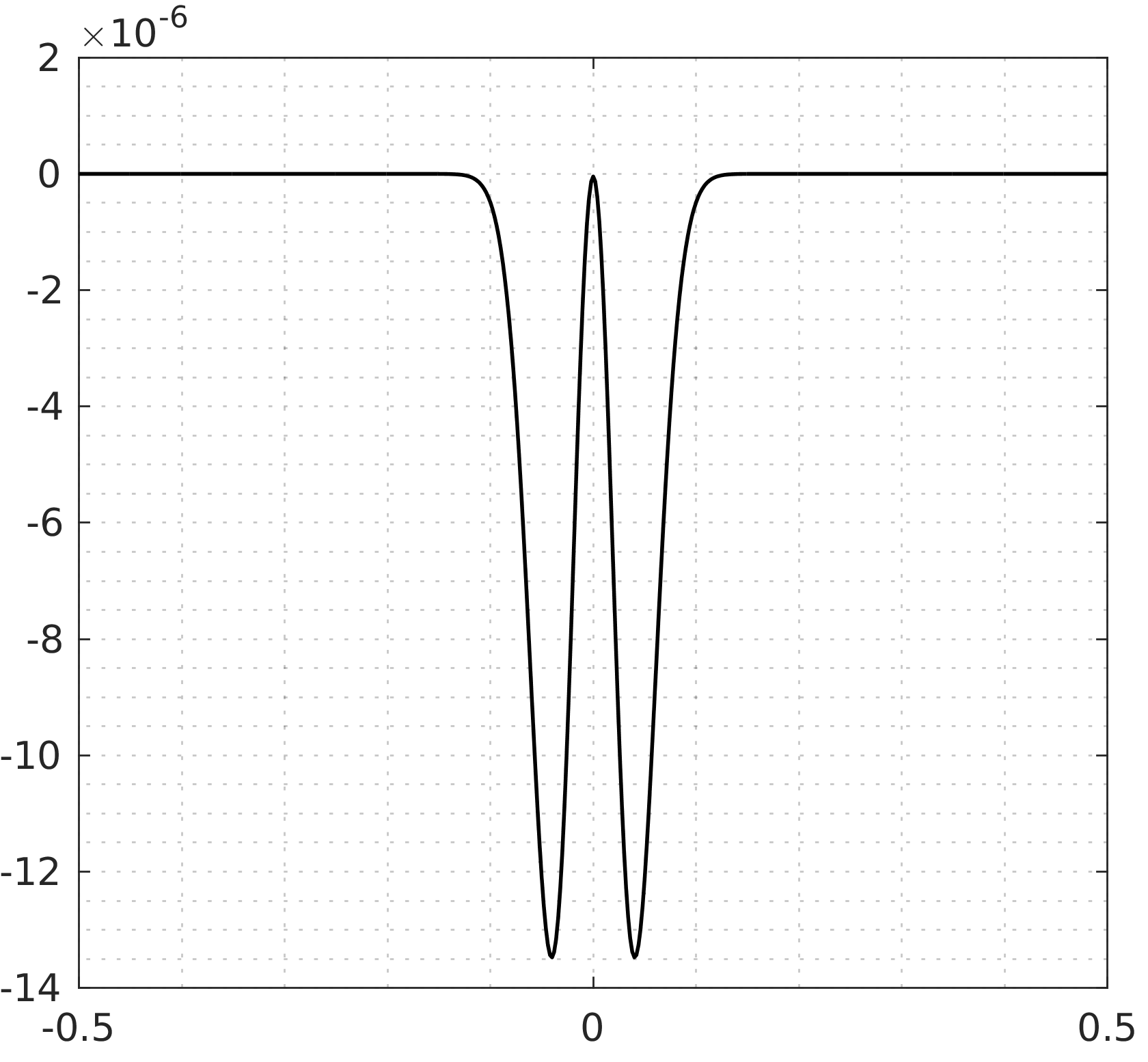}}
    \subfigure[$\hat{\mathcal{E}}^S_{u_{np}+a_p}$]{\includegraphics[scale = 0.7]{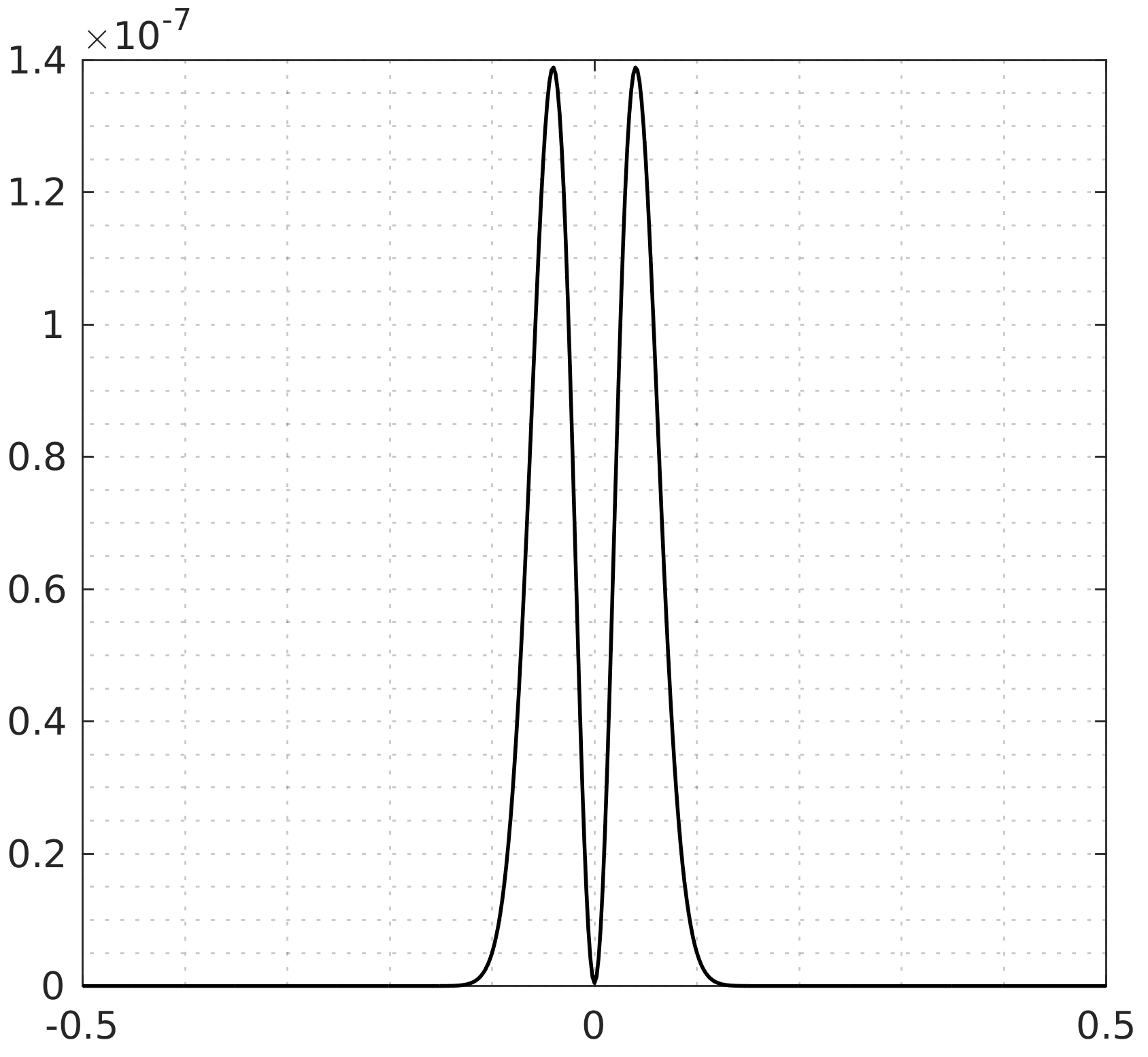}}
    \subfigure[$\hat{\mathcal{E}}^S_{u_{np}-a_p}$]{\includegraphics[scale = 0.7]{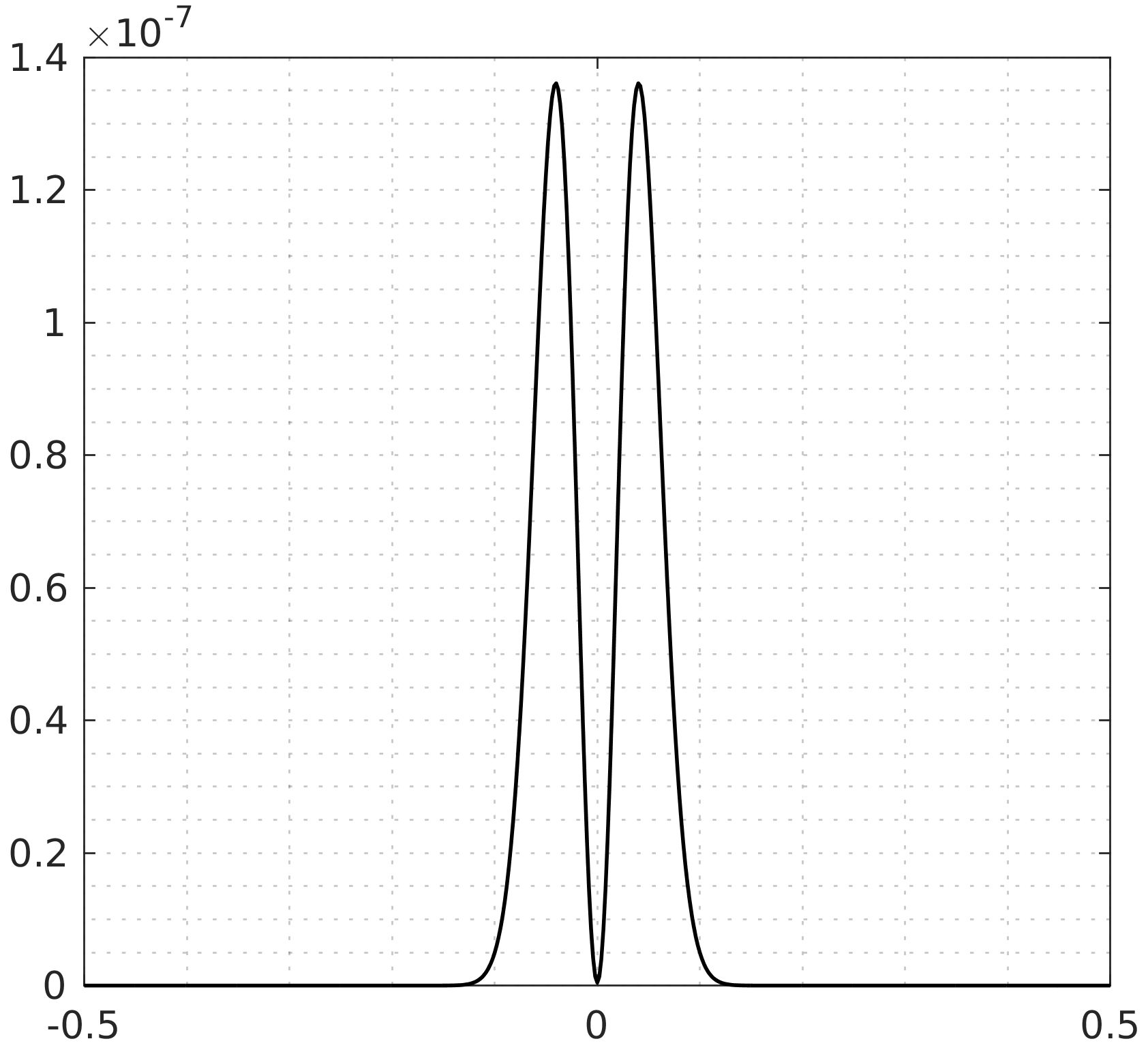}}
    \subfigure[$\Delta\hat{\mathcal{E}}_{p}$]{\includegraphics[scale = 0.7]{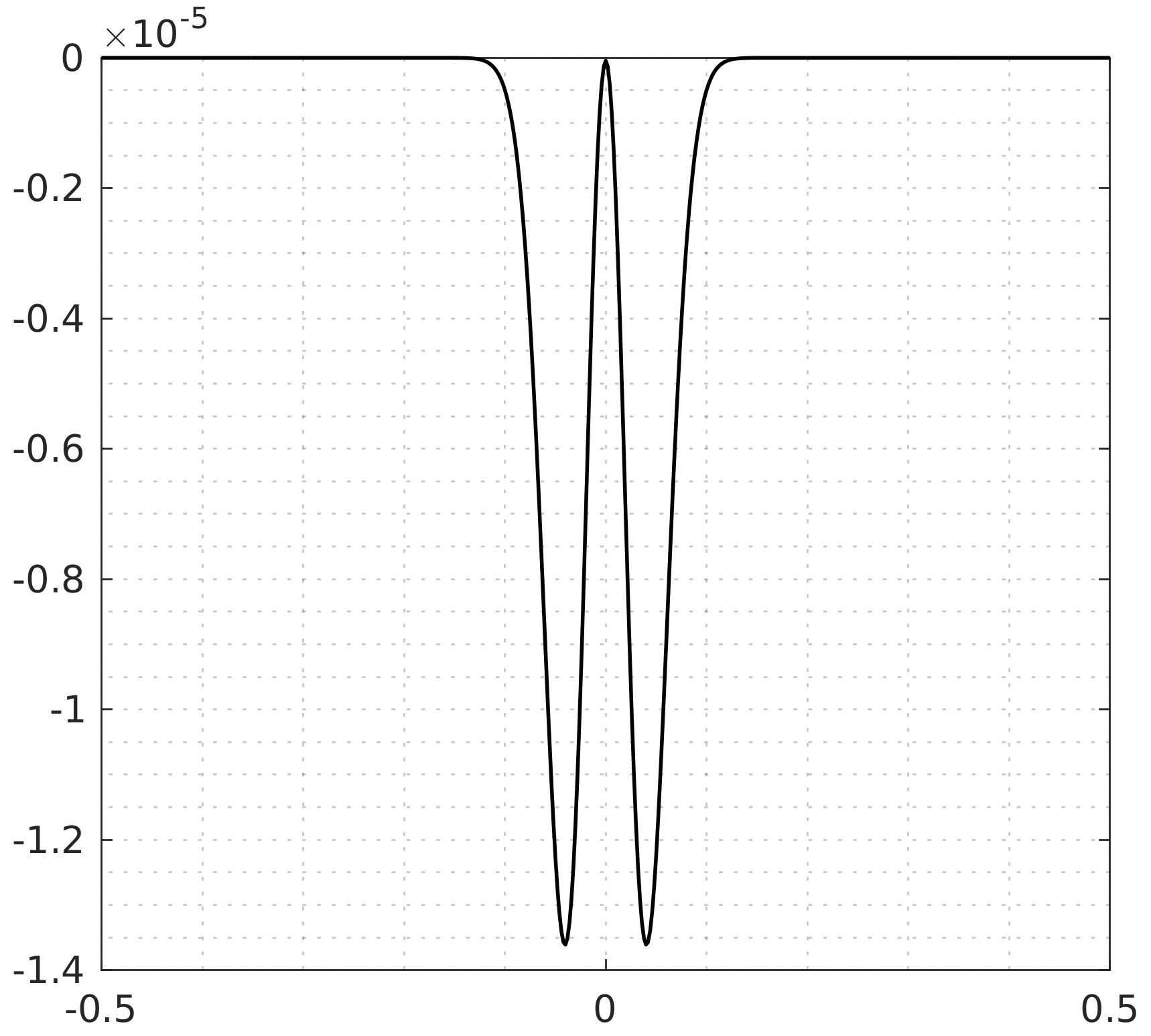}}
    \caption{Sound wave: Entropy production fields at $t = 0$ for the ES Miczek flux. $M_r = 10^{-2}$.}
    \label{fig:Sound_Mic_dS}
\end{figure}


\end{document}